\documentclass[10pt]{thesislanl}
\usepackage{amssymb}
\providecommand{\LyX}{L\kern-.1667em\lower.25em\hbox{Y}\kern-.125emX\@}
\advisor{Professor Richard E. Prange}
\chairtitle{Chairman/Advisor}
\committee{Assoc. Professor Steven Anlage \\
                Assoc. Professor Brian R. Hunt \\
                Asst. Professor Wolfgang Losert \\
                Professor Edward Ott }
\department{Department of Physics, University of Maryland}

\input tcilatex
\QQQ{Language}{
American English
}

\begin{document}

\title{Semiclassical surface of section \\
perturbation theory}
\author{Oleg A. Zaitsev}
\date{2001}

\dedication{\centerline{{\em To my parents}}}

\abstract{\ We derive analytic expressions for the wavefunctions and
energy levels in the semiclassical approximation for perturbed
integrable systems. We find that some eigenstates of such systems are
substantially different from any of the unperturbed states, which requires
some sort of a resonant perturbation theory. We utilize the semiclassical
surface of section method by Bogomolny that reduces the spatial dimensions
of the problem by one. Among the systems considered are the circular
billiard with a perturbed boundary, including the short stadium; the 
perturbed rectangular billiard, including the tilted square and the 
square in magnetic field; the bouncing ball states in the stadium and 
slanted stadium; and the whispering gallery modes. The surface of section 
perturbation theory is compared with the Born-Oppenheimer approximation, 
which is an alternative way to describe some classes of states in these 
systems. We discuss the derivation of the trace formulas from Bogomolny's 
transfer operator for the chaotic, integrable, and almost integrable 
systems.}

\comment{\ This is where I can make all my comments
on how the paper is going. }

\acknowledgements{ 
\hspace*{1cm}It is my pleasure to thank my advisor Prof.\ Richard
Prange for the inspiration he provided and patience while working with me
and answering my numerous questions. Needless to say, his ideas formed the
skeleton of this work. At the same time, he left me enough room to work on
my own initiatives, giving a good advice, when needed. 

\hspace*{1cm}I am indebted to my colleague Dr.\ Romanas Narevich for the
countless discussions and help. He co-authored many of the results
presented in this work and created many of the figures.
}

\prefacefile{PLACE-TAKER The main purpose of this work is to study the
perturbed integrable systems in the semiclassical regime
\cite{PraNarZai2}. The examples of such systems include a circular or
rectangular billiard with a distorted boundary, a rectangular billiard
with a weak magnetic field, coupled quartic oscillators --- any system
that is a ``small'' perturbation of an integrable system. In a classical
integrable system even an infinitesimal perturbation produces a
qualitatively big effect --- it changes the topology of the phase space
near the periodic orbits. Whether this change will affect the quantum
states depends on the energy of the particle. 

As will be explained in due course, these systems possess the special 
spatially localized states related to the classical periodic orbits. 
Localization may take place even for relatively small perturbations. For 
example, in the case of a distorted boundary even a perturbation 
smaller than the wavelength may lead to a strong effect. 

We use the Bogomolny $T$-operator method throughout this work. It provides
the semiclassical description for the Poincar\'e surface of section. With
this approach we reduce the dimensionality of the problem by one. In
particular, for the two-dimensional systems one can derive a
one-dimensional Schr\"odinger equation on the surface of section. It is
often easy to find the qualitative behavior of its solutions and thus
predict the localized states even without doing extensive calculations. 
In some cases the $T$-operator method produces the results similar to the 
Born-Oppenheimer approximation that we also consider. 

Here is a brief description of the work. Chapter 1 is of an introductory
character. We review the $T$-operator method in general for the reader's
convenience. We also demonstrate its connection with the boundary integral
method for billiards. The Maslov phases are discussed as well.

In Chapter 2 the perturbation theory is derived and discussed from various
aspects. Possible experiments are suggested. Although the theory is
formulated for the circular billiard with a perturbed boundary, the
general character of the derivation and results is emphasized. In the
following chapters we apply the method to other systems without detailed
explanations. 

In the first part of Chapter 3 we adapt the theory to the rectangular
billiard with a perturbed boundary. We consider a tilted square as an
example. In the rest of the chapter we study several non-perturbative
systems where the perturbation theory still can be used for special
classes of states, perhaps after some modification. Among such states are
the bouncing ball and the whispering gallery modes. We also show 
how the whispering gallery mode near the boundary with a point of 
zero curvature can be described in terms of scattering. 

In Chapter 4 we consider an example where the perturbation is 
not a distorted boundary, but a magnetic field. We analyze the square 
billiard in uniform magnetic field and with an off-center Aharonov-Bohm 
flux line. Experimental possibilities are also discussed. 

The Born-Oppenheimer approximation is a subject of Chapter 5. We revisit 
some of the examples of earlier chapters and analyze them with this 
method. We compare this approach with the $T$-operator perturbation 
theory. 

In Chapter 6 we derive the trace formulas for chaotic, integrable, and 
almost integrable systems starting with the $T$-operator. We consider an 
example of coupled quartic oscillators. Finally, in Chapter 7 we 
summarize the results of this work.}

\makefrontmatter

\chapter{Introductory chapter: Elements of the semiclassical theory}

We begin with a review of the semiclassical methods that provide the
framework for the perturbation theory developed in the subsequent
chapters. The starting point for most of our calculations will be the
Bogomolny $T$-operator equation, which is essentially a semiclassical
Green's function method adapted to the Poincar\'e surface of section. In
billiards the procedure is related to the boundary integral method. 

\section{Stationary phase approximation}

Almost any semiclassical theory takes advantage of the \emph{stationary
phase} \emph{approximation }$(S\Phi )$, which provides the most direct
connection with the underlying classical behavior. Since it is used
consistently throughout this work it seems necessary to remind the reader
of the application of this method. The problem that often arises is to
evaluate an integral of the form
\begin{equation} I=\int dxe^{iS(x)/\hbar} 
\end{equation} 
asymptotically for $\hbar \rightarrow 0$. It is argued that under certain
conditions the main contribution comes from the neighborhoods of the
points where the derivative of $S(x)$ vanishes, the so called
\emph{stationary points}. Near such point $x_{\mathrm{st}}$ we can expand
\begin{equation} 
S(x)=S(x_{\mathrm{st}})+\frac 12S^{\prime \prime
}(x_{\mathrm{st}})(x-x_{\mathrm{st}})^2+O\left[
(x-x_{\mathrm{st}})^3\right]
\end{equation} 
and the integral is approximately equal to
\begin{equation} 
I\approx \sum_{x_{\mathrm{st}}}\sqrt{\frac{2\pi \hbar }{-iS^{\prime \prime
}(x_{\mathrm{st}})}}e^{iS(x_{\mathrm{st}})/\hbar }.  \label{SPhiint}
\end{equation} 
The square root is analytically continued from the real positive numbers.
Equation (\ref{SPhiint}) is a good approximation to the integral if the third
order terms that we left out are small within the region $\left|
x-x_{\mathrm{st}}\right| <\sqrt{\hbar /\left| S^{\prime \prime
}(x_{\mathrm{st}})\right| }$ that contributes the most to the integral.
This translates into the condition
\begin{equation} 
\left| S^{\prime \prime }(x_{\mathrm{st}})\right| ^{3/2}\gg \hbar
^{1/2}\left| S^{\prime \prime \prime }(x_{\mathrm{st}})\right|
\end{equation} 
which is normally satisfied for small $\hbar $ unless $\left| S^{\prime
\prime }(x_{\mathrm{st}})\right| $ is unusually small. 

\section{Boundary integral method}

The perturbation theory that is the subject of this work is based on the
\emph{semiclassical} \emph{surface of section transfer operator} (or
$T$\emph{-operator}) method developed in generality by Bogomolny
\cite{Bog} and discussed below. This method is especially convenient in
billiards where the surface of section can be associated with the
boundary. It is instructive to see how the $T$-operator formulation
follows explicitly from a version of the \emph{boundary integral method
}\cite{Boa,GeoPra} designed to solve the Helmholtz equation in billiards. 

\begin{figure}[tbp]
{\hspace*{2.7cm} \psfig{figure=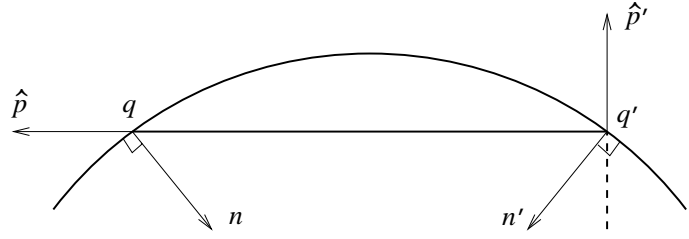,height=3cm,width=9cm,angle=0}}
{\vspace*{.13in}}
\caption[The classical orbit from point $q^\prime$ to point $q$ on the   
boundary.]
{The classical orbit from point $q^\prime$ to point $q$ on the
boundary. The dashed line shows the orbit incident at $q^\prime$.
\label{1_1}}
{\vspace{1.2 cm}} 
\end{figure}

The wavefunction $\Psi (r)$ for a particle of mass $m$ and energy $E$
moving freely inside a two-dimensional domain $B$ with the impenetrable
boundary $\partial B$ is determined by the Schr\"odinger equation
\begin{equation}
\left( \nabla ^{\prime 2}+k^2\right) \Psi (r^{\prime })=0  \label{Helm1}
\end{equation}
with the Dirichlet boundary conditions 
\begin{equation}
\Psi (r^{\prime })\mid _{\partial B}=0.
\end{equation}
Here the wavenumber $k=\sqrt{2mE}/\hbar $ and $\nabla ^{\prime 2}$ is the
two-dimensional Laplacian acting on $r^{\prime }$. The \emph{free-space}
Green function $G_0(r^{\prime },r)$ (that does not satisfy the boundary
conditions on $\partial B$) is defined by the equation 
\begin{equation}
\left( \nabla ^{\prime 2}+k^2\right) G_0(r^{\prime },r)=\delta (r^{\prime
}-r)  \label{Helm2}
\end{equation}
which is solved by the Hankel function \cite{AbrSte} 
\begin{equation}
G_0(r^{\prime },r)=-\frac i4H_0^{(1)}\left( k\left| r^{\prime }-r\right|
\right) .
\end{equation}
Multiplying Eq.\ (\ref{Helm1}) by $-G_0(r^{\prime },r)$ and Eq.\
(\ref{Helm2}) by $\Psi (r^{\prime })$, adding them together and
integrating over the billiard's area we arrive to the integral equation
\begin{equation}
\Psi (r)=\int_Bd^2r^{\prime }\left[ \Psi (r^{\prime })\nabla ^{\prime
2}G_0(r^{\prime },r)-G_0(r^{\prime },r)\nabla ^{\prime 2}\Psi (r^{\prime
})\right] .
\end{equation}
With the help of Green's theorem we transform it to the boundary integral 
\begin{equation}
\Psi (r)=\oint_{\partial B}dq^{\prime }\left[ G_0(q^{\prime
},r)\frac{\partial \Psi (q^{\prime })}{\partial n^{\prime }}-\Psi
(q^{\prime })\frac{\partial G_0(q^{\prime },r)}{\partial n^{\prime
}}\right] \label{Psiint}
\end{equation}
where $q^{\prime }$ is the coordinate along the boundary and $n^{\prime }$
is the normal at point $q^{\prime }$ directed inside the boundary. The
second term vanishes due to the boundary conditions. Taking point $r=q$ on
the boundary and introducing the new function 
\begin{equation}
\mu (q)=\frac{\partial \Psi (q)}{\partial n}
\end{equation}
we can write the integral equation which is \emph{equivalent} to the
original Helmholtz equation (\ref{Helm1}), 
\begin{equation}
\mu (q)=\int dq^{\prime }\frac{\partial G_0(q^{\prime },q)}{\partial n}\mu
(q^{\prime }).  \label{G0neqn}
\end{equation}
Its kernel has a $\delta $-function type singularity at $q^{\prime }=q$.
Indeed, 
\begin{equation}
\frac{\partial G_0(q^{\prime },q)}{\partial n}=\frac{ik}4H_0^{(1)\prime
}\left[ kL(q,q^{\prime })\right] \frac{\partial L(q,q^{\prime })}{\partial n}
\label{bimH0}
\end{equation}
with 
\begin{equation}
\frac{\partial L(q,q^{\prime })}{\partial n}=\left\{ 
\begin{array}{l}
\left| \hat p\cdot n\right| ,\quad q^{\prime }\neq q \\ 
1,\quad q^{\prime }=q
\end{array}
\right.
\end{equation}
where $L(q,q^{\prime })$ is the length of the classical orbit (chord)
going from point $q^{\prime }$ to $q$ and $\hat p$ is a unit vector along
this orbit (or a unit momentum) (Fig.\ \ref{1_1}). Using the asymptotic
expansion for the derivative of Hankel's function for small argument
\cite{AbrSte} we find
\begin{equation}
\frac{\partial G_0(q^{\prime },q)}{\partial n}\sim \left\{ 
\begin{array}{l}
-\frac{\left| \hat p\cdot n\right| }{2\pi L(q,q^{\prime })}\sim -\frac
1{4\pi },\quad q^{\prime }\rightarrow q \\ 
\infty ,\quad q^{\prime }=q
\end{array}
\right. .
\end{equation}
The $-1/4\pi $ is written under the assumption of a convex boundary with
non-vanishing curvature, but this assumption is not essential for extracting
the $\delta $-singularity. So, we may assume that the boundary is locally a
straight line. Then, for any $q$ and $q^{\prime }$, $\partial L(q,q^{\prime
})/\partial n=\stackunder{n\rightarrow 0}{\lim }n/\sqrt{\left( q-q^{\prime
}\right) ^2+n^2}$ and the singular part of the kernel 
\begin{equation}
\left( \frac{\partial G_0(q^{\prime },q)}{\partial n}\right)
_{\mathrm{sin}\text{\textrm{g}}}=\stackunder{n\rightarrow 0}{\lim }\frac
n{2\pi \left[ \left( q-q^{\prime }\right) ^2+n^2\right] }=\frac 12\delta
\left( q-q^{\prime }\right) . 
\end{equation}
The integral equation (\ref{G0neqn}) may now be rewritten as 
\begin{equation}
\mu (q)=\int dq^{\prime }K(q,q^{\prime })\mu (q^{\prime })  \label{Keqn}
\end{equation}
with the regular kernel 
\begin{equation}
K(q,q^{\prime })=2\left( \frac{\partial G_0(q^{\prime },q)}{\partial
n}\right) _{\mathrm{reg}}. 
\end{equation}
Thus the problem is reduced to the one-dimensional integral equation (\ref
{Keqn}) for $\mu (q)$. This function contains most of the information
about the state and $\det (1-K)=0$ is the quantization condition. [The second
quantization condition comes from the periodicity requirements on $\mu
(q)$.] If needed, the wavefunction $\Psi (r)$ can be subsequently
determined from Eq.\ (\ref{Psiint}). 

Our goal is to develop a semiclassical approximation to this scheme. When
$kL\gg 1$ we may use the large argument asymptotics for the Hankel
function \cite{AbrSte} and the kernel now becomes
\begin{eqnarray}
K(q,q^{\prime }) &\approx &-\sqrt{\frac k{2\pi L(q,q^{\prime })}}\left| \hat
p\cdot n\right| e^{ikL(q,q^{\prime })-i\pi /4}  \nonumber \\
\ &=&-\sqrt{\frac 1{2\pi }\frac{\partial ^2S(q,q^{\prime })}{\partial
q\partial q^{\prime }}\frac{\left| \hat p\cdot n\right| }{\left| \hat
p^{\prime }\cdot n^{\prime }\right| }}e^{iS(q,q^{\prime })-i\pi /4}.
\end{eqnarray}
We introduced the reduced action (measured in units of $\hbar $) 
\begin{equation}
S(q,q^{\prime })=kL(q,q^{\prime })
\end{equation}
for the classical orbit $q^{\prime }\mapsto q$. Clearly, $K(q,q^{\prime
})$ is not symmetric. The semiclassical approximation allows to symmetrize
it. First, notice that, in spite of the notation, both $\left| \hat p\cdot
n\right| $ and $\left| \hat p^{\prime }\cdot n^{\prime }\right| $ are the
functions of two coordinates, $q$ and $q^{\prime }$. The essential part of
the semiclassical approach, however, is to evaluate all the integrals
containing the fast changing action at the exponent in the stationary
phase approximation. It will be clear from the following chapters (and is
well known) that this approximation amounts to extracting those $q$ and
$q^{\prime }$ that are connected by a classical orbit. Thus we may
consider, say, $\left| \hat p\cdot n\right| $ to be a function of $q$ only
(as well as the two fixed parameters related to the initial conditions,
one of which is the total energy). Similarly, $\left| \hat p^{\prime
}\cdot n^{\prime }\right| $ will be a function of $q^{\prime }$. 

Define a function 
\begin{equation}
\psi (q)=\frac{\mu (q)}{\sqrt{\left| \hat p\cdot n\right| }}.  \label{psimu}
\end{equation}
Then the semiclassical analog to Eq.\ (\ref{Keqn}) is 
\begin{equation}
\psi (q)=\int_{\partial B}dq^{\prime }T(q,q^{\prime })\psi (q^{\prime })
\label{psiTpsi}
\end{equation}
where the $T$-operator 
\begin{equation}
T(q,q^{\prime };E)=K(q,q^{\prime };E)\sqrt{\frac{\left| \hat p^{\prime
}\cdot n^{\prime }\right| }{\left| \hat p\cdot n\right| }}=-\sqrt{\frac
1{2\pi i}\frac{\partial ^2S(q,q^{\prime };E)}{\partial q\partial q^{\prime
}}}e^{iS(q,q^{\prime };E)}.  \label{Tbil}
\end{equation}
Unlike $K(q,q^{\prime })$, the $T$-operator is symmetric and unitary
(semiclassically). See Sec.\ \ref{propT} for more details. Clearly, Eq.\
(\ref {psiTpsi}) has a solution if $\det (1-T)=0$. The original
wavefunction $\Psi (r)$ can be found from $\psi (q)$ (see Sec.\
\ref{Tgen}). We emphasize once again that the stationary phase
approximation is expected when dealing with the $T$-operator. In the next
section we discuss the relationship between the $T$-operator and the
classical surface of section map. 

\section{Surface of section map}

\label{sosm}

In the previous section we reduced the two-dimensional problem in a billiard
to a one-dimen\-sional problem on its boundary. The state of the system 
can be equivalently described by either 2D-wavefunction $\Psi (r)$ or
1D-wavefunction $\psi (q)$. In the classical language, we reduced the
continuous motion in the four-dimensional phase space of the billiard to a
map in the two-dimensional phase space of the boundary. It is an example of
a \emph{surface of section map}. Again, this map contains all the
information about the continuous motion.

Formally, the map $(p^{\prime },q^{\prime })\longmapsto (p,q)$ is implicitly
given by the equations 
\begin{equation}
p=\frac{\partial S(q,q^{\prime })}{\partial q},\quad p^{\prime
}=-\frac{\partial S(q,q^{\prime })}{\partial q^{\prime }}. 
\label{clasmap}
\end{equation}
The generalized momentum (in the units of $\hbar $) $p$ is the projection of
the total momentum onto the boundary. To connect the classical and quantum
descriptions consider a version of Eq.\ (\ref{psiTpsi}) 
\begin{equation}
\psi (q)=\int dq^{\prime }T(q,q^{\prime })\psi ^{\prime }(q^{\prime })
\label{Tpsiprime}
\end{equation}
describing the propagation of a wavepacket. We write the wavefunction $\psi
^{\prime }(q^{\prime })$ on a locally defined Lagrangian manifold $p^{\prime
}(q^{\prime })$ in the semiclassical form (Sec.\ 6.3.c in Ref.\ \cite{Tab}) 
\begin{equation}
\psi ^{\prime }(q^{\prime })=\sqrt{\frac{\partial ^2S(P,q^{\prime
})}{\partial P\partial q^{\prime }}}e^{i\int p^{\prime }(q^{\prime
})dq^{\prime }}=\sqrt{-\left( \frac{\partial Q}{\partial q^{\prime
}}\right) _P}e^{i\int p^{\prime }(q^{\prime })dq^{\prime }}
\end{equation}
where $S(P,q^{\prime })$ is the Legendre transform of $S(q,q^{\prime })$
to the canonical coordinates $(P,Q)$ such that $P$ is a local integral of
motion. We substitute it in Eq.\ (\ref{Tpsiprime}) and integrate by the
$S\Phi $. The stationary point $q_{\mathrm{st}}^{\prime }$ is determined
by the second of the Eqs.\ (\ref{clasmap}). In the neighborhood of point
$q$ we define the function $p(q)$ by
\begin{equation}
\int^qp(q^{\prime \prime })dq^{\prime \prime }=S\left[
q,q_{\mathrm{st}}^{\prime }(q)\right] +\int^{q_{\mathrm{st}}^{\prime
}(q)}p^{\prime }(q^{\prime \prime })dq^{\prime \prime }. 
\end{equation}
This definition satisfies the first of the Eqs.\ (\ref{clasmap}). Thus the
integral 
\begin{eqnarray}
&&\int dq^{\prime }T(q,q^{\prime })\psi ^{\prime }(q^{\prime })  
= -\sqrt{\frac{-\frac{\partial ^2S}{\partial q\partial q^{\prime }}\
\left( \frac{\partial Q}{\partial q^{\prime }}\right) _P}{\frac{\partial
^2S}{\partial q^{\prime 2}}+\frac{dp^{\prime }}{dq^{\prime }}}}e^{i\int
p(q)dq} \nonumber \\
&& =-\sqrt{\left( \frac{\partial q^{\prime }}{\partial q}\right)
_P\left( \frac{\partial Q}{\partial q^{\prime }}\right) _P}e^{i\int
p(q)dq}
 =\sqrt{\frac{\partial ^2S(q,P)}{\partial q\partial P}}e^{i\int
p(q)dq-i\pi }=\psi (q). 
\end{eqnarray}
(To prove the second equality differentiate the equation $\partial
S(q,q^{\prime })/\partial q^{\prime }+p^{\prime }(q^{\prime })=0$. The
Maslov phase $-\pi $ results from the reflection at the billiard's
boundary.) This shows that if a manifold $p^{\prime }(q^{\prime })$
evolves into the manifold $p(q)$ the $T$ -operator provides the evolution
of the wavefunction defined on this manifold. 

\section{$T$-operator: the general case}

\label{Tgen}

The phase space of billiard boundary is an example of a \emph{Poincar\'e
surface of section} (PSS). The above discussion makes it clear now why the
$T$-operator is called the surface of section transfer operator. In
general, PSS is a $(2N-2)$-dimensional manifold in the phase space of an
$N$-dimensional system crossed by all classical trajectories. Bogomolny
\cite {Bog} derived the expression for the $T$-operator in the case when
PSS is an $(N-1)$-dimensional manifold in the coordinate space of the
system together with its conjugate momenta. He assumed the Hamiltonian has
the form
\begin{equation}
H(\hat p,r)=\frac 12\hat p^2+V(r)
\end{equation}
but presumably his result is more general. The $T$-operator is now 
\begin{equation}
T(q,q^{\prime };E)=\frac 1{\left( 2\pi i\right) ^{\left( N-1\right)
/2}}\sum_{\mathrm{cl.\ tr.}}\left| \det \left[ \frac{\partial
^2S(q,q^{\prime };E)}{\partial q\partial q^{\prime }}\right] \right|
^{1/2}e^{iS(q,q^{\prime };E)-i\frac \pi 2\nu }.  \label{Tgen1}
\end{equation}
The sum is over all classical trajectories that go from $q^{\prime }$ to
$q$ on PSS and correspond to \emph{one }Poincar\'e mapping. All
trajectories should leave the PSS with the positive, say, normal component
of the momentum. The reduced action $S(q,q^{\prime };E)$ is still measured
in the units of $\hbar $. The \emph{Maslov index} $\nu $ is determined by
the number and type of caustics encountered by the trajectory
\cite{BerMou}. For example, a regular caustic of dimension $(N-1)$
increases $\nu $ by 1; a hard wall increases $\nu $ by 2. We discuss the
origin of Maslov phase in Sec.\ \ref{mpat}. In a billiard with its
boundary as a PSS\footnote{In the billiard problems it is sometimes
convenient to choose PSS infinitesimally close to the hard wall, not on
the hard wall.} $\nu =2$ and there is only one trajectory $q^{\prime
}\longmapsto q$. Then Eq.\ (\ref{Tbil}) follows. 

A surface of section wavefunction is determined from the integral equation 
\begin{equation}
\psi (q)=\int_{\mathrm{PSS}}dq^{\prime }T(q,q^{\prime })\psi (q^{\prime })
\label{TgenBog}
\end{equation}
that we will call \emph{Bogomolny's equation }and the quantization condition
is 
\begin{equation}
\det (1-T)=0. \label{totgdet}
\end{equation}
The original wavefunction can be found by propagating $\psi (q)$, 
\begin{equation}
\Psi (r)=\int_{\mathrm{PSS}}dq\tilde G(r,q;E)\psi (q)  \label{PsiGpsi}
\end{equation}
where 
\begin{equation}
\tilde G(r,q;E)=\frac{i^{1/2}}{\left( 2\pi i\right) ^{\left( N-1\right)
/2}}\sum_{\mathrm{cl.\ tr.}}\left| \frac 1{v(r)}\det \left[ \frac{\partial
^2S(r,q;E)}{\partial r_{\perp }\partial q}\right] \right|
^{1/2}e^{iS(r,q;E)-i\frac \pi 2\nu }.  \label{Gmix}
\end{equation}
The trajectories start at point $q$ on PSS with the positive normal
component of the momentum and end at point $r$ of the interior; $v(r)$ is
the modulus of the velocity at point $r$; $r_{\perp }$ is the direction
perpendicular to the trajectory at point $r$. In the case of a billiard
one can show that Eqs.\ (\ref{PsiGpsi}) and (\ref{Gmix}) agree with Eqs.\
(\ref{Psiint}) and (\ref {psimu}) up to normalization if the asymptotic
form of Hankel's function is used. 

\section{Properties of the $T$-operator}

\label{propT}

In this section we review some of the properties of the $T$-operator \cite
{Bog}. As was noticed by Bogomolny, $T$-operator is similar to the
time-dependent semiclassical Green's function [cf.\ Eq.\
(\ref{Tpsiprime})]. It is therefore not surprising that they share some of
the properties. To simplify notation we assume the system to be
two-dimensional and omit the Maslov indices. 

First, note that the $T$-operator vanishes for short trajectories, i.e.\
$T(q,q^{\prime })\rightarrow 0$ as $q\rightarrow q^{\prime }$. Indeed, in
this case $S(q,q^{\prime })\sim p\left| q-q^{\prime }\right| $, so the
prefactor in Eq.\ (\ref{Tgen1}) vanishes. 

As was mentioned before, the $T$-operator, unlike the exact kernel
$K(q,q^{\prime })$, is unitary in semiclassical approximation. To see this
we evaluate the integral
\begin{equation}
\left( T^{\dagger }T\right) (q_1,q_2)=\frac 1{2\pi }\int dq\sum \sum
\left[ \frac{\partial ^2S(q,q_1)}{\partial q\partial q_1}\frac{\partial
^2S(q,q_2)}{\partial q\partial q_2}\right] ^{1/2}e^{i\left[
S(q,q_2)-S(q,q_1)\right] }
\end{equation}
by the $S\Phi $. The stationary phase condition $\partial
S(q,q_1)/\partial q=\partial S(q,q_2)/\partial q$ requires that the orbits
$q_1\longmapsto q$ and $q_2\longmapsto q$ have the same final momentum in
the PSS. For a deterministic map it means they are the same orbit, i.e.\
$q_1=q_2$. Expand $S(q,q_2)\cong S(q,q_1)+(q_2-q_1)\partial
S(q,q_1)/\partial q_1$. Then the integral becomes
\begin{equation}
\left( T^{\dagger }T\right) (q_1,q_2)=\frac 1{2\pi }\int dq\left(
\frac{\partial p_1}{\partial q}\right) _{q_1}e^{ip_1(q_2-q_1)}\cong \delta
(q_2-q_1). 
\end{equation}
The sum over classical orbits disappeared because the integration over $dp_1$
takes care of all the orbits leaving $q_1$. The $\delta $-function is not
ideal since $\left| p_1\right| $ is limited by the fixed total energy but it
is a good approximation semiclassically. The unitarity of the $T$-operator
leads to the resurgence of the spectral determinant \cite{GeoPra}.

Another useful property, the $n$th power of the $T$-operator has the form of
the $T$-operator for $n$ Poincar\'e mappings, that is $T^n(q,q^{\prime };E)$
is given by Eq.\ (\ref{Tgen1}) but the orbits now cross the PSS $n$ times.
Consider, for example, 
\begin{equation}
T^2(q,q^{\prime })=\frac 1{2\pi }\int dq^{\prime \prime }\sum \sum \left[
-\frac{\partial ^2S(q,q^{\prime \prime })}{\partial q\partial q^{\prime
\prime }}\frac{\partial ^2S(q^{\prime \prime },q^{\prime })}{\partial
q^{\prime \prime }\partial q^{\prime }}\right] ^{1/2}e^{i\left[
S(q,q^{\prime \prime })+S(q^{\prime \prime },q^{\prime })\right] }. 
\end{equation}
The $S\Phi $ condition $\partial S(q,q^{\prime \prime })/\partial
q^{\prime \prime }+\partial S(q^{\prime \prime },q^{\prime })/\partial
q^{\prime \prime }=0$ ensures that the orbits $q^{\prime }\longmapsto
q^{\prime \prime }$ and $q^{\prime \prime }\longmapsto q$ have the same
momentum at $q^{\prime \prime }$. In other words, it selects the classical
orbits $q^{\prime }\longmapsto q^{\prime \prime }$ $\longmapsto q$ that
cross the PSS twice. To complete the proof we need to show that
\begin{equation}
\frac{\frac{\partial ^2S(q,q^{\prime \prime })}{\partial q\partial
q^{\prime \prime }}\frac{\partial ^2S(q^{\prime \prime },q^{\prime
})}{\partial q^{\prime \prime }\partial q^{\prime }}}{\frac{\partial
^2S_2(q,q^{\prime })}{\partial q^{\prime \prime 2}}}=\frac{\partial
^2S_2(q,q^{\prime })}{\partial q^{\prime \prime }\partial q^{\prime }}
\end{equation}
where $S_2(q,q^{\prime })=S(q,q^{\prime \prime })+S(q^{\prime \prime
},q^{\prime })$. For this we differentiate the $S\Phi $ condition by $q$ and
by $q^{\prime }$. The composition property is important for the derivation
of the trace formula using the $T$-operator (see Sec.\ \ref{gdot}) and 
for the semiclassical Fredholm theory \cite{GeoPra,GeoPra2,FisGeoPra}.

The $T$-operator in momentum representation is semiclassically a finite
matrix if the PSS has a finite length $\mathcal{L}$. Suppose the PSS is a
closed line. We define a complete set of the momentum eigenfunctions 
\begin{equation}
\phi _p(q)=\frac 1{\sqrt{\mathcal{L}}}e^{ipq}
\end{equation}
where $p=(2\pi /\mathcal{L})\times \mathrm{integer}$. The $T$-operator
matrix is 
\begin{equation}
T_{pp^{\prime }}=\frac 1{\mathcal{L}}\int dqdq^{\prime }T(q,q^{\prime
})e^{i\left( p^{\prime }q^{\prime }-pq\right) }.
\end{equation}
The $S\Phi $ condition requires that $p$ and $p^{\prime }$ be equal to the
projection of the classical momentum at points $q$ and $q^{\prime }$,
respectively. But this projection is limited by $\sqrt{2mE}$ by the
absolute value. Hence the size of the matrix is no greater than
$\mathcal{L}\sqrt{2mE}/\pi $ or $2\mathcal{L}/\lambda $ where $\lambda $
is the de Broglie wavelength. 

\section{Maslov phase and the change of variables}

\label{mpat}

A neighborhood of a caustic in a multidimensional system, like a turning
point in one dimension, is the region where the semiclassical approximation
breaks down. It turns out that the $T$-operator and the wavefunctions
acquire additional phases, called \emph{Maslov's phases}, in the regions of
validity \cite{MasFed}. The Maslov phases will change, in general, the
overall probability density and the quantization conditions and thus it is
important to understand their origin and be able to properly account for
them.

The central point in Bogomolny's derivation of the $T$-operator is the use
of the semiclassical approximation to the Green function in the energy
representation \cite{Gut} 
\begin{eqnarray}
G(r,r^{\prime };E)=\frac 1{i\left( 2\pi i\right) ^{\left( N-1\right)
/2}}\sum_{\mathrm{cl.\ tr.}} &&\left| \frac 1{v(r)v(r^{\prime })}\det \left[ 
\frac{\partial ^2S(r,r^{\prime };E)}{\partial r_{\perp }\partial r_{\perp
}^{\prime }}\right] \right| ^{1/2}  \nonumber \\
&&\times e^{iS(r,r^{\prime };E)-i\frac \pi 2\nu }.  \label{mpatG}
\end{eqnarray}
It satisfies $(E-H)G(r,r^{\prime };E)=\delta (r-r^{\prime })$ and is not
valid for short trajectories, at least in two dimensions. $r_{\perp }$,
$r_{\perp }^{\prime }$ are the local coordinates perpendicular to the
velocity at $r$, $r^{\prime }$, respectively. $v=\left| \partial
H/\partial p\right| $ is the modulus of the velocity (actions and momenta
are defined in the units of $\hbar $). The Maslov index $\nu $ in the
$T$-operator (\ref {Tgen1}) is inherited from this Green's function. 

Let us follow a particular orbit starting from $r^{\prime }$. At some
points along the orbit one or more eigenvalues of the matrix $\partial
^2S/\partial r_{\perp }\partial r_{\perp }^{\prime }$ may become infinite
and change sign \cite{BerMou}. That is what happens when the orbit touches
a caustic (see below). If $m$ eigenvalues change sign (caustic of order
$m$) the determinant in Eq.\ (\ref{mpatG}) changes by $(-1)^m$. If we drop
the modulus and analytically continue the determinant in the complex space
``around'' the singularity \cite{Kel} as in the one-dimensional case
\cite{LanLif} we will find that the Green function acquired the phase
$-m\pi /2$. One has to assume the function is exponentially small in the
classically forbidden region beyond the caustic. (This method does not
work near a hard wall.) The Maslov index of an orbit is the sum of the
contributions from all caustics the orbit encounters. Each caustic
increases $\nu $ by $m$, each hard wall increases $\nu $ by $2$. In
particular, if $r^{\prime }=q^{\prime }$ and $r=q $ lie on the surface of
section and the orbit makes one Poincar\'e mapping, this index enters the
$T$-operator. 

\begin{figure}[tbp]
{\hspace*{2.7cm} \psfig{figure=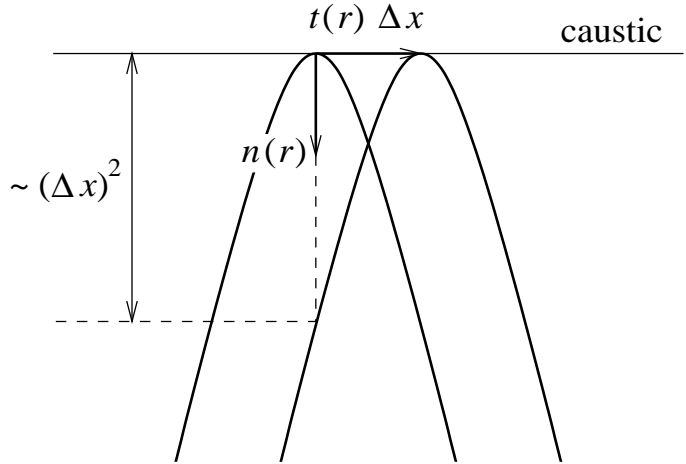,height=6.3cm,width=9cm,angle=0}}
{\vspace*{.13 in}} 
\caption{Two orbits that touch the caustic and lie almost in the same plane. 
\label{1_2}} {\vspace{1.2 cm}} 
\end{figure}

So far we defined the Maslov index in terms of singularities of the matrix
$\partial ^2S/\partial r_{\perp }\partial r_{\perp }^{\prime }$. Now we
give it the geometrical interpretation \cite{CreRobLit,BerMou}. Let us
surround the orbit of the previous paragraph (we call it the central
orbit) with a sufficiently narrow tube of trajectories with the same
energy $E$ all of which originate from point $r^{\prime }$. They form an
$(N-1)$-parameter family that can be parametrized by the vector $\partial
S/\partial r_{\perp }^{\prime }$, which depends on $r$. For a given orbit
in the family we will choose $r$ to be on the caustic. It is possible that
\emph{all} the trajectories within the tube (remember, the tube is narrow)
touch an $(N-m)$-dimensional surface which is independent of the shape of
the tube. This surface is called a \emph{caustic of order }$m$. Suppose
the central orbit touches the caustic surface at point $r$. The caustic
has $m-1$ dimensions less than the number of parameters in the tube. This
accounts for $m-1$ infinite eigenvalues of $\partial ^2S/\partial r_{\perp
}\partial r_{\perp }^{\prime }$. (When different trajectories touch the
caustic at the same point, the derivative of the parametrization vector
becomes infinite.) Without the loss of generality we can always assume
that the caustic is locally flat by making an appropriate coordinate
transformation. The orbits are parabolic near the caustic. Consider an
orbit that touches the caustic a small distance $\Delta x$ from the
central orbit and lies locally in almost the same plane (Fig.\ \ref{1_2}).
(To find this orbit note that the plane of the parabola of the central
orbit is fixed by two unit vectors: one is the tangent vector $t(r)$ at
the point of touch, another, $n(r)$, is orthogonal to the caustic at the
point of touch. If the neighboring orbit touches the caustic at
$r+t(r)\Delta x$, its tangent at this point is almost parallel to $t(r)$
by continuity. Its second vector $n\left[ r+t(r)\Delta x\right] $ belongs
to the $m$-dimensional complement to the caustic, so its direction can be
fixed by $m-1$ free parameters mentioned above.) Now, the difference in
parameters for these two orbits is of order $\Delta x$, but the distance
between them at point $r$ in the direction of $n(r)$ is of order $\left(
\Delta x\right) ^2$. This brings another singular eigenvalue of $\partial
^2S/\partial r_{\perp }\partial r_{\perp }^{\prime }$. Thus, this matrix
has $m$ singular eigenvalues in total.  

A special case is the caustic associated with the original point
$r^{\prime } $. Creagh \emph{et al}.\ \cite{CreRobLit} argue that each
negative eigenvalue of $-\partial ^2S/\partial r_{\perp }\partial r_{\perp
}^{\prime }$ for positive time $t=0+$ decreases the Maslov index by $-1$.
(Here one has to be careful to preserve the orientation of the local basis
$(\mathbf{e}_{\parallel }, \mathbf{e}_{\perp }^i)$, $i=1,\ldots ,N-1$,
along the orbit, where $\mathbf{e}_{\parallel }$ points in the direction
of propagation.) This apparently contradicts the statement in Ref.\
\cite{BerMou} that $\nu =0$ for short trajectories since there are no
caustics in between. Notice, however, that for the kinetic-plus-potential
systems all eigenvalues of $-\partial ^2S/\partial r_{\perp }\partial
r_{\perp }^{\prime }$ are positive \cite {CreRobLit}. In general, as the
example in the next section shows, this additional Maslov index is
important for the quantization conditions. 

\begin{sloppypar}
It is sometimes convenient to evaluate the classical action in the
canonical coordinates, other than the original ones. For example, in the
next section the action-angle variables are used. It would be desirable to
be able to construct the $T$-operator and solve for the wavefunction
directly in the new coordinates. Let us analyze how the Green function (\ref
{mpatG}) transforms under canonical transformation $(p,q)\longmapsto (P,Q)$.
We rewrite Eq.\ (\ref{mpatG}) in an equivalent form \cite{BerMou} 
\begin{eqnarray}
G(q,q^{\prime };E) &=&\sum_{\mathrm{cl.\ tr.}}G_1(q,q^{\prime };E)\equiv
\sum_{\mathrm{cl.\ tr.}}\frac{\sqrt{\Delta (q,q^{\prime };E)}}{i\left( 2\pi
i\right) ^{\left( N-1\right) /2}}e^{iS(q,q^{\prime };E)},  \label{mpatG2} \\
\Delta (q,q^{\prime };E) &=&-\left( \frac{\partial ^2S}{\partial
E^2}\right) ^{1-N}\det \left( \frac{\partial ^2S}{\partial
E^2}\frac{\partial ^2S}{\partial q\partial q^{\prime }}-\frac{\partial
^2S}{\partial E\partial q}\frac{\partial ^2S}{\partial E\partial q^{\prime
}}\right) . 
\end{eqnarray}
Note that we removed the modulus from under the square root together with
the Maslov index. For a given trajectory we can define the coordinate
transformations $Q=Q(q,q^{\prime })$ and $Q^{\prime }=Q^{\prime
}(q,q^{\prime })$ which can be multivalued. The probability density for a
semiclassical wavefunction is a sum of probabilities for individual orbits,
therefore the transformed Green's function 
\begin{equation}
G^{(Q)}(Q,Q^{\prime })=\sum_{\mathrm{cl.\ tr.}}G_1(q,q^{\prime
})\sqrt{\left| \det \frac{\partial q}{\partial Q}\right| \left| \det
\frac{\partial q^{\prime }}{\partial Q^{\prime }}\right| }
\end{equation}
where $q$ and $q^{\prime }$ are the functions of $Q$ and $Q^{\prime }.$
But $\Delta (q,q^{\prime })=\Delta (Q,Q^{\prime })$ $\det \left( \partial
Q/\partial q\right) $ $\det \left( \partial Q^{\prime }/\partial q^{\prime
}\right) .$ Finally, the Green function
\begin{equation}
G^{(Q)}(Q,Q^{\prime };E)=\sum_{\mathrm{cl.\ tr.}}\sqrt{\Delta (Q,Q^{\prime
};E)\ \mathrm{sgn}\left( \mathrm{\det }\frac{\partial Q}{\partial
q}\mathrm{\det }\frac{\partial Q^{\prime }}{\partial q^{\prime }}\right)
}\frac{e^{iS^{(Q)}(Q,Q^{\prime };E)}}{i\left( 2\pi i\right) ^{\left(
N-1\right) /2}}
\label{mpatG3}
\end{equation}
where $S^{(Q)}(Q,Q^{\prime };E)=S(q,q^{\prime };E).$ Equation (\ref{mpatG3})
has the same form as Eq.\ (\ref{mpatG2}) except for the signature of the
Jacobians. Let us fix $q^{\prime }$ and follow $q$ along the trajectory.
If $Q=Q(q)$ is a multivalued function, some eigenvalues of $\partial
Q/\partial q $ become singular and change sign as the orbit goes from one
sheet of $Q(q) $ to another. This happens if a caustic in $q$-space is
completely or partially removed in $Q$-space by making $Q(q)$ multivalued.
If, for instance, the caustic is removed completely, the number of
singular eigenvalues is equal to the order of the caustic. Suppose $k$
eigenvalues become singular. Then the matrix inside $\Delta (Q,Q^{\prime
};E)$ has $k$ singular eigenvalues less then the matrix inside $\Delta
(q,q^{\prime };E).$ Thus the total number of singular eigenvalues, and,
consequently, the Maslov index, is invariant under the coordinate
transformation.\footnote{If there is a new caustic of order $m$ in
$Q$-space that did not exist in $q$-space then $m$ eigenvalues of
$\partial Q/\partial q$ go through zero and change sign but in the
opposite direction than the corresponding $m$ singular eigenvalues of
$\Delta (Q,Q^{\prime };E).$ Thus, again, the Maslov index does not
change.} The important conclusion from this discussion is that the
semiclassical Green function or the $T$-operator can be evaluated with
Eqs.\ (\ref{mpatG}) or (\ref{Tgen1}) in any set of canonical coordinates,
apart from the Maslov index which is determined by the topology of the
orbits in the \emph{physical space}. This result is quite different from
the classical assumption that all canonical coordinates are equivalent. 
\end{sloppypar}

\section{Two-dimensional separable system}

\label{tdss}

As an illustration of the $T$-operator technique we apply it to a
two-dimensional system which is separable in the Cartesian coordinates. We
will solve the problem in the action-angle (AA) variables. They change the
topology of the orbits and thus bring certain complications. It will be
helpful to understand them before moving on to the perturbation theory of
the following chapters. The results are, of course, well known from simpler
methods. The quantization of the rotationally invariant integrable systems
in two and three dimensions using Bogomolny's $T$-operator was done by
Goodings and Whelan \cite{GooWhe}.

\begin{figure}[tbp]
{\hspace*{2.7cm} \psfig{figure=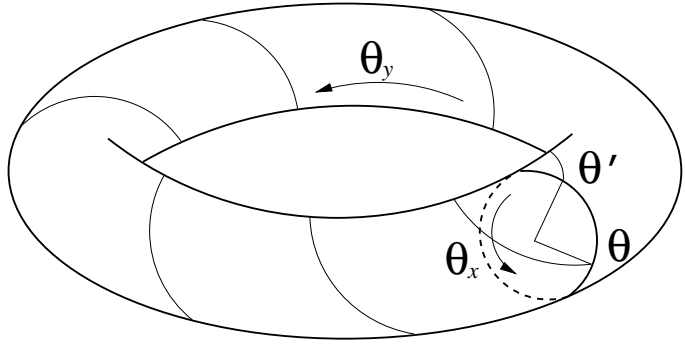,height=4.5cm,width=9cm,angle=270}}
{\vspace*{.13in}}
\caption{Poincar\'e map from $\theta_x = \theta^\prime$ to $\theta_x = 
\theta$ on the surface of section $\theta_y = \mathrm{const}$.
\label{1_3}}
{\vspace{1.2 cm}}
\end{figure}

For a separable system we can define the action-angle variables
$(I_x,\theta _x)$ and $(I_y,\theta _y)$ in $x$ and $y$ directions,
respectively. We choose the $x$ axis ($y=0$) as the PSS which corresponds
to $\theta _y=\mathrm{const}$. One Poincar\'e mapping $x^{\prime
}\longmapsto x$ is described as $\theta _x^{\prime }\longmapsto \theta
_x,$ $\theta _y\longmapsto \theta _y+2\pi $ in the AA variables
(Fig.\ \ref{1_3}). Note that by definition $\theta _x>\theta _x^{\prime };$
the actions $I_i=(2\pi )^{-1}\oint p_idx_i$ ($i=x,y$) are integrals of
motion; the functions $x(\theta _x)$ and $y(\theta _y)$ have period $2\pi
$. We made a transformation to the AA variables because in these
coordinates the $T$-operator has a particularly simple form: 
\begin{equation} T(\theta ,\theta ^{\prime };E)=\left( \frac 1{2\pi
i}\left| \frac{\partial ^2S}{\partial \theta \partial \theta ^{\prime
}}\right| \right) ^{1/2}e^{iS(\Delta \theta ;E)-i\frac \pi 2\nu (\theta
,\theta ^{\prime };E)} \label{tdssT} \end{equation} where we dropped the
subscript ``$x$'' in $\theta _x$ and, apart from the Maslov index, $T(\theta
,\theta ^{\prime };E)$ depends only on the difference $\Delta \theta
=\theta -\theta ^{\prime }$. We assumed there is only one trajectory
connecting $\theta ^{\prime }$ and $\theta .$ If there are more than one
trajectory with the same energy, they will have different $I$'s, and the
PSS wavefunctions $\psi \left( \theta \right) \propto e^{iI_x\theta }\
$can be treated independently in the semiclassical approximation. 

The action 
\begin{equation}
S(\Delta \theta ;E)=\Delta \theta I_x+2\pi I_y  \label{tdssS}
\end{equation}
where the $I$'s should be expressed in terms of $\Delta \theta $ and $E.$
One relation is the Hamiltonian 
\begin{equation}
H(I_x,I_y)=E.  \label{tdssH}
\end{equation}
Suppose it can be solved for $I_y,$ 
\begin{equation}
I_y=g_E(I_x).  \label{tdssge}
\end{equation}
Define the frequency ratio 
\begin{equation}
\alpha \equiv \alpha _E(I_x)\equiv \frac{\omega _x}{\omega _y}=\frac{\left(
\partial H/\partial I_x\right) _{I_y}}{\left( \partial H/\partial I_y\right)
_{I_x}}
\end{equation}
where $I_y$ was substituted from Eq.\ (\ref{tdssge}). Then the second
relation is 
\begin{equation}
\alpha _E(I_x)=\frac{\Delta \theta }{2\pi }.  \label{tdssal}
\end{equation}
Eqs.\ (\ref{tdssge}) and (\ref{tdssal}) allow one to write the action
$S(\Delta \theta ;E)$.\footnote{\label{tdssosc}The $T$-operator is not well
defined for a two-dimensional harmonic oscillator where $\Delta \theta
=\mathrm{const}.$} Note a useful formula \cite{UllGriTom}
\begin{equation}
\alpha =-g_E^{\prime }(I_x)  \label{tdssag}
\end{equation}
which can be derived by differentiating Eqs.\ (\ref{tdssH}) and 
(\ref{tdssge}).

The Maslov index $\nu (\theta ,\theta ^{\prime };E)$ counts the number of
caustics encountered by the orbit in the Cartesian coordinates. Suppose
there are two turning points in both $x$- and $y$-directions that produce
the first order caustics. The Maslov index 
\begin{equation}
\nu (\theta ,\theta ^{\prime })=\nu _0+\nu _x(\theta ,\theta ^{\prime })+\nu
_y,\quad \nu _y=2.
\end{equation}
Here $\nu _x,$ $\nu _y$ are the number of times the orbit goes through the
turning points in $x$- and $y$-directions, respectively, and $\nu
_0=-\Theta \left( \partial ^2S/\partial r_{\perp }\partial r_{\perp
}^{\prime }\mid _{t\rightarrow 0}\right) $ is related to the starting
point caustic (see the previous section), $\Theta $ is the Heaviside
function. Assume for the definitiveness that the turning points in $x$ are
located at $\theta \limfunc{mod}2\pi =0,\pi $ (they must be separated by
$\pi $ because of the time-reversal symmetry). Suppose $\pi n<\Delta
\theta <\pi (n+1),$ for some $n=0,1,\ldots $~. Then, depending on the end
point $\theta ,$ the orbit encounters $n$ or $n+1$ turning points in $x$.
It is easy to find
\begin{equation}
\nu _x(\theta ,\Delta \theta )=\left\{ 
\begin{array}{l}
n+1,\quad 0<\theta <\Delta \theta -\pi n \\ 
n,\quad \Delta \theta -\pi n<\theta <\pi
\end{array}
\right.
\end{equation}
continued in $\theta $ with period $\pi .$ The additional Maslov index $\nu
_0$ can be conveniently expressed as \cite{UllGriTom} 
\begin{equation}
\nu _0=-\Theta \left[ g_E^{\prime \prime }(I_x)\right] .
\end{equation}

\begin{figure}[tbp]
{\hspace*{2.7cm} \psfig{figure=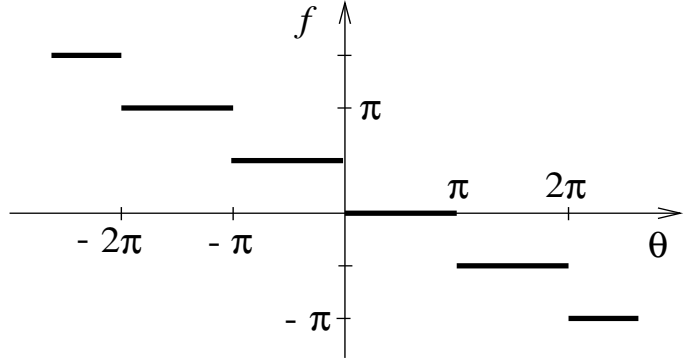,height=4.9cm,width=9cm,angle=0}}
{\vspace*{.13in}}
\caption{Step-wise phase $f(\theta)$ from Eq.\ (\ref{tdsspsi}).
\label{1_4}}
{\vspace{1.2 cm}}
\end{figure}

To solve Bogomolny's equation (\ref{TgenBog}) we make an ansatz 
\begin{equation}
\psi \left( \theta \right) =e^{iI\theta +if(\theta )}  \label{tdsspsi}
\end{equation}
where $I=\mathrm{const}$ to be determined and $f(\theta )$ is a
step-function. With $T$ given by Eq.\ (\ref{tdssT}) Bogomolny's equation
becomes 
\begin{equation}
e^{if(\theta )}=\int d\left( \Delta \theta \right) \left[ \frac{\left|
S^{\prime \prime }\left( \Delta \theta \right) \right| }{2\pi i}\right]
^{1/2}e^{i\left[ S\left( \Delta \theta \right) -I\Delta \theta +f(\theta
-\Delta \theta )-\frac \pi 2\nu (\theta ,\Delta \theta )\right] }
\label{tdssBog}
\end{equation}
where the integration variable has been changed from $\theta ^{\prime }$
to $\Delta \theta $ and $e^{iI\theta }$ was canceled on both sides. The
stationary phase condition is
\begin{equation}
I=S^{\prime }\left( \Delta \theta _{\mathrm{st}}\right) =I_x(\Delta \theta
_{\mathrm{st}},E). 
\end{equation}
The second equality follows from Eq.\ (\ref{tdssS}) taking into account
that $ \left( \frac {\partial I_y} {\partial \Delta \theta} \right) _E=
g_E^{\prime }(I_x)\left( \frac {\partial I_x} {\partial \Delta \theta}
\right) _E$ $=-\frac {\Delta \theta} {2\pi} \left( \frac {\partial I_x}
{\partial \Delta \theta} \right) _E$ [see Eqs.\ (\ref {tdssal}) and
(\ref{tdssag})]. Note that $f$ and $\nu $ do not change the stationary
point. We expand $S\left( \Delta \theta \right) $ near the stationary
point and integrate. Equation (\ref{tdssBog}) becomes
\begin{equation}
e^{if(\theta )}=\left[ \frac{\left| S^{\prime \prime }\left( \Delta \theta
_{\mathrm{st}}\right) \right| }{S^{\prime \prime }\left( \Delta \theta
_{\mathrm{st}}\right) }\right] ^{1/2}e^{i\left[ 2\pi I_y(\Delta \theta
_{\mathrm{st}},E)+f(\theta -\Delta \theta _{\mathrm{st}})-\frac \pi 2\nu
(\theta ,\Delta \theta _{\mathrm{st}})\right] }.  \label{tdssexp}
\end{equation}
Applying another chain of equalities $S^{\prime \prime }\left( \Delta
\theta \right) =\left( \frac {\partial I_x} {\partial \Delta \theta}
\right) _E=\frac 1 {2\pi} \left( \frac {\partial I_x} {\partial \alpha}
\right) _E$ $=-\frac 1 {2\pi g_E^{\prime \prime }}$ we find that the
pre-exponential factor cancels the Maslov index $\nu _0.$ Let us require
that
\begin{equation}
f(\theta )-f(\theta -\Delta \theta _{\mathrm{st}})+\frac \pi 2\nu _x(\theta
,\Delta \theta _{\mathrm{st}})=0  \label{tdssf}
\end{equation}
for all $\theta .$ Then Eq.\ (\ref{tdssexp}) is solved if 
\begin{equation}
I_y(\Delta \theta _{\mathrm{st}},E)=n_y+\frac 12
\end{equation}
for some $n_y=0,1,\ldots $~. This is, of course, a well known
Einstein-Brillouin-Keller (EBK) quantization of an action variable. One can
check that the step-wise function (Fig.\ \ref{1_4}) 
\begin{equation}
f(\theta )=0\ \mathrm{if}\ 0<\theta <\pi ;\quad f(\theta +\pi )-f(\theta
)=-\frac \pi 2
\end{equation}
satisfies Eq.\ (\ref{tdssf}). The second quantization condition 
\begin{equation}
I_x(\Delta \theta _{\mathrm{st}},E)=n_x+\frac 12,\quad n_x=0,1,\ldots
\end{equation}
comes from the $2\pi $-periodicity of $\psi \left( \theta \right) $ Eq.\
(\ref {tdsspsi}). (This requirement ensures that $\psi \left( \theta
\right) $ is uniquely defined on the cross-section of an invariant torus
$0\leq \theta \leq 2\pi ,$ $\theta _y=\mathrm{const}$.)

Thus, after all this trouble, we arrive to the familiar EBK expression for
the energy 
\begin{equation}
E_{n_x,n_y}=H\left( I_x=n_x+\frac 12,I_y=n_y+\frac 12\right)
\end{equation}
with the PSS wavefunction 
\begin{equation}
\psi \left( \theta \right) =e^{i\left( n_x+\frac 12\right) \theta +if(\theta
)}.  \label{tdsspsi2}
\end{equation}
This function is discontinuous at $\theta =\pi n.$ This may seem to be
unphysical, but remember that the semiclassical approximation breaks down
near the turning points. Therefore Eq.\ (\ref{tdsspsi2}) should not be used
near $\theta =\pi n.$ The exact wavefunction would smoothly join the
discontinuity. If there were hard walls instead of caustics the actions
would be integral and $f(\theta )$ would have the steps of size $\pi .$

The wavefunction in $x$-representation can be recovered by the standard
procedure $\psi \left( x\right) =\psi \left( \theta \right) \sqrt{\left|
\partial \theta /\partial x\right| }.$ Note that $\partial \theta /\partial
x=\omega _x/v_x,$ where $v_x$ is the velocity, and $I_x\theta =\int p_xdx.$
Suppose $x(\theta =0)$ is the turning point on the left. Then $\psi \left(
x\right) $ has two branches: 
\begin{eqnarray}
\ &&\psi _1\left( x\right) \propto \frac 1{\sqrt{\left| v_x\right|
}}e^{i\int_{x(0)}^x\left| p_x\right| dx^{\prime }},\quad 0<\theta <\pi ,
\nonumber \\
\ &&\psi _2\left( x\right) \propto \frac 1{\sqrt{\left| v_x\right|
}}e^{-i\int_{x(0)}^x\left| p_x\right| dx^{\prime }+i\frac \pi 2},\quad
-\pi <\theta <0,
\end{eqnarray}
which makes 
\begin{equation}
\psi \left( x\right) \propto \frac 1{\sqrt{\left| v_x\right| }}\cos \left[
i\int_{x(0)}^x\left| p_x\right| dx^{\prime }-i\frac \pi 4\right] ,
\end{equation}
the one-dimensional bound state wavefunction \cite{LanLif}.

Finally, we note that the algorithm we used to solve the semiclassical
problem in the AA coordinates (or other coordinates that change the topology
of the trajectories) may be not unique. Both in this and in the previous
section, where we considered the change of coordinates in the $T$-operator,
we assumed that the wavefunctions do not acquire Maslov's index on caustics
--- we always took the \emph{absolute value }of the Jacobian: $\psi \left(
q\right) =\psi \left( Q\right) \sqrt{\left| \partial Q/\partial q\right| }$
. Alternatively, we could drop the modulus or add the Maslov phase. Then the
Maslov phase of the $T$-operator would be determined by the caustics in the
current coordinates (not physical space) and the wavefunction $\psi $ would
not have the unnatural discontinuities as in the above example. However, the
straightforward quantization conditions would be incorrect, for instance,
the actions would be integral instead of half-integral. Thus the special
formulation of the quantization conditions would be necessary, again giving
significance to the physical coordinates. Whichever method is used, the
physically important quantities, the wavefunction $\psi \left( x\right) $ and
the quantized energy, will be the same.

\section{Conclusions}

Bogomolny's $T$-operator is a powerful tool that combines the
semiclassical approximation and Poincar\'e's surface of section.
Bogomolny's equation determines the surface of section wavefunction and
the energy levels. The full wavefunction can be reconstructed from the
surface of section wavefunction with the help of a semiclassical
propagator. In billiards the method is related to the boundary integral
method. The properties of $T$-operator include unitarity, the composition
property, and the finite size in the momentum representation, all within
the stationary phase approximation. The $T$-operator propagates the 
semiclassical wavepacket consistently with the classical surface of section 
map. The regions where the semiclassical approximation breaks down, like 
caustics and walls, are responsible for the additional (Maslov) phases in 
the $T$-operator and wavefunctions. One should be careful, when making 
coordinate transformations with singular points, not to change the Maslov 
phase. For the integrable systems the $T$-operator method is consistent 
with the EBK quantization. 

\chapter{Perturbation theory}

\label{perth}

In this chapter we present a systematic derivation of the perturbation
theory that provides a semiclassical description for almost integrable
systems \cite{PraNarZai}. In a theory of this type there are two competing
quantities: the large action (compared to the Planck constant) and the
small perturbation from integrability. Their interplay determines to what
degree the structure of classical phase space is reflected in the quantum
results. 

It is a well-known fact of the classical theory \cite{LicLie} that the
topology of invariant tori near the periodic orbits changes under
perturbation. One needs a resonant classical perturbation theory in order
to describe it. On the other hand, the standard EBK quantization procedure
relies on the classical invariant tori, which means that a resonant
semiclassical theory might be necessary for some classes of states. Here
is a simple estimate. For a periodic orbit of action $I$ the perturbation
of size $\epsilon$ changes the topology of the tori within the layer
$\delta I \sim \sqrt {\epsilon} I$. If $\delta I \gtrsim \hbar$, several
quantum levels become mixed, i.e.\ the resonant theory is required. 

For a billiard of linear size $L$ with the perturbed boundary the
condition becomes $kL \sqrt {\epsilon} \gtrsim 1$ where $k=2\pi/ \lambda$
is the wavenumber. This means that, unless the perturbation $\delta L \ll
\lambda^2 /L$, the billiard will have strongly perturbed states.\footnote
{In a billiard the level spacing $\hbar^2 /mL^2$ should be compared to the
shift of a given level $\hbar^2 k^2 \delta L/ mL$, which results in the
same criterion. (Noted by M. Sieber.)} Note that even the distortion of
the boundary shorter than the wavelength may be strong enough to mix
several energy levels. 

In the following sections we derive the analytic expressions for the
wavefunctions and energy levels of the states associated with the
classical resonances. Although our theory is effectively a resonant
perturbation theory, the diagonalization of unperturbed states is not
explicit. The Bogomolny equation allows to express the results in a 
simple, easy to visualize form. 

We use the circle billiard with a perturbed boundary as an example.
Several papers published in recent years \cite{BorCasLi,FraShe,Bor,CasPro}
discuss the localization and diffusion in angular momentum space of this 
system. We too find the angular momentum localized, with the degree of 
localization depending on the smoothness of the perturbation. 

There are several other methods that deal with almost integrable systems
that are sometimes similar, sometimes complementary to the following
theory. For instance, the quantization of Birkhoff-Gustavson normal form
\cite{Bir,Gus,Rob} is useful for the perturbed harmonic oscillators ---
the case to which our theory does not directly apply. The Born-Oppenheimer
approximation (Ch.\ \ref{qboa}) is similar to the perturbation theory in
some cases. The perturbed Berry-Tabor formula (Ch.\ \ref{trform})
expresses the contributions of the periodic orbits to the density of
states consistently with the perturbation theory.

\section{Perturbed integrable systems}

\label{pis}

An $N$-dimensional Hamiltonian system is called \emph{integrable }if it
has $N$ independent integrals of motion in involution \cite{Tab}. Its
motion in the phase space is confined to an $N$-dimensional manifold that
has the topology of a torus. It is called an \emph{invariant torus.} The
integrals of motion can be chosen to be the action variables $I_i,$
$i=1,\ldots ,N$. Then the angle variables $\theta _i$ (identified with
$\theta _i+2\pi $) provide the natural coordinates on the torus. The
Hamiltonian of the system depends only on the actions, $H_0(I)=E.$ The
equations of motion $\dot \theta _i=\partial H_0/\partial I_i\equiv \omega
_i(I)$ can be easily integrated. 

In the case $N=2$ we can choose the Poincar\'e surface of section (PSS)
as, say, $\theta _2=0$ (Fig.\ \ref{1_3}). The dynamics of the system
induces the PSS map $(I^{\prime },\theta ^{\prime })\longmapsto (I,\theta
)$ (we dropped the subscript ``1'') simply as $I=I^{\prime },$ $\theta
=\theta ^{\prime }+2\pi \omega _1/\omega _2.$ If all points $(I,\theta )$
belonging to a certain trajectory are plotted $I$ \emph{vs} $\theta ,$
$0<\theta <2\pi ,$ they all will lie on a horizontal line
$I=\mathrm{const}.$ The line will either be covered densely if $\omega
_1/\omega _2$ is irrational or have a finite number of points if it is
rational. We may say that the line $I=\mathrm{const}$ is an intersection
of an invariant torus with the PSS. The reduced action $S(\theta ,\theta
^{\prime })=S_0(\theta -\theta ^{\prime })=(\theta -\theta ^{\prime
})I_1+2\pi I_2$ [cf.\ Eq.\ (\ref{tdssS})] is a generating function for
this map: $I=\partial S/\partial \theta ,$ $I^{\prime }=-\partial
S/\partial \theta ^{\prime }$ (Sec.\ \ref{sosm}). Note that $S(\theta
,\theta ^{\prime })$ depends only on the difference $\theta -\theta
^{\prime }.$ The action variables must be regarded as functions of $\theta
-\theta ^{\prime }$ and $E$ (see Sec.\ \ref{tdss} for details). 

The Hamiltonian of a \emph{perturbed integrable system} 
\begin{equation}
H=H_0(I_1,I_2)+\epsilon H_2(I_1,I_2,\theta _1,\theta _2)+\epsilon
^2H_4+\cdots  \label{pish}
\end{equation}
differs from an integrable Hamiltonian by the terms proportional to a small
parameter $\epsilon .$ Likewise, the action of such system 
\begin{equation}
S(\theta ,\theta ^{\prime })=S_0(\theta -\theta ^{\prime })+\epsilon
S_2(\theta ,\theta ^{\prime })+\epsilon ^2S_4+\cdots  \label{pisS}
\end{equation}
is expanded in powers of $\epsilon .$ (We reserve the odd subscripts for
half-integral powers of $\epsilon $ that will appear in other expansions.)
The classical perturbation theory \cite{LicLie}, in principle, allows one to
find the corrections to the action if the Hamiltonian is known, for example, 
$S_2=-\int H_2dt$. Since only short orbits are involved, the problem of
small divisors does not arise. In the case of a billiard with a perturbed
boundary, $S(\theta ,\theta ^{\prime })$ can be easily deduced directly.

Of course, when the perturbation is on, $I$'s are no longer integrals of
motion. Consequently, the orbits on the PSS are no longer confined to the
lines $I=\mathrm{const}$. One can imagine several possibilities. The lines
could be slightly deformed, they could change their topology, or disappear
completely (i.e.\ the orbit would densely cover a two-dimensional area in the
PSS). As a matter of fact, all three cases may be present within one system,
depending on the region in the phase space.

\begin{sloppypar}
According to the Kolmogorov-Arnold-Moser (KAM) theorem, for small enough
$\epsilon $ and smooth enough perturbation most of the tori continue to
exist and remain close to the unperturbed tori. The measure of the
destroyed or considerably modified tori tends to zero as $\epsilon
\rightarrow 0.$ The destroyed tori are located near the \emph{rational,}
or \emph{resonant,} original tori. These are the tori that have a rational
winding number $\omega _1/\omega _2$, i.e.\ they support a \emph{periodic
orbit}, or \emph{resonance}. Consider a $(p,q)$ resonance, i.e.\ the
winding number is $p/q,$ an irreducible fraction. Then all points on the
line $I=\mathrm{const}$ are the fixed points of $\mathcal{T}^q,$ where
$\mathcal{T}$ is one Poincar\'e mapping. The Poincar\'e-Birkhoff fixed
point theorem \cite{Tab} states that under the perturbation only an even
number, a multiple of $2q$, of fixed points will remain. They will
alternate between stable and unstable. 
\end{sloppypar}

\begin{figure}[tbp]
{\hspace*{2.7cm} 
\psfig{figure=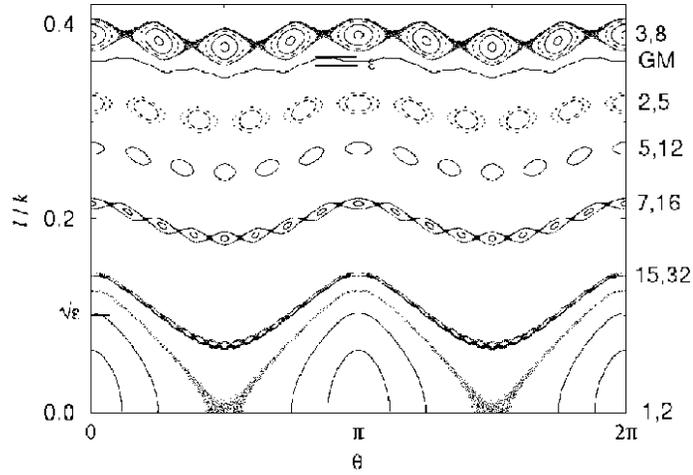,height=6.26cm,width=9cm,angle=0}}
{\vspace*{.13in}}
\caption[The surface of section (boundary phase space) of a perturbed
circular billiard.]
{The surface of section (boundary phase space) of a perturbed
circular billiard. $l/k$ is an angular momentum normalized to its maximal
value and $\theta$ is an angular coordinate along the boundary. The
numbers on the right denote the orbits near the $(p,q)$ resonances
perturbed by order $\sqrt{\epsilon}$. GM is an orbit with the golden mean
winding number perturbed by order $\epsilon$. The secondary resonances and
the chaotic regions near the separatrices, that are exponentially small
when $\epsilon \rightarrow 0$, can also be observed. Here the perturbation
is the ``smoothed stadium'' with the parameters $\epsilon = 0.1$, $\eta =
0.19$ defined in Sec.\ 2.2.1. 
\label{2_1}} 
{\vspace{1.2 cm}} 
\end{figure}

As Fig.\ \ref{2_1} illustrates, the tori form resonant islands around the
fixed points. The size of the islands is proportional to
$\sqrt{\epsilon}.$ Thus most of the tori near a rational torus change
their topology under the perturbation and form a new system of tori. The
new tori, in turn, form secondary resonance chains (as in the example of
$(2,5)$ resonance in the figure) although they are exponentially smaller
than the primary resonances \cite{LicLie}. There is a chaotic region near
the separatrix formed by numerous intersections of stable and unstable
manifolds. It too is exponentially small \cite{Laz}. The size of the
resonant islands reduces rapidly with $q$ (see Sec.\ \ref{rnr}), so the
total measure of the strongly modified phase space is finite and
proportional to $\sqrt{\epsilon }.$ The rest of the tori are slightly
perturbed on the scale of $\epsilon .$ Figure \ref{2_1} shows the torus
with the ``most irrational'' golden mean (GM) winding number. It is
supposedly the last torus to be destroyed with the increase of the
perturbation, although this is not a subject of this work. The figure also
illustrates how the rational numbers with large $q$ tend to become
``closer'' to irrational numbers. If one disregards the small island
chains in $(7,16)$ or $(15,32)$ resonances the overall curve is similar to
a perturbed irrational torus. Its oscillation is of order of $\epsilon .$
This effect is relevant to the semiclassical theory since the quantization
misses those details of the phase space that have a typical area less than
$2\pi \hbar $ (see Secs. \ref{ClasInt} and \ref{rnr}). 

\begin{figure}[tbp]
{\hspace*{2.7cm} \psfig{figure=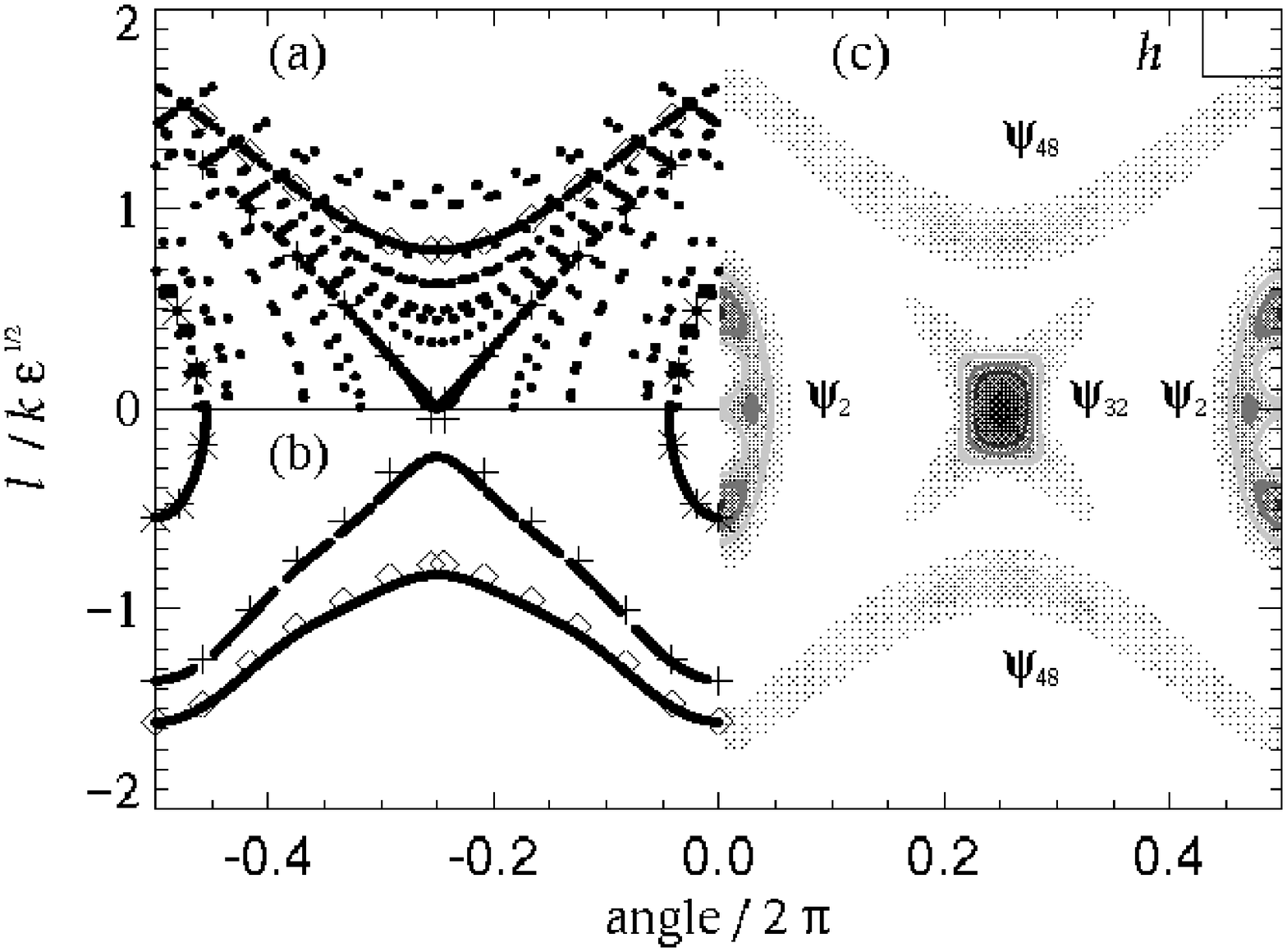,height=7cm,width=9cm,angle=0}}
{\vspace*{.13in}} 
\caption[The surface of section for (a) ``short stadium'' and (b)
``smoothed stadium.'' (c) Husimi plots for the numerical states of Fig.\
2.4.]
{The surface of section for (a) ``short stadium'' and (b)
``smoothed stadium'' of Sec.\ 2.2.1 with $\epsilon = 0.01$, $\eta =
0.15$. $l/k$ as in Fig.\ 2.1. The points $\Diamond$, $+$, and $\ast$
denote the intersection of the invariant tori with the surface of section,
$l_{\mathrm{inv}} (\theta)$, as given by the perturbation theory (see
Sec.\ 2.3.1), for the non-resonant, separatrix, and resonant
orbits, respectively. In (a), where KAM fails, orbits started at symbols
were iterated forward and backward 15 times (appear as dots $\bullet$).
Only short time structure is regular. In (b), where KAM applies, three
orbits, each iterated 1000 times, coalesce into solid lines. (c) Husimi
plots for the numerical states of Fig.\ 2.4, as well as the scar
state $\psi_{32}$, for the short stadium. The square has area $h$. 
\label{2_2}} 
{\vspace{1.2 cm}} 
\end{figure}

If the perturbation is not smooth enough (so that the KAM theorem does not
apply) the invariant tori may not exist at all [Fig.\ \ref{2_2} (a)]. In
this case the semiclassical perturbation theory needs diffraction
corrections, although the simplest version still reflects the main
features of the states for small $\epsilon$. 

\section{Formulation of the theory}

\label{fott}

\subsection{Perturbed circular billiard}

\label{pcb}

The perturbation theory we are about to present is based on the
$T$-operator and requires a careful choice of coordinates and surface of
section. From the theoretical point of view the natural choice would be
the action-angle (AA) variables where the unperturbed action depends only
on the difference of the angles. In practice working in the AA variables
is not always physically transparent and may require a complicated
transformation from the original coordinates and back. On the other hand,
billiards with a perturbed boundary have an advantage that the reduced
action is just a geometrical quantity proportional to the length of the
orbit. A perturbed circular billiard whose boundary in the polar
coordinates is $r(\theta )=R_0+\epsilon \Delta R(\theta )$ is a
particularly convenient system since the angle $\theta $ together with the
(modulus of) angular momentum $l$ are, in fact, the AA variables (when
$\epsilon =0$, of course). (We can always make the angular average of
$\Delta R$ vanish and assume $\Delta R\sim R_0$.) We will formulate the
perturbation theory for this system, which allows the simple mathematical
description and direct physical interpretation. There should be no
principal difficulty to generalize the theory or tailor it to other
systems, as we do in the subsequent chapters. We would like to point out
some limitations that will become more apparent as we proceed. First, the
theory is the most useful in two-dimensional systems, where the PSS has a
one-dimensional space component. Second, the unperturbed action $S_0$
should depend only on the difference of coordinates in the PSS which may
force one to use the AA variables. Finally, the theory is not suitable for
the perturbed two-dimensional harmonic oscillator since $S_0$ is not well
defined (see the footnote on p. \pageref{tdssosc}). 

\begin{figure}[tbp]
{\hspace*{2.7cm} \psfig{figure=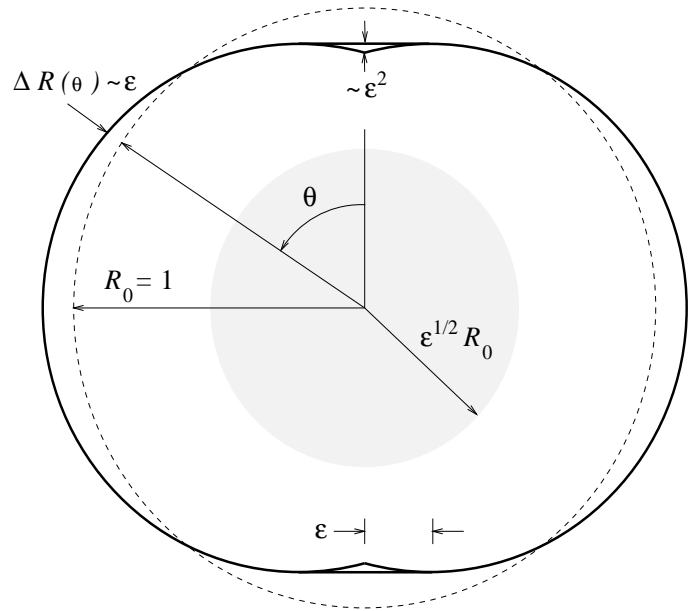,height=8cm,width=9cm,angle=270}}
{\vspace*{.13in}}
\caption[Short Bunimovich stadium $R(\theta) = R_0 + \epsilon \Delta R
(\theta)$ with $\epsilon = 0.3$.]
{Short Bunimovich stadium $R(\theta) = R_0 + \epsilon \Delta R
(\theta)$ with $\epsilon = 0.3$. The straight segments have length
$2\epsilon$. The dashed line denotes the circle of radius $R_0 = 1$. The
difference between the actual stadium and the approximation for $\Delta
R(\theta)$ that we use is of order $\epsilon^2$. The classical orbits that
pass through the shaded region are affected by the $l=0$ resonance. 
\label{2_3}}
{\vspace{1.2 cm}}
\end{figure}

From now on we assume $R_0=1.$ A wavenumber $k$ also becomes
dimensionless. (To restore the proper units one should substitute $k$ by
$kR_0$.) A perturbation of the form $\Delta R(\theta )=\left| \sin \theta
\right| -2/\pi $ models the short Bunimovich stadium that recently
received a lot of attention \cite{BorCasLi,Bor,CasPro}. The straight
segments in this stadium have length $2\epsilon .$ Our perturbed circle
approximates it to the order of $\epsilon $ and has a discontinuous
derivative (Fig.\ \ref{2_3}) [we neglected the straight segments when
deriving $\Delta R(\theta ),$ which thus describes the outer boundary of
two circles with their centers being $2\epsilon $ apart]. Although the
invariant tori do not exist, our theory still works if $k\epsilon
^{3/2}\ll 1.$ We also consider a ``smoothed stadium'' $\Delta R(\theta
)=\sqrt{\sin ^2\theta +\eta ^2}-C_\eta ,$ where $C_\eta $ makes the
angular average vanish. If $\epsilon $ is sufficiently small compared to
$\eta $ the invariant tori exist [Fig.\ \ref{2_2} (b)]. We choose $\eta
\sim \sqrt{\epsilon }$ in our numerical examples. 

We associate the PSS with the boundary of the billiard. We parametrize it by
the angle $\theta $ with the conjugate momentum $l.$ Our goal is to
construct the $T$-operator (\ref{Tbil}) and solve Bogomolny's equation (\ref
{psiTpsi}) perturbatively. For convenience we assume $\hbar =1$ and the mass 
$m=1/2.$ Then the energy $E=k^2.$ The action 
\begin{equation}
S(\theta ,\theta ^{\prime })=kL(\theta ,\theta ^{\prime })=k\left[
L_0(\theta -\theta ^{\prime })+\epsilon L_2(\theta ,\theta ^{\prime
})+\cdots \right]  \label{pcbS}
\end{equation}
where $L(\theta ,\theta ^{\prime })$ is the chord length between points
$\theta $ and $\theta ^{\prime }$ on the boundary. For a perfect circle we
find
\begin{equation}
L_0(\theta -\theta ^{\prime })=2\left| \sin \frac{\theta -\theta ^{\prime
}}2\right| \label{pcbL0}
\end{equation}
and the correction 
\begin{equation}
L_2(\theta ,\theta ^{\prime })=\left| \sin \frac{\theta -\theta ^{\prime
}}2\right| \left[ \Delta R(\theta )+\Delta R(\theta ^{\prime })\right] . 
\label{pcbL2}
\end{equation}
We always assume $k\gg 1$ and $\epsilon \ll 1$ as general requirements for
the semiclassical and perturbation theories, respectively.

An example of the PSS is shown on Fig.\ \ref{2_1}. Note that in a perfect
circle of radius $1$ the angular momentum is bounded, $\left| l\right|
\leq k.$ The resonances correspond to the periodic trajectories in a
circle. For instance, the $(1,2)$ resonance is related to the $l=0$ orbit
passing through the center of the circle. The $(1,3)$ resonance is a
triangle shaped periodic orbit, etc. Our theory is designed to find the
quantum states located primarily within a given resonance. 

\subsection{Simplest case: $(1,2)$ resonance, $k\epsilon \ll 1$}

\label{r12}

The theory simplifies significantly for the states near $(1,2)$ resonance,
partly due to the symmetry of the perturbation $L_2(\theta ,\theta
^{\prime }).$ We also assume the perturbation to be sufficiently weak,
$k\epsilon \ll 1,$ a requirement that will be later relaxed. It was
explained in the chapter's introduction that as long as $k \sqrt{\epsilon
}\gtrsim 1$ the resonant perturbation theory is needed. The relevant
classical orbits have the angular momentum $l\sim k\sqrt{\epsilon }$ and
pass through the shaded region in Fig.\ \ref{2_3}. 

In a perfect circle the resonant trajectory maps point $\theta ^{\prime }$
to $\theta =\theta ^{\prime }+\pi $. We therefore expand the $T$-operator
in $\delta \theta ^{\prime }=\theta ^{\prime }-\theta +\pi ,$ as well as
in $k\epsilon .$ In this case Eq.\ (\ref{Tbil}) becomes
\begin{equation}
T(\theta ,\theta ^{\prime })\simeq -\sqrt{\frac k{4\pi i}(1+\cdots
)}e^{i\left( 2k-\frac k4\delta \theta ^{\prime 2}+\cdots \right) }\left[
1+ik\epsilon V(\theta )+\cdots \right] \label{r12Texp}
\end{equation}
where $V(\theta )=L_2(\theta ,\theta -\pi )=\Delta R(\theta )+\Delta
R(\theta -\pi ).$ We also expand the wavefunction 
\begin{equation}
\psi (\theta ^{\prime })=\psi (\theta -\pi )+\delta \theta ^{\prime }\psi
^{\prime }(\theta -\pi )+\frac{\delta \theta ^{\prime 2}}2\psi ^{\prime
\prime }(\theta -\pi )+\cdots
\end{equation}
that solves Bogomolny's equation $\psi =T\psi $. We evaluate the integral
$\int T(\theta ,\theta ^{\prime })$ $\psi (\theta ^{\prime })d\theta
^{\prime }$ in the stationary phase approximation ($S\Phi $) thus
justifying the smallness of $\delta \theta ^{\prime }.$ The equation is
reduced to
\begin{equation}
\psi (\theta )=e^{i\left( 2k+\frac \pi 2\right) }\left[ \psi (\theta -\pi
)-\frac ik\psi ^{\prime \prime }(\theta -\pi )+ik\epsilon V(\theta )\psi
(\theta -\pi )\right] .
\end{equation}
Assume we can expand $\psi (\theta - \pi) \simeq e^{-i \left(2k + \frac 
\pi 2 \right)} \psi (\theta) - ik \epsilon E_m \psi (\theta - \pi)$, making 
$\psi (\theta -\pi )=e^{-i\omega }\psi (\theta )$ where 
\begin{equation}
\omega (k)\simeq 2k+\frac \pi 2+k\epsilon E_m.  \label{r12om}
\end{equation}
The constant $E_m$ is to be determined. We find that $\psi (\theta )$ has to
satisfy the ordinary differential equation 
\begin{equation}
\psi ^{\prime \prime }(\theta )+k^2\epsilon \left[ E_m-V(\theta )\right]
\psi (\theta )=0,  \label{r12Sch}
\end{equation}
which is a one-dimensional Schr\"odinger equation. We will refer to
$V(\theta )$ as a ``potential'' and to $E_m$ as a ``(surface of section)
energy'' when it does not cause a confusion. Note that the strength of the
potential is scaled as $\left( k\sqrt{\epsilon }\right) ^2$. 

The eigenstates $\psi _m(\theta )$ of Eq.\ (\ref{r12Sch}) with the
eigenenergy $E_m$ solve Bogomolny's equation, $m$ is a quantum number that
labels the states. Since $V(\theta )$ has period $\pi $ we can always find
the $2\pi $-periodic solutions such that $\psi _m(\theta )=\pm \psi
_m(\theta -\pi ).$ The second quantization condition is then 
\begin{equation}
\omega (k)=2\pi n+\omega _m  \label{r12qc}
\end{equation}
where $\omega _m=0$ or $\pi $ (depending on $m$) and $n$ is integer. This
provides an equation for $k$ and the total energy $E_{nm}$ 
\begin{equation}
k_{nm}=\sqrt{E_{nm}}=\frac{2\pi \left( n-\frac 14\right) +\omega
_m}{2-\epsilon E_m}. 
\end{equation}
Note that $E_m$ and $\psi _m$ weakly depend on $n$ via the parameter
$k^2\epsilon .$

Eq.\ (\ref{r12Sch}) can be solved by standard methods either numerically or
analytically. If $k\sqrt{\epsilon }\gg 1$ we can approximate its solution in
WKB: 
\begin{equation}
\psi _m(\theta )\propto \frac 1{\sqrt{f^{\prime }(\theta
)}}e^{ik\sqrt{\epsilon }f(\theta )} \label{r12psi}
\end{equation}
where 
\begin{equation}
f(\theta )=\pm \int d\theta \sqrt{E_m-V(\theta )}.
\end{equation}
If $E_m>V(\theta )$ for all $\theta ,$ we may quantize $E_m$ by the
condition 
\begin{equation}
k\sqrt{\epsilon }\int_0^\pi d\theta \sqrt{E_m-V(\theta )}=\omega _m+2\pi m.
\end{equation}
We call this the ``rotational'' case since all $\theta $ are in the
``classically allowed region.'' As will be clarified later, these states
quantize the classical orbits in the surface of section that lie near the
resonance chain though \emph{outside} of the separatrix (Fig.\ \ref{2_1}). 
In the KAM language, the topology of the tori did not change as a result
of the perturbation, but they are strongly distorted by the nearby
resonance. If we compare a picture of a resonance chain with the phase
portrait of a pendulum, these tori would be analogous to the rotational
trajectory of a pendulum. 

\begin{figure}[tbp]
{\hspace*{2.7cm}
\psfig{figure=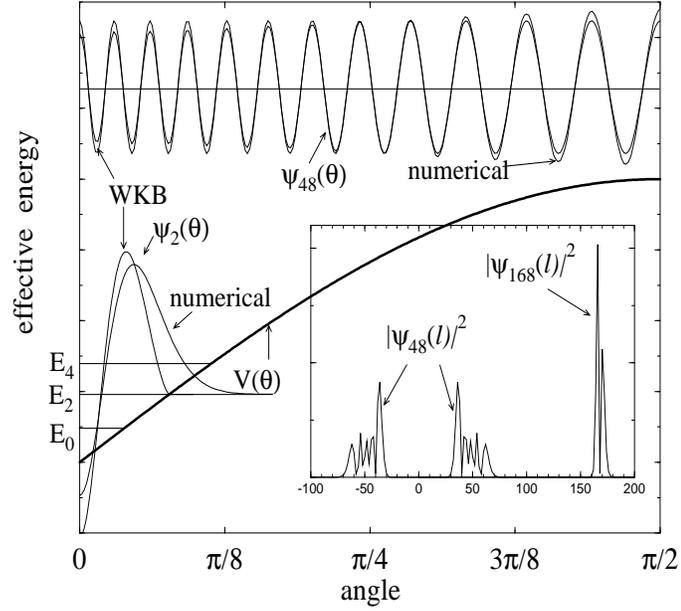,height=8.2cm,width=9cm,angle=270}}
{\vspace*{.13in}} 
\caption[Examples of the low angular momentum states
$\psi_m (\theta)$ in the short stadium.] 
{Examples of the low angular momentum states $\psi_m (\theta)$ in the
short stadium. The effective potential $V(\theta)$ is shown as a thick
line. The WKB solutions of Eq.\ (2.10) are compared with the numerical
solutions of Bogomolny's equation $\psi = T \psi$ (see Sec.\ 2.3.5). The
zero axis for each state is its WKB effective energy $E_m$. For the
librational state the simplistic boundary condition $\psi_2 (\theta) = 0$
at the turning point was taken in the WKB case. The states and the
potential are symmetric at $\theta = 0$. $k \sqrt{\epsilon} =42.3$ is
fixed.  Inset: the angular momentum representation of the rotational
states $m= 48$ and $m = 168$ (Sec.\ 2.3.4). 
\label{2_4}} 
{\vspace{1.2 cm}} 
\end{figure}

\begin{figure}[tbp]
{\hspace*{2.7cm} 
\psfig{figure=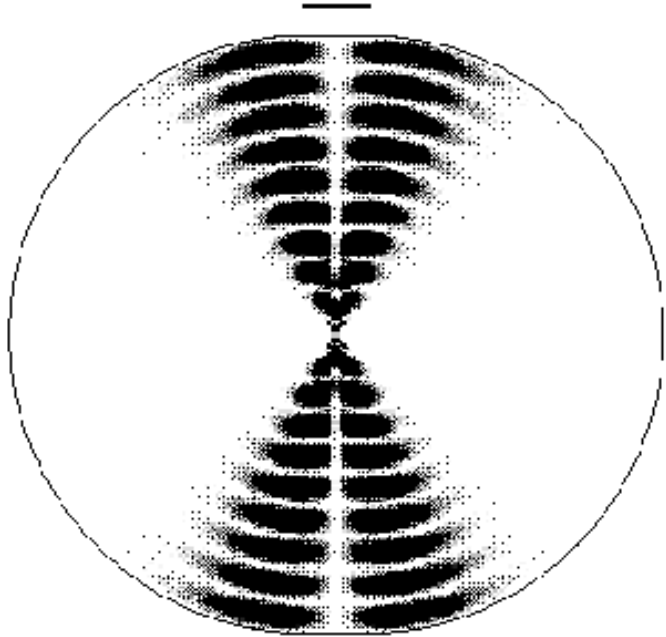,height=9.09cm,width=9cm,angle=0}}
{\vspace*{.13in}} 
\caption[The low angular momentum state $n=10$, $m=2$
for the short stadium with $\epsilon =0.1$.] 
{The low angular momentum state $n=10$, $m=2$ for the short stadium with
$\epsilon =0.05$. The absolute square of the wavefunction is calculated
from the perturbation theory. The theoretical wavefunction is inaccurate
near the center of the billiard (see Sec.\ 2.4). The line segment indicates
the length and position of the straight parts of the boundary. 
\label{2_5}} 
{\vspace{1.2 cm}}
\end{figure}

\begin{figure}[tbp] {\hspace*{2.7cm}
\psfig{figure=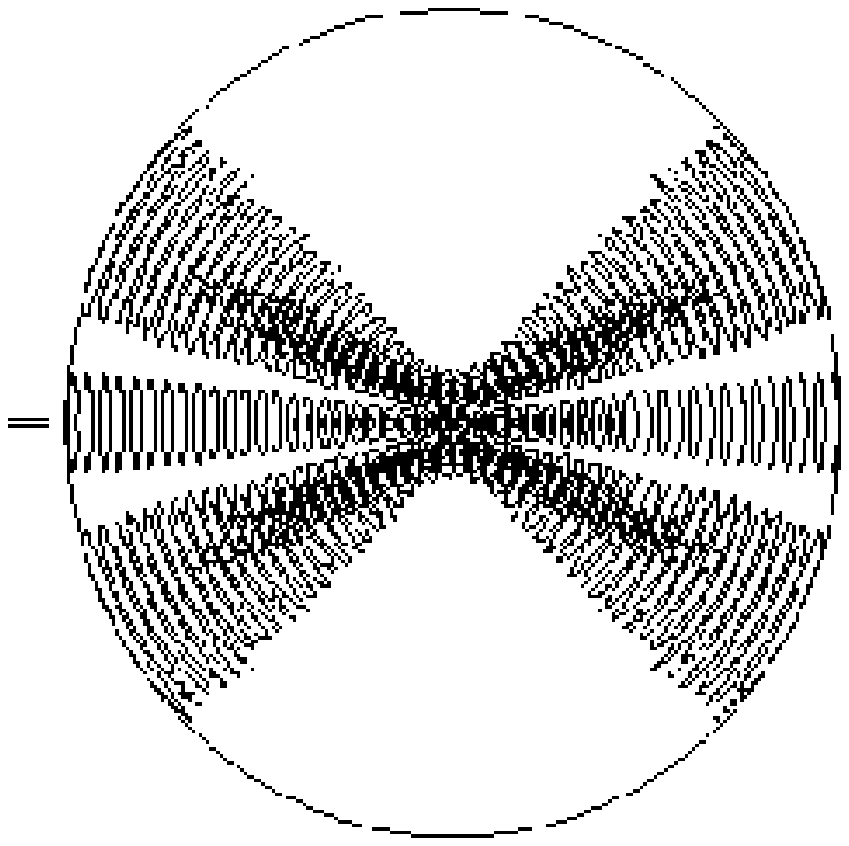,height=9.25cm,width=9cm,angle=0}}
{\vspace*{.13in}} 
\caption[Contour plot of the numerical low angular momentum state $n=25$, 
$m=3$ for the short stadium with $\epsilon =0.01$.]
{Contour plot of the numerical low angular momentum state $n=25$, $m=3$
for the short stadium with $\epsilon =0.01$. The parallel lines on the
left indicate the length and position of the straight segments of the
boundary. 
\label{2_6}}
{\vspace{1.2 cm}}
\end{figure}

Likewise, if $E_m<\max V(\theta )$ the motion will be librational. These
are the tori that lie inside the separatrix and surround the stable
periodic orbit. Their topology is changed by the perturbation. The
potential $V(\theta )$ has at least two wells between $0$ and $2\pi .$ If
the tunneling between the wells can be neglected the degenerate levels
will be given by the Bohr-Sommerfeld conditions \cite{LanLif}
\begin{equation} 
k\sqrt{\epsilon }\int_{\theta _{m-}}^{\theta _{m+}}d\theta
\sqrt{E_m-V(\theta )}=\pi \left( m+\frac 12\right) . 
\end{equation} 
Here the limits of the integration are the classical turning points. The
wavefunction inside the well is
\begin{equation} 
\psi _m(\theta )=\left[ E_m-V(\theta )\right] ^{-1/4}\sin \left(
\int_{\theta _{m-}}^\theta d\theta ^{\prime }\sqrt{E_m-V(\theta ^{\prime
})} +\frac \pi 4\right) .  \label{r12loc}
\end{equation} 
Figures \ref{2_2} (c) and \ref{2_4} show the examples of the rotational
and librational states. The two-dimensional wavefunctions for the
librational states with $m=2,3$ are shown in Figs. \ref{2_5}, \ref{2_6},
respectively. 

\subsection{General case: $(p,q)$ resonance, $k\epsilon ^{3/2}\ll 1$}

\label{gc}

In this section we quantize the orbits near an arbitrary $(p,q)$
resonance. We also lift the restriction $k\epsilon \ll 1.$ Instead, we
construct the perturbation series $\psi (\theta )=\exp \left\{ ik\left[
f_0(\theta )+bf_1(\theta )+\right. \right.$ $\left. \left. b^2f_2(\theta
)+\cdots +b^Mf_M(\theta )\right] \right\} ,$ where $b=\sqrt{\epsilon }$
and $f_i^{\prime }\sim 1.$ We call $M$ the order of the perturbation
theory. The large and small parameters $k$ and $b$ should satisfy the
conditions $kb^{M-1}\gg 1$ and $kb^{M+1}\ll 1.$ Equation (\ref {r12psi}),
for example, is the case $M=1.$ The first condition ensures the validity
of this WKB-type\ ansatz by making the phase fast varying. In other words,
the quantum system will reflect the classical structure to order $M.$ The
second condition allows to neglect the terms of order $M+1$. In practice,
the procedure becomes very tedious for $M>2$ and the results are not
interesting qualitatively. In this case one can simply solve the problem
numerically. Although we derive the expressions for up to $M=3$ we do not
go beyond $M=2$ in the examples studied in this work. 

In a perfect circle the change of angle after one mapping $\theta -\theta
^{\prime }=2\pi p/q\equiv \Theta _{pq}$ for a $(p,q)$-resonant orbit. (By
definition, the particle makes $p$ times around the billiard after $q$
mappings.) The angular momentum $l_{pq}=\left( \mathrm{sgn\ }p\right) k\cos
\left( \Theta _{pq}/2\right) .$ For the perturbed circle we make an
ansatz 
\begin{equation}
\psi (\theta )=e^{il_{pq}\theta +ik\left[ bf_1(\theta )+b^2f_2(\theta
)+\cdots \right] }.  \label{gcans}
\end{equation}
The function should also have a slowly varying prefactor but we disregard
it for now. We retain terms to order $kb^2$ in the phase of the
$T$-operator and the lowest order term in the prefactor. The phase of
$T(\theta ,\theta ^{\prime })$ and $\psi (\theta ^{\prime })$ can be
expanded in $\delta \theta ^{\prime }=\theta ^{\prime }-\theta +\Theta
_{pq}.$ We will see that $\delta \theta ^{\prime }\sim b.$ Introducing the
notation $-d=L_0^{\prime \prime }(\Theta _{pq})$ for the second derivative
of the unperturbed chord length (\ref{pcbL0}) and $W_{pq}(\theta
)=L_2(\theta ,\theta -\Theta _{pq})$ we can write the integral
\begin{eqnarray}
&&\ \int d\theta ^{\prime }T(\theta ,\theta ^{\prime })\psi (\theta ^{\prime
})\simeq \int d\theta ^{\prime }\sqrt{\frac{kd}{2\pi i}}e^{i\left[
kL_0(\Theta _{pq})-\frac{kd}2\delta \theta ^{\prime 2}+kb^2W_{pq}(\theta
)-\pi \right] }  \nonumber \\
&&\ \ \ \times e^{il_{pq}(\theta -\Theta _{pq})+ik\left[ bf_1(\theta -\Theta
_{pq})+bf_1^{\prime }(\theta -\Theta _{pq})\delta \theta ^{\prime
}+b^2f_2(\theta -\Theta _{pq})\right] }  \label{gcTpsi}
\end{eqnarray}
where we have taken into account that $l_{pq}=S_0^{\prime }(\Theta _{pq}).$
The integral can be evaluated in the $S\Phi $ approximation. The stationary
point $\delta \theta _{\mathrm{st}}^{\prime }=bf_1^{\prime }(\theta -\Theta
_{pq})/d$ $\sim b.$

After elementary manipulations equation $\psi =T\psi $ becomes 
\begin{eqnarray}
\ &&\ e^{ik\left[ bf_1(\theta )+b^2f_2(\theta )\right] }=  \nonumber
\label{gcbal} \\
&&\ \ e^{ikL_0(\Theta _{pq})-il_{pq}\Theta _{pq}+i\frac \pi 2+ikbf_1(\theta
-\Theta _{pq})+ikb^2\left[ f_2(\theta -\Theta _{pq})+W_{pq}(\theta
)+f_1^{\prime 2}/2d\right] }.  \label{gcbal}
\end{eqnarray}
We may try to proceed in the same fashion as in the $(1,2)$ resonance
case. However the assumption $\psi (\theta -\Theta _{pq})=e^{-i\omega
}\psi (\theta )$, in general, is not self-consistent. Indeed, it would
mean that $f_1^{\prime }$ has period $\Theta _{pq}$ in this approximation
and $W_{pq}(\theta )+f_1^{\prime 2}/2d=\mathrm{const.}$ But $W_{pq}(\theta
)$, in general, does not have period $\Theta _{pq},$ so the latter
equation cannot be satisfied. The correct conditions would be
\begin{eqnarray}
f_1(\theta )-f_1(\theta -\Theta _{pq}) &=&\mathrm{const,}  \label{gcdf} \\
f_2(\theta )-f_2(\theta -\Theta _{pq}) &=&W_{pq}(\theta )+f_1^{\prime
2}/2d-E_m  \label{gcf2}
\end{eqnarray}
where $E_m$ is the constant to be determined. The first equation makes
$f_1^{\prime }$ $\Theta _{pq}$-periodic. The second equation defines $f_2$
to compensate for the lack of periodicity in $W_{pq}(\theta ).$ Thus it is
essential to include $f_2$ in the wavefunction when $W_{pq}$ does not have
period $\Theta _{pq}.$

Define a $q$-average of $W_{pq}$ 
\begin{equation}
\bar V_q(\theta )=\frac 1q\sum_{r=1}^qW_{pq}(\theta +r\Theta _{pq}).
\end{equation}
$\bar V_q(\theta )$ has a period $\Theta _{1q}=2\pi /q.$ (This follows
from the $\Theta _{pq}$-periodicity and the existence of an integer $s$
such that $s\Theta _{pq}=2\pi N+\Theta _{1q}.$) Performing the $q$-average
on Eq.\ (\ref {gcf2}) we can determine
\begin{equation}
f_1^{\prime }(\theta )=\pm \sqrt{2d\left[ E_m-\bar V_q(\theta )\right] }.
\label{gcf1}
\end{equation}
Thus $f_1^{\prime }$ also has a period $2\pi /q.$ This leaves us with
equation 
\begin{equation}
f_2(\theta )-f_2(\theta -\Theta _{pq})=W_{pq}(\theta )-\bar V_q(\theta
)\equiv V_{pq}(\theta )
\end{equation}
that determines $f_2$ up to a $\Theta _{1q}$-periodic (or $q$-periodic)
function, which appears in the next order of the perturbation theory. This
equation has an explicit solution 
\begin{equation}
\tilde f_2(\theta )\equiv f_2-\bar f_{2q}=-\frac
1q\sum_{r=1}^{q-1}rV_{pq}(\theta -r\Theta _{pq})
\end{equation}
where $\bar f_{2q}$ is a $q$-average. In the future we will, as a rule,
leave out the tilde. If the perturbation is expanded in the Fourier series
$W_{pq}(\theta )=\sum_l\tilde W_le^{il\theta }$ then $\bar V_q(\theta
)=\sum_l\tilde W_{ql}e^{iql\theta }$ and $\tilde f_2(\theta
)=\sum_l^{\prime }\left( 1-e^{-il\Theta _{pq}}\right) ^{-1}\tilde
W_le^{il\theta }.$ The prime indicates that integers $l$ divisible by $q$
are not included in the sum. 

\begin{figure}[tbp]
{\hspace*{2.7cm} \psfig{figure=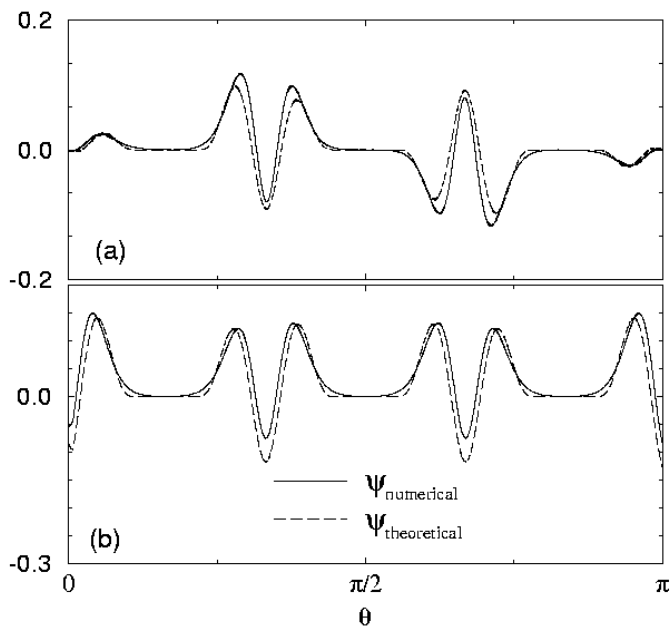,height=8.4cm,width=9cm,angle=0}}
{\vspace*{.13in}}
\caption[Numerical and theoretical surface of section wavefunctions for 
the $(1,3)$ resonance in the short stadium.]
{Numerical and theoretical surface of section wavefunctions for the
$(1,3)$ resonance in the short stadium with $k=4032$, $\epsilon=6.79
\times 10^{-4}$. The theoretical wavefunction is $\psi = \cos (kbf_1) \sin
(l\theta + kb^2 f_2)$. The fast dependence on $l \theta$ is removed by
locally averaging (a) $\psi \sin l\theta$ and (b) $\psi \cos l\theta$. 
The effective potential $\bar V_3(\theta)$ has a nominal period $2\pi/3$,
but there are two symmetric wells per period for this perturbation. The
angular momentum $l=k \cos \frac \pi 3$ is integer, so the function $\hat
\psi(\theta)$ is $2\pi/3$-periodic (see Sec.\ 2.2.4). In the current
example this function is the even combination of the second excited states
in each well. 
\label{2_7}}
{\vspace{1.2 cm}}
\end{figure}

\begin{figure}[tbp]
{\hspace*{2.7cm} 
\psfig{figure=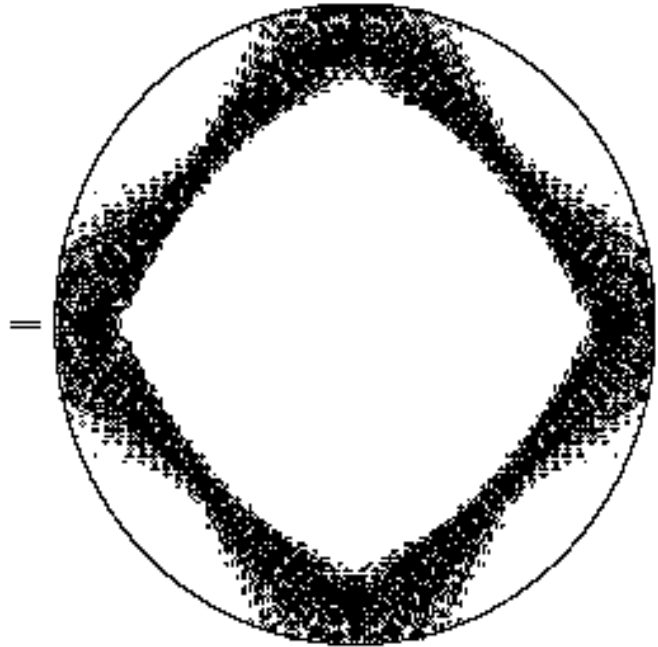,height=9.17cm,width=9cm,angle=0}}
{\vspace*{.13in}}
\caption[Contour plot of a numerically obtained state for the $(1,4)$
resonance in the short stadium.]
{Contour plot of a numerically obtained state for the $(1,4)$ resonance in
the short stadium with $k = 280.54$, $\epsilon = 0.01$. The parallel lines
on the left indicate the length and position of the straight segments
of the boundary. 
\label{2_8}}
{\vspace{1.2 cm}}
\end{figure}

Apart from function $f_2$ the wavefunction $\psi (\theta )$ is of WKB form
for the potential $\bar V_q(\theta ):$
\begin{equation}
\psi _m(\theta )=\left[ E_m-\bar V_q(\theta )\right] ^{-1/4}e^{il_{pq}\theta
\pm ik\sqrt{\epsilon }\int d\theta \sqrt{2d\left[ E_m-\bar V_q(\theta
)\right] }+ik\epsilon f_2(\theta )}.  \label{gcWKB}
\end{equation}
It is shown in Sec.\ \ref{pref} that $\psi (\theta )$ has a standard WKB
prefactor $\left( f_1^{\prime }\right) ^{-1/2}$ in the lowest order (which
has been added now by hand). This form is not valid near the turning points
and is not convenient in the classically forbidden region. Hence it would be
helpful to have a differential equation, similar to Eq.\ (\ref{r12Sch}). Let 
\begin{equation}
\psi (\theta )=e^{il_{pq}\theta +ikb^2f_2(\theta )}\hat \psi (\theta ).
\label{gchat}
\end{equation}
Then the new function satisfies the equation $\hat \psi =\hat T\hat \psi $
where the $T$-operator 
\begin{equation}
\hat T(\theta ,\theta ^{\prime })\simeq \sqrt{\frac{kd}{2\pi i}}e^{i\left[
S_0(\Theta _{pq})-\frac{kd}2\delta \theta ^{\prime 2}-\pi \right]
}e^{ikb^2\left[ W_{pq}(\theta )+f_2(\theta -\Theta _{pq})-f_2(\theta
)\right] }.
\end{equation}
The second exponential is $\exp \left[ ikb^2\bar V_q(\theta )\right] .$ So,
if $k\epsilon \left| E_m-\bar V_q(\theta )\right| \ll 1$ we can repeat the
argument in the previous section and derive a differential equation 
\begin{equation}
\hat \psi ^{\prime \prime }+2dk^2\epsilon \left[ E_m-\bar V_q(\theta
)\right] \hat \psi =0.  \label{gcSch}
\end{equation}
This equation is valid either when $k\epsilon \ll 1$ or near the turning
points. If $k\epsilon \sim 1$ and $\theta $ is far from the turning points
$\hat \psi $ can be shown to satisfy this equation approximately by the
direct substitution of the WKB form (\ref{gcWKB}). Thus the first order
wavefunction $\hat \psi $ can \emph{always} be determined from Eq.\ (\ref
{gcSch}). An example of an eigenfunction for the $(1,3)$ resonance is
shown in Fig.\ \ref{2_7}. Figure \ref{2_8} depicts a two-dimensional 
wavefunction for the $(1,4)$ resonance. 

\subsection{Quantization conditions}

\label{qc}

The eigenenergies $E_m$ are found after imposing the periodicity condition
$\psi _m(\theta +2\pi )=\psi _m(\theta ).$ For $\hat \psi (\theta )$ it
translates to
\begin{equation}
\hat \psi (\theta +2\pi )=e^{-i2\pi \delta }\hat \psi (\theta )  \label{gcbc}
\end{equation}
where $\delta $ is the fractional part of $l_{pq}\neq 0.$ (For negative
$l_{pq}$ we define $[l_{pq}]$ as the closest integer from \emph{above }and
$\delta =l_{pq}-[l_{pq}]<0.$) If $\psi (\theta )$ has a WKB form we may
again distinguish between the rotational [$E_m>\max \bar V_q(\theta )$]
and librational [$E_m<\max \bar V_q(\theta )$] cases. In the former case
the condition is
\begin{equation}
kb\int_0^{2\pi }d\theta \sqrt{2d\left[ E_m-\bar V_q(\theta )\right] }=2\pi
(m\mp \delta )
\end{equation}
where $\delta $ is the fractional part of $l_{pq}.$ The double sign reflects
the two possibilities in Eq.\ (\ref{gcf1}). In the librational case, if the
tunneling between the wells can be neglected, 
\begin{equation}
kb\int_{\theta _{m-}}^{\theta _{m+}}d\theta \sqrt{2d\left[ E_m-\bar
V_q(\theta )\right] }=\pi \left( m+\frac 12\right) .  \label{qclib}
\end{equation}
The wavefunction $\hat \psi $ within a well is given by Eq.\ (\ref{r12loc})
up to a phase. Assuming $\bar V_q(\theta )$ has $q$ wells, the phase shift
between the wells $\hat \psi (\theta +\Theta _{1q})/\hat \psi (\theta )=\exp
(-i\delta \Theta _{1q})$ [cf.\ Eq.\ (\ref{gcbc})]. Note that $-\delta \Theta
_{pq}/kb$ is the constant in Eq.\ (\ref{gcdf}).

The second quantization condition comes from Eq.\ (\ref{gcbal}). With the
help of Eqs.\ (\ref{gcdf}) and (\ref{gcf2}) we find 
\begin{equation}
\omega _m(k)\equiv kL_0(\Theta _{pq})-[l_{pq}]\Theta _{pq}+k\epsilon
E_m+\frac \pi 2=2\pi n,\quad l_{pq}\neq 0.  \label{qck}
\end{equation}
The energy levels $E_{nm}=k_{nm}^2$ are degenerate under $l_{pq}\rightarrow
-l_{pq}.$ This is a consequence of the time-reversal symmetry. The real
degenerate eigenstates are the even and odd combinations of the states (\ref
{gcWKB}) with $\pm p.$ Since $l$ is large the states with $\pm l$ do not
overlap. However the overlap with the states from other resonances can
create a transition between $l$ and $-l.$ We call it a \emph{resonance
assisted tunneling}. With this tunneling the states (\ref{gcWKB}) are no
longer the eigenstates, while their even and odd combinations are. The
degeneracy between them will be removed. We will disregard this effect.

\subsection{Third order theory: $k\epsilon ^2\ll 1\sim k\epsilon ^{3/2}$}

If $k\epsilon ^{3/2}\gtrsim 1$ one has to keep terms to order $kb^3$ in the
ansatz (\ref{gcans}) and $T$-operator. The additional terms in the phase of
Eq.\ (\ref{gcTpsi}) are 
\begin{eqnarray}
&&\ \ k\left[ -\frac 16L_0^{\prime \prime \prime }(\Theta _{pq})\delta
\theta ^{\prime 3}+b^2\frac{\partial L_2(\theta ,\theta -\Theta
_{pq})}{\partial \theta ^{\prime }}\delta \theta ^{\prime }\right. 
\nonumber \\
&&\ \ \left. +\frac b2f_1^{\prime \prime }(\theta -\Theta _{pq})\delta
\theta ^{\prime 2}+b^2f_2^{\prime }(\theta -\Theta _{pq})\delta \theta
^{\prime }+b^3f_3(\theta -\Theta _{pq})\right] .
\end{eqnarray}
Although the stationary point $\delta \theta _{\mathrm{st}}^{\prime
}=bf_1^{\prime }(\theta -\Theta _{pq})/d$ receives a $b^2$-order correction,
it does not enter explicitly in the result. The above terms should be
evaluated at $\delta \theta _{\mathrm{st}}^{\prime }$ and included in the
phase of Eq.\ (\ref{gcbal}). The balance of the $kb^3$-terms yields 
\begin{equation}
f_3(\theta )-f_3(\theta -\Theta _{pq})=-\frac{L_0^{\prime \prime \prime
}(\Theta _{pq})f_1^{\prime 3}}{6d^3}+\frac{f_1^{\prime }\left[ f_2^{\prime
}+\partial L_2/\partial \theta ^{\prime }\right] }d+\frac{f_1^{\prime
2}f_1^{\prime \prime }}{2d^2}+c  \label{totf3}
\end{equation}
where $f_i^{\prime },$ $f_i^{\prime \prime },$ $\partial L_2/\partial
\theta ^{\prime }$ are evaluated at $\theta ^{\prime }=\theta -\Theta
_{pq}$ and $c=\mathrm{const}.$ Taking the $q$-average of both sides, we
have an equation determining $\bar f_{2q},$ the $q$-periodic part of
$f_2,$
\begin{equation}
\bar f_{2q}^{\prime }=\frac{L_0^{\prime \prime \prime }(\Theta
_{pq})f_1^{\prime 2}}{6d^2}-\frac{cd}{f_1^{\prime }}+\frac{L_0^{\prime
}(\Theta _{pq})}{L_0(\Theta _{pq})}\bar V_q  \label{totf2q}
\end{equation}
where we used Eq.\ (\ref{pcbL2}). This expression must have a vanishing
angular average, since $\bar f_{2q}^{\prime }$ is the derivative of a
periodic function, which determines $c.$ Now $f_3$ can be found from Eq.\
(\ref{totf3}) up to a $q$-periodic function which is determined in the
next order of the perturbation theory. Note that if $f_1$ is
double-valued, so is $f_3.$

\subsection{Non-resonant case}

The non-resonant tori that are not affected by the nearby resonances do not
change their topology. They get perturbed proportionally to $\epsilon .$
Unlike the resonant case, $\sqrt{\epsilon }$ does not enter the classical
picture. As a consequence (see Sec.\ \ref{ClasInt}), the odd powers of $b$
drop out of the expansion (\ref{gcans}). The unperturbed state has an
integer angular momentum $l,$ and after one mapping the angle changes by a
non-resonant amount $\theta -\theta ^{\prime }=\Theta _l\equiv 2\left( 
\mathrm{sgn\ }l\right) \cos ^{-1}\left| l/k\right| .$

\begin{figure}[tbp]
{\hspace*{2.7cm} \psfig{figure=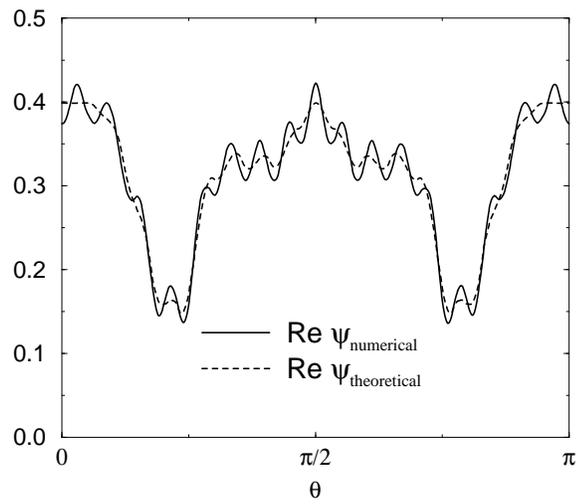,height=7.4cm,width=9cm,angle=270}}
{\vspace*{.13in}}
\caption[Numerical and theoretical surface of section wavefunctions for the
non-resonant state on the golden mean torus.]
{Numerical and theoretical surface of section wavefunctions for the
non-resonant state on the golden mean torus. The system is a smoothed
stadium with $\epsilon = 10^{-4}$, $\eta = 2 \times 10^{-2}$. For this
torus $\Theta_l = \pi (\sqrt 5 -1)$. The factor $e^{il\theta}$ has been
removed and the real part of the wavefunction is shown. 
\label{2_9}}
{\vspace{1.2 cm}}
\end{figure}

We start with an ansatz $\psi (\theta )=\exp \left[ il\theta +ik\epsilon
f_2(\theta )\right] $ and proceed as in the resonant case. Similar to Eq.\
(\ref{gcbal}) we find
\begin{equation}
\ e^{ik\epsilon f_2(\theta )}=\ e^{ikL_0(\Theta _l)-il\Theta _l+i\frac \pi
2+ik\epsilon \left[ f_2(\theta -\Theta _l)+W_l(\theta )\right] }
\end{equation}
where $W_l(\theta )=L_2(\theta ,\theta -\Theta _l).$ If we require 
\begin{equation}
f_2(\theta )-f_2(\theta -\Theta _l)=W_l(\theta )
\end{equation}
we find the quantization condition 
\begin{equation}
kL_0(\Theta _l)-l\Theta _l=2\pi \left( n-\frac 14\right) .
\end{equation}
Thus, at this order of the calculation, there is no shift in energy levels.
(Note that the above expression is the Debye approximation to the roots of
Bessel's function $J_l(k)$ for large $k$ and fixed $l/k.$) Expanding the
perturbation in the Fourier series $W_l(\theta )=\sum_r\tilde W_re^{ir\theta
}$ we find 
\begin{equation}
f_2(\theta )=\sum_{r\neq 0}\frac{\tilde W_re^{ir\theta }}{1-e^{-ir\Theta
_l}}.  \label{nrcf2}
\end{equation}
Note that $\tilde W_{r=0}=0$ since, by definition, the perturbation has a
vanishing angular average. Equation (\ref{nrcf2}) would not be valid in a
resonant case, when some denominators vanish, or if $\Theta _l$ is close to
a rational $\Theta _{pq}$. The same is true if $\tilde W_r$ does not drop
off with $r$ sufficiently fast, since $\exp \left( -ir\Theta _l\right) $ can
be arbitrary close to $1$ for some $r$. This is related to the \emph{problem
of small denominators} in classical mechanics \cite{Tab,LicLie}. If the
series does not converge, the resonant solution is needed. Figure \ref{2_9}
illustrates the wavefunction associated with the GM torus.

\section{Comments and discussion}

\subsection{Classical interpretation}

\label{ClasInt}

In Sec.\ \ref{sosm} we discussed the relation between the $T$-operator and
the surface of section map. The map $(l^{\prime },\theta ^{\prime
})\longmapsto (l,\theta )$ is given by the equations $l^{\prime
}=-\partial S(\theta ,\theta ^{\prime })/\partial \theta ^{\prime }$ and
$l=\partial S(\theta ,\theta ^{\prime })/\partial \theta .$ We have shown
that if a curve $l^{\prime }(\theta ^{\prime })$ is mapped into $l(\theta
)$ then the $T$-operator maps the wavefunction $\psi ^{\prime }(\theta
^{\prime })\sim e^{i\int l^{\prime }(\theta ^{\prime })d\theta ^{\prime
}}$ into $\psi (\theta )=T\psi ^{\prime }(\theta )\sim e^{i\int l(\theta
)d\theta }.$ The eigenstates of Bogomolny's equation $\psi =T\psi $ are
invariant under the map. Hence the curve $l=l_{\mathrm{inv}}(\theta )$
associated with an eigenstate is mapped on itself. We call this curve an
\emph{invariant loop}. Comparing with Eq.\ (\ref{gcans}) we conclude that
our method gives a perturbation expansion for this loop
\begin{equation}
l_{\mathrm{inv}}(\theta )=l_0+k\left[ bf_1^{\prime }(\theta )+b^2f_2^{\prime
}(\theta )+\cdots \right]  \label{cil}
\end{equation}
and similar in the non-resonant case. Of course, this expansion can be
derived by purely classical methods (see also Sec.\ \ref{dqoc}). If we
neglect terms of order $kb^3$ and higher, 
\begin{equation}
\tilde l_{\mathrm{inv}}=l_0\pm kb\sqrt{2d\left[ E_m-\bar V_q(\theta )\right] 
}+kb^2f_2^{\prime }(\theta )  \label{cilinv}
\end{equation}
will be an approximation to the invariant loop. Under the Poincar\'e map
it will be transformed into a new loop $l_1$. The area enclosed between
the curves $\tilde l_{\mathrm{inv}}$ and $l_1$ will be proportional to
$k\epsilon ^{3/2.}.$ If this area is smaller than $2\pi \hbar ,$ i.e.\
$k\epsilon ^{3/2.}\ll 1$, this approximation is good for the purposes of
quantum mechanics. This is true even if no classical invariant loop
exists. 

Clearly, $l_{\mathrm{inv}}(\theta )$ is an intersection of an (approximate)
invariant torus with the PSS. If we start with a point $(l,\theta )$ and
propagate it under the surface of section map its images will lie on an
invariant loop going through this point. Fig.\ \ref{2_1} shows some of the
invariant loops. There are no invariant loops in a stochastic region near
the separatrix, but if its typical area is less than $2\pi \hbar ,$ the
separatrix will be an invariant curve for a semiclassical wavepacket. The
same is true for the secondary resonances.

The three terms in Eq.\ (\ref{cilinv}) play different roles. $l_0$
specifies the unperturbed torus. If this torus is close to a resonance a
perturbation will transform its neighborhood into a resonant chain. The
resonant islands have a size $\sqrt{\epsilon }$ and are described by the
$q$-periodic function $f_1^{\prime }\propto \pm \sqrt{E_m-\bar V_q(\theta
)}.$ The orbits are labeled by $E_m,$ which is therefore an approximate
constant of the classical motion. ($E_m=\max \bar V_q$ for a separatrix.)
The islands, in general, do not lie along a horizontal line. Instead, they
form a wave of size $\epsilon .$ This wave is given by the function
$f_2^{\prime },$ more precisely, by the non-$q$-periodic part $\tilde
f_2^{\prime }$ determined in the second order perturbation theory. The
$q$-average $\bar f_{2q}^{\prime }$ distorts the shape of the islands but
does not shift them with respect to each other. In principle, higher order
corrections can be found. In the non-resonant case $f_1^{\prime }=0,$ and
$f_2^{\prime }$ describes the distortion of the torus, which is of order
$\epsilon .$ In the special case of $(1,2)$ resonance the islands are not
shifted by symmetry, so $f_2^{\prime }=0.$

\begin{figure}[tbp]
{\hspace*{2.7cm} 
\psfig{figure=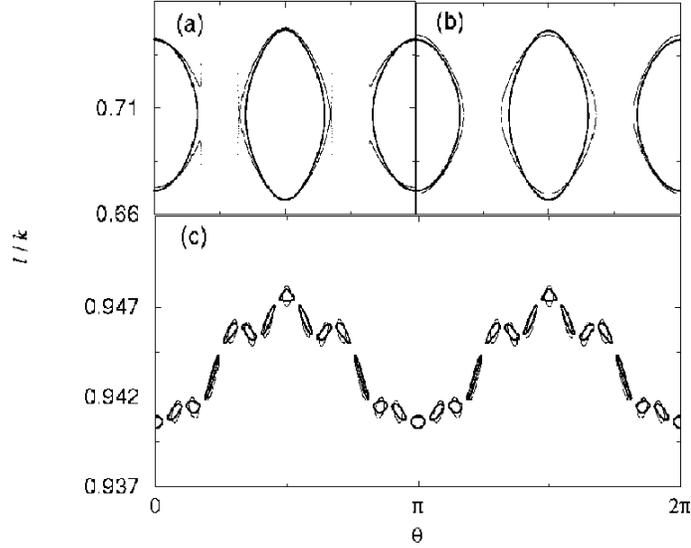,height=7.27cm,width=9cm,angle=0}}
{\vspace*{.13in}}
\caption[Exact and approximate invariant loops near the $(1,4)$ and
$(1,28)$ resonances in the smoothed stadium.]
{Exact and approximate invariant loops near the (a),(b) $(1,4)$ and (c)
$(1,28)$ resonances in the smoothed stadium. The exact loops (thick line)
are obtained by the propagation of the classical map. The approximate
loops (thin line) are $l_{\mathrm{inv}}(\theta)$ that includes terms to
order $b^3$ in (a) and $b^2$ in (b),(c). The parameters are $\epsilon =
0.05$, $\eta = 0.7$ in (a),(b), and $\epsilon = 0.01$, $\eta = 0.19$ in
(c). 
\label{2_10}}
{\vspace{1.2 cm}}
\end{figure}

\begin{figure}[tbp]
{\hspace*{2.7cm} 
\psfig{figure=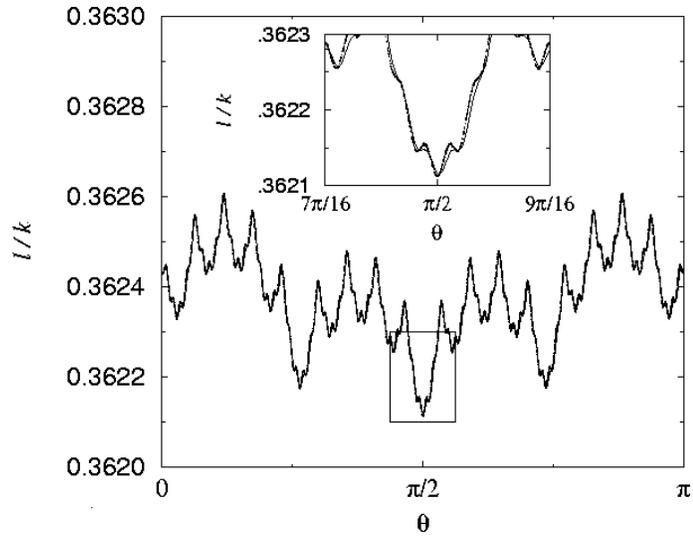,height=7.06cm,width=9cm,angle=0}}
{\vspace*{.13in}}
\caption[Exact and approximate golden mean torus for the smoothed stadium.]
{Exact and approximate golden mean torus for the smoothed stadium
(parameters as in Fig.\ 2.9). The numerical classical map (dots, 20,000
iterations) is approximated by $\cos \frac 1 2 \Theta_l + \epsilon
f_2^\prime (\theta)$ (solid line). The inset enlarges a portion of the
figure to demonstrate the degree of precision of the approximation. 
Presumably, this is the last torus to disappear as $\eta$ is decreased. 
\label{2_11}}
{\vspace{1.2 cm}}
\end{figure}

We calculated the invariant loops for the smoothed stadium from Eq.\
(\ref{cilinv}). For the $(1,2)$ resonance they are shown as discrete
symbols in Fig.\ \ref{2_2} (b). The orbits follow these curves closely and
the deviation must be due to the higher order corrections. We can also
formally calculate the loops for the short stadium [Fig.\ \ref{2_2} (a)]
although the orbits stay close to the loops only for a short time.
Moreover, the expansion (\ref{cil}) breaks down in the higher orders,
because $\bar V_q(\theta )$ has a singular second derivative.
Nevertheless, there are quantum states localized near these loops [Fig.\
\ref{2_2} (c)].  Fig.\ \ref{2_10} compares $l_{\mathrm{inv}}(\theta )$
with the orbits near the $(1,4)$ and $(1,28)$ resonances in the smoothed
stadium. Figure \ref{2_11} shows the same for the non-resonant GM torus. 

\subsection{Resonant or non-resonant?}

\label{rnr}

When $q$ becomes large the resonant islands become small (e.g. resonance
$(15,32)$ in Fig.\ \ref{2_1}). When the area of the islands becomes
smaller then $2\pi \hbar $ the resonant structure is disregarded by the
quantum state. Thus, on the bigger scale, the $(p,q)$-chain with large $q$
looks similar to a perturbed non-resonant torus. This agrees with the
intuitive, as well as the number-theoretical \cite{Khi} notion that the
irrational numbers can be approximated by rationals with large
denominators with the increasing precision. 

Clearly, the resonant solution is needed if the librational state $E_m<\max
\bar V_q$ exists. This requires [Eq.\ (\ref{qclib})] 
\begin{equation}
\bar V_q>\frac{q^2}{k^2\epsilon d}.  \label{rnrVq}
\end{equation}
We assumed that the perturbation $W_{pq}(\theta ),$ as well as $\bar V_q,$
has vanishing angular average. As $q$ increases, $\bar V_q,$ the
$q$-average of $W_{pq},$ becomes closer to the angular average and drops
off. Another way to say it, $\bar V_q$ is of order of the $q$th Fourier
coefficient $\tilde W_q$ of $W_{pq}(\theta )=\sum_l\tilde W_le^{il\theta
}.$ But $\tilde W_l$ decreases with $l,$ usually exponentially for
analytic perturbations or as $l^{-s}$ if $W_{pq}(\theta )$ has $s-2$
continuous derivatives. If $\bar V_q\ll q^2/k^2\epsilon d$ then
$kbf_1^{\prime }\approx \pm k\sqrt{2\epsilon dE_m}$ shifts the constant
part of the angular momentum so the nontrivial expansion begins with
$k\epsilon f_2$ as in the non-resonant case. 

Condition (\ref{rnrVq}) also means that the area of a resonant island
$k\sqrt{\epsilon d\bar V_q}/q>1$ as we implied above. Note that when
$k\sqrt{\epsilon }\ll 1$ all phase space of the system can be treated with
the non-resonant perturbation theory in agreement with our earlier
estimates. 

\begin{sloppypar}
The area of \emph{all} resonances in the system can be estimated as
$\sum_{p,q}k\sqrt{\epsilon d\bar V_q}<k\sqrt{\epsilon d}\sum_qq\sqrt{\bar
V_q}.$ So if $\bar V_q$ drops off faster than $q^{-4}$ (i.e.\ the
perturbation has three continuous derivatives) the total affected area
will vanish as $\epsilon \rightarrow 0$, as in the KAM theorem 
\cite{LicLie}. 
\end{sloppypar}

The perturbation $\Delta R(\theta )=\left| \sin \theta \right| -2/\pi $
introduced in Sec.\ \ref{pcb} has a discontinuous derivative, so $\bar
V_q\sim q^{-2}.$ In this case the KAM theorem breaks down completely
\cite{Bun}. We expect our leading order solution, which does not
include the derivatives, to be valid. Our theory contains the singular
second derivatives of $\Delta R$ in the $k\epsilon ^2$-order. We assumed
that $k\epsilon ^2$ is small. However, multiplied by an infinite $\bar
V_q^{\prime \prime }$ it may bring finite corrections. We call them the
\emph{diffraction corrections}. Figure \ref{2_4} compares the theoretical
and numerically exact wavefunctions in this case. 

It is easy to show that if $\Delta R(\theta )$ has only one Fourier
harmonic, say $\Delta R(\theta )=\cos \left( l\theta \right) $ for some
integer $l,$ then $\bar V_q\neq 0$ only if $q$ is a divisor of $l.$ (See
also Sec.\ \ref{fep}.) For example, $l=1$ and $l=2$ are to first
approximation a shifted circle and an integrable ellipse, respectively. Thus 
$\Delta R$ can be transformed to a perturbation about an integrable system
with perturbation parameter $\epsilon ^2$ rather than $\epsilon .$ In the
case of ellipse $\bar V_2\neq 0$ is the bouncing ball state which does not
exist in a perfect circle. The states with the larger angular momentum are
similar to the states in the circle.

\subsection{Close to a large resonance}

Another important question is whether a given unperturbed torus $l$
(resonant or non-resonant) will be affected by a neighboring resonance
$l_{pq}$ once the perturbation is on. This is the case if the distance
$\left| l-l_{pq}\right| <k\sqrt{\epsilon d\bar V_q},$ the size of the
resonance. We can estimate $\left| l-l_{pq}\right| \sim \left| \partial
l/\partial \Delta \theta \right| \left| \Theta _l-\Theta _{pq}\right| $
where $\Delta \theta $ is the change of angle after one mapping. Note that
$\left| \partial l/\partial \Delta \theta \right| =kd$ and $\left| \Theta
_l-\Theta _{pq}\right| >q^{-2},$ which is a generic precision when a
number is approximated by the rationals \cite{Khi}. Thus if $\bar
V_q>d/\epsilon q^4 $ the torus $l$ may be affected by a $(p,q)$ resonance.
If $\bar V_q$ drops off faster than $q^{-4}$ only finite number of
resonances may affect the torus. 

If the torus \emph{is} affected by a larger resonance, two cases are
possible. If $\left| l-l_{pq}\right| $ is sufficiently small, the topology
of torus $l$ will be changed, i.e.\ the torus will disappear and a new
librational torus will appear inside the separatrix. In this case,
obviously, torus $l$ cannot be used as a starting point of the
perturbation theory; instead, the expansion should be done near the
resonant torus $l_{pq.}$ On the other hand, if $\left| l-l_{pq}\right| $
is large enough, so that the topology is not changed, torus $l$ still
exists outside the separatrix, although it may be strongly distorted. In
this case torus $l$ may be considered either as a rotational trajectory
for the resonance $(p,q)$ or the perturbation expansion can be done
directly near the torus $l.$ In fact, the larger $\left| l-l_{pq}\right|
$, the better is the expansion near $l$ and the worse is the expansion
near $l_{pq.}$ Note that $l$ can be another resonance $(p^{\prime
},q^{\prime })$ with $q^{\prime }\gg q.$ Consider, for example, the
resonance $(15,32)$ in Fig.\ \ref{2_1}. It is a secondary resonance for
the large $(1,2)$ resonance. The perturbation expansion near the $(1,2)$
resonance yields only $f_2$ part of the $(15,32)$ resonance, i.e.\ it
cannot describe the loops. However, for small $\epsilon $ the loops are
small and will not be reflected in the quantum state anyway. 

Suppose $l$ is a non-resonant torus. The above discussion suggests that if
$ l $ lies outside of the separatrix of a large resonance $l_{pq},$ we
should be able to describe this torus by both the non-resonant
perturbation theory for torus $l$ and the resonant perturbation theory for
torus $l_{pq}$ to some approximation. For a rotational state far enough
from the resonance we can assume $\left| \bar V_q\right| \ll E_m$ and
expand $bf_1^{\prime }\simeq \pm \sqrt{2\epsilon dE_m}\left( 1-\bar
V_q/2E_m\right) .$ We are going to show that
\begin{equation}
l_{pq}\pm k\sqrt{2\epsilon dE_m}\left( 1-\bar V_q/2E_m\right)
+kb^2f_2^{\prime }=l+kb^2f_{2l}^{\prime }  \label{ctlreq}
\end{equation}
where $f_{2l}^{\prime }$ is a non-resonant function. The constant part of
$f_1^{\prime }$ shifts the unperturbed angular momentum, so $l_{pq}\pm
k\sqrt{2\epsilon dE_m}\equiv l_{pq}+\delta l\simeq l.$ According to
Eq.\ (\ref{nrcf2})
\begin{equation}
f_{2l}^{\prime }(\theta )=\sum_{r\neq 0}ir\frac{L_0\left( \Theta _l\right)
\left( 1+e^{-ir\Theta _l}\right) \Delta \tilde R_re^{ir\theta }}{2\left(
1-e^{-ir\Theta _l}\right) }  \label{ctlrf2}
\end{equation}
where $\Delta \tilde R_r$ is a Fourier component of $\Delta R(\theta )$. We
can expand Eq.\ (\ref{ctlrf2}) in $\delta \Theta =\Theta _l-\Theta
_{pq}\propto \sqrt{\epsilon }.$ The denominator becomes small when $r$ is a
multiple of $q.$ We separate those terms. Then 
\begin{equation}
f_{2l}^{\prime }(\theta )\approx \tilde f_2^{\prime }(\theta
)+\sum_r\frac{L_0\left( \Theta _{pq}\right) +L_0^{\prime }\delta \Theta
}{\delta \Theta }\Delta \tilde R_{rq}e^{irq\theta } \label{ctlrsum}
\end{equation}
where $\tilde f_2^{\prime }(\theta )$ is the part of $f_2^{\prime }$ that
vanishes under the $q$-average. The sum is $\left( \delta \Theta
^{-1}+L_0^{^{\prime }}/L_0\right) \bar V_q.$ With $\delta \Theta
^{-1}=-kd/\delta l+L_0^{\prime \prime \prime }/2L_0^{\prime \prime }+O\left(
\delta l\right) $ the leading term is $\mp \bar V_q\sqrt{d/2\epsilon E_m},$
which takes care of the respective term in Eq.\ (\ref{ctlreq}), when
multiplied by $kb^2.$ The next order terms give $\bar f_{2q}^{\prime },$ the 
$q$-average of $f_2^{\prime },$ since, according to Eq.\ (\ref{totf2q}), 
\begin{equation}
\bar f_{2q}^{\prime }=\left[ \frac{L_0^{\prime }(\Theta _{pq})}{L_0(\Theta
_{pq})}+\frac{L_0^{\prime \prime \prime }(\Theta _{pq})}{2L_0^{\prime \prime
}(\Theta _{pq})}\right] \bar V_q.
\end{equation}
We thus confirmed that a rotational torus near a large resonance can be
chosen as a starting point for the perturbation series in the semiclassical
approximation.

\subsection{Localization in angular momentum}

\label{liam}

As we explained in Sec.\ \ref{ClasInt}, the derivative of the phase of the
eigenfunction $\psi (\theta )$ approximates the invariant loop
$l_{\mathrm{inv}}(\theta )$ in the PSS phase space [see Eq.\ (\ref{cil})].
This suggests that the angular momentum spectrum of a given eigenstate is
concentrated within the range of $l$ covered by its invariant loop.
Formally, we define the wavefunction in the angular momentum
representation as
\begin{equation}
\psi _l=\int d\theta \left[ E_m-\bar V_q(\theta )\right]
^{-1/4}e^{i\int^\theta d\theta ^{\prime }l_{\mathrm{inv}}(\theta ^{\prime
})-l\theta }.
\end{equation}
If $k\sqrt{\epsilon }$ is large, the $S\Phi $ can be employed. The
stationary phase condition can be satisfied if $l_{\mathrm{inv}}(\theta
)=l$ for some $\theta .$ In particular, for a librational state near the
$(1,2)$ resonance the spectrum range is $0\leq \left| l\right|
<k\sqrt{\epsilon }\sqrt{\max \left( E_m-V\right) }.$ It has much overlap
with zero angular momentum. In a higher resonance the range is centered
near the resonant angular momentum $l_{pq}$ and spreads by
$k\sqrt{\epsilon }\sqrt{2d\max \left( E_m-\bar V_q\right) }$ in both
directions. In a non-resonant state the spectrum is highly localized near
the unperturbed angular momentum, the spread being of order $k\epsilon .$
The numerical examples are shown in Fig.\ \ref{2_4}. 

Outside of the range of $l_{\mathrm{inv}}(\theta ),$ the angular momentum
components $\left| \psi _l\right| ^2$ decay exponentially for smooth $\bar
V_q(\theta )$. In this case the range of $l_{\mathrm{inv}}(\theta )$ defines
the \emph{localization length}. In the stadium case $\Delta R(\theta )\sim
\left| \sin \theta \right| $ (Sec.\ \ref{pcb}), on the other hand, they decay
as a power law, that makes it necessary to define the localization
length more precisely \cite{CasPro}.

Note that $\psi _l$ can be calculated by diagonalization of $T$-operator
in the angular momentum representation. Consider the matrix elements
$T_{ll^{\prime }}=(2\pi )^{-1}$ $\int d\theta d\theta ^{\prime }T(\theta
,\theta ^{\prime })e^{i\left( l^{\prime }\theta ^{\prime }-l\theta \right)
}. $ The stationary phase conditions require $l=\partial S/\partial \theta
$ and $l^{\prime }=-\partial S/\partial \theta ^{\prime }$ where the
action $S(\theta ,\theta ^{\prime })$ is given by Eq.\ (\ref{pcbS}). It
follows that for $\left| l-l^{\prime }\right| >k\epsilon $ the matrix
elements are small, i.e.\ $T_{ll^{\prime }}$ is a band diagonal matrix. 

\subsection{Numerical computations}

\label{nc}

To verify our theory we conducted several numerical checks. Essentially, we
compared the wavefunctions and energy levels obtained in Sec.\ \ref{fott}
(``theoretical'' solution) with the numerical solution of Bogomolny's
equation $\psi =T\psi $, which we presume to be semiclassically exact. The
problem can be separated into two parts. First we solve a more general
equation 
\begin{equation}
e^{i\omega _m(k)}\psi =T(k)\psi  \label{ncepsi}
\end{equation}
assuming $k$ continuous (it follows from the semiclassical unitarity of
$T$). Then we can find the allowed $k$'s from the quantization condition
$\omega _m(k)=2\pi n.$ Unless we are interested in the energy
quantization, we may consider only the general eigenvalue problem
(\ref{ncepsi}) with fixed $k$. This is done in some numerical examples to
study the approximation of the wavefunctions. 

Equation (\ref{ncepsi}) can be solved both theoretically and numerically.
The theory of Sec.\ \ref{fott} remains unchanged but one should not equate
$\omega _m(k)$ to $2\pi n$ in Eq.\ (\ref{qck}). To solve the equation
numerically \cite{FisGrePra} we choose an ansatz $\psi (\theta )$ and
evaluate a discrete function $F(n)=\left( T^n\psi \right) \left( \theta
_0\right) $ at some $\theta =\theta _0.$ We then find its Fourier image
$\bar F\left( \omega \right) ,$ which consists of the peaks near the
eigenphases $\omega _m(k)$ (see below). We choose $\omega _m$ of the
highest peak. The eigenstate $\psi _m(\theta )=\left\langle \left(
e^{-i\omega _m}T\right) ^n\psi (\theta )\right\rangle _{n\rightarrow
\infty }$ is the average over $n.$

To justify this method expand the ansatz $\psi (\theta )=\sum_mc_m\psi
_m(\theta )$ in the eigenstates of the $T$-operator. Then $F(n)\approx
\sum_mc_me^{i\omega _mn}\psi _m(\theta _0).$ The Fourier transform $\bar
F\left( \omega \right) =\sum_nF(n)e^{-i\omega n}\approx (2\pi
)^{-1}\sum_mc_m\delta \left( \omega -\omega _m\right) \psi _m(\theta _0)$
has peaks at the eigennumbers $\omega _m$. If the ansatz is close to some
eigenstate $\psi _{m_0}(\theta ),$ the respective peak will have the largest
weight (for a generic $\theta _0$). In the combination $\left( e^{-i\omega
_{m_0}}T\right) ^n\psi (\theta )=c_{m_0}\psi _{m_0}(\theta )+\sum_m^{\prime
}c_me^{i\left( \omega _m-\omega _{m_0}\right) n}\psi _m(\theta )$ the sum
over $m\neq m_0$ disappears after the $n$-average is taken.

Figure \ref{2_4} compares the theoretical (``WKB'') and numerical
wavefunctions for the short stadium (Sec.\ \ref{pcb}). The librational and
rotational states near the $(1,2)$ resonance are shown. The considerable
difference in the bound state is due to the simplistic WKB\ solution of
Eq.\ (\ref{r12Sch}) with the condition that $\psi (\theta )$ vanishes at
the turning point. This one-dimensional equation can be solved, of course,
numerically, but even the simple approach renders the main features of the
state. Figure \ref{2_2} (c) contains the Husimi plots of the numerical
wavefunctions. In addition to the librational and rotational states, we
show the state near the separatrix. This state has $E_m$ just greater than
the maximum $V(\theta ),$ which means that the wavefunction has an
excessive weight near an unstable periodic orbit (due to the WKB
prefactor). This explains the observed ``scars'' of unstable orbits
\cite{Hel}. Note that scars do not appear in the previously developed
perturbation theories that quantize the neighborhood of a stable orbit
(minimum $V$). It is necessary to know the potential $V$ near its maximum,
not minimum. Figure \ref{2_7} shows the wavefunctions for the $(1,3)$
resonance. In this case we remove the fast dependence on $l\theta $ by
local averaging. The wavefunctions for the non-resonant GM torus in a
smoothed stadium are shown in Fig.\ \ref{2_9}. The examples of the
two-dimensional wave-functions are given in Figs.\ \ref{2_5}, \ref{2_6},
and \ref{2_8}. 

The numerical and theoretical eigenphases modulo $2\pi $, as defined by
Eq.\ (\ref{ncepsi}), are compared in Table \ref{Table1}. We consider the
states with different $m$, but all belonging to the lowest $(1,2)$
resonance in the short stadium billiard. The wavenumber $k$ is fixed. One
can see that the errors in $\omega _{\mathrm{theor}}$ are small compared
with the average spacing, $\omega _{m+1}-\omega _m,$ within the resonance.
The spacing is of order [cf.\ Eqs.\ (\ref{r12om}) and (\ref{qck})]
\begin{equation}
k\epsilon \left( E_{m+1}-E_m\right) \simeq \sqrt{\epsilon }.
\end{equation}

The energy levels $\left( k_{nm}\right) ^2$ are given by the condition
$\omega _m(k)=2\pi n.$ Table \ref{Table2} shows numerical and theoretical
wavenumbers for fixed $m$. The error is much smaller then the differences
\begin{equation}
k_{n+1,m}-k_{n,m}\simeq \pi \quad \mathrm{and}\quad k_{n,m+1}-k_{n,m}\simeq 
\sqrt{\epsilon }
\end{equation}
given by Eq.\ (\ref{r12qc}). Note, however, that our theory gives the energy
levels grouped by the resonances they belong to. In this example we have
found the levels close to the lowest resonance. There are levels coming from
other resonances (i.e the states with higher angular momentum) in the same
energy range. The mean spacing of \emph{all }levels in the billiard in terms
of $k$ is $2/k$, which is of the order of the errors committed.
Nevertheless, the results are still useful since the interaction between the
states belonging to different resonances is small.
\TeXButton{Table1}
{\begin{table}[tbp]
\vspace{0.5cm}
\centering
\begin{tabular}{cccc}
$m$ & $E_m+4/\pi $ & $\omega_{\rm{num}}$ & $\omega_{\rm{theor}}$\\ 
\hline 
$1$ & $0.2083$ & $5.6199$ & $5.6310$ \\ 
$3$ & $0.4321$ & $5.7701$ & $5.7680$ \\ 
$5$ & $0.6053$ & $5.8726$ & $5.8741$ \\ 
$7$ & $0.7544$ & $5.9661$ & $5.9654$ \\ 
$9$ & $0.8878$ & $6.0465$ & $6.0471$ \\ 
$11$ & $1.0097$ & $6.1222$ & $6.1218$ \\ 
$13$ & $1.1223$ & $6.1904$ & $6.1908$ \\ 
$15$ & $1.1272$ & $6.2553$ & $6.2550$ \\ 
$17$ & $1.3254$ & $0.0317$ & $0.0319$ \\ 
$19$ & $1.4175$ & $0.0885$ & $0.0884$ \\ 
$21$ & $1.5040$ & $0.1411$ & $0.1413$ \\ 
$23$ & $1.5852$ & $0.1912$ & $0.1911$ \\ 
$25$ & $1.6614$ & $0.2376$ & $0.2378$ \\ 
$27$ & $1.7327$ & $0.2814$ & $0.2814$ \\ 
$29$ & $1.7989$ & $0.3217$ & $0.3220$ \\ 
$31$ & $1.8599$ & $0.3592$ & $0.3593$ \\ 
$33$ & $1.9152$ & $0.3924$ & $0.3932$ \\ 
$35$ & $1.9638$ & $0.4243$ & $0.4229$ \\ 
$37$ & $2.0010$ & $0.4461$ & $0.4458$\end{tabular}
\vspace*{0.5cm}
\caption[Numerical ($\omega_{\rm{num}}$) and theoretical
($\omega_{\rm{theor}}$) eigenphases modulo $2\pi$, compared for states
with different $m$, but all belonging to the same low angular momentum
resonance.]
{Numerical ($\omega_{\rm{num}}$) and theoretical ($\omega_{\rm{theor}}$)
eigenphases modulo $2\pi$, compared for states with different $m$, but all
belonging to the same low angular momentum resonance.  $\Delta R$
corresponds to the stadium billiard. WKB ``energy'' parameter
$E_m-V_{\min}$ is also given, $k=1000$ and $\epsilon =6.1\times 10^{-4}$
are fixed. 
\label{Table1}}
\end{table}}
\TeXButton{Table2}
{\begin{table}[tbp]
\vspace{0.5cm}
\centering
\begin{tabular}{cccc}
$n$ & $E_m+4/\pi $ & $k_{\rm{num}}$& $k_{\rm{theor}}$ \\ 
\hline 
$319$ & $1.0108$ & $998.3207$ & $998.3213$ \\ 
$318$ & $1.0128$ & $995.1784$ & $995.1789$ \\ 
$317$ & $1.0148$ & $992.0358$ & $992.0364$ \\ 
$316$ & $1.0169$ & $988.8934$ & $988.8940$ \\ 
$315$ & $1.0190$ & $985.7509$ & $985.7515$ \\ 
$314$ & $1.0210$ & $982.6085$ & $982.6090$ \\ 
$313$ & $1.0231$ & $979.4660$ & $979.4666$ \\ 
$312$ & $1.0252$ & $976.3235$ & $976.3241$ \\ 
$311$ & $1.0273$ & $973.1810$ & $973.1816$ \\ 
$310$ & $1.0294$ & $970.0387$ & $970.0392$\end{tabular}
\vspace{0.5cm}
\caption[Energies $k$ with different quantum numbers $n,$ but same $m=11$,
computed numerically and found solving Eq.\ (\ref{qck}).]
{Energies $k$ with different quantum numbers $n,$ but same $m=11$,
computed numerically and found solving Eq.\ (\ref{qck}). $\epsilon$ as in 
Table 2.1.} \label{Table2}
\end{table}}

\subsection{Possible experiments}

\label{pe}

A number of experimental techniques could, in principle, be used to verify
the theoretical results (see also Sec.\ \ref{pe2}). For example, using a
scanning tunnel microscope (STM) individual iron atoms can be positioned on
a copper surface to form a boundary of a two-dimensional domain called a
quantum corral \cite{CroLutEig}. The two-dimensional electron gas on the
surface will be partially confined within the corral, and the spatial images 
of the electron density standing waves can be produced with STM 
\cite{HelCroLut}.

Another possible system would be a shallow container of liquid which is
vibrated to produce standing surface waves \cite{KudAbrGol,AgaAlt}. The
dissipation may limit the selection of eigenstates that can be observed.

It is also feasible to utilize the analogy between a wavefunction and an
electromagnetic wave in a resonator. Both the wavefunction in a billiard and
the electric field in a cavity satisfy the Helmholtz equation with the
Dirichlet conditions. A technique was developed \cite{GokWuBri} that allows
to measure the spatial distribution of the magnitude of electric field in a
microwave cavity.

\section{Two-dimensional wavefunctions}

\label{tdw}

Although the surface of section wavefunction $\psi (\theta )$ together with
the quantization conditions contain most information about the state, it
might be occasionally necessary to reconstruct the actual two-dimensional
wavefunction $\Psi \left( r,\theta \right) $ in the billiard. The procedure
is straight-forward and involves evaluation of the integral (\ref{PsiGpsi})
by the $S\Phi .$

\begin{figure}[tbp]
{\hspace*{2.7cm} \psfig{figure=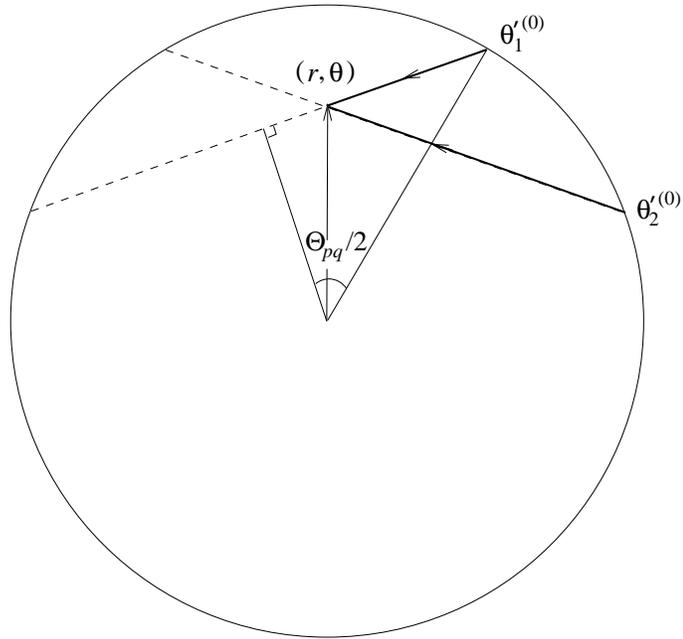,height=8.47cm,width=9cm,angle=0}}
{\vspace*{.13in}}
\caption[Two orbits with angular momentum $l_{pq}$ arriving at point $(r,
\theta)$ in the circular billiard.]
{Two orbits with angular momentum $l_{pq} = (\mathrm{sgn}) k \cos
(\Theta_{pq} / 2)$ in the circular billiard arriving at point $(r,
\theta)$ from points $\theta_{1,2}^{\prime (0)}$ on the boundary. 
\label{2_12}}
{\vspace{1.2 cm}}
\end{figure}

We start with the PSS wavefunction (or one of its branches) 
\begin{equation}
\psi (\theta )=\frac 1{\sqrt{f_1^{\prime }\left( \theta \right)
}}e^{il_{pq}\theta +ik\left[ bf_1(\theta )+b^2f_2(\theta )\right] }
\end{equation}
where $b=\sqrt{\epsilon }.$ Without worrying about normalization, and thus
leaving out constant factors, we write the Green function (\ref{Gmix}) 
\begin{equation}
\tilde G(\mathbf{r},\theta ^{\prime })=\left| \frac{\partial
^2L(\mathbf{r},\theta ^{\prime })}{\partial r_{\perp }\partial \theta
^{\prime }}\right| ^{1/2}e^{ikL(\mathbf{r},\theta ^{\prime })} 
\label{tdwGtil} 
\end{equation}
where $L(\mathbf{r},\theta ^{\prime })$ is the length of a direct orbit
between point $\theta ^{\prime }$ on the boundary and point
$\mathbf{r=}\left( r,\theta \right) $ inside the billiard. In the
unperturbed billiard there are two such orbits of angular momentum
$l_{pq}$ for any given $\mathbf{r}$ (Fig.\ \ref{2_12}). They start at the
points
\begin{equation}
\theta _{1,2}^{\prime (0)}=\theta -\frac{\Theta _{pq}}2\pm \cos ^{-1}\left( 
\frac{l_{pq}}{kr}\right)
\end{equation}
on the boundary. These points will satisfy the lowest order stationary phase
condition in the integral $\int d\theta ^{\prime }\tilde G(\mathbf{r};\theta
^{\prime })\psi (\theta ^{\prime })$. We therefore can expand the phase of
the integrand in $\delta \theta ^{\prime }=\theta ^{\prime }-\theta
_{1,2}^{\prime (0)}\sim b.$ We keep terms up to $b^2.$ In the prefactor we
keep only the lowest order. It can be shown that at these points 
\begin{equation}
\frac{\partial ^2L}{\partial r_{\perp }\partial \theta ^{\prime
}}=-\frac{\gamma (1)}{L_{1,2}^{(0)}}
\end{equation}
where 
\begin{equation}
L_{1,2}^{(0)}=L^{(0)}\left( \mathbf{r},\theta _{1,2}^{\prime (0)}\right)
=\gamma (1)\mp \gamma (r)
\end{equation}
is the length of the orbits for an unperturbed circle and 
\begin{equation}
\gamma (r)=\sqrt{r^2-\left( l_{pq}/k\right) ^2}.
\end{equation}
Integrating over $\delta \theta ^{\prime }$ near the two stationary points
we find the two-dimensional wavefunction (up to a constant factor) 
\begin{eqnarray}
\ &&\Psi \left( \mathbf{r}\right) =  \nonumber \\
\ &&\sum_{1,2}\frac{\psi \left( \theta _{1,2}^{\prime (0)}\right)
}{\sqrt{k\gamma (r)}}e^{ikL_{1,2}^{(0)}+ikb^2\left[ \gamma (1)\Delta
R\left( \theta _{1,2}^{\prime (0)}\right) \mp \frac{f_1^{\prime 2}\left(
\theta _{1,2}^{\prime (0)}\right) L_{1,2}^{(0)}}{2\gamma (1)\gamma
(r)}\right] \pm i\frac \pi 4},\quad r>\frac{l_{pq}}k, \label{tdwPsi} \\
\ &&\Psi \left( \mathbf{r}\right) \approx 0,\quad r<\frac{l_{pq}}k.
\end{eqnarray}

Note that the theory breaks down when $r\rightarrow \left( l_{pq}/k\right) .$
This is easy to understand in the case of a perfect circle. Indeed, $r=l/k$
is a caustic for the trajectories with angular momentum $l.$ The
semiclassical approximation normally fails near caustics. There are no
classical trajectories in the region $r<l/k,$ which means that the
wavefunction should be exponentially small there. This is the case for the
Bessel function $J_l(kr)$ for $r<l/k$ and $l\gg 1$ \cite{AbrSte}.

The explicit form of wavefunction (\ref{tdwPsi}) is rather cumbersome in the
general case. It simplifies significantly for the $l=0$ resonance. In this
case $\theta _1^{\prime (0)}=\theta $ and $\theta _2^{\prime (0)}=\theta
-\pi $ and the wavefunction becomes 
\begin{eqnarray}
\ &&\Psi \left( \mathbf{r}\right) =\frac{\cos \Gamma }{\sqrt{kr}}\frac{\exp
ikbf_1\left( \theta \right) }{\sqrt{f_1^{\prime }\left( \theta \right) }},
\label{tdwl0} \\
\Gamma &\equiv &kr+kb^2\left[ \frac{f_1^{\prime 2}}{2r}-\frac{\Delta R\left(
\theta \right) -\Delta R\left( \theta -\pi \right) }2\right] -\pi \left(
\frac m2+\frac 14\right) .
\end{eqnarray}
We assumed that $\omega _m=\pi m$ in Eq.\ (\ref{r12qc}) and dropped all
constant factors. The non-$\pi $-periodic correction to the phase
$kb^2\tilde f_2\left( \theta \right) =0$ for the lowest resonance and the
$\pi $-periodic $kb^2\bar f_2\left( \theta \right) $ can be included in
the phase, if needed. Note that the first factor in Eq.\ (\ref{tdwl0}) is
the Debye expansion \cite{AbrSte} for a linear combination of Bessel's
functions $\alpha \left( \theta \right) J_l\left( kr\right) +\beta \left(
\theta \right) N_l\left( kr\right) $ with $l\equiv kbf_1^{\prime }$ when
\begin{equation}
kr\gg l\gg 1.  \label{tdwnu}
\end{equation}
This relation will reemerge in Sec.\ \ref{bita} where we solve the same
problem by the Born-Oppen\-heimer approximation. Equation (\ref{tdwnu}) sets
the limits of \emph{our} theory, as well. Indeed, if $\mathbf{r}$ is close
to the center of the billiard, the stationary phase trajectory connecting
this point with the boundary is not well-defined.

It is instructive to show that the two-dimensional wavefunction
(\ref{tdwPsi}) satisfies Dirichlet's conditions on the boundary. When we
substitute $r=1+b^2\Delta R\left( \theta \right) ,$ we find
\begin{eqnarray}
\ &&\theta _1^{\prime (0)}\cong \theta +\frac{b^2\Delta R\left( \theta
\right) }{\gamma (1)}\frac{l_{pq}}k,\quad \theta _2^{\prime (0)}\cong
\theta -\Theta _{pq}-\frac{b^2\Delta R\left( \theta \right) }{\gamma
(1)}\frac{l_{pq}}k, \\
\ &&\gamma (r)\cong \gamma (1)+\frac{b^2\Delta R\left( \theta \right)
}{\gamma (1)}, \\
\ &&L_1^{(0)}\cong -\frac{b^2\Delta R\left( \theta \right) }{\gamma
(1)},\quad L_2^{(0)}\cong 2\gamma (1)+\frac{b^2\Delta R\left( \theta
\right) }{\gamma (1)}. 
\end{eqnarray}
We use these expressions in Eq.\ (\ref{tdwPsi}) keeping terms up to order
$b^2 $ in the phase and the lowest order in the prefactor. Note that the
wavefunction $\Psi \left( \mathbf{r}\right) $ $\propto \cos \left[ \left(
\phi _1-\phi _2\right) /2\right] ,$ where $\phi _i$ are the phases of the
two terms in Eq.\ (\ref{tdwPsi}). Using Eqs.\ (\ref{gcdf}),\footnote{The
constant in this equation is $-\delta \Theta _{pq}/kb,$ where $\delta $ is
the fractional part of $l_{pq},$ as explained in Sec.\ \ref{qc}} (\ref
{gcf2}), and (\ref{qck}) one can show that the cosine argument is $\pi
/2-\pi n,$ completing the proof. 

The examples of the two-dimensional wavefunctions in the short stadium 
are shown in Figs.\ \ref{2_5}, \ref{2_6}, and \ref{2_8}. 

\section{Additional remarks}

\subsection{The prefactor}

\label{pref}

It was noted in Sec.\ \ref{gc} that the PSS wavefunction $\psi (\theta )$ has
a standard WKB\ prefactor $\left[ f_1^{\prime }\left( \theta \right) \right]
^{-1/2}$ if $k\epsilon \left| E_m-\bar V_q(\theta )\right| \ll 1.$ Now we
lift this restriction. We solve Bogomolny's equation 
\begin{equation}
\psi (\theta )=\int d\theta ^{\prime }T\left( \theta ,\theta ^{\prime
}\right) \psi (\theta ^{\prime })  \label{tppsi}
\end{equation}
with the ansatz in general form 
\begin{equation}
\psi (\theta )=\exp \left[ il_{pq}\theta +ik\sum_{n=1}^\infty b^nf_n(\theta
)\right]
\end{equation}
where $b=\sqrt{\epsilon }.$ We will be interested in the imaginary part of
the phase. To evaluate the integral (\ref{tppsi}) by $S\Phi $ we expand the
phase of the integrand in $\delta \theta ^{\prime }=\theta ^{\prime }-\theta
+\Theta _{pq}.$ This phase is 
\begin{equation}
\phi \left( \theta ,\theta ^{\prime }\right) =kL_0\left( \theta ,\theta
^{\prime }\right) +l_{pq}\theta +k\sum_{n=1}^\infty b^nF_n\left( \theta
,\theta ^{\prime }\right)
\end{equation}
where $F_n\left( \theta ,\theta ^{\prime }\right) =L_n\left( \theta ,\theta
^{\prime }\right) +f_n(\theta ).$ Here $L_n\left( \theta ,\theta ^{\prime
}\right) $ is the term in the expansion of the chord length in Eq.\ (\ref
{pcbS}), $L\left( \theta ,\theta ^{\prime }\right) =\sum_{n=0}^\infty
b^nL_n\left( \theta ,\theta ^{\prime }\right) ,$ $L_n=0$ for odd $n$.

The integral (\ref{tppsi}) is equal to 
\begin{equation}
\left[ 1+\left( \sum_{n=1}^\infty b^nF_n^{\prime \prime }/L_0^{\prime
\prime }\right) \right] ^{-1/2}\exp i\left[ \phi \left( \theta ,\theta
-\Theta _{pq}\right) -k\frac{\left( \sum_{n=1}^\infty b^nF_n^{\prime
}\right) ^2}{L_0^{\prime \prime }+\sum_{n=1}^\infty b^nF_n^{\prime \prime
}}+\frac \pi 2\right] \label{tpint}
\end{equation}
where all functions are evaluated at $\theta ^{\prime }=\theta +\Theta _{pq}$
and all the derivatives are with respect to $\theta ^{\prime }.$ The
prefactor can be moved to the exponent by the transformation 
\begin{equation}
\left[ 1+\left( \sum_{n=1}^\infty b^nF_n^{\prime \prime }/L_0^{\prime
\prime }\right) \right] ^{-1/2}=\exp \left[ -\frac{bf_1^{\prime \prime
}}{2L_0^{\prime \prime }}\right] +O\left( b^2\right) . 
\end{equation}
It adds the imaginary component to the phase in Eq.\ (\ref{tpint}). This
phase should be equal to the phase in the \emph{l.h.s.}\ of Eq.\
(\ref{tppsi}). The imaginary part of the resulting equation is
\begin{equation}
\mathrm{Im}\left\{ k\sum_{n=1}^\infty b^n\left[ f_n(\theta )-f_n(\theta
-\Theta _{pq})\right] +k\frac{\left( \sum_{n=1}^\infty b^nF_n^{\prime
}\right) ^2}{L_0^{\prime \prime }+\sum_{n=1}^\infty b^nF_n^{\prime \prime
}}\right\} =\frac{bf_1^{\prime \prime }}{2L_0^{\prime \prime }}+O\left(
b^2\right) .  \label{tpim}
\end{equation}
We need to keep terms only to the order $b.$ Suppose we can choose $n_0$
such that $kb^{n_0+1}\sim b.$ Let us make a self-consistent assumption that
there is no variable imaginary part in $f_n(\theta )$ for $n<n_0.$
Collecting terms of order $kb^{n_0+1}$ we can write 
\begin{equation}
\mathrm{Im}\left[ f_{n_0+1}(\theta )-f_{n_0+1}(\theta -\Theta
_{pq})+\frac{f_1^{\prime }f_{n_0}^{\prime }}{L_0^{\prime \prime }}\right]
=\frac{f_1^{\prime \prime }}{2kb^{n_0}L_0^{\prime \prime }}. 
\end{equation}
After the $q$-average we obtain the equation 
\begin{equation}
kb^{n_0}\mathrm{Im}\bar f_{n_0}^{\prime }=\frac{f_1^{\prime \prime
}}{2f_1^{\prime }}=\left( \ln \sqrt{f_1^{\prime }}\right) ^{\prime }
\end{equation}
which means that the prefactor in function $\psi (\theta )$ is $e^{-\ln 
\sqrt{ f_1^{\prime }}+O\left( b^2\right) }=1/\sqrt{f_1^{\prime }}+O\left(
b^2\right) .$ Note that $\mathrm{Im}f_{n_0}$ is $q$-periodic since,
according to Eq.\ (\ref{tpim}), $\mathrm{Im}\left[ f_{n_0}(\theta
)-\right.$ $\left. f_{n_0}(\theta -\Theta _{pq})\right] $ $=0.$

\subsection{Fourier expanded perturbation}

\label{fep}

Consider a system described by the perturbed Hamiltonian (\ref{pish}).
Suppose the perturbation is expanded in Fourier series 
\begin{equation}
H_2\left( \mathbf{I},\mathbf{\theta }\right)
=\sum_{\mathbf{m}}H_{\mathbf{m}}\left( \mathbf{I}\right)
e^{i\mathbf{m\cdot \theta }}. 
\end{equation}
It is known in the classical perturbation theory \cite{Tab} that each
Fourier component produces a resonance when $\mathbf{I}$ satisfies the
equation 
\begin{equation}
\mathbf{\omega (I)\cdot m}=0
\end{equation}
where $\mathbf{\omega (I)}=\partial H_0\left( \mathbf{I}\right) /\partial 
\mathbf{I}$ are the unperturbed frequencies. Below we show that the
semiclassical theory agrees with this condition.

Suppose the perturbation consists of one component $H_{\mathbf{m}}\left(
\mathbf{I}\right) e^{i\left( m_1\theta _1+m_2\theta _2\right) }.$ We
choose the surface of section $\theta _2=0,$ as in Sec.\ \ref{pis}. We
show that the effective potential for the $(p,q)$ resonance $\bar
V_q\left( \theta _1\right) =0$ unless $\omega _1/\omega _2=$
$p/q=-m_2/m_1.$ The perturbed action $S_2\left( \theta ,\theta ^{\prime
};E\right) =-\int H_2dt,$ where the integral is taken over the unperturbed
trajectory between two consecutive crossings of the PSS at $\theta
_1=\theta ^{\prime }$ and $\theta _1=\theta .$ With the substitution
$\theta _1=\theta ^{\prime }+\omega _1t$ and $\theta _2=\omega _2t$ we
find
\begin{equation}
S_2\left( \theta ,\theta ^{\prime };E\right)
=-\frac{H_{\mathbf{m}}e^{im_1\theta ^{\prime }}}{\omega _2}\int_0^{2\pi
}\exp \left[ i\left( m_1\frac{\omega _1}{\omega _2}+m_2\right) \theta
_2\right] d\theta _2.
\end{equation}
Note that $H_{\mathbf{m}}\left( \mathbf{I}\right) $ and $\omega _i\left( 
\mathbf{I}\right) $ depend on $\theta -\theta ^{\prime }$ and $E.$ The
effective potential 
\begin{equation}
\bar V_q(\theta )=\left\langle S_2\left( \theta ,\theta -\Theta
_{pq};E\right) \right\rangle _q\propto \sum_{r=0}^{q-1}e^{2\pi im_1rp/q}=0
\end{equation}
unless $m_1p/q\equiv l$ is integer. But in this case $\int_0^{2\pi }\exp
\left[ i\left( m_1\frac{\omega _1}{\omega _2}+m_2\right) \theta _2\right]
d\theta _2=0,$ except when $l=-m_2,$ i.e.\ $\mathbf{\omega \cdot m}=0.$ In
the latter case $\bar V_q(\theta )=-\frac{2\pi }{\omega
_2}H_{\mathbf{m}}e^{im_1\theta }.$ Note that the current definition of
$\bar V_q(\theta )$ differs from one in Sec.\ \ref{gc} by a factor of $k.$

\subsection{Direct quantization of classical perturbation theory}

\label{dqoc}

In Sec.\ \ref{ClasInt} we have mentioned that some results of the classical
perturbation theory follow from our semiclassical theory. In fact, the
connection between the two theories is even closer. In this section we are
going to show how one could develop the semiclassical theory in the
action-angle variables solely on the grounds of the classical \emph{resonant}
perturbation theory, without employing the $T$-operator formalism.

\subsubsection{First order theory}

We begin with the classical Hamiltonian (\ref{pish}) keeping terms to
order $\epsilon .$ We will quantize the orbits near the $(p,q)$ resonance,
for which the ratio of unperturbed frequencies $\omega _1/\omega _2=$
$p/q.$ Following the recipe \cite{LicLie}, we make a canonical
transformation to the rotating frame $\left( \mathbf{I},\mathbf{\theta
}\right) \rightarrow ( \hat{\mathbf{I}},\hat{\mathbf{\theta
}}) $ defined by the equations
\begin{eqnarray}
\ &&I_1=q\hat{I}_1,\quad I_2=\hat{I}_2-p\hat{I}_1, \\
\ &&\hat{\theta }_1=q\theta _1-p\theta _2,\quad \hat{\theta
}_2=\theta _2. 
\end{eqnarray}
The new Hamiltonian is $\hat H( \hat{\mathbf{I}},\hat{\mathbf{\theta }})
=H\left( \mathbf{I}( \hat{\mathbf{I}}) ,\mathbf{\theta }(
\hat{\mathbf{\theta }}) \right) .$ Like the original Hamiltonian, it can
be expanded in the perturbation series
\begin{equation}
\hat H ( \hat{\mathbf{I}},\hat{\mathbf{\theta }}) =\hat H_0(
\hat{\mathbf{I}}) +\epsilon \hat H_2( \hat{\mathbf{I}},\hat{\mathbf{\theta
}}) . 
\end{equation}

The new variable $\hat{\theta }_1$ is ``slow'', i.e.\ $\dot
{\hat{\theta }}_1\simeq q\omega _1-p\omega _2\ll \dot
{\hat{\theta }}_2.$ It describes the slow deviation from the
resonance. Under this condition one can make an \emph{infinitesimal}
canonical transformation $( \hat{\mathbf{I}},\hat{\mathbf{\theta }})
\rightarrow \left( \overline{\mathbf{I}},\overline{\mathbf{\theta
}}\right) = ( \hat{\mathbf{I}},\hat{\mathbf{\theta }}) +O\left( \epsilon
\right) $ that would eliminate the $\hat{\theta }_2$-dependence from
the Hamiltonian. The transformed Hamiltonian has the form
\begin{equation}
\bar H( \hat{I}_1,\hat{\theta }_1;\hat{I}_2) =\hat H_0( \hat{\mathbf{I}})
+\epsilon \bar H_2( \hat{I}_1,\hat{\theta }_1;\hat{I}_2)
\end{equation}
where the perturbed part is simply the average of $\hat H_2$ over
$\hat{\theta }_2,$
\begin{equation}
\bar H_2( \hat{I}_1,\hat{\theta }_1;\hat{I}_2) =\frac 1{2\pi
q}\int_0^{2\pi q}\hat H_2( \hat{\mathbf{I}},\hat{\mathbf{\theta }})
d\hat{\theta }_2. 
\end{equation}
(When $\hat{\theta }_1$ is fixed, the period of $\hat H_2$ in variable
$\hat{\theta }_2$ is $2\pi q.$) At this stage we neglect the
difference between the ``hatted'' and ``barred'' variables. 

Now the problem becomes essentially one-dimensional.
$\hat{I}_2=\hat{I}_{20}=\mathrm{const}$ is an integral of motion.
We define $\hat{I}_{10} $ by the resonance condition
\begin{equation}
\left. \frac{\partial \hat H_0}{\partial \hat{I}_1}\right|
_{\hat{I}_1=\hat{I}_{10}}=0.  \label{dqocres}
\end{equation}
(The meaning of this definition will become clear later, when we transform
back to the original variables.) The energy conservation $\bar H=E$ requires 
\begin{equation}
\frac{\hat H_0^{\prime \prime }}2( \Delta \hat{I}_1)
^2+\epsilon \bar H_2( \hat{\mathbf{I}}_{\mathbf{0}},\hat{\theta 
}_1) =E-E_0.  \label{dqocE}
\end{equation}
Here $\Delta \hat{I}_1=\hat{I}_1-\hat{I}_{10},$ $E_0=\hat
H_0( \hat{\mathbf{I}}_0) ,$ and $\hat H_0^{\prime \prime
}=( \partial ^2\hat H_0/\partial \hat{I}_1^2)
_{\hat{\mathbf{I}}=\hat{\mathbf{I}}_0}.$ Remarkably, Eq.\
(\ref{dqocE}) is similar to the energy conservation law for a
one-dimensional particle of mass $\hat H_0^{\prime \prime }$ with momentum
$\Delta \hat{I}_1$ moving in a potential $\epsilon \bar H_2(
\hat{\mathbf{I}}_{\mathbf{0}},\hat{\theta }_1) .$ We will
quantize it by changing the variables to the appropriate operators, but
first we will express the constant parameters in terms of the original
actions $\mathbf{I}_{\mathbf{0}}.$

To begin with, note that 
\begin{equation}
\frac{\partial \hat H_0}{\partial \hat{I}_1}=q\omega _1\left(
\mathbf{I}\right) -p\omega _2\left( \mathbf{I}\right)
\end{equation}
so that condition (\ref{dqocres}) means that
$\hat{\mathbf{I}}_{\mathbf{0}}$ is a resonant torus. To calculate the
second derivative, it is convenient to express $( \partial /\partial
\hat{I}_1) _{\hat{I}_2}$ in terms of $\left( \partial
/\partial I_1\right) _E.$ Using the notation from Sec.\ \ref{tdss},
$I_2=g_E\left( I_1\right) $ and $\alpha \left( \mathbf{I}\right) =\omega
_1\left( \mathbf{I}\right) /\omega _2\left( \mathbf{I}\right)
=-g_E^{\prime }\left( I_1\right) ,$ we can write
\begin{equation}
\frac 1q\left( \frac \partial {\partial \hat{I}_1}\right)
_{\hat{I}_2}=\left( \frac \partial {\partial I_1}\right) _{I_2}-\alpha
\left( \mathbf{I}\right) \left( \frac \partial {\partial I_2}\right)
_{I_1}=\left( \frac \partial {\partial I_1}\right) _E. 
\end{equation}
Then the second derivative becomes 
\begin{equation}
\hat H_0^{\prime \prime }=q\left( \frac{\partial \omega _2\left( q\alpha
-p\right) }{\partial I_1}\right) _E=\left. -q^2\omega _2g_E^{\prime \prime
}\right| _{\mathbf{I=I}_{\mathbf{0}}}
\end{equation}
where we took into account that $q\alpha -p=0$ at the resonance. If the PSS
is chosen at $\theta _2=0,$ we can express $\hat H_0^{\prime \prime }$ in
terms of the unperturbed action (\ref{pisS}). We have shown in Sec.\ \ref
{tdss} that $S_0^{\prime \prime }\left( \Theta _{pq}\right) =-\left( 2\pi
g_E^{\prime \prime }\right) ^{-1},$ therefore 
\begin{equation}
\hat H_0^{\prime \prime }=\frac{q^2\omega _2}{2\pi S_0^{\prime \prime
}\left( \Theta _{pq}\right) }.  \label{dqocH0}
\end{equation}

\begin{sloppypar}
Now let us take care of the potential term in Eq.\ (\ref{dqocE}). Returning
to the original variables we write 
\begin{equation}
\bar H_2( \hat{\mathbf{I}}_{\mathbf{0}},\hat{\theta
}_1) =\frac{\omega _2}{2\pi q}\int_0^{2\pi q/\omega _2}H_2\left(
\mathbf{I}_{\mathbf{0}},\theta _1=\frac{\hat{\theta }_1}q+\omega
_1t,\theta _2=\omega _2t\right) dt. 
\end{equation}
Note that the correction to the action between two crossings of the surface
of section $S_2\left( \theta _1,\theta _1-\Theta _{pq}\right) =-\int_0^{2\pi
/\omega _2}H_2dt$ where the integral is taken along the unperturbed orbit.
If we define the effective potential as a $q$-average $\bar V_q(\theta
_1)=\left\langle S_2\left( \theta _1,\theta _1-\Theta _{pq}\right)
\right\rangle _q,$ we find that 
\begin{equation}
\bar H_2( \hat{\mathbf{I}}_{\mathbf{0}},\hat{\theta
}_1) =-\frac{\omega _2}{2\pi }\bar V_q( \theta
_1=\hat{\theta }_1/q) .  \label{dqocH2}
\end{equation}
\end{sloppypar}

Equation (\ref{dqocE}) can be transformed to the Schr\"odinger equation by
changing $\Delta \hat I_1\rightarrow -i\partial /\partial \hat \theta _1.$
With the help of Eqs.\ (\ref{dqocH0}) and (\ref{dqocH2}), we have 
\begin{equation}
\frac 1{2S_0^{\prime \prime }\left( \Theta _{pq}\right) }\frac{\partial
^2\hat \psi }{\partial ( \hat{\theta }_1/q) ^2}+\epsilon \bar
V_q( \hat{\theta }_1/q) \hat \psi =\epsilon \mathcal{E}\hat
\psi  \label{dqocSch}
\end{equation}
where 
\begin{equation}
\epsilon \mathcal{E}=-\left( 2\pi /\omega _2\right) \left( E-E_0\right) .
\label{dqocen}
\end{equation}
It is easy to construct the complete wavefunction, which is trivial in
$\hat{\theta }_2,$
\begin{equation}
\Psi \left( \theta _1,\theta _2\right) =\hat \psi \left(
\frac{\hat{\theta }_1}q\right)
e^{i\hat{\mathbf{I}}_{\mathbf{0}}\cdot \hat{\mathbf{\theta
}}}=\hat \psi \left( \theta _1-\frac pq\theta _2\right)
e^{i\mathbf{I}_{\mathbf{0}}\cdot \mathbf{\theta }}.  \label{dqoc2d}
\end{equation}
Setting $\theta _2=0$ we obtain the one-dimensional wavefunction that
would also follow from the first order perturbation theory based on the
$T$-operator. In fact, Eq.\ (\ref{dqocSch}) is similar to Eq.\
(\ref{gcSch}) when $\theta _1=$ $\hat{\theta }_1/q.$ Thus both
theories are equivalent to this order. 

\subsubsection{Quantization conditions}

The quantization conditions follow from the requirement that $\Psi \left(
\theta _1,\theta _2\right) $ be $2\pi $-periodic in both variables up to the
Maslov phases. With 
\begin{equation}
\hat \psi \left( \theta +2\pi \right) =\hat \psi \left( \theta \right) \exp
\left( -i2\pi \delta \right) ,  \label{dqocbc}
\end{equation}
where $0\leq \delta <1$ is to be determined, we obtain two equations, 
\begin{eqnarray}
\ &&I_{10}=l+\nu _l+\delta , \\
\ &&I_{20}=n+\nu _n-\frac pq\delta .
\end{eqnarray}
Here $l$ and $n$ are integer and $\nu _l$ and $\nu _n$ are the Maslov
phases. Since $I_{10}$ and $I_{20}$ are related by the resonance condition
(\ref{dqocres}), the above equations allow us to express $\delta ,$
$I_{10},$ and $I_{20}$ in terms of $l$ and $n.$ Note that $l$ and $n$ are
not independent. Usually $l,n\gg \delta ,$ so they are related by the
resonance condition on $I_{10}$ and $I_{20}.$ In the perturbed circle, for
example, $l$ is an integer part of $l_{pq}.$

The Schr\"odinger equation (\ref{dqocSch}) with Eq.\ (\ref{dqocbc}) as the
boundary condition provides the quantized ``energy'' levels $\mathcal{E}_m.$
The total energy 
\begin{equation}
E=H_0\left( \mathbf{I}_{\mathbf{0}}\right) -\epsilon
\mathcal{E}_m\frac{\omega _2\left( \mathbf{I}_{\mathbf{0}}\right) }{2\pi }
\label{dqocq}
\end{equation}
is found from Eq.\ (\ref{dqocen}).

Let us compare this result with the quantization condition given by the
$T$-operator [cf.\ Eq.\ (\ref{qck})],
\begin{equation}
S_0\left( \Theta _{pq},E\right) -\left( l+\nu _l\right) \Theta
_{pq}+\epsilon \mathcal{E}_m=2\pi \left( n+\nu _n\right) .
\end{equation}
Here $S_0\left( \Theta _{pq},E\right) $ is the action for an unperturbed
system with energy $E$ between two crossings of the PSS $\theta _2=0$ with
the difference in $\theta _1$ equal $\Theta _{pq}.$ By inverting the
function $S_0$ one finds the energy 
\begin{equation}
E=H_0\left[ \Delta \theta _1=\Theta _{pq},S=\left( l+\nu _l\right) \Theta
_{pq}+2\pi \left( n+\nu _n\right) -\epsilon \mathcal{E}_m\right] .
\end{equation}
It can be shown that $\left( \partial H_0/\partial S\right) _{\Delta
\theta _1}=\omega _2/2\pi .$ Hence expanding the above equation in
$\epsilon $ we obtain Eq.\ (\ref{dqocq}), taking into account that
$H_0\left( \mathbf{I}_{\mathbf{0}}\right) =H_0\left[ \Delta \theta
_1=\Theta _{pq},\right. $ $\left. S=\Theta _{pq}I_{10}+2\pi I_{20}\right]
.$

\subsubsection{Second order theory}

To find the next order correction to the wavefunction (\ref{dqoc2d}) we
need to take into account the difference between the coordinates $(
\hat{\mathbf{I}},\hat{\mathbf{\theta }}) $ and $\left(
\overline{\mathbf{I}},\overline{\mathbf{\theta }}\right) $ defined above.
In particular, $\Delta \hat{I}_1$ in Eq.\ (\ref{dqocE}) should be
changed to
\begin{equation}
\Delta \overline{I}_1=\Delta \hat{I}_1-\epsilon \frac{\partial
F}{\partial \hat{\theta }_1}
\end{equation}
where the function $F(
\hat{\mathbf{I}}_{\mathbf{0}},\hat{\mathbf{\theta }}) $ can
be determined from the equation \cite{LicLie}
\begin{equation}
\omega _2\frac{\partial F}{\partial \hat{\theta }_2}\cong \bar H_2( 
\hat{\mathbf{I}}_{\mathbf{0}},\hat{\theta }_1) -\hat H_2( 
\hat{\mathbf{I}}_{\mathbf{0}},\hat{\mathbf{\theta }}) .
\label{dqocF}
\end{equation}
The solution of the Schr\"odinger equation will be modified by a factor
$\exp \left( i\epsilon F\right) .$

The two-dimensional wavefunction becomes 
\begin{equation}
\Psi \left( \theta _1,\theta _2\right) =\hat \psi \left(
\frac{\hat{\theta }_1}q\right)
e^{i\hat{\mathbf{I}}_{\mathbf{0}}\cdot \hat{\mathbf{\theta
}}+i\epsilon F\left(
\hat{\mathbf{I}}_{\mathbf{0}},\hat{\mathbf{\theta }}\right) }. 
\end{equation}
Function $F$ should satisfy Eq.\ (\ref{dqocF}) and have period $2\pi $ in
$\theta _1$ and $\theta _2.$ Hence the energy quantization does not
change. Explicitly,
\begin{equation}
F( \hat{\theta }_1,\hat{\theta }_2) =\frac 1{\omega
_2}\int_0^{\hat{\theta }_2}\left[ \bar H_2( \hat{\theta
}_1) -\hat H_2( \hat{\theta }_1,\hat{\theta }_2^{\prime
}) \right] d\hat{\theta }_2^{\prime }+f_2\left( \theta
_1=\frac{\hat{\theta }_1}q\right)
\end{equation}
where we dropped argument $\hat{\mathbf{I}}_{\mathbf{0}}$ to shorten
notation. The function $f_2\left( \theta _1\right) $ comes from the
$T$-operator theory. It is the second order term in the phase of the PSS\
wavefunction $\psi \left( \theta _1\right) =\hat \psi \left( \theta
_1\right) \exp \left[ iI_{10}\theta _1+i\epsilon f_2\left( \theta
_1\right) \right] $ and it should satisfy the equation
\begin{eqnarray}
\ &&f_2\left( \theta _1+\Theta _{pq}\right) -f_2\left( \theta _1\right)
=S_2\left( \theta _1+\Theta _{pq},\theta _1\right) -\bar V_q(\theta _1) 
\nonumber  \label{dqocf2} \\
\ &=&\frac 1{\omega _2}\int_0^{2\pi }\left[ \bar H_2( \hat{\theta
}_1) -\hat H_2( \hat{\theta }_1,\hat{\theta }_2)
\right] _{\hat{\theta }_1=q\theta _1}d\hat{\theta }_2. 
\label{dqocf2}
\end{eqnarray}

The $2\pi $-periodicity in $\theta _1$ for $F(
\hat{\mathbf{\theta }}) $ is trivial. To show that it is $2\pi
$-periodic in $\theta _2$ consider
\begin{eqnarray}
&&\int_0^{\hat{\theta }_2+2\pi }\hat H_2( \hat{\theta
}_1-2\pi p,\hat{\theta }_2^{\prime }) d\hat{\theta
}_2^{\prime }=\int_0^{\hat{\theta }_2+2\pi }\hat H_2(
\hat{\theta }_1,\hat{\theta }_2^{\prime }-2\pi )
d\hat{\theta }_2^{\prime } \nonumber \\
&=&\int_0^{\hat{\theta }_2}\hat H_2( \hat{\theta
}_1,\hat{\theta }_2^{\prime }) d\hat{\theta }_2^{\prime
}+\int_{-2\pi }^0\hat H_2( \hat{\theta }_1,\hat{\theta
}_2^{\prime }) d\hat{\theta }_2^{\prime }. 
\end{eqnarray} 
Then it follows from Eq.\ (\ref{dqocf2}) that $F( \hat{\theta
}_1-2\pi p,\hat{\theta }_2+2\pi ) =F( \hat{\theta
}_1,\hat{\theta }_2) ,$ which is the desired result. 

Note that $F( \hat{\mathbf{\theta }}) $ is defined up to an
arbitrary function $F_1( \hat{\theta }_1) .$ This function is
constant along the unperturbed resonant trajectory. This trajectory crosses
the PSS $\theta _2=0$ at $q$ points separated by angle $\Delta \theta
_1=2\pi /q.$ Since $F_1( \hat{\theta }_1) $ must be
single-valued on the torus $\left( \theta _1,\theta _2\right) ,$ it has to
be $2\pi $\emph{-periodic in }$\hat{\theta }_1.$

The above results are in agreement with the $T$-operator perturbation
theory. To see this set $\hat{\theta }_2=0.$ Then $F\left( q\theta
_1,0\right) =f_2\left( \theta _1\right) $, defined up to a $2\pi
/q$-periodic function $F_1\left( q\theta _1\right) .$

In conclusion, we have shown that the semiclassical resonant perturbation
theory can be constructed on the basis of the classical theory without
employing the $T$-operator. Both theories are equivalent, at least to the
second order. The theory discussed in this section automatically gives the
two-dimensional wavefunction. Unfortunately, the use of action-angle
variables is rather important. The $T$-operator method can be more
convenient in practice, since it is not restricted to a specific set of
coordinates. Still, even here the unperturbed action should depend on the
difference of the surface of section coordinates, which, in effect, limits
the choice to the coordinates that are similar to action-angle.

\subsection{Perturbed spherical billiard}

\label{psb}

It is easy to generalize our theory in the case of low angular momentum
resonance in a perturbed spherical billiard. The billiard is a
three-dimensional cavity with the boundary $r\left( \theta ,\varphi \right)
=1+\epsilon \Delta R\left( \theta ,\varphi \right) .$ We choose the PSS to
coincide with the boundary, which is now two-dimensional, as well as the
surface of section wavefunction $\psi \left( \theta ,\varphi \right) .$

Following the method of Sec.\ \ref{r12} we expand the $T$-operator near
$\theta ^{\prime }=\pi -\theta ,$ $\varphi ^{\prime }=\varphi -\pi .$ With
the notation $\mathbf{r}=\left( \theta ,\varphi \right) $ and
$\mathbf{r}^{\prime }=\left( \theta ^{\prime },\varphi ^{\prime }\right) $
it becomes
\begin{equation}
T\left( \mathbf{r},\mathbf{r}^{\prime }\right) \simeq -\frac{k\sin \theta
}{4\pi i}\exp \left[ i2k\left( 1-\frac{\delta \theta ^{\prime
2}}8-\frac{\delta \varphi ^{\prime 2}}8\sin ^2\theta \right) \right]
\left[ 1+ik\epsilon V\left( \mathbf{r}\right) \right]
\end{equation}
where the effective potential $V\left( \mathbf{r}\right) =\Delta R\left(
\mathbf{r}\right) +\Delta R\left( \mathbf{-r}\right) .$ The wavefunction
is also expanded up to the quadratic terms. The integral $\int T\left(
\mathbf{r},\mathbf{r}^{\prime }\right) \psi \left( \mathbf{r}^{\prime
}\right) d\left( \delta \theta ^{\prime }\right) d\left( \delta \varphi
^{\prime }\right) $ is evaluated in the $S\Phi $ approximation. It should
be equal to $\psi \left( \mathbf{r}\right) :$
\begin{equation}
e^{i2k}\left\{ \psi \left( \mathbf{-r}\right) \left[ 1+ik\epsilon V\left(
\mathbf{r}\right) \right] +\frac 1{ik}\left[ \frac{\partial ^2\psi
}{\partial \theta ^2}+\frac 1{\sin ^2\theta }\frac{\partial ^2\psi
}{\partial \varphi ^2}\right] \right\} =\psi \left( \mathbf{r}\right) . 
\end{equation}
Substitute $\psi \left( \mathbf{-r}\right) =e^{-i\omega }\psi \left(
\mathbf{r}\right) $ where $\omega =2k+k\epsilon E_{lm}.$ Expanding in
$k\epsilon $ we obtain the differential equation
\begin{equation}
-\left[ \frac{\partial ^2\psi }{\partial \theta ^2}+\frac 1{\sin ^2\theta
}\frac{\partial ^2\psi }{\partial \varphi ^2}\right] +k^2\epsilon \left[
V\left( \mathbf{r}\right) -E_{lm}\right] \psi =0. 
\end{equation}
Solving this equation we find possible $E_{lm}$'s. Since $V\left(
\mathbf{r}\right) $ is symmetric the resulting wavefunctions can be made
symmetric or antisymmetric. Hence the quantization condition for $k$ is
\begin{equation}
2k+k\epsilon E_{lm}=2\pi n+\omega _{lm}
\end{equation}
where $\omega _{lm}=0$ or $\pi .$

The problem of low angular momentum resonance can also be solved by the
Born-Oppen\-heimer approximation (Sec.\ \ref{bita}). So far we have not
succeeded in treating high angular momentum resonances.

\vspace*{-.15cm}
\section{Conclusions}

\label{conc}
\vspace*{-.1cm}
In this chapter we developed the semiclassical resonant perturbation
theory. The essence of the theory is to quantize the classical motion near
the resonances. The Poincar\'e surface of section method and its
semiclassical analogue, Bogomolny's $T$-operator, effectively reduce the
dimensionality of the problem. The number of terms entering the
perturbation series is determined by relationship between the parameters
$k$ and $\epsilon .$ It would be impractical to use this theory if one
needs to find \emph{all} states in a certain energy range. Here the states
are classified by the regions of phase space (resonances) they belong to.
Each resonance is treated separately. We have found the expressions for
the PSS wavefunctions and the energy levels within any given resonance.
For two-dimensional systems the wavefunctions are the solutions of
one-dimensional Schr\"odinger equation. So, as soon as the effective
potential is known, it is easy to predict how the states will look like,
before a numerical solution is obtained. The effective potential contains
information about all orbits on a perturbed torus, not just the
neighborhood of a stable orbit. This, in particular, explains scars near
unstable periodic orbits. 

It should be noted that the states described by our theory are not
necessarily close to the true eigenmodes of the system. The states
associated with one resonance can mix appreciably with nearly degenerate
states of other resonances. Although the mixing can, in principle, be
estimated and the proper eigenmodes can be constructed, it is hard do it
systematically. This observation, however, does not completely invalidate
the theory. While the approximate states are not stationary, they may exist
for a long time before leaking to other degenerate states. Such states are
called the \emph{quasimodes }\cite{Arn}.

There are a number of other methods that deal with perturbed integrable
systems. The quantization of the Birkhoff-Gustavson normal form \cite
{Bir,Gus,Rob} can be considered complementary to the present method.
Birkhoff-Gustavson is applied to the systems that are, essentially,
perturbed harmonic oscillators. Our theory works for strongly anharmonic
systems, as was remarked earlier. In addition, the Birkhoff-Gustavson
method requires numerous algebraic transformations, which in practice
makes it numerical. The adiabatic methods based on the Born-Oppenheimer
approximation (BOA) will be discussed in Ch.\ \ref{qboa}. These methods are
equivalent to our perturbation theory in certain cases. The BOA is rather
straightforward and easy to understand since it\ is applied directly to
the Schr\"odinger equation. It does not require the surface of section and
automatically produces the two-dimensional wavefunction. Unfortunately,
its application is limited to the cases where a slowly changing parameter
can be found. For example, only the low angular momentum resonance in the
perturbed circle can be studied by the BOA, while the perturbation theory
does all of them. 

\chapter{Rectangular billiard and other systems}

In the previous chapter we developed the resonant perturbation theory for a
perturbed circular billiard. Now we apply this theory to the perturbed
rectangular billiard. In the second half of the chapter we consider examples
of the systems without an explicit small parameter. It is sometimes possible
to tailor the $T$-operator approach to these systems on the case-by-case
basis in order to study certain classes of states, like bouncing ball or
whispering gallery modes. The surface of section method can also be used 
for the scattering problems. 

\section{Rectangular billiard --- general case}

\label{rbgc}

This section has mostly an auxiliary purpose. We consider a rectangular
billiard with slightly perturbed sides. A number of particular cases of
this model has been studied such as the tilted billiard, the long stadium,
or the square in magnetic field. We will use the general results of this
chapter when discussing these examples. The last case is particularly
interesting since the perturbation here is the magnetic field (that breaks
the time-reversal symmetry) rather than the shape of the boundaries. A
separate chapter is devoted to this example. 

\subsection{Classification of trajectories}

\begin{figure}[tbp]
{\hspace*{2.7cm} \psfig{figure=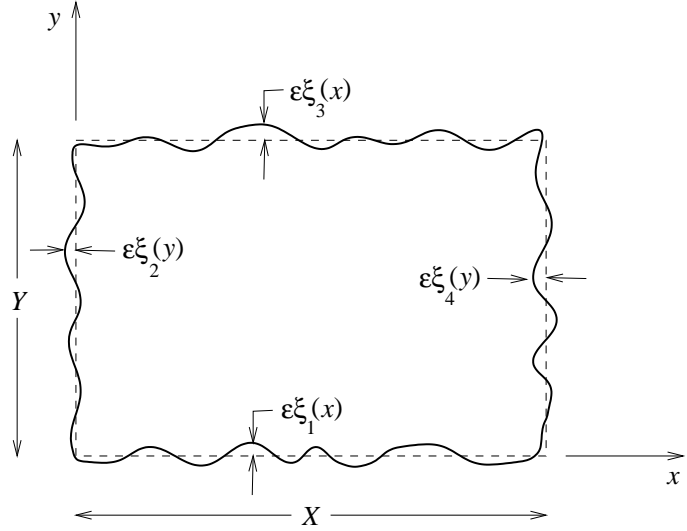,height=7cm,width=9cm,angle=0}}
{\vspace*{.13in}}
\caption{Perturbed rectangular billiard. 
\label{3_1}}
{\vspace{1.2 cm}}
\end{figure}

Consider a rectangle defined by its vertices $(0,0),$ $(X,0),$ $(X,Y),$
and $(0,Y),$ where $X,Y>0.$ The boundary shape, starting with the lower
boundary and going clockwise, is described by the functions (Fig.\
\ref{3_1})
\begin{eqnarray}
y &=&\epsilon \xi _1\left( x\right) ,\quad 0<x<1,  \nonumber \\
x &=&\epsilon \xi _2\left( y\right) ,\quad 0<y<1,  \nonumber \\
y &=&Y-\epsilon \xi _3\left( x\right) ,\quad 0<x<1,  \nonumber \\
x &=&X-\epsilon \xi _4\left( y\right) ,\quad 0<y<1.  \label{cotxy}
\end{eqnarray}
In the following we assume $\epsilon \ll 1,$ while $\left| \xi _i\right|
\sim \left| \xi _i^{\prime }\right| \sim X\sim Y\sim 1.$ The averages
$\left\langle \xi _i\right\rangle =0.$ Note that, by definition, $\xi
_i>0$ inside the unperturbed billiard. All the conclusions concerning the
relationship between the dimensionless wavenumber $k$ and the perturbation
parameter $\epsilon $ made for the circle billiard remain in force. In
particular, if $k\epsilon ^{3/2}\ll 1,$ it suffices to continue the
perturbation expansion up to the second order (assuming there are no
extremely large higher derivatives of $\xi _i$). 

\begin{figure}[tbp]
{\hspace*{1cm} \psfig{figure=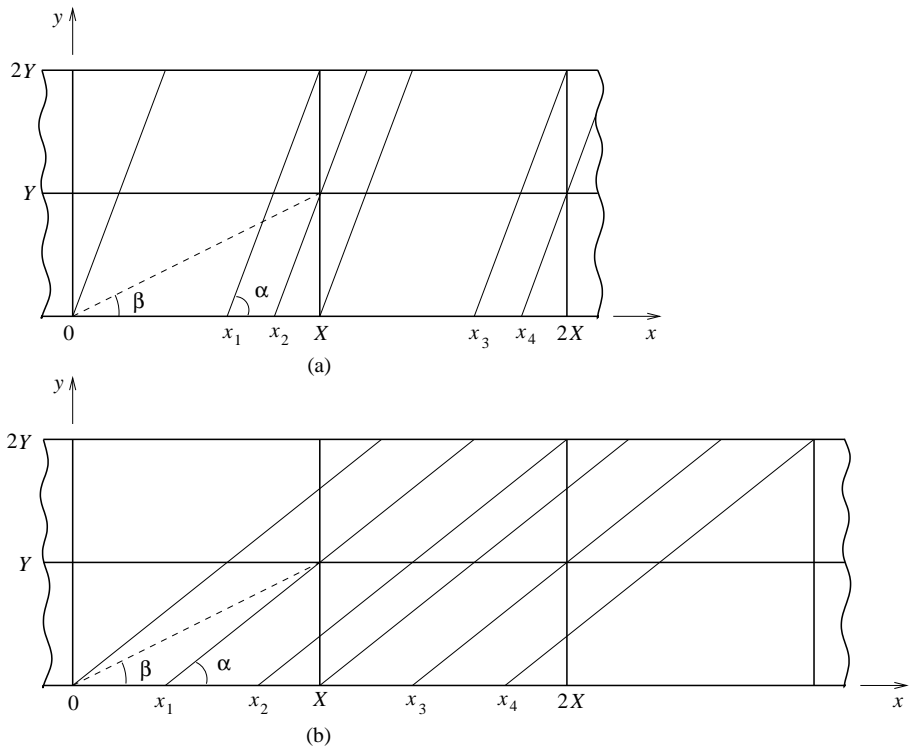,height=9.8cm,width=12cm,angle=0}}
{\vspace*{.13in}}
\caption[Method of images for the rectangular billiard.]
{Method of images for the rectangular billiard. (a) $\tan \alpha >2 \tan
\beta$, (b) $\tan \beta < \tan \alpha <2 \tan \beta$, where $\alpha$ is the
angle that the orbits make with the $x$ axis, $\tan \beta = Y/X$. The
orbits starting within the intervals defined by points $x_i$ and the
vertices reflect from the same sequences of the walls. 
\label{3_2}}
{\vspace{1.2 cm}}
\end{figure}

We choose the lower boundary of the billiard as the Poincar\'e surface of
section (PSS). Thus, an orbit might bounce off three other walls several
times between the two consecutive mappings. We will classify the orbits
according to their topology in the unperturbed billiard. But, first, we
introduce an equivalent representation for the trajectories, using the
method of images. Assuming for the moment there is no perturbation, we
reflect the billiard about its upper side and attach the image to the
original. Then we reflect the resulting billiard about its right side to
obtain a $2X\times 2Y$ billiard. After that we continue this billiard
periodically in $x$-direction. Thus, we have an infinite strip of width
$2Y$ (Fig.\ \ref{3_2}). The PSS $y=0$ is identified with $y=2Y.$ All
orbits go from the bottom to the top. 

Suppose an orbit makes angle $\alpha $ with the $x$-axis. We can always
assume $0<\alpha <\pi /2,$ i.e.\ the $x$-projection of the momentum
$p_x>0,$ in this extended scheme. The region $X<x<2X$ is responsible for
the negative $p_x$ in the original billiard. Similarly, $Y<y<2Y$
represents the negative $p_y.$ With these definitions the extended
unperturbed billiard has the topology of a torus, and the coordinates and
momenta are similar to the action-angle variables. 

We will consider only the orbits with $\tan \alpha \geq \tan \beta =Y/X,$
where $\beta $ is the angle between the $x$-axis and the diagonal. In the
opposite case we can switch the $x$- and $y$-directions. The trajectories
can be divided in two classes depending on whether $\tan \alpha $ is greater
(class I) or less (class II) than $2\tan \beta .$ Each family of orbits with
given $\alpha $ divides the stretch of PSS from $0$ to $2X$ into six
intervals between the points $0<x_1<x_2<X<x_3<x_4<2X$. Orbits starting
within the same interval reflect from the same sequence of walls as they
propagate. Although the borders of these intervals depend on $\alpha ,$ the
sequences of walls remain the same within the class. Points $x_i$ are
defined in Fig.\ \ref{3_2} (a) for class I and (b) for class II. Analytic
expressions can also be written down. For example, $x_1=X-2Y/\tan \alpha $
for class I. The orbits beginning in the interval $(0,x_1)$ reflect off side
3 (as defined in Fig.\ \ref{3_1}) before returning to the PSS. For the orbits
beginning in $(x_1,x_2)$ these are sides 3 and 4, etc.

\subsection{Perturbation theory}

We begin with constructing the $T$-operator for the system, 
\begin{equation}
T\left( x,x^{\prime }\right) =\sqrt{\frac k{2\pi i}\left| \frac{\partial
^2L}{\partial x\partial x^{\prime }}\right| }e^{ikL\left( x,x^{\prime
}\right) -i\frac \pi 2\nu \left( x,x^{\prime }\right) }. 
\end{equation}
$\nu \left( x,x^{\prime }\right) $ is the Maslov index for the orbit that
goes from point $x^{\prime }$ to point $x$ on the PSS. It increases by 2
on each reflection (in the original billiard). Note that each trajectory
reflects off the upper and lower boundaries once, which changes the phase
by $2\pi .$ Thus, one may count in $\nu \left( x,x^{\prime }\right) $ only
reflections from the side walls. As was explained in Sec.\ \ref{mpat}, the
Maslov phase remains, even when we work in the extended scheme. $L\left(
x,x^{\prime }\right) $ is the length of the orbit. With the intent to
calculate the two-dimensional wavefunction later, we first find
$\mathcal{L}\left( x,y;x^{\prime }\right) ,$ the length of the orbit that
starts at $x^{\prime }$ on the PSS and ends at point $\left( x,y\right) $
inside the billiard. 

We argue that the shift of the point of reflection \emph{along} the
boundary due to the perturbation can be neglected when calculating the
length. Here is what we mean by that. Consider, for example, the right
wall. It is parametrized by $y$ [see Eq.\ (\ref{cotxy}) for wall 4].
Suppose an orbit with fixed end points reflects at point $y_0$ in an
unperturbed billiard. If there is a perturbation, the orbit with the same
end points will reflect at point $y_0+\Delta y$ where the length is
extremal, that is $\partial L/\partial y=0.$ Since $\left| \xi _4^{\prime
}\right| \sim 1,$ the shift $\Delta y\sim \epsilon .$ The correction to
the length, however, is $\frac 12\left( \partial ^2L/\partial y^2\right)
\left( \Delta y\right) ^2\sim \epsilon ^2$ and can be neglected at this
order of the calculation. As a result, all the corrections will depend on
functions $\xi _i$ evaluated at the reflection points for unperturbed
trajectories. 

It is convenient to express the corrections to the length
$\mathcal{L}\left( x,y;x^{\prime }\right) $ in terms of auxiliary
functions $\eta _i\left( x,y;\alpha \right) ,$ $i=1,\ldots ,4,$ where
$\tan \alpha =y/\left( x-x^{\prime }\right) .$ To define them, consider
the trajectory from family $\alpha $ that arrives at point $\left(
x,y\right) $ of the extended billiard. On its way this trajectory may hit
the $i^{\mathrm{th}}$ wall. In this case $\eta _i$ is equal to the value
of $\xi _i$ at the point of reflection. Otherwise, $\eta _i=0.$ For
example, for class I trajectories, when $0\leq x\leq 2X$ and $0\leq y\leq
2Y,$
\begin{equation}
\eta _2\left( x,y;\alpha \right) =\left\{ 
\begin{array}{l}
0,\quad 0\leq y\leq x\tan \alpha \leq 2X\tan \alpha , \\ 
\xi _2\left( y-x\tan \alpha \right) ,\quad x\tan \alpha \leq y\leq Y+x\tan
\alpha , \\ 
\xi _2\left( 2Y+x\tan \alpha -y\right) ,\quad Y+x\tan \alpha \leq y\leq 2Y.
\end{array}
\right.  \label{tdwet}
\end{equation}
$\eta _i$'s have period $2X$ in $x.$ Note that $\eta _2$ is discontinuous at
wall 2, similar for other $\eta _i$'s. What complicates things even more,
the extended scheme is only $\emph{approximate}$ in the perturbed billiard,
in which the domains may slightly overlap near the boundaries. $\eta _i$ is
double valued near the $i^{\mathrm{th}}$ wall where $\xi _i<0.$ It is equal
there either $\xi _i$ or $0$, depending on which sheet the point is located.
Equation (\ref{tdwet}) can still be used in the boundary regions, but the
inequalities are incorrect there.

Now the length becomes 
\begin{eqnarray}
\ &&\mathcal{L}\left( x,y;x^{\prime }\right) =  \nonumber \\
\ &&\sqrt{\left\{ x-x^{\prime }-2\epsilon \left[ \eta _2\left( x,y\right)
+\eta _4\left( x,y\right) \right] \right\} ^2+\left\{ y-\epsilon \left[ \eta
_1\left( x,y\right) +2\eta _3\left( x,y\right) \right] \right\} ^2}= 
\nonumber \\
&&\ \ \mathcal{L}_0\left( y,\alpha \right) +\epsilon \mathcal{L}_2\left(
x,y;\alpha \right)
\end{eqnarray}
where 
\begin{equation}
\mathcal{L}_0\left( y,\alpha \right) =\frac y{\sin \alpha },\quad
\mathcal{L}_2\left( x,y;\alpha \right) =-2\cos \alpha \left( \eta _2+\eta
_4\right) -\sin \alpha \left( \eta _1+2\eta _3\right) . 
\end{equation}
When the end point is on the surface of section, i.e.\ substituting
$y=2Y-\epsilon \xi _1\left( x\right) =2Y-\epsilon \eta _1\left( x,0\right)
,$ we find the length that enters the $T$-operator
\begin{equation}
L\left( x,x^{\prime }\right) =\frac{2Y}{\sin \alpha }+\epsilon L_2\left(
x,\alpha \right)
\end{equation}
where $\tan \alpha =2Y/\left( x-x^{\prime }\right) $ and 
\begin{eqnarray}
L_2\left( x,\alpha \right) &=&-2\cos \alpha \left[ \eta _2\left( x,2Y\right)
+\eta _4\left( x,2Y\right) \right]  \nonumber \\
&&\ \ -\sin \alpha \left[ \eta _1\left( x,0\right) +\eta _1\left(
x,2Y\right) +2\eta _3\left( x,2Y\right) \right] .
\end{eqnarray}

We will be interested in the states near a $(p,q)$ resonance. The resonant
trajectory maps point $x^{\prime }$ to $x=x^{\prime }+x_{pq}$ on the upper
part of the PSS, where $x_{pq}=\left( p/q\right) 2X.$ Hence we expand the
phase of $T$-operator in $\Delta x^{\prime }=x^{\prime }-x+x_{pq}\sim
\sqrt{\epsilon }.$ By assumption, only the case when $p\leq q$ is
considered. The solution of Bogomolny's equation $\psi =T\psi $ has the
form [cf.\ Eq.\ (\ref {gchat})]
\begin{equation}
\psi \left( x\right) =\hat \psi \left( x\right) e^{i\left[ p_xx+k\epsilon
f_2\left( x\right) +g\left( x\right) \right] }.
\end{equation}
Here $p_x$ satisfies the resonance condition $p_x/p_y=x_{pq}/2Y$ or
$p_x=k/\sqrt{1+\left( \frac{qY}{pX}\right) ^2}.$ The step-function
$g\left( x\right) $ compensates for the change in the Maslov phase each
time the orbit crosses a domain boundary at $x=0,$ $\pm X,\pm 2X,\ldots \
.$ It satisfies the condition [cf.\ Eq.\ (\ref{tdssf})]
\begin{equation}
g\left( x\right) -g\left( x-x_{pq}\right) =-\frac \pi 2\nu \left(
x,x-x_{pq}\right)
\end{equation}
that has a solution 
\begin{equation}
g\left( x\right) =-\pi \left[ x/X\right] _{\mathrm{int}}
\end{equation}
where $\left[ \ldots \right] _{\mathrm{int}}$ is an integer part. Function
$\hat \psi \left( x\right) $ satisfies the Schr\"odinger equation
\begin{equation}
\hat \psi ^{\prime \prime }+2dk^2\epsilon \left[ E_m-\bar V_q\left( x\right)
\right] \hat \psi =0,  \label{ptSch}
\end{equation}
where $d=\left( \sin \alpha \right) ^3/2Y,$ with the effective potential 
\begin{equation}
\bar V_q\left( x\right) =-\left\langle L_2\left( x,\alpha \right)
\right\rangle _q.
\end{equation}
The angle $\alpha $ has the resonant value: $\tan \alpha _{pq}=qY/pX.$ Note
that $d$ and $\bar V_q$ are defined with the opposite sign from the circular
billiard. When the walls are deformed towards outside of the billiard, the
potential tends to decrease, thus supporting a localized state. In the
circular billiard, on the other hand, the states concentrate where the
billiard is deformed inwards.

The Schr\"odinger equation should be solved with the boundary condition
$\hat \psi \left( x+2X\right) =\hat \psi \left( x\right) e^{-i2\pi \delta
},$ where $\delta $ is the fractional part of $2Xp_x/2\pi .$ This
condition makes $\psi \left( x\right) $ periodic with period $2X.$ Once
$E_m$'s are known, the total energy can be found from the second
quantization condition
\begin{equation}
\frac{2Yk}{\sin \alpha }-2\pi \frac pq\left[ \frac{2Xp_x}{2\pi }\right]
_{\mathrm{int}}-k\epsilon E_m=2\pi n.  \label{ptqc}
\end{equation}
The part of $f_2(x)$ that vanishes under $q$-average is 
\begin{equation}
\tilde f_2(x)=\frac 1q\sum_{r=1}^{q-1}r\left[ \left\langle L_2\left(
x,\alpha \right) \right\rangle _q-L_2\left( x-rx_{pq},\alpha \right) \right]
.
\end{equation}
The $q$-periodic part can be found from the equation similar to Eq.\ (\ref
{totf2q}), but this time it will include the derivative of the perturbation,
since $L_2$ is not of the special form (\ref{pcbL2}).

\subsection{Two-dimensional wavefunction}

The wavefunction $\Psi \left( \mathbf{r}\right) =\int dx^{\prime }\tilde
G\left( \mathbf{r},x^{\prime }\right) \psi \left( x^{\prime }\right)$, where
the integral is evaluated in the stationary phase approximation ($S\Phi $).
The kernel 
\begin{equation}
\tilde G\left( \mathbf{r},x^{\prime }\right) =\frac 1{2\sqrt{2\pi }}\left|
\frac{\partial ^2\mathcal{L}\left( \mathbf{r};x^{\prime }\right)
}{\partial r_{\perp }\partial x^{\prime }}\right|
^{1/2}e^{ik\mathcal{L}\left( \mathbf{r};x^{\prime }\right) -i\frac \pi
2\nu \left( \mathbf{r};x^{\prime }\right) }
\label{tdw2G}
\end{equation}
where $r_{\perp }$ is the direction perpendicular to the trajectory at point 
$\mathbf{r.}$ The Maslov phase $\nu \left( \mathbf{r};x^{\prime }\right) $
changes by 2 on every boundary. The lowest order stationary phase condition
selects one point 
\begin{equation}
x^{\prime }\left( \mathbf{r}\right) =x-y\cot \alpha _{pq}  \label{tdw2x}
\end{equation}
indicating that only one classical trajectory contributes to the integral.
This is the trajectory that approximately makes angle $\alpha _{pq}$ with
axis $x$ and arrives at point $\mathbf{r.}$ Keeping terms to order $\epsilon 
$ in the phase and dropping the constant factors we find the two-dimensional
wavefunction 
\begin{equation}
\Psi (x,y)=\varphi (x,y)e^{ig\left( x^{\prime }\right) -i\frac \pi 2\nu
\left( \mathbf{r};x^{\prime }\right) }
\end{equation}
where 
\begin{equation}
\varphi (x,y)=\hat \psi \left( x^{\prime }\right) e^{i\mathbf{p\cdot
r}+ik\epsilon \left\{ f_2\left( x^{\prime }\right) -\frac y{2Y}\left[
E_m-\bar V_q\left( x^{\prime }\right) \right] +\mathcal{L}_2\left(
\mathbf{r};\alpha _{pq}\right) \right\} }
\end{equation}
and $x^{\prime }$ is given by Eq.\ (\ref{tdw2x}). $\Psi (x,y)$ has period 
$2X$ in $x$-direction.

\begin{figure}[tbp]
{\hspace*{2.7cm} \psfig{figure=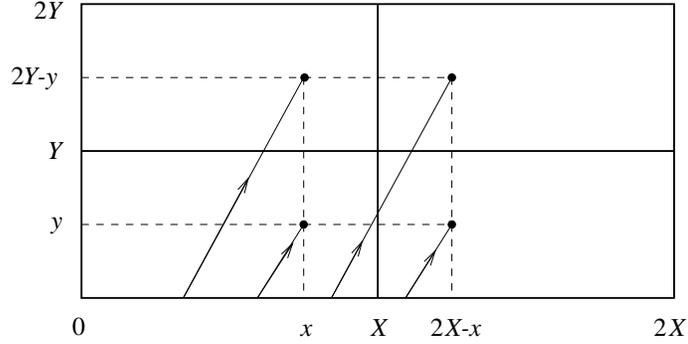,height=4.48cm,width=9cm,angle=0}}
{\vspace*{.13in}}
\caption{Four classical orbits that contribute to the wavefunction in a 
rectangle. 
\label{3_3}}
{\vspace{1.2 cm}}
\end{figure}

Each point in the original $X\times Y$ domain has four images in the
extended billiard (up to a translation by $2X$). Hence the physical
wavefunction $\Psi _{\mathrm{ph}}(x,y)$ is the sum of four respective
terms $\Psi (x,y)$ (Fig.\ \ref{3_3}). The combination $g\left( x^{\prime
}\right) -\frac \pi 2\nu \left( \mathbf{r};x^{\prime }\right) $ is
constant within one domain but differs by $\pi $ between neighboring
domains. Therefore the complete wavefunction is
\begin{equation}
\Psi _{\mathrm{ph}}(x,y)=\varphi (x,y)-\varphi (x,2Y-y)-\varphi
(2X-x,y)+\varphi (2X-x,2Y-y).  \label{tdw2Ps}
\end{equation}

This function vanishes on the physical boundary of the billiard to order
$\epsilon $. Consider, for example, the left border where $x=\epsilon \xi
_2(y).$ In the first term of Eq.\ (\ref{tdw2Ps}) $x^{\prime }\simeq -y\cot
\alpha .$ In the third term $x^{\prime }\simeq 2X-y\cot \alpha .$ The
phase difference between these terms is $p_x\left[ 2\epsilon \xi
_2(y)-2X\right] +2\pi \delta -2k\epsilon \cos \alpha \left[ \eta _2\left(
0^{+},y\right) -\eta _2\left( 2X^{-},y\right) \right] ,$ which is a
multiple of $2\pi .$ [The last term comes from $k\epsilon \mathcal{L}_2;$
$\eta _2\left( 0^{+},y\right) =\xi _2(y)$ and $\eta _2\left(
2X^{-},y\right) =0$]. Thus the first and third terms in Eq.\
(\ref{tdw2Ps}) cancel. The second and fourth terms cancel as well.
Similarly, the wavefunction vanishes on the other sides of the rectangle.
In particular, in order for it to vanish on the lower side, the
quantization condition (\ref{ptqc}) must be satisfied. 

\begin{sloppypar}
Setting $\epsilon =0,$ we find the familiar result for an unperturbed
billiard: $\Psi _{\mathrm{ph}}(x,y)$ $\propto \sin \left( \pi mx/X\right)
\sin \left( \pi ny/Y\right) $ and $k_{mn}=\sqrt{\left( \pi m/X\right)
^2+\left( \pi n/Y\right) ^2}.$ For generic $X$ and $Y$ the resonance
condition $p_x/p_y=pX/qY$ is incompatible with the quantized momenta
$p_x=\pi m/X$ and $p_y=\pi n/Y$ for any integer $p$ and $q$. This is of no
surprise because the resonances are irrelevant for the quantum unperturbed
billiard. 
\end{sloppypar}

\subsection{Example: tilted square}

\label{ets}

We illustrate the results of previous sections with the example of tilted
square. This is a trapezoid resulted from tilting one side of the square
by a small amount. This billiard was studied in a different context \cite
{KapHel}, but no specific states were mentioned. For definitiveness,
suppose the vertices of the trapezoid are located at points $(0,0),$
$(0,1),$ $(1,1)$, and $(1,\epsilon ).$ The lower side $y=\epsilon x$,
$0\leq x\leq 1,$ is chosen as the PSS. We will be interested in the states
near $(1,1)$ resonance, which are characterized by the diamond shaped
classical orbits. The PSS wavefunction $\psi \left( x\right) =\exp \left(
i\kappa x\right) \hat \psi \left( x\right) ,$ where $\kappa =k/\sqrt{2},$
indicates that the unperturbed orbits make angle $\alpha =45^{\circ }$
with the sides of the square. Note that, as in any $(1,1)$ resonance,
function $f_2$ appears only in the third order theory. 

The perturbed part $\hat \psi \left( x\right) $ satisfies the
Schr\"odinger equation (\ref{ptSch}) with the effective potential in the
extended scheme $V(x)=\sqrt{2}\left| x\right|$ ($\left| x\right| \leq 1$).
$V(x)$ is repeated with period 2. The requirement for $\psi \left(
x\right) $ to be of period 2 makes $\hat \psi \left( x+2\right)
=e^{i2\kappa }\hat \psi \left( x\right) .$ Equation (\ref{ptSch}) is a
piece-wise Airy equation. For the well-localized states deep inside the
wells the solutions are approximately
\begin{equation}
\hat \psi _m\left( x\right) =\mathrm{Ai}\left[ \left(
\frac{\sqrt{2}}{\mathcal{L}}k^2\epsilon \right) ^{1/3}x-z_m\right] ,\quad
0\leq x\leq 1,
\end{equation}
where $\mathcal{L}=\sqrt{8}$ is the length of the unperturbed orbit, $z_m$
is an extremum (zero) of $\mathrm{Ai}(-z)$ for even (odd) $m$ and $\hat
\psi _m\left( -x\right) =(-1)^m\hat \psi _m\left( x\right) .$ The
eigenenergy $E_m=z_m \left(2 \mathcal{L}/ k^2\epsilon \right) ^{1/3}$ and
the total energy
\begin{equation}
E_{nm}=k^2\simeq \left( \frac{2\pi n}{\mathcal{L}}\right) ^2\left[
1+z_m\left( \frac{2\epsilon }{\pi n}\right) ^{2/3}\right] .
\end{equation}
A state can be considered deep inside the well if $E_m\ll \max V(x),$ i.e.\
$z_m\ll \left( \sqrt{2}k^2\epsilon /\mathcal{L}\right) ^{1/3}.$

\begin{figure}[tbp]
{\hspace*{2.7cm} 
\psfig{figure=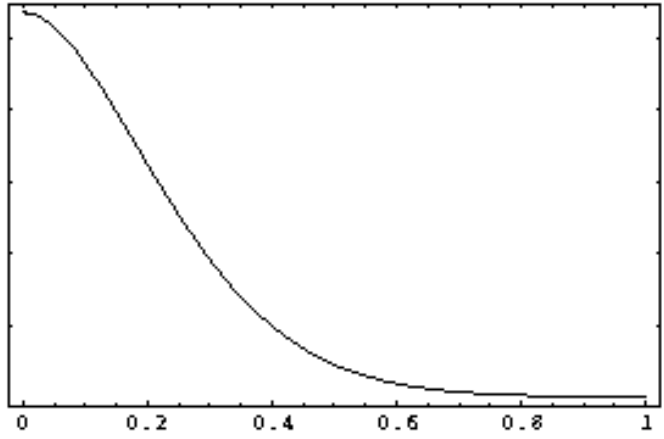,height=6.12cm,width=9cm,angle=0}}
{\vspace*{.13in}}
\caption[Function $\hat \psi_0 (x)$ for the tilted square.]
{Function $\hat \psi_0 (x)$ for the tilted square with $\epsilon
= 0.03$, state $n=50$, $m=0$. 
\label{3_4}}
{\vspace{1.2 cm}}
\end{figure}

\begin{figure}[tbp]
{\hspace*{2.7cm} 
\psfig{figure=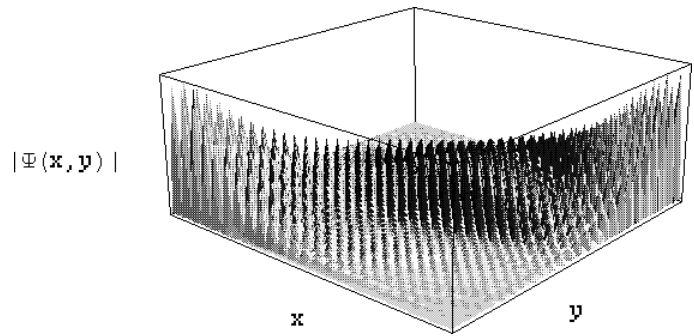,height=4.44cm,width=9cm,angle=0}}
{\vspace*{.13in}}
\caption[Two-dimensional wavefunction in the tilted square.] 
{The absolute value of the theoretical wavefunction to order
$k\sqrt{\epsilon}$ in the tilted square is plotted for the parameters of
Fig.\ 3.4. 
\label{3_5}}
{\vspace{1.2 cm}}
\end{figure}

\begin{figure}[tbp]
{\hspace*{2.7cm} 
\psfig{figure=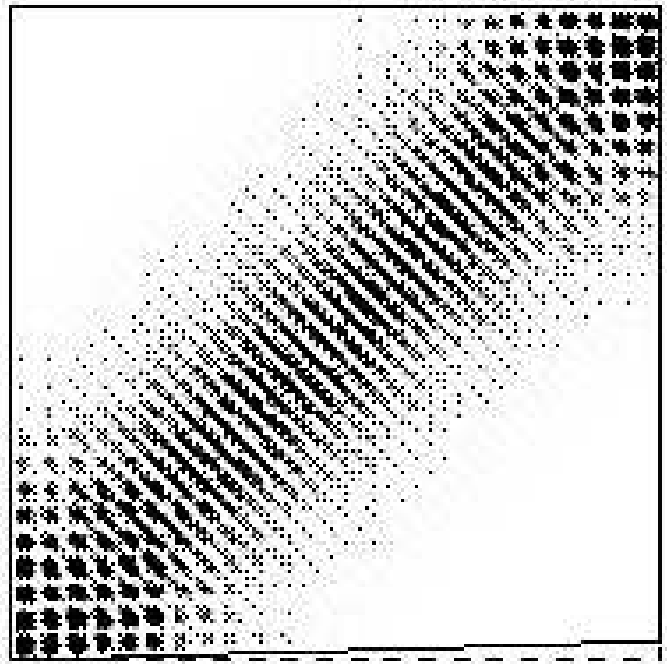,height=9.38cm,width=9cm,angle=0}}
{\vspace*{.13in}}
\caption[Density plot for the wavefunction of Fig.\ 3.5.]
{Density plot for the wavefunction of Fig.\ 3.5. The dashed line 
is $y=0$.
\label{3_6}}
{\vspace{1.2 cm}}
\end{figure}

\begin{figure}[tbp]
{\hspace*{2.7cm} \psfig{figure=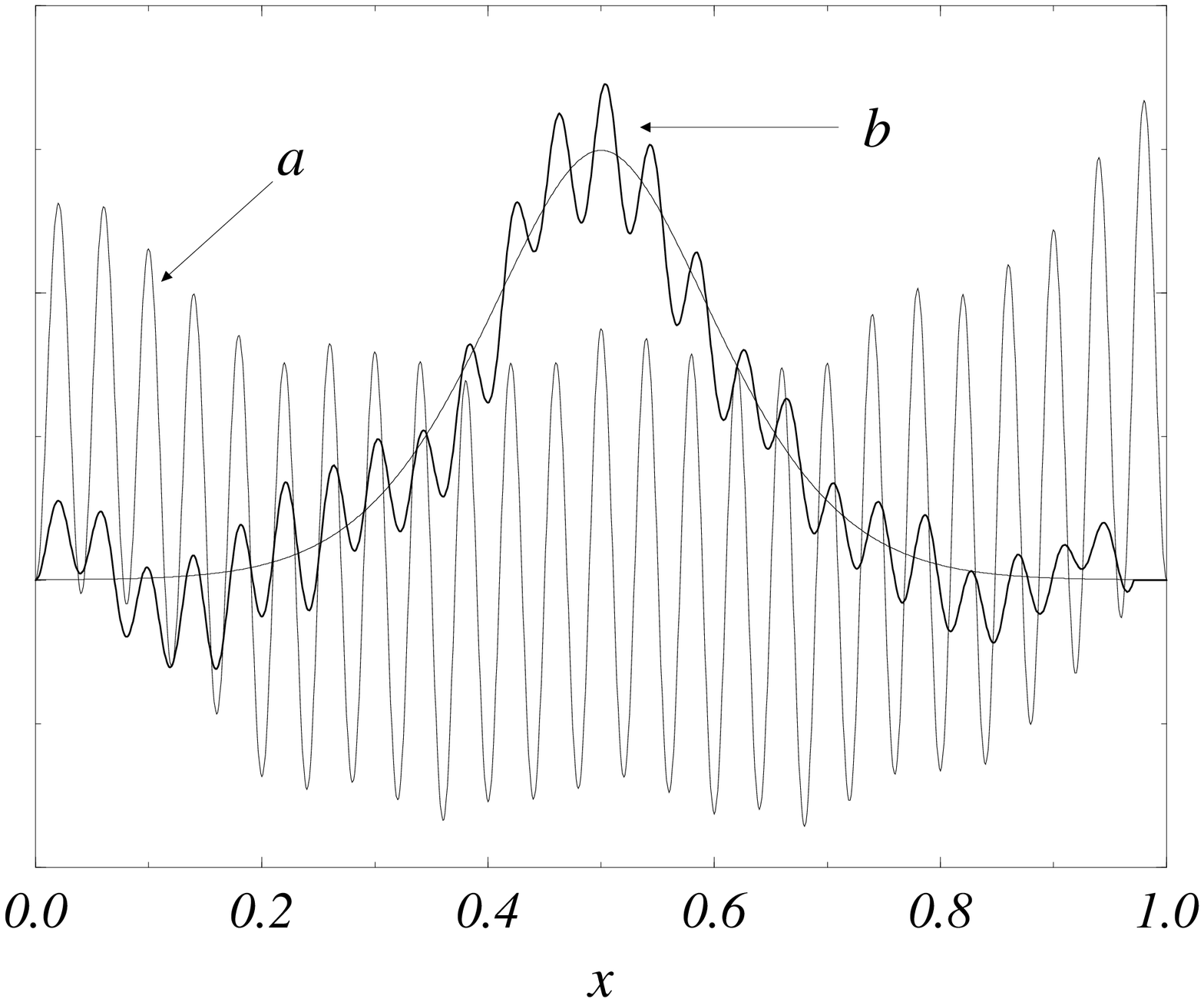,height=7.6cm,width=9cm,angle=270}}
{\vspace*{.13in}}
\caption[The cross-section of the numerical wavefunction for the state of
Figs.\ 3.4-3.6 along the diagonals.]
{The cross-section of the numerical wavefunction for the state of Figs.\
3.4-3.6 along the diagonals. (a) $\Psi (x,x)$, (b) $\Psi (x,1-x)$. The
theoretical wavefunction, which is smoother, is also shown in (b). The
diagonal of the unperturbed square $y = 1-x$ crosses the boundary at $x =
1/(1 + \epsilon)$, which explains the cut-off in (b). 
\label{3_7}}
{\vspace{1.2 cm}}
\end{figure}

Figure \ref{3_4} shows the lowest surface of section state $\hat \psi
_0\left( x\right) .$ It is localized near $x=0,$ which for the $\left(
1,1\right) $ resonance means that the two-dimensional state is
concentrated along the diagonal $y=x.$ To check this one looks at the
two-dimensional wavefunction in the physical domain that has the form
\begin{eqnarray}
&&\Psi (x,y)=e^{i\kappa (x+y)+ik\epsilon \sqrt{2}\left[
-yE_m+(y-1)V(x-y)\right] }\hat \psi _m(x-y)  \nonumber \\
\ \ \ &&+e^{i\kappa (-x-y)+ik\epsilon \sqrt{2}\left[ yE_m+(1-y)V(x-y)\right]
}\hat \psi _m(-x+y)  \nonumber \\
\ \ &&\ -e^{i\kappa (-x+y)+ik\epsilon \sqrt{2}\left[
-yE_m+(y-1)V(x+y)\right] }\hat \psi _m(-x-y)  \nonumber \\
\ &&\ -e^{i\kappa (x-y)+ik\epsilon \sqrt{2}\left[ yE_m+(1-y)V(x+y)\right]
}\hat \psi _m(x+y),\ 0\leq x,y\leq 1.
\end{eqnarray}
In Figs.\ \ref{3_5} and \ref{3_6} we plotted a simplified version of this
function where we neglected terms of order $k\epsilon $ and set $\kappa
=\pi n/2.$ To make the resulting function satisfy the boundary conditions
we made a substitution $y \rightarrow (y - \epsilon x)/(1 - \epsilon x)$,
which does not change the order of approximation. The reduced function
keeps the gross features of the exact function. The cross-section of the
state along the diagonal
\begin{equation}
\Psi (x,x)\simeq \left\{ 
\begin{array}{l}
\hat \psi _m(2x)-\hat \psi _m(0)\cos (\pi nx),\ m\ \mathrm{even} \\ 
0,\ m\ \mathrm{odd}
\end{array}
\right. \text{\quad (}0\leq x\leq 1\text{)}
\end{equation}
is shown in Fig.\ \ref{3_7} for a numerical wavefunction, as well as the
section across another diagonal $y=1-x$
\begin{equation}
\Psi (x,1-x)\simeq \left\{ 
\begin{array}{l}
\hat \psi _m(2x-1),\ m,n\ \mathrm{both\ even\ or\ both}\ \mathrm{odd} \\ 
0,\ \mathrm{otherwise}
\end{array}
\right. ,
\end{equation}
both numerical and theoretical. The line $y=1-x$ is the diagonal of the
unperturbed square, not the trapezoid. It crosses its boundary at
$x=1/(1+\epsilon )$ which explains the cut-off in the figure. 

Note that the neglect of the $k\epsilon $ order terms in $\kappa $ leads to
the symmetries $\hat \psi _m\left( -x\right) =(-1)^m\hat \psi _m\left(
x\right) $ and $\hat \psi _m\left( x+2\right) =(-1)^n\hat \psi _m\left(
x\right) $ for \emph{all }$x\in \left( -\infty ,+\infty \right) .$ This is
consistent with the fact that the potential $V(x)$ is even at $x=0$ and $x=1$
and the Schr\"odinger equation (\ref{ptSch}) is solved with the real
boundary condition. In general, however, the complex condition $\hat \psi
\left( x+2\right) =e^{i2\kappa }\hat \psi \left( x\right) $ leaves a weaker
symmetry $\hat \psi _m\left( -x\right) =(-1)^m\hat \psi _m^{*}\left(
x\right) .$

We have shown that even with a small perturbation of the side of the
square the strongly perturbed localized states can exist, as long as
$k\sqrt{\epsilon }\gg 1.$ By constructing an effective potential it is
easy to see that there are also states concentrated along the side $x=0$
corresponding to $(0,1)$ resonance. We will study these states using the
Born-Oppenheimer approximation when we revisit this billiard in Sec.\
\ref{bbs}. It is also clear that there are no diagonal states in the
symmetric trapezoid with sides $x=0,1,$ $y=\epsilon x,1-\epsilon x,$ but
there are states near $x=0.$ For the parallelogram billiard with sides
$x=0,1,$ $y=\epsilon x,1+\epsilon x $ there are states along the long
diagonal but no states near the edges. 

\section{Billiards without small parameter}

\label{bwsp}

In some cases the perturbation theory that we developed can be applied to a
system that is not a small perturbation of an integrable system. However,
this system may possess certain states that are close to the states in an
integrable system. Then these states could be studied using the perturbation
theory where the small parameter would depend on the state itself.

\begin{figure}[tbp]
{\hspace*{2.7cm} \psfig{figure=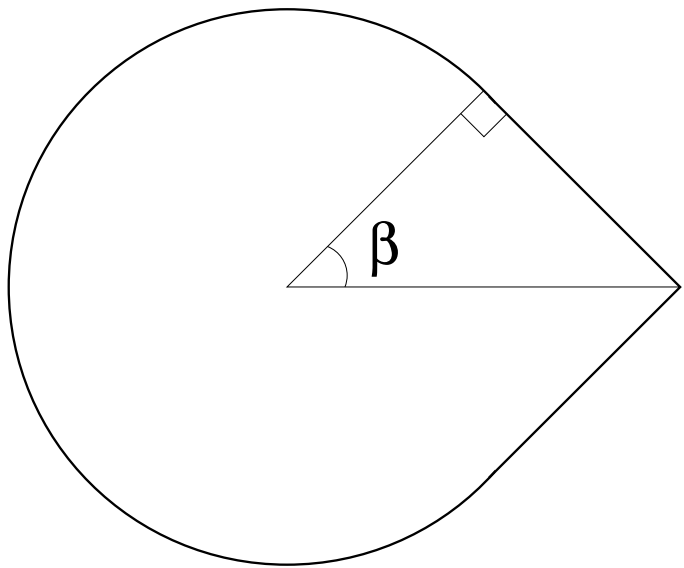,height=7.45cm,width=9cm,angle=270}}
{\vspace*{.13in}}
\caption{Ice-cream cone billiard.
\label{3_8}}
{\vspace{1.2 cm}}
\end{figure}

As a first example, consider the ice cream cone billiard (Fig.\ \ref{3_8}),
which is a unit circle for $\pi >\left| \theta \right| >\beta $ and a
triangular shaped region for $\left| \theta \right| <\beta .$ For
the surface of section states that are concentrated in the circular
region the perturbation theory for the circle can be used. The
perturbation of the boundary, $\Delta R\left( \theta \right) =1/\cos
\left( \beta -\theta \right) $ for $\left| \theta \right| <\beta $ and
zero elsewhere, grows as $\theta $ goes into the triangular region.
Consequently, the effective potential $\bar V_q\left( \theta \right) $ is
large near $\theta =0$, repeated with period $ 2\pi /q.$ The PSS
wavefunction obeys the Schr\"odinger equation (\ref{gcSch}) with
$\epsilon =1.$ Clearly, the state will not significantly penetrate the
non-circular region as long as $E_m\ll \max \bar V_q=q^{-1}\sin \left(
\Theta _{pq}/2\right) /\cos \beta $ ($2\pi /q>$ $2\beta $). This defines
the small parameter that depends on $E_m.$ For example, the states of the
low angular momentum resonance with small $E_m$ are concentrated near the
diameter at $\theta =\pm \pi /2.$ There is also an upper limit on the
denominator $q$: the potential becomes shallower when $2\pi /q$ is smaller
than $2\beta .$

\begin{figure}[tbp]
{\hspace*{2.7cm} \psfig{figure=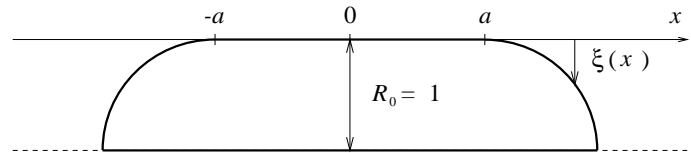,height=2cm,width=9cm,angle=270}}
{\vspace*{.13in}}
\caption[Bunimovich stadium billiard.]
{The Bunimovich stadium with the straight sides of length $2a$ and the
endcaps of radius $R_0 = 1$ reduced to the upper half by symmetry.  The
deviation of the upper boundary from the straight line is $\xi (x)$. 
\label{3_9}}
{\vspace{1.2 cm}}
\end{figure}

\begin{figure}[tbp]
{\hspace*{2.7cm} \psfig{figure=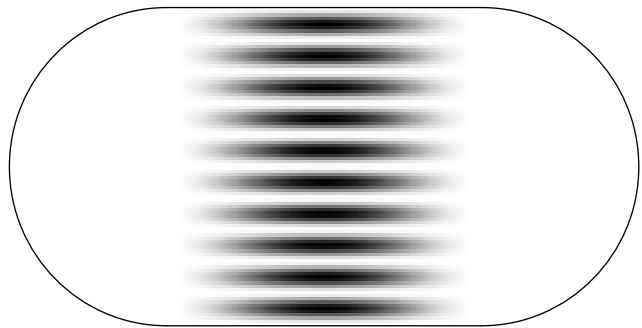,height=4.94cm,width=9cm,angle=0}}
{\vspace*{.13in}}
\caption[Bouncing ball mode in the Bunimovich stadium.]
{Square of the theoretical wavefunction for the bouncing ball state 
$n=10$, $m=1$ in the Bunimovich stadium.  
\label{3_10}}
{\vspace{1.2 cm}}
\end{figure}

\begin{figure}[tbp]
{\hspace*{2.7cm} 
\psfig{figure=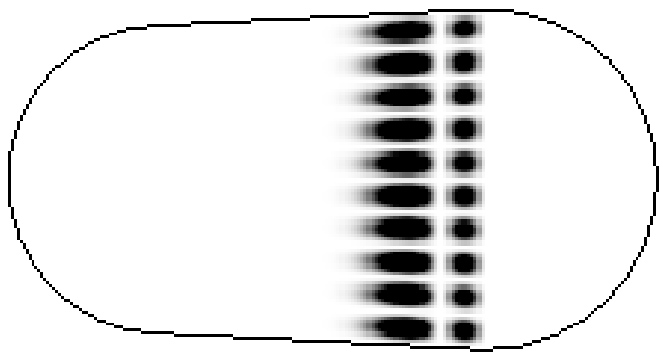,height=5.07cm,width=9cm,angle=0}}
{\vspace*{.13in}}
\caption[Bouncing ball mode in the slanted stadium.]
{Bouncing ball state in the slanted stadium. The radii differ by $\Delta
R/R = 0.1$. The quantum numbers are $n=10$, $m=2$. 
\label{3_11}}
{\vspace{1.2 cm}}
\end{figure}

\begin{figure}[tbp]
{\hspace*{2.7cm} 
\psfig{figure=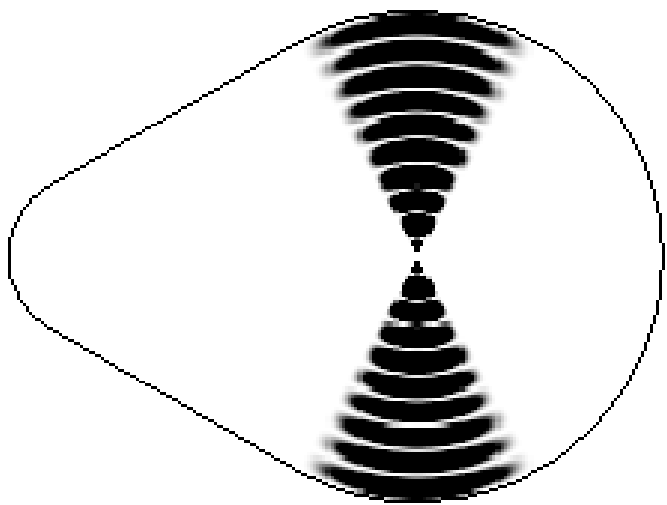,height=7.06cm,width=9cm,angle=0}}
{\vspace*{.13in}}
\caption[Localized state in the ``baseball'' stadium.]
{Localized state in the ``baseball'' stadium. The cap radii are $R_1 = (1/
\sqrt {20}) \ell$, $R_2 = (3/ \sqrt {20}) \ell$, where
$\ell$ is the distance between them. If treated as a perturbed
circle the quantum numbers are $n=10$, $m=1$. The theoretical wavefunction
is inaccurate near the center of the billiard (cf.\ Sec.\ 2.4). 
\label{3_12}}
{\vspace{1.2 cm}}
\end{figure}

Another example are the ``bouncing ball'' (BB) states in a stadium
billiard \cite{BacSchSti,BaiHosSte,Tan,KelRub}. The billiard has the
straight segments of length $2a$ and the endcaps of radius 1, and we can
reduce the billiard to its upper half by symmetry (Fig.\ \ref{3_9}). We
are looking for the states that have low momentum parallel to the sides
and large perpendicular momentum. If such states are concentrated outside
of the endcaps, they will not be too different from the localized
$(0,1)$-states in a perturbed infinite channel or a long rectangle. The
deviation of the upper boundary from the straight line is described by the
function $\xi (x)=0,$ $\left| x\right| <a,$ $\xi (x)\approx \left( \left|
x\right| -a\right) ^2/2,$ $\left| x\right| >a.$ The effective potential
$V(x)=2\xi (x)$ is multiplied by a large factor of $k^2$ in the
Schr\"odinger equation, thus it is almost a square well with $V\rightarrow
\infty $ for $\left| x\right| >a.$ Therefore the states are contained in
the straight region if $E_m\ll 1.$ Since $E_m\sim \left(m/
ka\right) ^2,$ the billiard should be longer than $m/k$ to apply the
method. (The opposite case is a perturbed circle.) The ratio of the
parallel and perpendicular momenta $p_x/p_y\approx \sqrt{E_m}$ is small,
as expected. Since we consider only half of the billiard, we need to find
the states with both Dirichlet's and Neumann's conditions on the lower
boundary. In the latter case the $T$-operator has an additional Maslov
phase $\pm \pi ,$ which should be added to the quantization condition
(\ref{ptqc}). The theoretical state $n=10,$ $m=1$ is shown in Fig.\ 
\ref{3_10}. The similar results for the BB states can be obtained by the
Born-Oppenheimer approximation \cite{BaiHosSte}. The theory can be
generalized, of course, to include the perturbation of the boundary of the
stadium. For example, the radii of the endcaps can be slightly different
so that the straight segments are tilted \cite{PriSmi}. Then, from what we
know, the BB states will be shifted towards the wider end (Fig.\ \ref{3_11}).
When the sides are strongly tilted we have an ice cream cone billiard
(Fig.\ \ref{3_12}). 

Kudrolli \emph{et al.\ }\cite{KudAbrGol} observed the surface wave
patterns in a Bunimovich stadium filled with liquid. The modes were excited
by oscillating the container vertically with an appropriate frequency. Only
three types of all possible modes were actually observed, including the BB
mode. It was argued \cite{AgaAlt} that the whispering gallery mode and the
excitations of other types do not appear due to strong boundary dissipation.

\begin{figure}[tbp]
{\hspace*{2.7cm} \psfig{figure=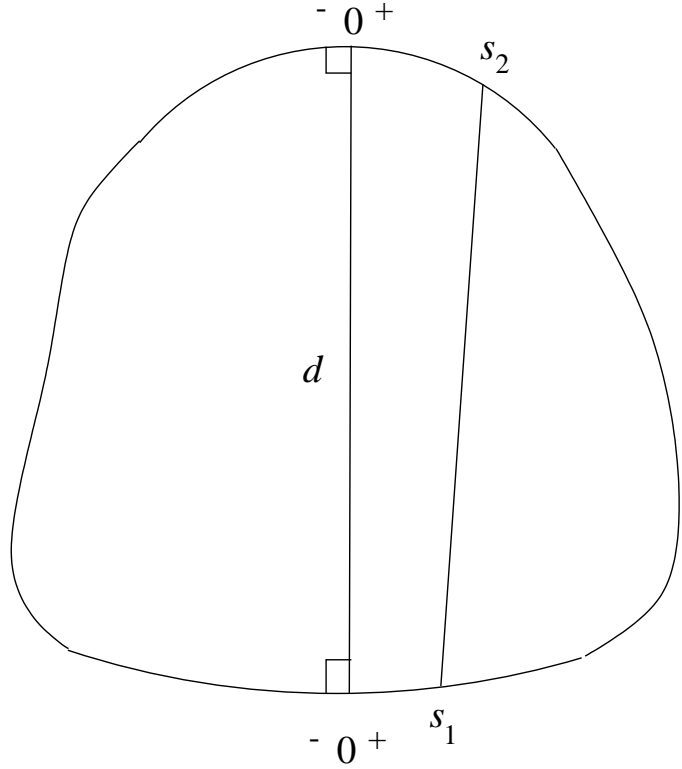,height=10.1cm,width=9cm,angle=0}}
{\vspace*{.13in}}
\caption[Short periodic orbit and a nearby orbit in a generic billiard.]
{Short periodic orbit of length $2d$ and a nearby orbit between points
$s_1$ and $s_2$ in a generic billiard. The positive and negative directions 
are indicated. 
\label{3_13}}
{\vspace{1.2 cm}}
\end{figure}

The notion of BB states can be generalized to the states near a short
periodic orbit that connects opposite sides in a generic billiard \cite
{Nar,BoyGor} (Fig.\ \ref{3_13}). Assume constant radii of curvature $R_1$
and $R_2$ in the neighborhoods of the points joined by the periodic orbit.
Then the length of a nearby chord in terms of the local coordinates along
the boundary is
\begin{equation}
L(s_1,s_2)\cong d-\frac{s_1^2}{2R_1}-\frac{s_2^2}{2R_2}+\frac{\left(
s_1-s_2\right) ^2}{2d}  \label{bwspL}
\end{equation}
where $2d$ is the length of the periodic orbit and $s_1$ and $s_2$ measure
the distance from the periodic orbit along the boundary with
counterclockwise and clockwise positive direction, respectively (see Fig.\
\ref{3_13}). Let the PSS coincide with the boundary. We define the PSS
wavefunctions locally on each side and assume that they are well
localized. Then we need to solve a couple of Bogomolny's equations
\begin{eqnarray}
\psi _1\left( s_1\right) &=&\int T\left( s_1,s_2\right) \psi _2\left(
s_2\right) ds_2,  \nonumber \\
\psi _2\left( s_2\right) &=&\int T\left( s_2,s_1\right) \psi _1\left(
s_1\right) ds_1.
\end{eqnarray}
The formula \cite{GraRyz} 
\begin{equation}
\int_{-\infty }^\infty e^{-\left( x-y\right) ^2}H_m\left( ax\right)
dx=\sqrt{\pi }\left( 1-a^2\right) ^{m/2}H_m\left(
\frac{ay}{\sqrt{1-a^2}}\right)
\end{equation}
suggests looking for the solution in the form 
\begin{equation}
\psi _i\left( s_i\right) =\sqrt{\alpha _i}\exp \left( -\frac k{2d}\alpha
_is_i^2\right) H_m\left( \sqrt{\frac kd\beta _i}s_i\right) \quad \left(
i=1,2\right)
\end{equation}
where $H_m\left( x\right) $ is a Hermite's polynomial. The constants $\alpha
_i$ and $\beta _i$ are determined after substituting the ansatz in the
Bogomolny equations. They are 
\begin{eqnarray}
&&\alpha _1=\sqrt{\frac{\frac d{R_1}-1}{\frac d{R_2}-1}\left[ 1-\left( \frac
d{R_1}-1\right) \left( \frac d{R_2}-1\right) \right] },  \nonumber \\
&&\beta _1=\sqrt{\frac{\frac d{R_1}-1}{\frac d{R_2}-1}}\alpha _1
\end{eqnarray}
and similar for $\alpha _2$ and $\beta _2.$ This approach is valid only if
$0<\left( \frac d{R_1}-1\right) \left( \frac d{R_2}-1\right) <1.$ Also the
localization lengths $\sqrt{d/k\alpha _i}$ should be smaller than the
characteristic scale of the boundary, including $R_i$. The quantization
condition is
\begin{equation}
kd=\pi \left( 2n+\frac m2+\frac 54\right) +\varphi \left( 2m+1\right)
\end{equation}
where 
\begin{equation}
\varphi =\frac 12\mathrm{sgn}\left( \frac d{R_i}-1\right) \arcsin
\sqrt{\left( \frac d{R_1}-1\right) \left( \frac d{R_2}-1\right) }. 
\end{equation}

In the case of equal radii $R_1=R_2=R$ we have the same functions with
$\alpha =\beta =\sqrt{\frac dR\left( 2-\frac dR\right) }$ on each side.
The phase $\varphi =\frac 12\arcsin \left( \frac dR-1\right) .$ Note that
when $d=2R$ the state is delocalized, as expected for a perfect circle.
Coincidentally, we have the correct WKB quantization in this limit,
although the theory obviously fails. In the confocal case $d=R_1=R_2$ the
solution is $\psi \left( s\right) =\exp \left( -\frac k{2d}s^2\right)
H_m\left( \sqrt{\frac kd}s\right) $ with the quantization condition
$kd=\pi \left( 2n+\frac m2+\frac 54\right) .$

\section{Whispering gallery modes}

\label{wgm}

A wide range of billiards can support the states that are concentrated
close to the boundary and called the \emph{whispering gallery} modes
(WGMs). This effect was first discussed by G. B. Airy and Lord Rayleigh
\cite{Ray}, and Keller and Rubinow \cite{KelRub} obtained the quantization
conditions. Apart from the trivial cases the WGMs should be regarded as
quasimodes (see Sec.\ \ref{conc}). The WGMs in the asymmetric resonant
optical cavities with emission have a long life-time and might be useful
in lasers \cite{NocSto}. There is a recent experimental observation of the
chaos-assisted tunneling between the WGMs in a superconducting microwave
cavity \cite{DemGraHei}, as well as of the excitation of WGMs by a vortex
in a Josephson junction \cite{WalUstKur}. 

\begin{figure}[tbp]
{\hspace*{2.7cm} 
\psfig{figure=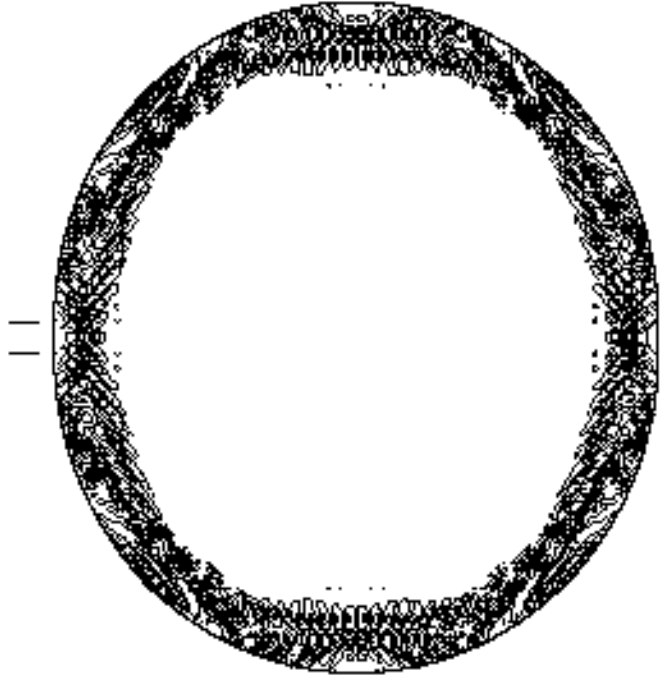,height=9.48cm,width=9cm,angle=0}}
{\vspace*{.13in}}
\caption[Whispering gallery mode in the short stadium.]
{Contour plot of the whispering gallery mode in the short stadium with
$\epsilon = 0.05$ (see Sec.\ 2.2.1). The numerically obtained wavefunction
with $k=242.7611$ is shown. There are no caustics in this case and almost
no classical orbits that stay close to the boundary for a long time. The
parallel lines on the left indicate the length and position of the
straight segments of the boundary. 
\label{3_14}}
{\vspace{1.2 cm}}
\end{figure}

The standard problem can be solved by various analytical methods giving
essentially the same leading order result. Among them the ray method by
Keller and Rubinow, the parabolic equation and the etalon methods are
reviewed in Ref.\ \cite{BabBul}; the Born-Oppenheimer approximation is
discussed in Sec.\ \ref{wgm2}; and the $T$-operator is used in this
section. All these methods are based on the adiabatic assumption that the
curvature of the boundary is slowly varying (the estimate is given below)
and never vanishes. Under these conditions a classical (non-periodic)
orbit is proven \cite{Laz} to have a caustic and an adiabatic invariant,
meaning that the invariant tori can be constructed in the perturbation
theory. If a boundary has at least one point of zero curvature, there are
trajectories infinitely close to the boundary that reverse themselves and
thus do not have a caustic \cite{Mat}. A wider variety of classical
wispering gallery trajectories is found in a magnetic billiard
\cite{RobBer}. It should be noted, however, that the existence of
classical tori is a sufficient but not necessary condition for a quantum
state to exist. For example, Fig.\ \ref{3_14} shows a state localized near
the boundary in a short stadium billiard. We discuss a possible way to
deal with such systems in Sec.\ \ref{sp}. 

To apply the Bogomolny's equation method we need to find the length of the
chord between two points on the boundary. The classical trajectory stays
close to the boundary, so, typically, the two points will be close to each
other and the boundary between them can be approximated by an arc of a
circle. Let $s$ be the distance along the boundary. It can be shown that the
chord length 
\begin{eqnarray}
L\left( s,s^{\prime }\right) &=&\left| s-s^{\prime }\right| -\frac{\left|
s-s^{\prime }\right| ^3}{24\left[ R\left( \frac{s+s^{\prime }}2\right)
\right] ^2}+O\left( \left| s-s^{\prime }\right| ^5\right) \nonumber \\ \
&\cong &\left| s-s^{\prime }\right| -\frac{\left| s-s^{\prime }\right|
^3}{24\left[ R\left( s\right) \right] ^2}-\frac{\left| s-s^{\prime
}\right| ^4R^{\prime }\left( s\right) }{24\left[ R\left( s\right) \right]
^3}
\label{wgmL}
\end{eqnarray}
where $R\left( s\right) $ is the radius of curvature. Although the best
estimate for the radius is at the middle-point, it is more convenient to
express it locally at point $s.$ We make an ansatz for the wavefunction 
\begin{equation}
\psi \left( s\right) = C(s) e^{ik\left[ s-f\left( s\right) \right]}. 
\label{wgmpsi}
\end{equation}
The wavefunction can be found using the first two terms in Eq.\
(\ref{wgmL}). One can show that the inclusion of the (small) third term in 
this form does not result in the $R^\prime /R$-order correction in the 
wavefunction.
Using the stationary phase approximation ($S\Phi $) in $\int T\psi ds$ we
find 
\begin{equation}
f^{\prime }\left( s\right) =\left[ \frac{3\sqrt{2}\pi \left( n-1/4\right) 
}{4kR\left( s\right) }\right] ^{2/3}  \label{wgmf}
\end{equation}
and the prefactor \cite{Zai}
\begin{equation}
C(s) \propto R^{-1/3} (s) \propto  \sqrt {f^\prime (s)}.
\end{equation}
The quantization condition comes from the periodicity of $\psi \left(
s\right) $ over the perimeter of the boundary $\mathcal{L}$, 
\begin{equation}
k\left[ \mathcal{L}-f\left( \mathcal{L}\right) +f\left( 0\right) \right]
=2\pi m.
\end{equation}
Quantum number $m$ indicates the momentum along the boundary and $n$
reflects the motion in the perpendicular direction.

The last term in Eq.\ (\ref{wgmL}) is small if 
\begin{equation}
\left| s-s^{\prime }\right| _{\mathrm{st}}\frac{\left| R^{\prime }\right|
}R\ll 1 \label{wgms}
\end{equation}
where the $S\Phi $ value $\left| s-s^{\prime }\right|
_{\mathrm{st}}=\sqrt{8f^{\prime }}R\sim n^{1/3}R^{2/3}k^{-1/3}$ is a
typical hop the classical orbit makes near point $s.$ Condition
(\ref{wgms}) becomes
\begin{equation}
\left| R^{\prime }\right| \ll \left( kR/n\right) ^{1/3}  \label{wgmapp}
\end{equation}
thus restricting the pace of variation of the curvature. 

\begin{figure}[tbp]
{\hspace*{2.7cm} \psfig{figure=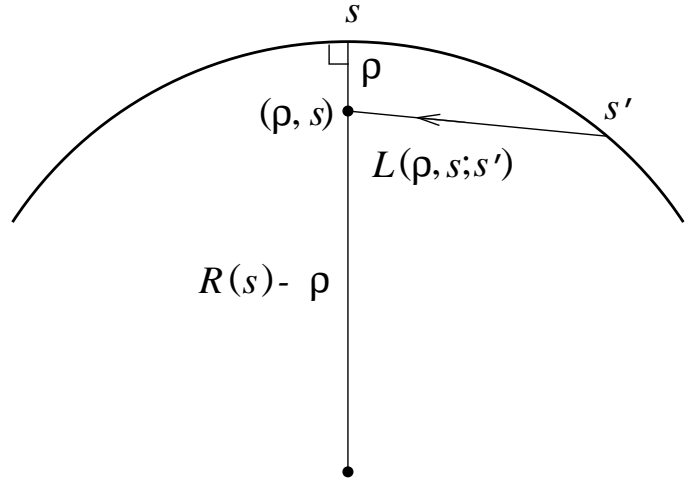,height=6.36cm,width=9cm,angle=0}}
{\vspace*{.13in}}
\caption[Classical orbit of length $L(\rho , s; s^\prime)$ from point
$s^\prime$ on the boundary to point $(\rho , s)$ inside the billiard.]
{Classical orbit of length $L(\rho , s; s^\prime)$ from point $s^\prime$ on
the boundary to point $(\rho , s)$ inside the billiard. The radius of
curvature at point $s$ is $R(s)$. 
\label{3_15}}
{\vspace{1.2 cm}}
\end{figure}

It is convenient to describe the motion in two dimensions in the $\left(
\rho ,s\right) $ coordinates where $\rho $ is the distance from the boundary
positive inside the billiard. These coordinates are well-defined close to
the boundary, where $\rho \ll R(s).$ The two-dimensional function can be
found by the standard technique, 
\begin{equation}
\Psi \left( \rho ,s\right) =\int \tilde G\left( \rho ,s;s^{\prime }\right)
\psi \left( s^{\prime }\right) ds^{\prime }
\end{equation}
where $\tilde G(\rho ,s;s^{\prime })\propto \exp \left[ ikL\left( \rho
,s;s^{\prime }\right) \right] .$ The distance $L\left( \rho ,s;s^{\prime
}\right) $ between point $s^{\prime }$ on the boundary and point $\left(
\rho ,s\right) $ inside the billiard is shown in Fig.\ \ref{3_15}. As in
the case of a circle (Sec.\ \ref{tdw}), there are two classical orbits
(for fixed $m$ and $n$) arriving at point $\left( \rho ,s\right) $ that
satisfy the $S\Phi $ condition. They leave the boundary at points
\begin{equation}
s_{1,2}^{\prime }=s-\sqrt{2}R\left( \sqrt{f^{\prime }}\pm \sqrt{f^{\prime
}-\frac \rho R}\right) .
\end{equation}
When we sum the contributions from these two stationary points we get the
wavefunction 
\begin{equation}
\Psi \left( \rho ,s\right) =\frac{\psi \left( s\right)
}{\left(f^{\prime }-\frac \rho R \right)^{1/4}}\cos \left[
\frac{2\sqrt{2}}3kR\left( f^{\prime }-\frac \rho R\right) ^{3/2}-\frac \pi
4\right] .  \label{wgmP}
\end{equation}
The function is defined in the classically allowed
region $\rho <f^{\prime }R.$ Clearly, $\rho =f^{\prime }R$ is an equation
for the caustic. Note that $\Psi \left( \rho ,s\right) $ vanishes on the
boundary, when $\rho =0.$

The $\rho $-dependent part of $\Psi \left( \rho ,s\right) $ is an
asymptotic form of the Airy function \cite{AbrSte} $\mathrm{Ai}\left(
-z\right) $ for $z=2^{1/3}\left( kR\right) ^{2/3}\left( f^{\prime }-\rho
/R\right) \gg 1,$ i.e.\ not too close to the caustic. Function (\ref{wgmP})
can be rewritten approximately as
\begin{equation}
\Psi \left( \rho ,s\right) =\left( kR\right) ^{-1/6}\mathrm{Ai}\left(
\frac{2^{1/3}k^{2/3}}{R^{1/3}}\rho -z_n\right) \exp \left[ i\int
l\left( s\right) ds/R\right] 
\label{wgmAi}
\end{equation}
where $l\left( s\right)
=kR\left( 1-f^{\prime }\right) $ is an angular momentum about the local
center of curvature and $z_n$ is the $n$th root of $\mathrm{Ai}\left(
-z\right) .$ Alternatively, the Airy function can be expressed in terms of
the Bessel function $J_{l\left( s\right) }\left[ k\left( R-\rho \right)
\right] $ with variable index $l\left( s\right) \gg 1.$ Hence, Eq.\ (\ref
{wgmAi}) can be interpreted locally as an eigenfunction for a circle of
radius $R\left( s\right).$ We will return to this point in Sec.\ \ref{wgm2}
where the Bessel function will follow directly from the Born-Oppenheimer
approximation.

\section{Scattering problem}

\label{sp}

As was mentioned in the previous section, the billiards that do not
support classical whispering gallery orbits may still have the quantum
states that are localized near the boundary (perhaps for finite energy).
The candidates for this behavior are the boundaries that have short
regions of zero, or even negative, curvature. These regions act like
scatterers on the whispering gallery waves that can still exist along the
remaining parts of the boundary. The scattering mixes the waves with
different transverse quantum numbers $n$ defined in the previous section.
As a consequence, the proper stationary states (or, strictly speaking,
quasimodes \cite{Arn}) are the linear combinations of states with
different $n$'s.  If only a limited number of $n$'s contribute to an
eigenstate, this eigenstate can be localized near the boundary, and thus
look like a whispering gallery mode. 

We illustrate these ideas with a model of a boundary that has one point of
zero curvature. The work is still in progress. Specifically, we assume 
that the curvature in the neighborhood of this point, say $s=0$, can be 
described by the simplest analytic expression 
\begin{equation}
\frac 1 {R(s)} = \frac {\mu s^2} 2 
\end{equation}
where $\mu$ is a dimensional parameter. [In the Cartesian coordinates this
boundary is locally $y(x) = (\mu/ 4!) x^4$ where zero curvature is at
$x=0$.] The boundary is assumed closed in a regular fashion for $|s|$
greater than all relevant scales. 

\subsection{Classical theory}

Suppose a classical whispering gallery orbit is incident on the singular
region from $s < 0$. This orbit generates a map in the phase space of the
boundary $(\epsilon, s)$, where $\epsilon$ is a (small) angle between the
orbit and the boundary at the point of reflection $s$. In order to derive
the map equations, we introduce, following the recipe in Ref.\ 
\cite{BabBul}, the local Cartesian coordinates with the origin at a 
reflection point $s_i$. The $x$ axis touches the boundary such that $x$ 
increases with $s$ and the $y$ axis points inside the boundary. 

Now we can write the parametric equation of the boundary
\begin{eqnarray}
x(s) &=& \int_{s_i}^s \cos \left[ \int_{s_i}^{s^\prime} \frac {d \xi} 
{R(\xi)} \right] ds^\prime = s - s_i + O\left(s^7 \mu^2\right), \nonumber \\
y(s) &=& \int_{s_i}^s \sin \left[ \int_{s_i}^{s^\prime} \frac {d \xi} 
{R(\xi)} \right] ds^\prime \nonumber \\
&=& \frac \mu 6 \left[ \frac 1 4 \left(s^4 - s_i^4 \right) - s_i^3 \left(s -
s_i \right) \right]+ O\left(s^{10} \mu^3\right). 
\end{eqnarray}
From this we find the angle of reflection
\begin{eqnarray}
\epsilon_i &\approx& \tan \epsilon_i = \frac {y (s_{i+1})} {x (s_{i+1})} 
\nonumber \\
&=& \frac \mu {24} \left[ s_{i+1}^3 + s_{i+1}^2 s_i + s_{i+1} s_i^2 -3 
s_i^3 \right] + O\left(s^9 \mu^3 \right) \label{ctmap1}
\end{eqnarray}
where $s_{i+1}$ is the next reflection point. Another equation results 
from the elementary geometry 
\begin{equation}
\epsilon_{i+1} + \epsilon_i = \int_{s_i}^{s_{i+1}} \frac {ds} {R(s)} = 
\frac \mu 6 \left[ s_{i+1}^3 - s_i^3 \right]. \label{ctmap2}
\end{equation}
Equations (\ref{ctmap1}) and (\ref{ctmap2}) provide the approximate 
classical map. The first is the cubic equation for $s_{i+1}$. Once 
$s_{i+1}$ is known, it can be used in the second equation to obtain 
$\epsilon_{i+1}$. 

Far from the region of small curvature, where
\begin{equation}
\left| \Delta s_i \right| \equiv \left| s_{i+1} - s_i \right| \ll s_i,
\label{cts}
\end{equation}  
we can expand the map equations in $\Delta s_i$ to express
\begin{equation}
\Delta \epsilon_i \equiv \epsilon_{i+1} - \epsilon_i \approx \frac \mu 6 
s_i \Delta s_i^2 \approx \frac 2 3 \epsilon_i \frac {\Delta s_i} {s_i}.
\end{equation}
Treating this as a differential equation we deduce that $\epsilon (s) 
\propto s^{2/3} \propto R^{-1/3} (s)$. This confirms the well-known 
result that 
\begin{equation}
I = \epsilon R^{1/3}
\end{equation}
is an adiabatic invariant asymptotically for $\epsilon \rightarrow 0$ if 
$R$ is bounded. 

\begin{figure}[tbp]
{\hspace*{2.7cm} 
\psfig{figure=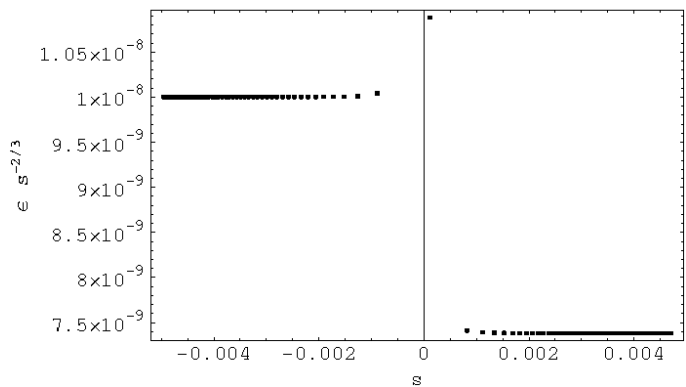,height=5.23cm,width=9cm,angle=0}}
{\vspace*{.13in}}
\caption[$\epsilon s^{-2/3}$ at the bounce points $s_i$ for a classical 
orbit.]
{$\epsilon s^{-2/3}$ at the bounce points $s_i$ for a classical orbit 
($\mu = 1$). The quantity is constant outside of the singular region near 
$s=0$. 
\label{3_16}}
{\vspace{1.2 cm}}
\end{figure}

In the present case, however, the adiabatic invariant is broken when the
trajectory enters the region of small curvature (however small the initial
$\epsilon$ may be). Figure \ref{3_16} shows the quantity $\epsilon_i
s_i^{-2/3}$, which is proportional to $I$, for a certain orbit passing
through the singular region. Clearly, it is almost constant on the
approach, then it changes significantly within a couple of bounces, after
which it stays constant again but at a different value. This observation
agrees with Eq.\ (\ref{cts}) that estimates the size of the critical region
as two jumps (i.e.\ only one reflection point of any orbit lies in the
critical region for this orbit). The critical region thus depends on the
orbit. Explicitly, the size of the jump outside of the critical region is
$\Delta s (s) \simeq \epsilon (s) R(s) \sim I/ (\mu s^2)^{2/3}$. Taking $s
\sim \Delta s$ we estimate the size of the critical region as 
\begin{equation}
|\Delta s| \sim I^{3/7}/ \mu^{2/7}. \label{ctdel}
\end{equation}
This estimate will reemerge in the semiclassical regime. 

\begin{figure}[tbp]
{\hspace*{2.7cm} 
\psfig{figure=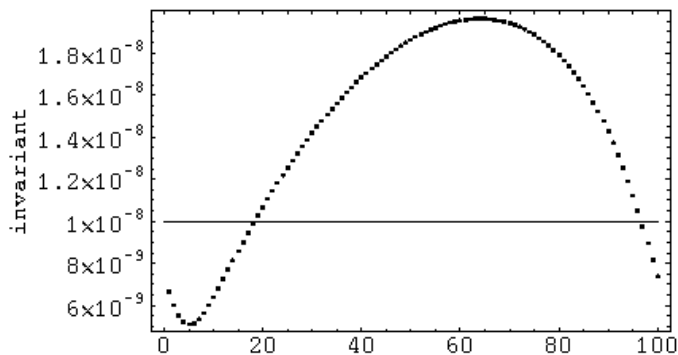,height=4.89cm,width=9cm,angle=0}}
{\vspace*{.13in}}
\caption[Values of adiabatic invariant for 100 orbits after they pass 
the singular region.]
{Values of adiabatic invariant $I$ for 100 orbits after they pass the
singular region. The orbits had the same initial $I$ indicated by the
horizontal line. 
\label{3_17}}
{\vspace{1.2 cm}}
\end{figure}

Now consider a family of orbits with the same $I$ that are incident on the
singular region from one side. We are interested in the ``spectrum'' of
$I$'s after the orbits pass through this region. The final values of $I$'s
for such family are shown in Fig.\ \ref{3_17}. We chose an arbitrary orbit
with a an initial adiabatic invariant $I_0$, shown as a horizontal line.
Then we started 100 orbits equidistantly from the interval $(s_0,s_1)$,
i.e.\ between the first two bounces of the original orbit, with the initial
$\epsilon = I_0 R^{-1/3}$. The figure shows the final $I$ \emph{vs.} the
orbit number. 

Clearly, there is a considerable spread in the final values, and thus the
family and the caustic associated with it no longer exist beyond the $s=0$
region.\footnote{A generic orbit going around the generic billiard will
have a certain statistical probability to increase $I$ on each return to
the $s=0$ region, i.e.\ it will make a kind of random walk in $I$-space. 
After some time it will eventually diffuse in the large $I$ region, where
the map equations are no longer valid. Thus there is no classical
localization near the boundary for a generic orbit.} In fact, numerically,
the maximal and minimal final values of $I$ are related to the initial
value $I_0$ by
\begin{equation}
I_{\max}/ I_0 \approx 1.96 , \;\; I_{\min}/ I_0 \approx 0.51. 
\label{ctImax}
\end{equation}
This agrees with the relation 
\begin{equation}
I_{\max} / I_0 = I_0 /I_{\min}, \label{ctI0}
\end{equation}
which is proven below. These results are important for the scattering 
theory in the next section. 

It should be pointed out that the right hand sides of Eq.\ (\ref{ctImax})
are valid for any $\mu$ and any sufficiently small $I_0$. This follows
from the scaling properties of the equations of motion (\ref{ctmap1}) and
(\ref{ctmap2}). For any trajectory $(\epsilon_i,s_i)$ there is a
trajectory $(\epsilon_i^\prime,s_i^\prime) = (\epsilon_i,s_i/ \alpha)$
that satisfies the equations of motion in a billiard with a new parameter
$\mu^\prime = \alpha^3 \mu$. The new adiabatic invariant $I^\prime = I / 
\alpha^{1/3} \propto I$. This proves the $\mu$-independence. Likewise, an 
orbit $(\alpha \epsilon_i, \alpha^{1/3} s_i)$ satisfies the equations of 
motion with the same $\mu$, but has the adiabatic invariant $\alpha^{7/9} 
I$. This shows the independence of $I_0$. 

Now we can justify Eq.\ (\ref{ctI0}). Suppose an orbit starts with $I_0$
and ends up with $I_1 = I_{\max} (I_0) = \eta I_0$ where $\eta \approx
1.96$. Since there are time-reversal and $s \rightarrow -s$ symmetries, we
can start an orbit with $I_1$ that ends up with $I_0 = \eta^{-1} I_1$.  We
claim that $I_0 = I_{\min} (I_1)$. Indeed, if there is a final $I_2 <
I_0$ then we can start an orbit with $I_2$ that ends up with $I_1$ and
$I_1/ I_2 > \eta$. But this contradicts the relation $I_{\max}(I)/ I =
\eta$, which is true for any $I$ according to the previous paragraph.  We
thus showed that $I_{\min} (I_1) = \eta^{-1} I_1$, which, again, must be 
true for any $I_1$. This completes the proof of Eq.\ (\ref{ctI0}).

\begin{figure}[tbp]
{\hspace*{2.7cm} 
\psfig{figure=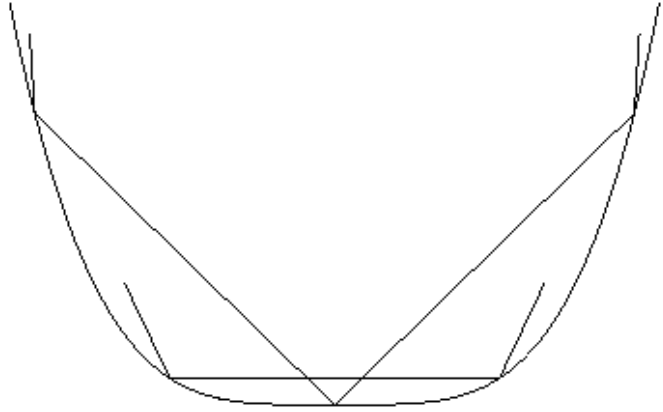,height=5.72cm,width=9cm,angle=0}}
{\vspace*{.13in}}
\caption{Two orbits that preserve the adiabatic invariant.
\label{3_18}}
{\vspace{1.2 cm}}
\end{figure}

Finally, we see from Fig.\ \ref{3_17} that there are two trajectories that 
preserve adiabatic invariant. These are the two orbits symmetric under $s 
\rightarrow -s$. One bounces at $s=0$, another at $s_{i+1} = - s_i$ for 
some $i$ (Fig.\ \ref{3_18}). 

\subsection{Semiclassical theory}

The destruction of caustic prevents us from applying the standard
quantization procedure, for example, the ray method. On the other hand,
the whispering gallery mode is well-defined outside the small curvature
region. So, before looking for the stationary states inside a closed
boundary, we may first consider the scattering problem of determining the
amplitudes of transitions between the modes due to the flat region. 

Suppose a mode with a transverse quantum number $n$ is incident from $s =
- \infty$. Its surface of section wavefunction $\psi_n (s)$ is given by
Eq.\ (\ref{wgmpsi}). With the open boundary the energy $k^2$ is not
quantized. It is conserved throughout the motion serving as an external
parameter. According to the Eqs.\ (\ref{wgms}) and (\ref{wgmapp}) the 
whispering gallery mode is well-defined for 
\begin{equation}
|s| \gg \left| s - s^\prime \right|_{\mathrm{st}} \sim \left( n/ \mu^2 k 
\right)^{1/7}. \label{stdel}
\end{equation}
Since the stationary phase quantity $\left| s - s^\prime
\right|_{\mathrm{st}}$ corresponds to the size of the classical jump, we
find that the strong inequality is similar to the classical condition
(\ref{cts}). In fact, the second estimate is associated with another
classical equation (\ref{ctdel}). To show this, we note that, since the
phase of $\psi_n (s)$ is a projection of the semiclassical momentum onto
the boundary, there is a relation $\cos \epsilon \simeq 1 - f_n^\prime$.
Here $\epsilon$ is the earlier defined angle between the classical orbit
and the boundary at the reflection point $s$, and $f_n^\prime$ is given by
Eq.\ (\ref{wgmf}). Then the adiabatic invariant can be expressed in terms 
of the quantum quantities as 
\begin{equation}
I = I_n (k) \simeq \sqrt {2 f_n^\prime} R^{1/3} = \frac {2^{1/3} \sqrt 
{z_n}} {k^{1/3}} \label{stIn}
\end{equation}
where $z_n \approx \left[ \frac 3 2 \pi \left(n - \frac 1 4 \right)
\right]^{2/3}$ is the $n$th root of $\mathrm{Ai} (-z)$ for large $n$. 
This connects Eqs.\ (\ref{stdel}) and (\ref{ctdel}). We see that the $n$th 
transverse mode is related to the classical family with the adiabatic 
invariant $I_n$. 

As we know, the whispering gallery wavefunction $\psi_n (s)$ is not valid
near $s=0$. It will be scattered into a number of transverse modes that
are well-defined for sufficiently large positive $s$. We are interested in
the transition amplitude between $\psi_n (s<0)$ and $\psi_{n^\prime}
(s>0)$. In order to apply the $T$-operator technique to the scattering
problem, the surface of section can be divided into an external (SSE) and
internal (SSI) pieces \cite{GeoPra2}, such that $\psi_n$ and
$\psi_{n^\prime}$ are well-defined on the SSE. From the previous 
discussion we conclude that the size of SSI is of order of two classical 
jumps. We will give an explicit definition later, but for now we 
assume only that the classical orbit that belongs to the family $I_n$ 
before the scattering and to the family $I_{n^\prime}$ after the 
scattering makes \emph{one and only one} bounce within the SSI. (In 
principle, we could extend the SSI to two or more bounces, which would 
make the theory more precise and also more complicated.) 

Suppose that the SSI is an interval $(-s_n, s_{n^\prime})$. The transfer 
operator between point $s < -s_n$ and point $s^\prime > s_{n^\prime}$ is 
\begin{equation}
T^2 (s^\prime, s) \simeq \int_{-s_n}^{s_{n^\prime}} ds^{\prime\prime} 
T(s^\prime, s^{\prime\prime}) T(s^{\prime\prime}, s) \simeq \left[ \frac 
k {2\pi i} \left| \frac {\partial^2 L} {\partial s^\prime \partial s} 
\right| \right]^{1/2} e^{ikL (s^\prime, s)} 
\end{equation}
where $L (s^\prime, s)$ is the length of the classical orbit between $s$
and $s^\prime$ that makes one bounce within the SSI. Note that this
expression is valid only for the given $n$ and $n^\prime$ in the $S\Phi$,
because we did not include the orbits from other families that have a
different number of bounces in the SSI. The scattering amplitude 
\begin{equation}
S_{n^\prime n} = \left(T^2 \right)_{n^\prime n} = \int_{- \infty}^{-s_n} 
ds \int_{s_{n^\prime}}^{\infty} ds^\prime \psi_{n^\prime}^* (s^\prime) 
T^2 (s^\prime, s) \psi_n (s). \label{stsnn}
\end{equation}
When we evaluate this integral by the $S\Phi$, only those $s$ and
$s^\prime$ that belong to a certain classical orbit are selected. Namely,
this is the orbit that belongs to the family $I_n(k)$ before the
scattering and $I_{n^\prime}(k)$ after the scattering. The stationary
points $s_{\mathrm{st}}$ and $s_{\mathrm{st}}^\prime$ are the points of
the last bounce before the SSI and the first bounce after the SSI,
respectively. We conclude from Eq.\ (\ref{ctImax}) and Fig.\ \ref{3_17} that,
given $n$, the stationary points exist for those $n^\prime$ that satisfy
$0.51 I_n (k) < I_{n^\prime}(k) < 1.96 I_n (k)$, or, according to Eq.\
(\ref{stIn}),
\begin{equation}
0.26 z_n < z_{n^\prime} < 3.84 z_n, \label{stzn}
\end{equation} 
moreover, they come in pairs. Equation (\ref{stzn}) shows that within the
$S\Phi$ an incident mode scatters into a limited number of modes. For
example, if $n=1$ (the lowest mode) then $n^\prime = 1, \ldots, 5$. 

Clearly, if the $n \rightarrow n^\prime$ transition is possible then the
$n^\prime \rightarrow n$ transition is also possible because of the
symmetry. The same conclusion follows from Eq.\
(\ref{stzn}),\footnote{$0.26 = 1/3.84$} based on the classical property
(\ref{ctI0}), which is a consequence of the same symmetries. In addition,
the probabilities of these transitions must be the same. This is ensured 
by the unitarity of the scattering matrix (\ref{stsnn}). In the $S\Phi$ 
the $T$-operator is unitary and so is $S_{n^\prime n}$. 

The $S\Phi$ result is independent of the limits of integration $s_n$ and
$s_{n^\prime}$. In order to go beyond the $S\Phi$, one should, first of
all, improve the transfer operator to have it include orbits with a
different number of bounces. The boundaries of the SSI can be defined by
the orbit in Fig.\ \ref{3_18} that does \emph{not} bounce at $s=0$. Let
$s_n$ be the reflection point of this orbit that is the closest to the
origin. From the map equations we find
\begin{equation}
s_n \approx \left(36 I_n^3 \mu^{-2} \right)^{1/7} \approx \left[108 \pi 
(n - 1/4) k^{-1} \mu^{-2} \right]^{1/7}
\end{equation}
[cf.\ Eq.\ (\ref{stdel})]. With this definition the $S\Phi$ orbit makes one 
bounce within the SSI. 

If the boundary is closed, i.e.\ the states $\psi_n (s)$ are defined on
$s_{n^\prime} < s < - s_n$, then one can consider the propagator $(\Lambda
\cdot S)_{n^\prime n}$ around the circumference. Here $\Lambda_{n^\prime
n}$ is a diagonal unitary matrix that accounts for the phase change
outside of the singular region. Its elements are just $e^{ik
[\mathcal{L}_n - f_n (\mathcal{L}_n) + f_n (0)]}$ where $\mathcal{L}_n$ is
the length of the regular part of the boundary for the state $\psi_n (s)
\sim e^{ik [s - f_n (s)]}$. The product $(\Lambda \cdot S)_{n^\prime n}$
can be diagonalized. Its eigenvalues will be of the form $e^{i \lambda_j
(k)}$. Since $S_{n^\prime n} \approx 0$ for $n$ and $n^\prime$ outside of
the range (\ref{stzn}), we presume that the new eigenmodes will, in
general, include the number of original modes within this range (although
this qualitative statement should be verified). The wavenumber $k$ is
quantized by the condition $\lambda_j (k) = 2\pi n$. So, if a billiard
boundary contains a point of zero curvature, the whispering gallery modes
may still exist (Fig.\ \ref{3_14}) but they are no longer simple Airy
function type solutions. These conclusions need to be verified by the
direct numerical computations, which are in the plans for the future. 

\section{Conclusions}

In this chapter we considered several examples when the perturbation
theory can be applied. The perturbed rectangular billiard, like the
perturbed circle, is one of the few systems where the \emph{unperturbed}
action depends only on the difference of their \emph{natural} coordinates. 
It is straightforward to build the perturbation theory in its general form
in this case without changing to the action-angle variables. As we know, a
perturbed system may possess the eigenstates that are localized near the
(semiclassically) stable periodic orbits. We have an example of such state
localized near the long diagonal in a tilted square. To have a strong
localization, the perturbation of the boundary $\delta L$ must be much
greater than $\lambda^2 /L$, where $\lambda$ is the wavelength. 

We have discussed a number of non-perturbative cases in which the
$T$-operator method can be used to derive the analytic expressions for the
wavefunctions and energy levels. These include the localized states in the
ice-cream cone billiard, the bouncing ball states in a stadium and near a
period-2 orbit, the whispering gallery modes. Moreover, the theory allows
to include an additional perturbation to such systems, for example, to 
tilt the sides of the stadium. 

The standard whispering gallery modes result from the quantization of
classical families of orbits that are localized near the boundary. The
families lie on the invariant tori that exist for the smooth boundaries
with positive curvature in the limit of small sliding angle. When the
boundary contains a point or a short interval of zero curvature, the
classical tori do not exist but the localized quantum states are still
possible. The bound state problem involves finding the scattering
amplitudes, which is another area where the semiclassical surface of
section method can be used.

\chapter{Square billiard in magnetic field\label{sbim}}

In this chapter we consider a square billiard where the perturbation is
caused not by the boundary, but by the magnetic flux perpendicular to its
plane. We concentrate on two particular cases: a uniform magnetic field
and an Aharonov-Bohm flux line (ABFL). The latter system is purely quantal
--- the famous Aharonov-Bohm effect \cite{AhaBoh} results from the phase
interference and has no classical analogue. We will be interested
precisely in the effect of phase interference on the eigenstates in the
billiard, and therefore even in the uniform field case will neglect the
Lorentz force (below we estimate the field where it can be done). We know
from the previous chapters that a relatively small non-integrable
perturbation of an integrable square makes it possible for the strongly
localized states to exist (Figs.\ \ref{3_5}, \ref{3_6}). This is still the
case when the perturbation is the magnetic flux \cite{NarPraZai}, and
these states will be the primary focus of our discussion. 

Although we confine our attention to the weakly non-integrable billiard, the
ergodic behavior of the system for larger fields can be of interest. As in
the Sinai billiard, the magnetic field imposes its circular symmetry on the
square, which leads to chaos. Classically, depending on the field, mixed and
chaotic regimes are possible \cite{BerKun}. Quantally, one is interested in
the energy level statistics. The energy level distribution in a strongly
chaotic billiard depends on the symmetries of the system, in particular, the
time-reversal symmetry. The magnetic field breaks this symmetry, and, in
general, the statistics will be different from the non-magnetic chaotic
billiard \cite{BerRob}. (The square in uniform field still has an
antiunitary symmetry, however.) The spectral statistics of a rectangular
billiard with the ABFL at the center is discussed in Refs. \cite
{DatJaiMur,RahFis}.

The measurements of the orbital magnetic susceptibility in the array of GaAs
ballistic squares \cite{LevReiPfe} motivated a certain amount of theoretical
activity in the last few years \cite{RicUllJal}. The trace formula is a
powerful semiclassical approach that allows one to express the density of
states (and thus the susceptibility) in terms of the actions of the periodic
orbits. For the relevant temperatures only the short orbits need to be taken
into account. In the weak field regime, when the perturbed Berry-Tabor trace
formula can be used, our perturbation theory also allows to find the
susceptibility.

The main reason to use the perturbation theory, however, is to predict and
give the approximate expressions for the wavefunctions. Even the relatively
simple analytical formulas produce the states with the non-trivial
probability and current distributions. It is possible to have current loops
of opposite sense within one state. For a given field one finds the states
with the overall current ranging from paramagnetic to diamagnetic, including
those with the small overall current but strong local currents nearly
canceling each other.

\section{Uniform field}

\subsection{Limits of applicability}

\label{loa}

A free particle of charge $e$ moves in the uniform field $B$ along a
circular orbit of cyclotron radius $R_{\mathrm{c}}=mcv/eB$ where $v$ is
its speed. If confined by the square boundary the orbit will consist of
the respective arcs. We are interested in the regime when the Lorentz
force can be neglected in the lowest order and the orbits are made up of
nearly straight segments. This means that $\epsilon =L/R_{\mathrm{c}},$
where $L$ is the side of the square, is a small parameter. In this case
the classical momentum $mv>>eA/c,$ where $A$ is the vector potential, so
one can approximate $mv\simeq \hbar k$ in the semiclassical picture. Then
we can express $\epsilon =2\pi \phi /\phi _0kL,$ where $\phi =BL^2$ is the
flux through the billiard and $\phi _0=hc/e$ is the flux quantum. We shall
see that $\epsilon $ is the small parameter of the perturbation theory. 

In the units $\hbar =c=e=L=1$ we have $\epsilon =B/k\ll 1$ where the
dimensionless field $B=\phi =2\pi \phi /\phi _0.$ Without loss of
generality we assume $B>0.$ In the semiclassical approximation we take the
wavenumber $ k $ $\gg 1.$ As explained in the beginning of Ch.\
\ref{perth}, the resonant perturbation theory is required when
$k\sqrt{\epsilon }=\sqrt{kB}\gtrsim 1,$ i.e.\ when the magnetic field is
relatively large. The upper boundary for $B$ is given by the number of
terms kept in the perturbation series and by the smoothness of the
perturbation (which determines the onset of diffraction). If the terms of
order no higher than $M$ are kept then $kb^{M+1}\ll 1$ where $b=\sqrt{
\epsilon }.$ To keep the diffraction effects small we limit $kb^4\ll 1$
(see the next section), but if we do not go higher than the second order
then the stricter condition $kb^3\ll 1$ must be imposed. Note that if $k$
is fixed then $B$ may not be too big, if $B$ is fixed then $k$ should be
large enough, and if $\epsilon $ is fixed then the energy is bounded. With
the ABFL the diffraction is stronger, and $kb^2\ll 1$ in that case. 

\subsection{Effective potential and eigenstates}

\begin{figure}[tbp]
{\hspace*{2.7cm} \psfig{figure=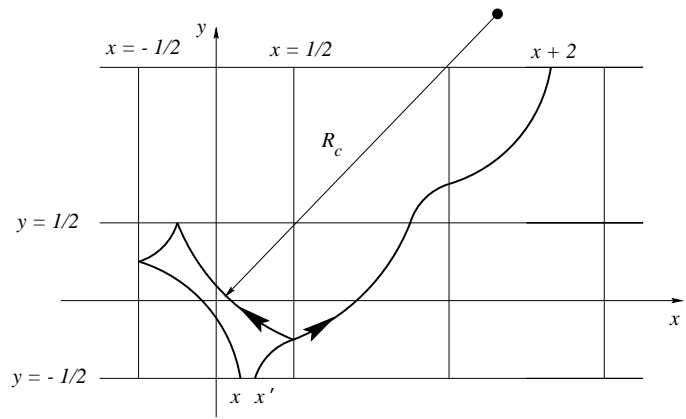,height=5.46cm,width=9cm,angle=270}}
{\vspace*{.13in}}
\caption[Method of images for the square in magnetic field.]
{Method of images for the square in magnetic field. The magnetic field has
different signs in the adjacent squares and the orbits are curved
differently. $R_{\mathrm{c}}$ is the cyclotron radius. The curvature is
exaggerated for clarity. The Poincar\'e surface of section is $y= -1/2$
identified with $y= 3/2$. The orbit shown is close to the $(1,1)$ periodic
orbit. It goes from $x^\prime$ to $x+2$ identified with point $x$ in the
original square. The square causes the particle to circulate in the
paramagnetic sense, which is opposite to the direction of a free orbit in
the field. 
\label{4_1}}
{\vspace{1.2 cm}}
\end{figure}

As far as the formal theory is concerned, the square in magnetic field is
similar to the square with the perturbed boundary (Sec.\
\ref{rbgc}).\footnote{The same lowest order results are given by the
channeling method (Sec.\ \ref {chap}).} Again we use the method of images
reflecting the square about its boundaries. The magnetic field has
opposite signs in the adjacent squares and the orbits are curved
differently (Fig.\ \ref{4_1}). Note that we chose the origin at the center
of the square. The Poincar\'e surface of section (PSS) is the bottom side
$y=-1/2$ identified with $y=3/2.$ We will be interested in the states near
resonances (periodic orbits) since they differ the most from the states of
the unperturbed billiard. Figure \ref{4_1}, for example, shows a
trajectory near the $(1,1)$ resonance. This trajectory is paramagnetic,
the square boundaries make it circulate in the opposite direction from a
free particle in the field. 

The action along the orbit in the extended scheme is 
\begin{equation}
S\left( x,x^{\prime }\right) =k\sqrt{4+\left( x-x^{\prime }\right) ^2}+\Phi
\left( x,x^{\prime }\right)
\end{equation}
where $\Phi =\left( e/c\right) \int \mathbf{A\cdot dr}$. The curvature of
the orbit can be neglected because it makes a correction of $\epsilon ^2.$
We will primarily consider the $(1,1)$ resonance since, as will become
clear, it is the most important. In this case the PSS wavefunction has the
form $\psi \left( x\right) =e^{i\kappa x}\hat \psi _m\left( x\right) $ where 
$\kappa =k\cos 45^{\circ }=k/\sqrt{2}.$ The slower varying function $\hat
\psi _m\left( x\right) $ satisfies the Schr\"odinger equation 
\begin{equation}
\hat \psi _m^{\prime \prime }+\left[ E_m-V\left( x\right) \right] \hat \psi
_m=0.  \label{epeSch}
\end{equation}
The effective potential $V\left( x\right) =-k\Phi \left( x,x-2\right)
/\mathcal{L}$ is proportional to the perturbed part of the action
evaluated along the unperturbed orbit of length $\mathcal{L}=\sqrt{8}$.
Note that $\Phi \left( x,x-2\right) $ is just the flux enclosed by the
unperturbed periodic orbit in the original square, equal $B$ times the
area of the loop with the positive sign for counter-clockwise orbits. The
potential is then
\begin{equation}
V\left( x\right) =\left\{ 
\begin{array}{l}
-Bk\left( \frac 12-2x^2\right) /\mathcal{L},\quad x\in \left[ -\frac
12,\frac 12\right] \\ 
Bk\left[ \frac 12-2\left( x+1\right) ^2\right] /\mathcal{L},\quad x\in
\left[ \frac 12,\frac 32\right]
\end{array}
\right.  \label{epeV}
\end{equation}
with the periodicity $V\left( x+2\right) =V\left( x\right) .$ The potential
is proportional to $k^2\epsilon ,$ which justifies the use of $\epsilon $ as
a small parameter. $V\left( x\right) $ consists of alternating parabolic
wells and barriers of height $Bk/2$ joined smoothly at $x=\pm 1/2,$ where
the potential is odd. Note that the potential for the tilted square of Sec.\ 
\ref{ets} for the same resonance is even on the boundary because of the
time-reversal symmetry.

The potential $V\left( x\right) $ has a discontinuous second derivative at
$x=\pm 1/2$ due to the diffraction from the corners. It will bring a
step-function to the $kb^4$-order of the perturbation theory. Since the
perturbation expansion takes place in the $\emph{phase}$ of Bogomolny's
equation, the discontinuity allows to shift the pieces of $V^{\prime
\prime }\left( x\right) $ independently by integer number of $2\pi .$ This
is the reason for the condition $kb^4\ll 1.$ (By the same token, we must
require $kb^3\ll 1$ in the tilted square.)

The potential is negative for the trajectories leaving the PSS with the
positive velocity projection $v_x.$ These trajectories go counter-clockwise
around the square. The minimum at $x=0$ indicates a stable periodic orbit.
In its neighborhood the potential is of the harmonic oscillator type, so for
the low-lying states the energies are approximately given by 
\begin{equation}
E_m=-\frac 12\frac{Bk}{\mathcal{L}}+\left( m+\frac 12\right)
\sqrt{\frac{8Bk}{\mathcal{L}}}.  \label{epeE}
\end{equation}
The formula holds for $m\ll \sqrt{Bk/\mathcal{L}}.$ The lowest
wavefunction $\hat \psi _0\left( x\right)
=e^{-\sqrt{\frac{Bk}{2\mathcal{L}}}x^2}$ is localized near $x=0.$ The
two-dimensional state is localized near the stable periodic orbit (Fig.\
\ref{4_4}). Not surprisingly, it maintains the paramagnetic current, as
explained below. The wavefunctions with higher $m$'s penetrate into the
region $1/2<\left| x\right| <1$ more and more. Hence these states include
the diamagnetic orbits. The states with large $m$ are diamagnetic. The
maximum at $x=1$ marks an unstable periodic orbit, which is almost a
time-reversed stable orbit. Our method accounts for the scars of unstable
orbits (Sec.\ \ref{nc}), i.e.\ the states with $E_m\approx V_{\max }$ (Fig.\
\ref{4_11}). 

In general, Eq.\ (\ref{epeSch}) should be solved with the periodic
condition $ \psi \left( x+2\right) =\psi \left( x\right) .$ It is
accompanied by the quantization
\begin{equation}
k_{nm}=2\pi n/\mathcal{L}+E_m\left( k\right) /k
\end{equation}
that follows from Bogomolny's equation. Thus $n$ is roughly the number of
wavelengths along the periodic orbit. In the first approximation $k$ in the 
\emph{r.h.s.} can be replaced by $2\pi n/\mathcal{L}.$ Note that $E_m/k$
weakly depends on $n$ [cf.\ Eq.\ (\ref{epeE})]. The energy 
\begin{equation}
E_{nm}=k^2=\left( 2\pi n/\mathcal{L}\right) ^2+2E_m  \label{epeEnm}
\end{equation}
depends on $B$ only via $E_m.$

\begin{sloppypar}
The two-dimensional wavefunction could be found similarly to the tilted
square (Sec.\ \ref{ets}). But here we do it by another method that takes into
account the symmetries of the problem. The PSS wavefunction $\psi \left(
x\right) =e^{i\kappa x}\hat \psi _m\left( x\right) $ can be propagated with
the Green's function (\ref{tdw2G}) to give a two-dimensional wavefunction 
\begin{equation}
\Psi _0\left( x,y\right) =e^{i\kappa \left( x+y\right) }\hat \psi _m\left(
x-y-\frac 12\right)  \label{epeP0}
\end{equation}
where we ignored the terms of order $k\epsilon $ in the phase. In this
approximation $\kappa =\pi n/2$ and $\hat \psi _m\left( x\right) $ solves
the Schr\"odinger equation with the boundary condition $\hat \psi _m\left(
x+2\right) =\left( -1\right) ^n\hat \psi _m\left( x\right) .$ By symmetry,
there are three more states with the same energy obtained from $\Psi _0$
by $90^{\circ }$ rotations. For example, $\Psi _1\left( x,y\right)
=\mathcal{R}\Psi _0\left( x,y\right) =\Psi _0\left( y,-x\right) $ where
$\mathcal{R}:\left( x,y\right) \rightarrow \left( y,-x\right) $ is a
rotation operator. The gauge is chosen to make the Hamiltonian invariant
under $\mathcal{R}.$ The eigenstates in the physical domain will be the
linear combinations of $\Psi _i$'s that are also the eigenvectors of
$\mathcal{R}$ with the eigenvalues $i^{-r}$, $r=0,\ldots ,3.$ The
eigenfunction of symmetry $r$ is
\begin{equation}
\Psi _{\left( r\right) }\left( x,y\right) =\left(
\sum_{s=0}^3i^{rs}\mathcal{R}^s\right) e^{i\kappa \left( x+y\right) }\hat
\psi _m\left( x-y-\frac 12\right) .  \label{epePsi}
\end{equation}
\end{sloppypar}

The eigenstate symmetry $r$ depends on the quantum numbers $n$ and $m.$ By
the boundary condition $\Psi _{\left( r\right) }\left( x,-1/2\right) $ has
to vanish. The $s=0,3$ terms are proportional to $e^{i\kappa x}$ and must
cancel each other. This implies $\hat \psi _m\left( -x\right)
=-i^{-3r}e^{-i\kappa }\hat \psi _m\left( x\right) .$ Since the potential
$V\left( x\right) $ is even and $\hat \psi _m$ satisfies real boundary
conditions we can choose $\hat \psi _m\left( -x\right) =\left( -1\right)
^m\hat \psi _m\left( x\right) .$ Then $n$ is related to $m$ and $r$ as
\begin{equation}
n\limfunc{mod}4=\left[ 2\left( 1-m\limfunc{mod}2\right) +r\right]
\limfunc{mod}4. 
\end{equation}
For fixed $m$ the successive values of $n$ cycle through the representations
of $\mathcal{R}.$ Note that $\hat \psi _m\left( x+2\right) =\left( -1\right)
^r\hat \psi _m\left( x\right) .$

These results are easy to generalize to the other resonances $\left(
p,q\right) .$ In the weak field approximation the orbits with even $pq$ do
not enclose any flux. These orbits can be subdivided into $pq$ smaller
loops, half of which are positive, and half are negative, so the total flux
is zero. (If the curvature is taken into account, there will be a small
flux). In particular, the bouncing ball $\left( 0,1\right) $ orbits are
non-magnetic, and $\left( 1,1\right) $ is the lowest resonance that responds
to the magnetic field. If $pq$ is odd, the effective potential is 
\begin{equation}
V_{pq}\left( x\right) =\frac qp\left[
\frac{\mathcal{L}_{11}}{\mathcal{L}_{pq}}\right] ^3V\left[ q\left( x+\frac
12\right) -\frac 12\right]
\end{equation}
where $\mathcal{L}_{pq}=2\sqrt{p^2+q^2}$ and $V$ is given by Eq.\
(\ref{epeV}). The potential has a period $2/q.$ If $kB\ll
pq\mathcal{L}_{pq}^3$, the potential does not support localized states and
can be ignored. The PSS states should satisfy the boundary condition $\hat
\psi _m\left( x+2\right) =e^{-i2\pi \delta }\hat \psi _m\left( x\right) $
where $\delta $ is the fractional part of $np/\left( p^2+q^2\right) .$ The
energy is
\begin{equation}
k_{nm}^2\approx \left( 2\pi n/\mathcal{L}_{pq}\right) ^2+\left(
1+\frac{p^2}{q^2}\right) E_m^{\left( q\right) }
\end{equation}
where we have to distinguish between $E_m^{\left( q\right) }$ for the
resonance $\left( p,q\right) $ and $E_m^{\left( p\right) }=\left( p/q\right)
^2E_m^{\left( q\right) }$ for the resonance $\left( q,p\right) .$ These
resonances produce nearly uncoupled states with the same energy that are
related by a $90^{\circ }$ rotation. The splitting of the levels is
negligible unless these resonances are close in the phase space. The latter
occurs when $p/q$ is close to 1 so that the tunneling can be assisted by the 
$\left( 1,1\right) $ resonance, but not too close to have $\left( p,q\right) 
$ and $\left( q,p\right) $ resonances destroyed by the $\left( 1,1\right) $
resonance.

\subsection{Magnetic response}

\label{mres}

The magnetic susceptibility at high temperatures is dominated by the Landau
term coming from the smooth part of the density of states. At low
temperatures the oscillatory corrections become large as the Fermi
wavenumber $k_F$ increases. This contribution to the susceptibility can be
found using the trace formula \cite{RicUllJal}, which does not require the
knowledge of the wavefunctions or the spectrum. When the magnetic field is
weak enough, the resonant perturbation theory can be used to calculate the
share of a particular resonance in the susceptibility. This method is
related to the perturbed Berry-Tabor formula (Sec.\ \ref{pbtf}).

We start with a grand potential for a system of non-interacting electrons 
\begin{equation}
\Omega \left( T,\mu ,B\right) =-k_BT\sum_a\ln \left[ 1+e^{-\left( E_a-\mu
\right) /k_BT}\right]
\end{equation}
where $\mu $ is the chemical potential and the sum is over the
single-electron states. If we are interested in the part of the
susceptibility coming from the $\left( 1,1\right) $ resonance, we include in
the sum the subset of states given by Eq.\ (\ref{epeEnm}). The 
magnetization $ \mathcal{M}=-\partial \Omega /\partial B$ is 
\begin{equation}
\mathcal{M}\left( T,\mu ,B\right) =-\sum_{nm}\frac{\partial E_{nm}}{\partial
B}f_D\left[ E_{nm}\left( B\right) \right] .
\end{equation}
Here $f_D$ is the Fermi-Dirac distribution function. The sum over $n$ can be
done using the Poisson sum formula. The result is 
\begin{equation}
\mathcal{M}\left( T,\mu ,B\right) =-\sum_m\sum_{s=1}^\infty \alpha
_m\frac{\mathcal{L}k_BT/2k_F}{\sinh \left( \pi
s\mathcal{L}k_BT/2k_F\right) }\sin \left[ \mathcal{L}s\left(
k_F-\frac{E_m}{k_F}\right) \right]
\end{equation}
where $\alpha _m=\partial E_m/\partial B$ evaluated at $k=k_F.$ For the
other resonances $\left( p,q\right) $ one should substitute
$E_m\rightarrow \left( 1+\frac{p^2}{q^2}\right) E_m^{\left( q\right) }/2$
if $pq$ is odd; the resonances with even $pq$ are non-magnetic. The
magnetization decreases exponentially as $\mathcal{L}$ becomes large.
Therefore the $\left( 1,1\right) $ resonance gives the strongest magnetic
response. If we neglect the higher resonances, we do not need to keep the
terms with $s>1$ either. In the trace formula approach $s$ gives the
number of repetitions of the primitive periodic orbit. The susceptibility
is given by $\chi =\partial \mathcal{M}/\partial B.$ It is exponentially
suppressed for high temperature giving place to the Landau susceptibility
that decays as a power law. Since $\chi $ is proportional $k_F^{3/2}$
($k_F$ coming from $\alpha _m$ and of order $\sqrt{k_F}$ terms in the sum
over $m$) it is greater than the $k_F$ independent Landau term at low
temperature, which is of order of one in our units. In the experiment
\cite{LevReiPfe} the temperature was about 10 level spacings, i.e.\
$k_BT\simeq 20\pi $. The squares contained $2\div 6\times 10^4 $ electrons
which makes $k_F\simeq 300\div 600.$ Then $\pi s\mathcal{L}k_BT/2k_F\simeq
0.5\div 1,$ and the exponential suppression is not too strong. If an
experiment is conducted on an array of squares the result should be
averaged over the distribution of sizes. Because of the oscillating sine
factor the result will be reduced. Reference \cite{RicUllJal} provides the
corrections. 

\begin{figure}[tbp]
{\hspace*{2.7cm} 
\psfig{figure=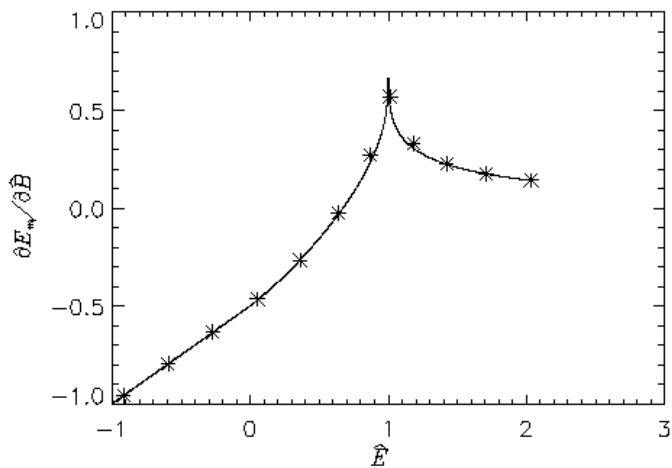,height=6.46cm,width=9cm,angle=0}}
{\vspace*{.13in}}
\caption[$\partial E_m / \partial \hat B$ as a function of $\hat E$.]
{$\partial E_m / \partial \hat B$ as a function of $\hat E$. The 
stars are at the $(1,1)$ resonance states $n = 62$, $m = 0,2, \ldots, 
m_{\max} =14, \ldots$ Magnetic field $B=25$. The continuous curve is 
given by Eq.\ (4.15).
\label{4_2}}
{\vspace{1.2 cm}}
\end{figure}

The magnetization depends on the response of individual states $\alpha _m.$
Note that $\partial E_m/\partial \hat B=\partial \left\langle H\right\rangle
_m/\partial \hat B=\left\langle \hat \psi _m\right| \hat V\left| \hat \psi
_m\right\rangle $ where $\hat B=Bk/2\mathcal{L},$ $\hat V\left( x\right)
=V\left( x\right) /\hat B,$ and $H$ is the effective Hamiltonian in the
Schr\"odinger equation (\ref{epeSch}). Semiclassically, 
\begin{equation}
\frac{\partial E_m}{\partial \hat B}=\frac{\int dx\hat V\left( x\right)
\left[ \hat E-\hat V\left( x\right) \right] ^{-1/2}}{\int dx\left[ \hat
E-\hat V\left( x\right) \right] ^{-1/2}}.
\end{equation}
If $\hat E=E_m/B$ is treated as a continuous variable, the values of
$\partial E_m/\partial \hat B$ as a function of $\hat E$ will fall on a
continuous curve independent of $B$ (Fig.\ \ref{4_2}). The integration is
between the turning points. The states with low $m$'s have negative
$\alpha _m$ and thus are paramagnetic. The wavefunction has more weight in
the region $x\in \left[ -\frac 1 2,\frac 1 2\right] $ which is classically
identified with the counter-clockwise orbits and where $\hat V$ is
negative. The states with higher $m$'s are diamagnetic. In this case the
weight shifts towards the turning points located in the region $\left|
x\right| \in \left[\frac 1 2,1\right] $ where $\hat V$ is positive. If
$\hat E=\hat V_{\max},$ the curve has a cusp. The simple WKB\ formula
fails in this case. The wavefunction is large near the unstable periodic
orbit, which is strongly diamagnetic. As $m$ gets even bigger, the
wavefunction becomes more uniformly distributed in $x$ with slightly
higher weight near $x=1.$ So the states remain diamagnetic but their
magnetization goes down. 

The quantized transverse ``energy'' is approximately given by the
Bohr-Som\-merfeld condition
\begin{equation}
\int dx\sqrt{\hat E_m-\hat V\left( x\right) }=\pi \left( m+\frac 12\right)
/\sqrt{\hat B}. 
\end{equation}
The number of transverse modes in the resonant island, that can be
estimated as $0.6\sqrt{\hat B},$ depends on $\hat B.$ In the example of
Fig.\ \ref{4_2} $k \approx 142$ and $B=25$, and there are 15 states inside
the well. With these parameters we have two special states: $m=14$ which
is close to the diamagnetic maximum and $m=10$ which is almost
non-magnetic. 

\subsection{Visual representation of eigenstates}

\label{vroe}

As we learned in the previous sections, the square in magnetic field can
support the eigenstates with different degree of localization and a wide
range of magnetic properties. Here we show some examples of the
wavefunctions and current distributions calculated using the perturbation
theory (referred to as ``theoretical'' results) and compared with the
numerical solution of the Schr\"odinger equation in the square
(``numerical'' results). We restricted our attention to the states of the
$\left( 1,1\right) $ resonance that are invariant under the $90^{\circ }$
rotation, i.e.\ having the symmetry $r=0$ [Eq.\ (\ref{epePsi})]. These are
the states with either even $n/2$ and odd $m$ or odd $n/2$ and even $m$. 

A couple of comments on the calculation procedures are in order. In the
perturbation theory the one-dimensional Schr\"odinger equation
(\ref{epeSch}) was solved numerically. The two-dimensional wavefunction
was constructed using Eq.\ (\ref{epePsi}). We call these results
``theoretical.'' 

\begin{figure}[tbp]
{\hspace*{2.7cm} 
\psfig{figure=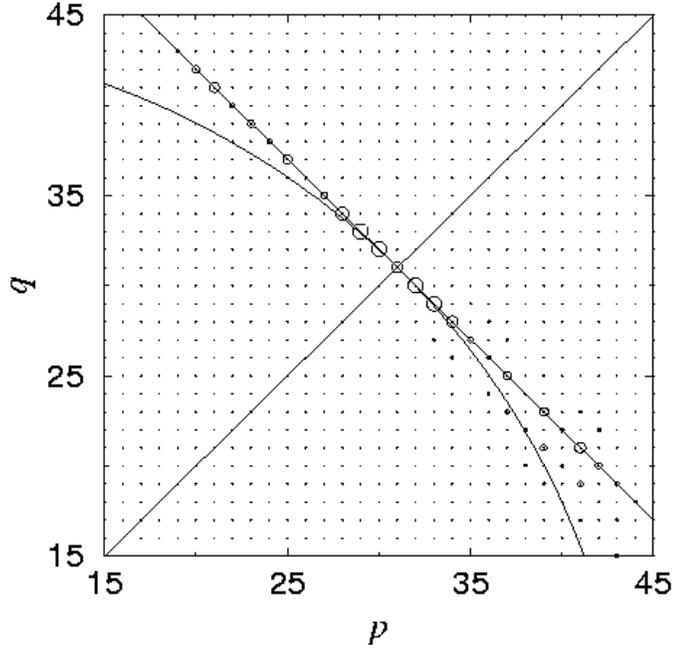,height=8.86cm,width=9cm,angle=0}}
{\vspace*{.13in}}
\caption[Decomposition of the state $m=14$, $n=62$, $B=25$ into $B=0$
basis states.]
{Decomposition of the state $m=14$, $n=62$, $B=25$ into $B=0$ 
basis states. The area of a circle is proportional to the square of the 
amplitude. The numerically exact results are shown below the diagonal 
$p=q$. Above the diagonal are the weights of theoretical wavefunction 
Eq.\ (4.8). A part of the constant energy circle $p^2 + q^2 = 
\mathrm{const}$ is shown as well. \label{4_3}}
{\vspace{1.2 cm}}
\end{figure}

In the ``numerical'' calculations the two-dimensional Schr\"odinger
equation was solved by diagonalization of the Hamiltonian in the basis of
states in the square without magnetic field. The states with the required
symmetry are
\begin{equation}
\Psi _{pq}\left( x,y\right) =\left\{ 
\begin{array}{l}
\sqrt{2}\left[ \cos \pi px\cos \pi qy+\cos \pi qx\cos \pi py\right] ,\ p\neq
q\ \mathrm{odd} \\ 
2\cos \pi px\cos \pi qy,\ p=q\ \mathrm{odd} \\ 
\sqrt{2}\left[ \sin \pi px\sin \pi qy-\sin \pi qx\sin \pi py\right] ,\ p\neq
q\ \mathrm{even}
\end{array}
\right. .
\end{equation}
The labels $p,q$ should not be confused with the resonances. Expand $\hat
\psi _m\left( x\right) =\sum \bar \psi _{m,l}e^{i\pi lx}$ in Eq.\ (\ref
{epePsi}) where $l$ is integer for $r$ even and half-odd integer for $r$
odd. Also $\bar \psi _{m,l}=\left( -1\right) ^m\bar \psi _{m,-l}.$ We
expect the largest weights in the state $\left( n,m\right) $ come from the
$B=0$ states with $p=\frac n2+l,$ $q=\frac n2-l.$ Figure \ref{4_3} shows
these states aligned along the diagonal in $\left( p,q\right) $ plane. The
energy of the basis states is $p^2+q^2=\frac{n^2}2+2l^2$ dropping a factor
of $\pi ^2.$ If $l^2\ll n,$ it is much closer to the base energy $n^2/2$
than the next base energy $\left( n\pm 2\right) ^2/2\approx \frac{n^2}2\pm
2n.$ Nevertheless, there are basis states with other base energies lying
near the constant energy circle (Fig.\ \ref{4_3}), for example, the states
with $p^{\prime }=\frac{n+2}2+l^{\prime },$ $q^{\prime
}=\frac{n+2}2-l^{\prime }$ where $l^{\prime 2}\simeq n.$ However, the
matrix elements of the Hamiltonian $\mathcal{H}_{pq,p^{\prime }q^{\prime
}}$ are small if the differences $\left| p-p^{\prime }\right| ,$ $\left|
p-q^{\prime }\right| ,$ etc., are large. When the condition $n\gg l^2\sim
m^2$ breaks down, the basis states with other base energies will be
contributing to the eigenstate. This is the case when the terms left out
in the phase of Eq.\ (\ref{epeP0}) become important. The terms that shift
$n$ are of order $\left( E_m-V_{\min }\right) /k\sim m^2/n.$ Thus, for
large $m$ our perturbation theory is not too accurate. 

The current density is given by 
\begin{equation}
\mathbf{j}\left( x,y\right) =2\mathrm{Re}\Psi ^{*}\left( x,y\right) \left[
-i\mathbf{\nabla }-\mathbf{A}\left( x,y\right) \right] \Psi \left(
x,y\right) . 
\label{vroej}
\end{equation}
The property $\mathbf{\nabla \cdot j}=0$ ensures that the streamlines are
the closed loops. However, because of the approximate nature of both
numerical and theoretical calculations, the small left-over divergence in
the regions of small current density may have a big effect. This makes it
problematic to construct a reasonable approximation for the streamlines.
To overcome this difficulty we \emph{imposed }the divergence-free
character of the current by representing $\mathbf{j}\left( x,y\right)
=\mathbf{\nabla }\times \mathbf{\hat z}\chi \left( x,y\right) $ where the
function $\chi \left( x,y\right) =\int_{1/2}^yj_x\left( x,y^{\prime
}\right) dy^{\prime }.$ Then the streamlines of $\mathbf{j}$ are the
contour lines of $\chi .$ In the exact problem $\chi \left( x,y\right)
=\chi \left( y,x\right) .$ To see this, note that the Hamiltonian has a
symmetry $\mathcal{H}\left( x,y\right) =\mathcal{H}^{*}\left( y,x\right)
,$ which means that for a non-degenerate state the wavefunction has the
same symmetry (up to a phase factor). From this follows the symmetry for
the current $j_x\left( x,y\right) =-j_y\left( y,x\right) .$ Then $\chi
\left( y,x\right) =-\int_{1/2}^xj_y\left( x^{\prime },y\right) dx^{\prime
}=\int_{1/2}^x\frac{\partial \chi \left( x^{\prime },,y\right) }{\partial
x^{\prime }}dx^{\prime }=\chi \left( x,y\right) ,$ using the condition
$j_x\left(\frac 12,y^{\prime }\right) =0.$ In practice, however, $\chi $ is
not symmetric. This is the pay-off for making the approximate current
non-divergent. Therefore in our calculations we used the symmetrized
version $\frac 12\left[ \chi \left( x,y\right) +\chi \left( y,x\right)
\right] .$ In the figures below we have a good agreement between thus
obtained streamlines and other types of current density representations. 

\begin{figure}[tbp]
{\hspace*{2.7cm} 
\psfig{figure=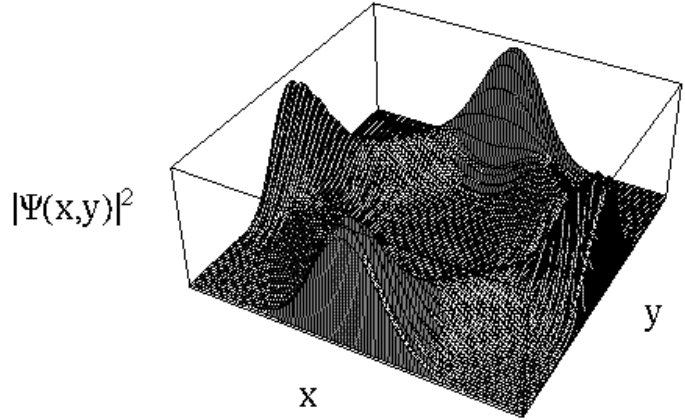,height=5.81cm,width=9cm,angle=0}}
{\vspace*{.13in}}
\caption[Absolute square of the numerical wavefunction for the localized 
paramagnetic state $m=0$, $n=62$, $B=31.4$.]
{Absolute square of the numerical wavefunction for the localized
paramagnetic state $m=0$, $n=62$, $B=31.4$. According to Eq.\ (4.19), the
profile along the side of the billiard is proportional to $| \hat
\psi_0 |^2$ and there are fast interference oscillations in the
transverse direction. 
\label{4_4}}
{\vspace{1.2 cm}}
\end{figure}

\begin{figure}[tbp]
{\hspace*{2.7cm} 
\psfig{figure=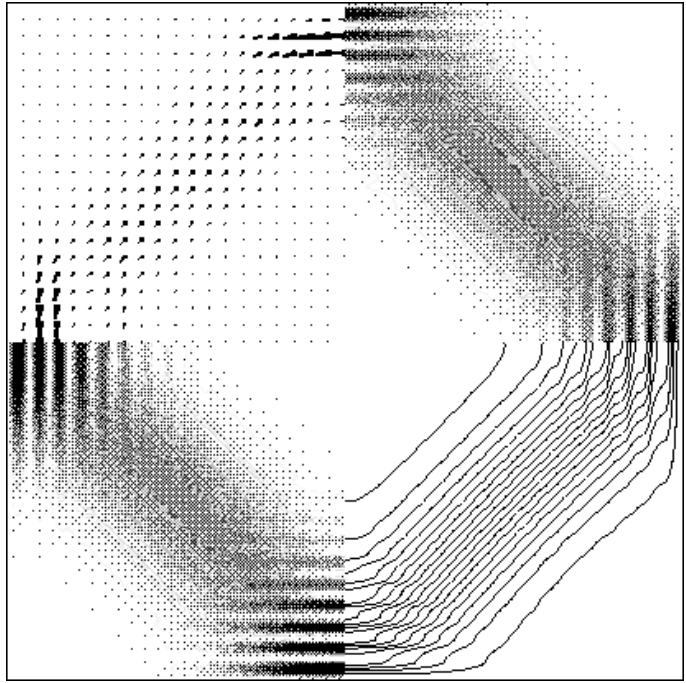,height=9.15cm,width=9cm,angle=0}}
{\vspace*{.13in}}
\caption[Numerical wavefunction and current for the state in Fig.\ 4.4.]
{Numerical wavefunction and current for the state in Fig.\ 4.4. The state
is symmetric under the $90^\circ$ rotation. Counter-clockwise from the
lower left: a density plot of $|\Psi|^2$, current streamlines, a density
plot of $| \mathbf j |$, and a vector field representation of the current
(the dot is at the calculated point, the size and direction of the stick
represents the current density magnitude and direction). 
\label{4_5}}
{\vspace{1.2 cm}}
\end{figure}

\begin{figure}[tbp]
{\hspace*{2.7cm} 
\psfig{figure=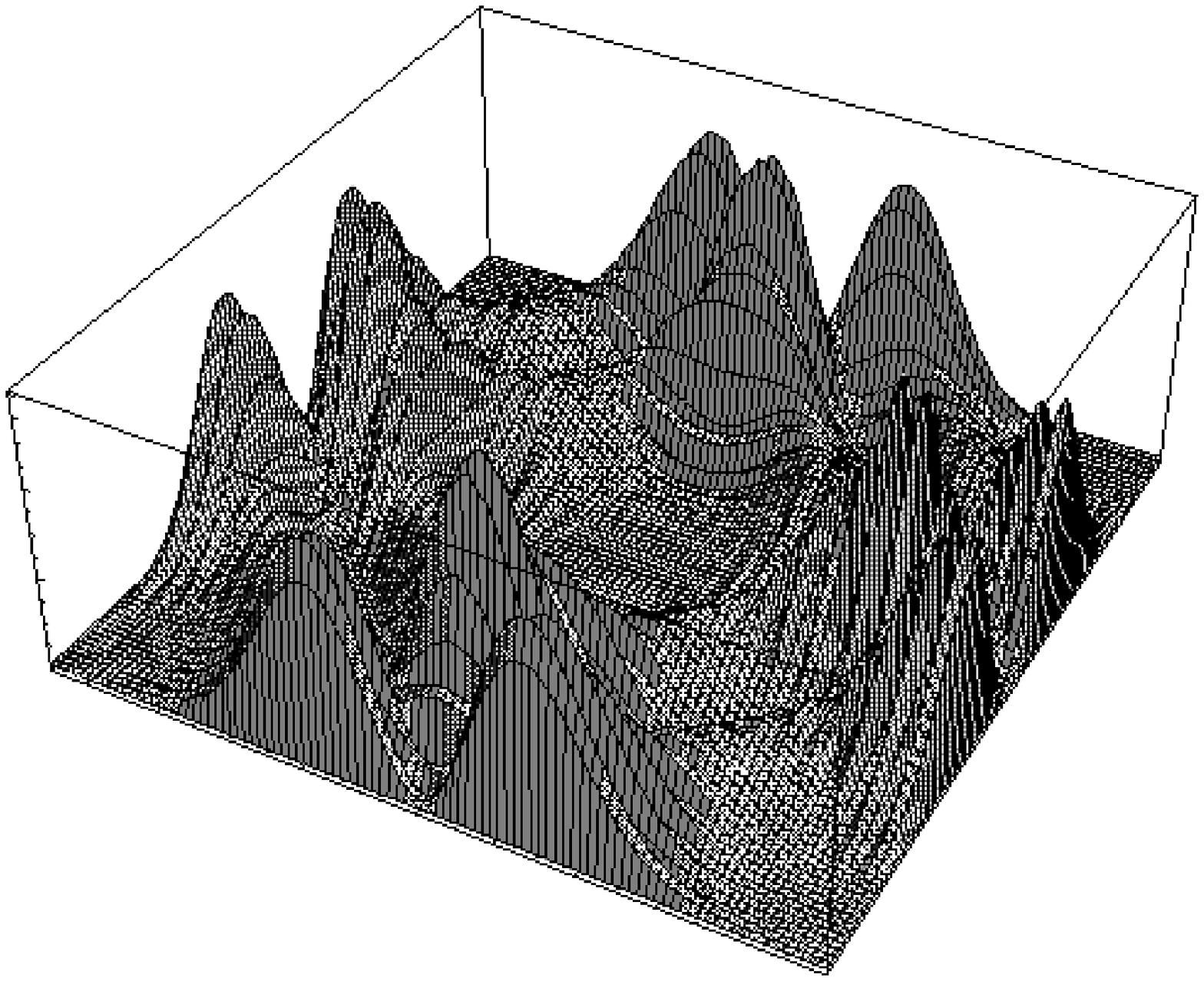,height=7.43cm,width=9cm,angle=0}}
{\vspace*{.13in}}
\caption{Absolute square of the numerical wavefunction for the state $m=1$,
$n=60$, $B=31.4$. 
\label{4_6}}
{\vspace{1.2 cm}}
\end{figure}

\begin{figure}[tbp]
{\hspace*{2.7cm} 
\psfig{figure=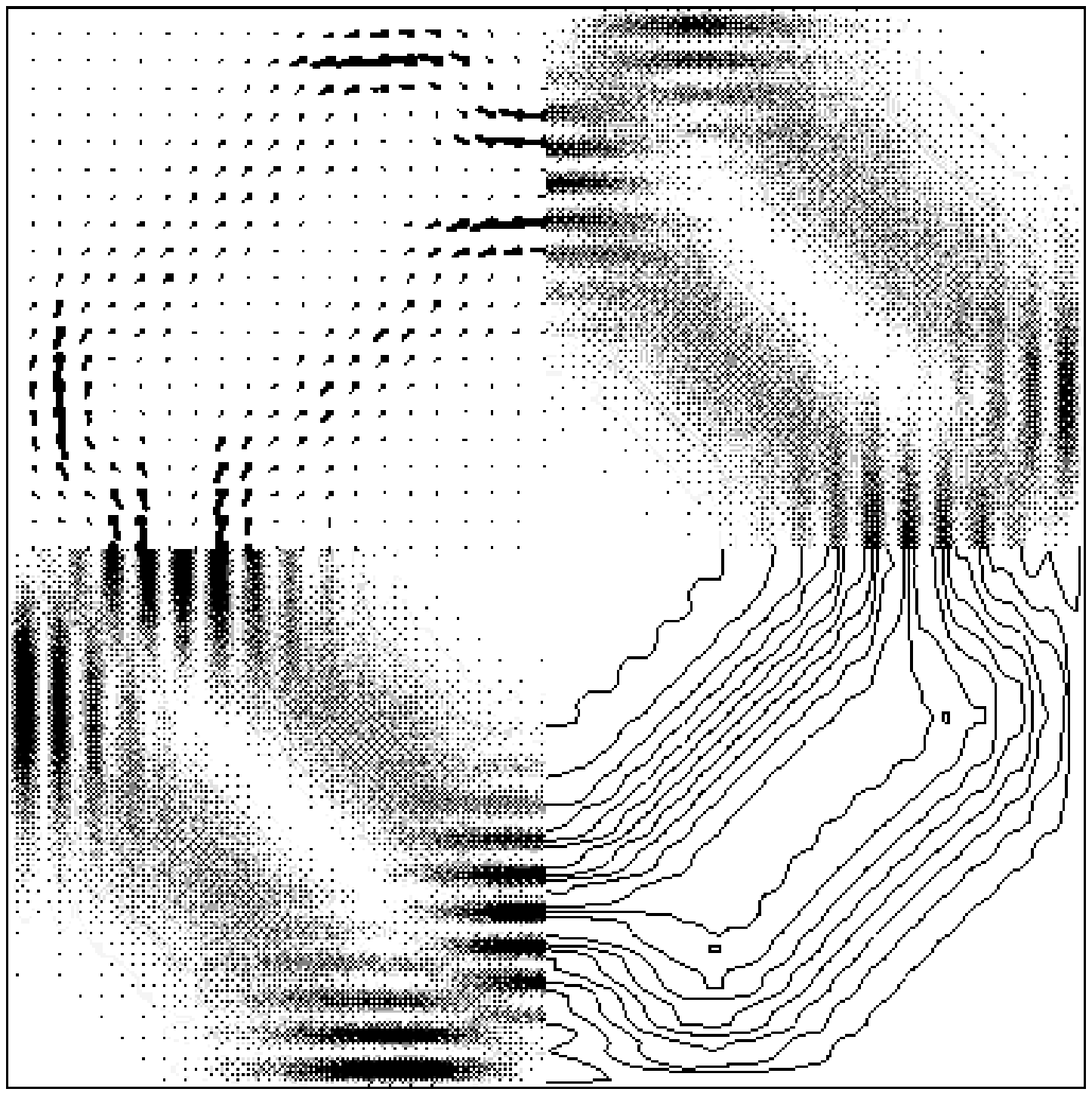,height=9.15cm,width=9cm,angle=0}}
{\vspace*{.13in}}
\caption[Numerical wavefunction and current for the state in Fig.\ 4.6.]
{Numerical wavefunction and current for the state in Fig.\ 4.6. 
The same representation as Fig.\ 4.5.
\label{4_7}}
{\vspace{1.2 cm}}
\end{figure}

\begin{figure}[tbp]
{\hspace*{2.7cm} 
\psfig{figure=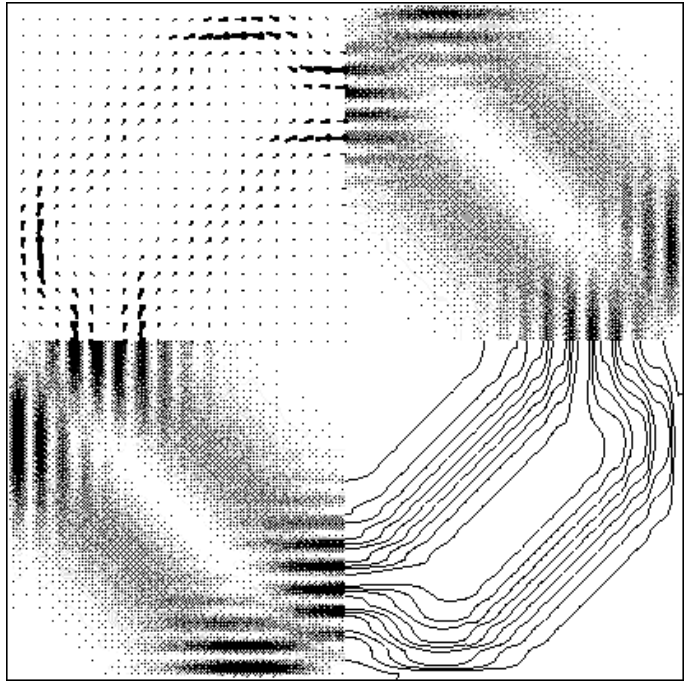,height=9.15cm,width=9cm,angle=0}}
{\vspace*{.13in}}
\caption[Theoretical wavefunction and current for the state in Fig.\ 4.6.]
{Theoretical wavefunction and current for the state in Fig.\ 4.6. The same
representation as Fig.\ 4.5. 
\label{4_8}}
{\vspace{1.2 cm}}
\end{figure}

Qualitatively, the $\left( 1,1\right) $ resonance states can be divided in
four classes. The first class consists of the states of the low transverse
modes with $E_m-V_{\min }\ll Bk.$ These states are strongly localized
along the stable orbit and have relatively simple paramagnetic current
patterns. In Figs.\ \ref{4_4}-\ref{4_8} the magnetic field is $B=31.4.$
Figure \ref{4_4} shows a three-dimensional representation of the numerical
wavefunction $n=62,$ $m=0.$ The theoretically estimated wavenumber is
$k\approx 2\pi 62/\mathcal{L}\approx 140.$ The parameter $k\sqrt{\epsilon
}=\sqrt{Bk}\approx 66$ is large indicating the strong localization due to
the resonance. Because of the localization only one of the four terms in
Eq.\ (\ref{epePsi}) dominates each side of the square periodic orbit. For
example, near $x=-y=1/4$ only the $s=0 $ term is appreciable. Here $\left|
\Psi \left( x,y\right) \right| ^2\approx \left| \hat \psi _0\left(
x-y-\frac 12\right) \right| ^2$ which is well approximated by a Gaussian.
Near the square boundary there is interference between two terms. For
example, near $y=-1/2$ the $s=0,3$ terms are large, so
\begin{eqnarray}
\left| \Psi \left( x,y\right) \right| ^2 &\approx &\left| e^{i\left( \pi
n/2\right) y}\hat \psi _0\left( x-y-\frac 12\right) +e^{-i\left( \pi
n/2\right) y}\hat \psi _0\left( -x-y-\frac 12\right) \right| ^2  \nonumber \\
\ &=&4\left| \hat \psi _0\left( x\right) \cos \left( \frac{\pi n}2y\right)
\right| ^2.  \label{vroeP}
\end{eqnarray}
The cosine factor produces interference fringes parallel to the border. At
the first maximum $\left| \Psi \left( 0,-\frac 12+ \frac 1n\right) \right|
^2$ is four times larger than $\left| \Psi \left(\frac 14,-\frac 14\right)
\right| ^2.$ The cross-section $\left| \Psi \left( x,-\frac 12+\frac
1n\right) \right| ^2$ is also a Gaussian. The interference pattern can be
seen in the density plot in the lower left quarter of Fig.\ \ref{4_5}. The
rest of the figure shows the current for this state. Starting on the lower
left and going counter-clockwise we show a density plot of $|\Psi|^2$, the
streamlines, the absolute value of the current density $\left|
\mathbf{j}\left( x,y\right) \right| ,$ and the vector field
$\mathbf{j}\left( x,y\right) $ (the dot is at the calculated point and the
stick size and direction represent the current density magnitude and
direction). Of course, close to the boundary the current is parallel to
the boundary. Its interference fringes mimic those of the wavefunction. 

Figures \ref{4_6}-\ref{4_8} show another state of the same class $n=60,$
$m=1.$ Its surface of section wavefunction is close to the first excited
state of a harmonic oscillator. Clearly, the probability and current
densities are very small near the point $x=-y=1/4.$ They are not exactly
zero, however, because the terms with $s=1$-$3$ in Eq.\ (\ref{epePsi})
have exponentially small tails there. Figures \ref{4_6}, \ref{4_7} were
obtained numerically and Fig.\ \ref{4_8} shows the results of theoretical
calculations. We see that the theory works well in the low $m$ case. 

\begin{figure}[tbp]
{\hspace*{2.7cm} \psfig{figure=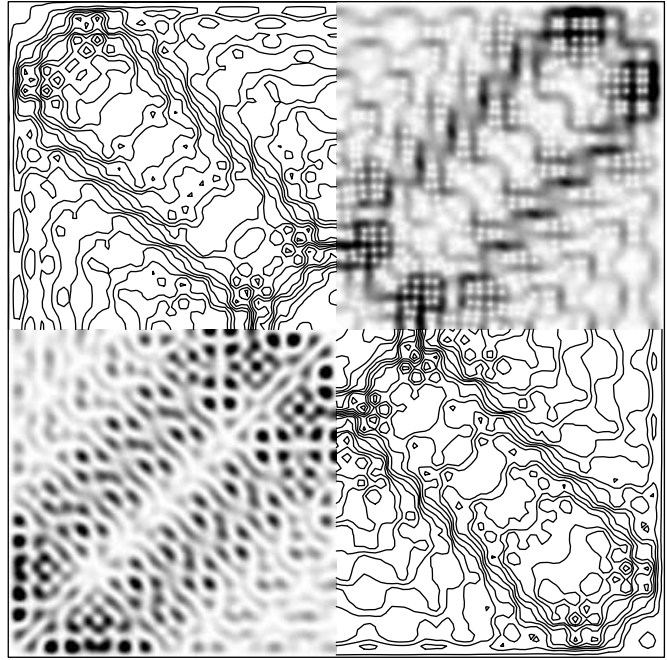,height=9cm,width=9cm,angle=0}}
{\vspace*{.13in}}
\caption[Almost non-magnetic state $m=10$, $n=62$, $B=25$.]
{Almost non-magnetic state $m=10$, $n=62$, $B=25$. The numerical results 
are shown as before, except for the upper-left corner, where the 
theoretical streamlines are shown. The theory and numerics disagree 
mainly in the low-current regions. The diamagnetic loops close to the 
diagonals nearly cancel the paramagnetic loops in the triangular wedges 
between the diagonals.
\label{4_9}}
{\vspace{1.2 cm}}
\end{figure}

\begin{figure}[tbp]
{\hspace*{2.7cm} 
\psfig{figure=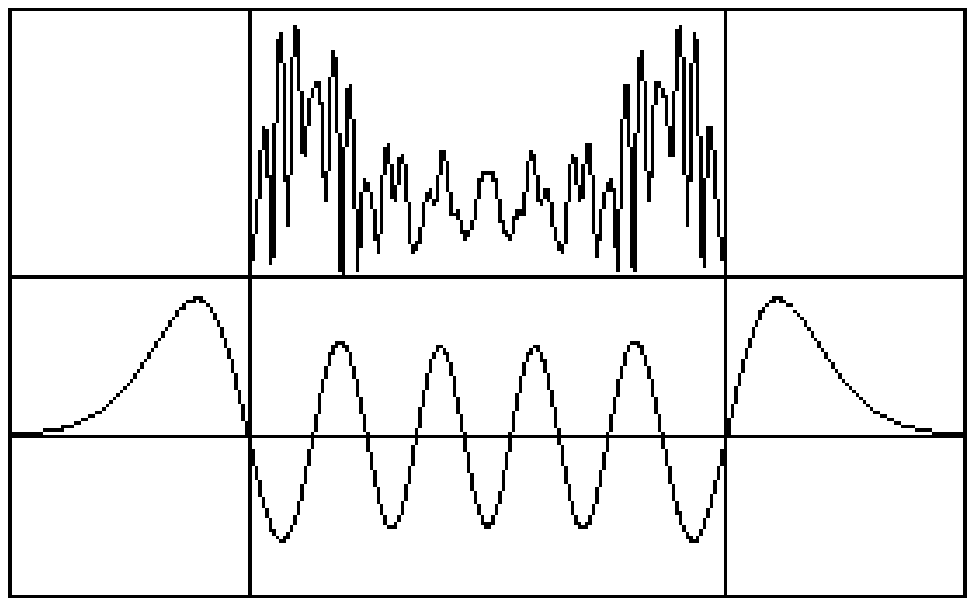,height=5.8cm,width=9cm,angle=0}}
{\vspace*{.13in}}
\caption[Normal derivative $| \partial \Psi_{nm} / \partial y|_{y=-1/2}$,
obtained numerically, and the theoretical surface of section function 
$\hat \psi_m (x)$ for the state of Fig.\ 4.9.]
{Normal derivative $| \partial \Psi_{nm} / \partial y|_{y=-1/2}$, obtained
numerically (upper plot), and the theoretical surface of section function
$\hat \psi_m (x)$ (lower plot) for the state of Fig.\ 4.9. When $m$ is
large, $\hat \psi_m (x)$ extends considerably outside the domain $\left[ -
\frac 1 2, \frac 1 2 \right]$ (indicated by the vertical lines). In order
to construct the normal derivative, the tails have to be ``folded'' back
with an additional phase factor. For small $m$ both functions are
proportional. 
\label{4_10}}
{\vspace{1.2 cm}}
\end{figure}

The states with $\left| E_m\right| \ll V_{\max }=Bk/2\mathcal{L}$ belong
to the second class. These states are delocalized and have rather
complicated probability and current distributions because all four terms
in Eq.\ (\ref {epePsi}) are important throughout the square. These states
lie on the border of para- and diamagnetism and are weakly magnetic.
Figure \ref{4_9} shows such state $n=62,$ $m=10,$ $B=25.$ It has very
small magnetization (Fig.\ \ref{4_2}) because the loops of rather strong
diamagnetic and paramagnetic currents nearly cancel each other. The figure
shows the numerical results as before, except for the upper left corner,
where the theoretical streamlines are shown. The theory and numerics
differ primarily in the lower current regions. To understand the
complexity of the higher $m$ states and the lack of localization consider
the normal derivative $\left| \partial \Psi _{62,10}/\partial y\right|
_{y=-1/2}$ on the boundary (Fig.\ \ref{4_10}). In the low $m$ state it
would be proportional to $| \hat \psi _m\left( x\right) | $ [cf.\ Eq.\
(\ref{vroeP})]. When $m$ becomes larger $\hat \psi _m\left( x\right) $
extends significantly beyond the physical domain $\left[ -\frac 12,\frac
12\right] $ (Fig.\ \ref{4_10}), so it has to be ``folded'' back to this
domain with an additional rapidly varying phase factor $e^{i\kappa x}.$
Formally, it means that the terms with $s=1,2$ in Eq.\ (\ref {epePsi})
cannot be neglected. 

\begin{figure}[tbp]
{\hspace*{2.7cm} 
\psfig{figure=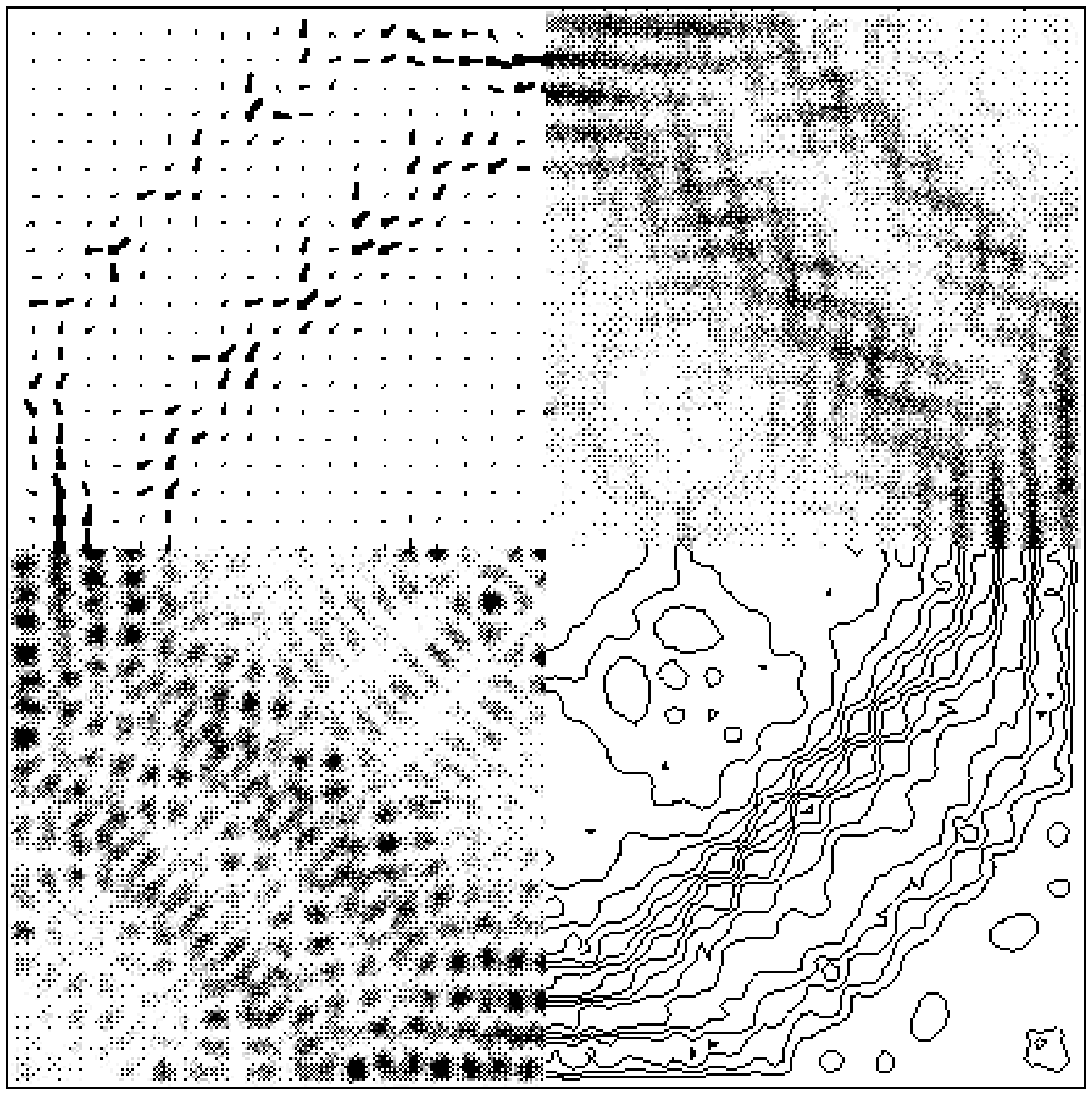,height=9.15cm,width=9cm,angle=0}}
{\vspace*{.13in}}
\caption[Numerical wavefunction and current for the state of maximum
diamagnetism $m=14$, $n=62$, $B=25$.]
{Numerical wavefunction and current for the state of maximum diamagnetism 
$m=14$, $n=62$, $B=25$. The same representation as Fig.\ 4.5.
\label{4_11}}
{\vspace{1.2 cm}}
\end{figure}

\begin{figure}[tbp]
{\hspace*{2.7cm} 
\psfig{figure=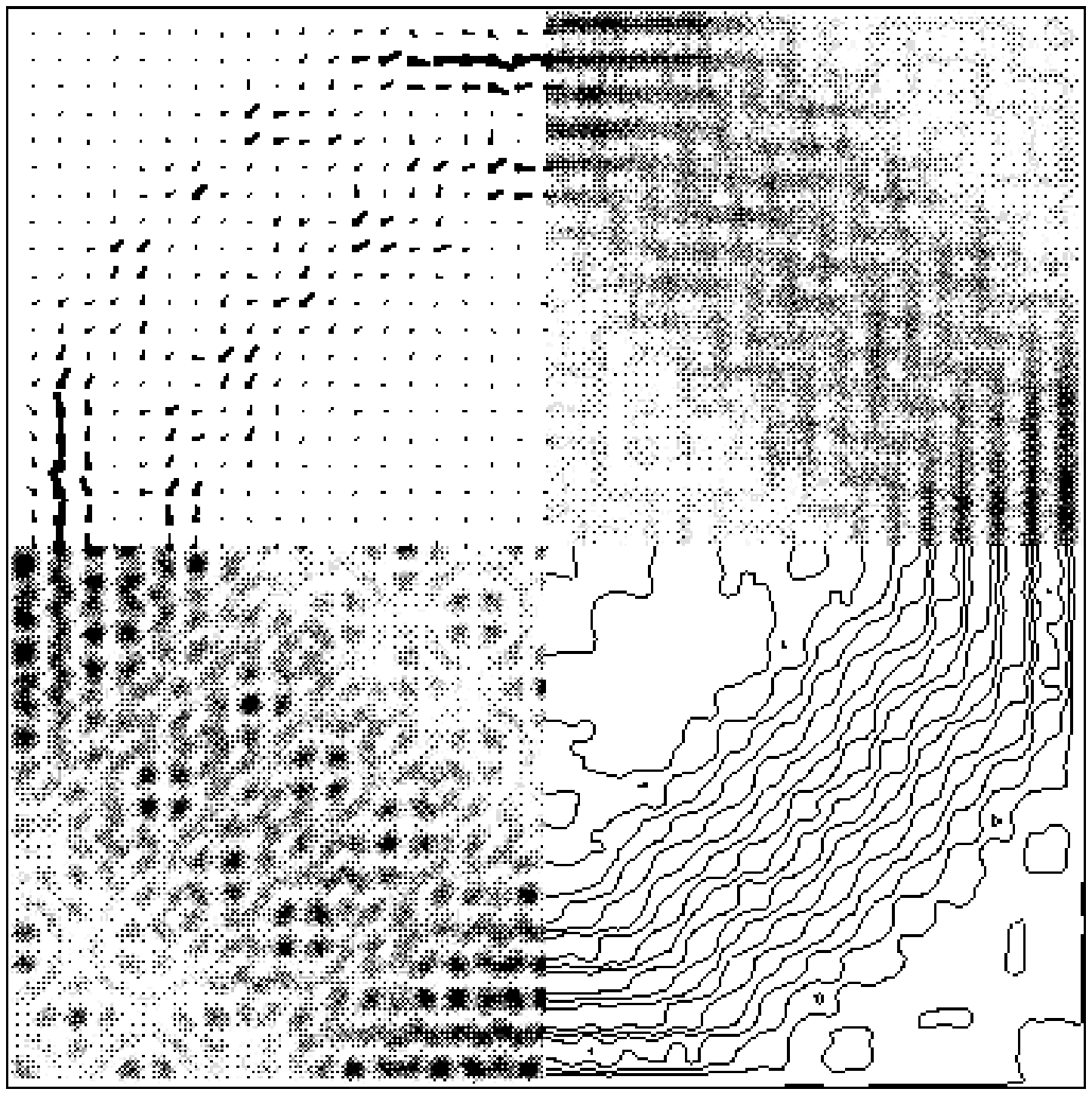,height=9.15cm,width=9cm,angle=0}}
{\vspace*{.13in}}
\caption[Theoretical wavefunction and current for the state in Fig.\ 4.11.]
{Theoretical wavefunction and current for the state in Fig.\ 4.11.  The
same representation as Fig.\ 4.5. 
\label{4_12}}
{\vspace{1.2 cm}}
\end{figure}

\begin{figure}[tbp]
{\hspace*{2.7cm} 
\psfig{figure=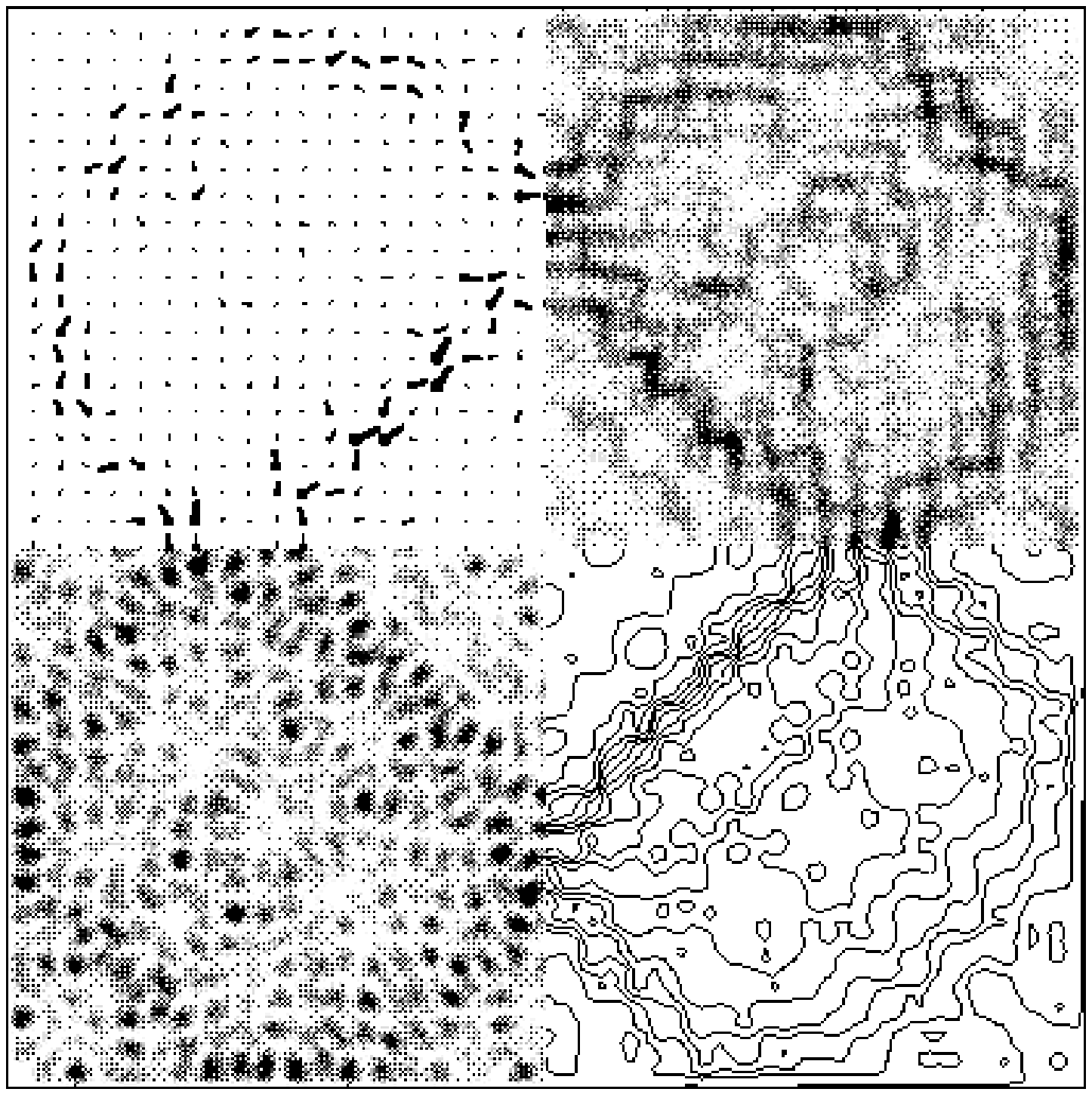,height=9.15cm,width=9cm,angle=0}}
{\vspace*{.13in}}
\caption[Numerical wavefunction and current for the diamagnetic state
$m=14$, $n=62$, $B=31.4$.]
{Numerical wavefunction and current for the diamagnetic state $m=14$,
$n=62$, $B=31.4$. The same representation as Fig.\ 4.5. 
\label{4_13}}
{\vspace{1.2 cm}}
\end{figure}

The third class includes the diamagnetic states with $E_m$ near the top of
the potential. The state at the maximum of the magnetization curve (Fig.\
\ref{4_2}) with $n=62,$ $m=14,$ $B=25$ is shown in Figs.\ \ref{4_11}
(numerics) and \ref{4_12} (theory). In the case of large $m$ the theory
does relatively poorly, although it is still qualitatively correct. The
unperturbed basis states beyond the diagonal $p+q=n$ appear in the
Hamiltonian matrix (Fig.\ \ref{4_3}, below the line $p=q$), while the
theory does not reflect this (Fig.\ \ref{4_3}, above the line $p=q$). This
state is localized near the unstable orbit, i.e.\ this is a scar state that
we predict. One can even see the two channels in Fig.\ \ref{4_11} because
if $E_m$ is just below $V_{\max }$ the wavefunction $\hat \psi _m\left(
x\right) $ has the excessive weight near the turning points on both sides
of the maximum of the potential at $x=1.$ In this case we find
$E_m=628.57$ and $V_{\max }=648.64.$ For comparison, we show the state
$n=62,$ $m=14,$ $B=31.4$ (Fig.\ \ref{4_13}). Here $E_m=703.4$ is much
lower than $V_{\max }=792.2,$ so the two channels are further apart. 

\begin{figure}[tbp]
{\hspace*{2.7cm} 
\psfig{figure=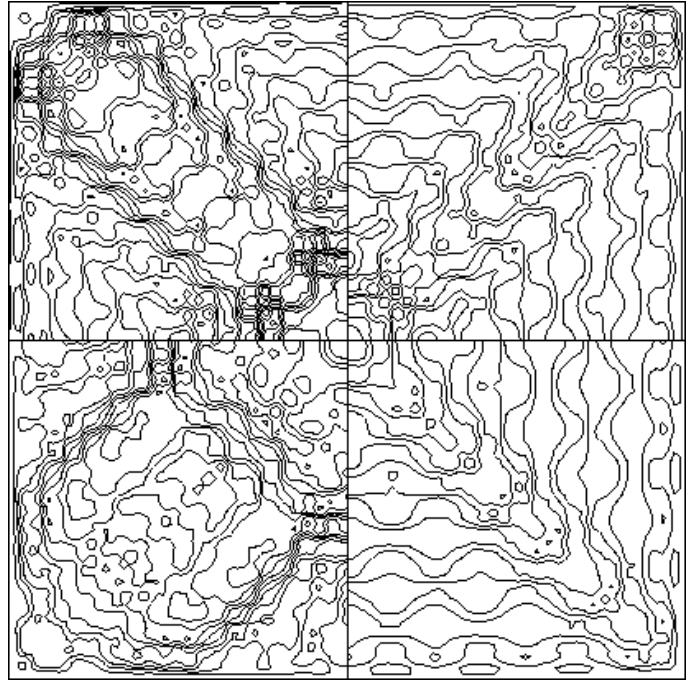,height=9.15cm,width=9cm,angle=0}}
{\vspace*{.13in}}
\caption[Streamlines for the sequence of states $m=6, 8, 10, 12$, $n=62$, 
$B=25$.]
{Streamlines for the sequence of states $m=6, 8, 10, 12$, $n=62$, $B=25$, 
counter-clockwise from the lower right.
\label{4_14}}
{\vspace{1.2 cm}}
\end{figure}

One can observe the change of the topology of the current from paramagnetic
to diamagnetic by looking at the sequence of states $m=6,8,10,12$, $n=62,$
and $B=25$ (Fig.\ \ref{4_14}). Similar effect could be achieved by the 
change of magnetic field.

The states with $E_m\gg V_{\max }$ belong to the fourth class. The magnetic
field is a small perturbation in this case and the states are weakly
diamagnetic. The expansion near $\left( 1,1\right) $ resonance will
eventually break down and higher resonances become involved.

\section{Aharonov-Bohm flux line}

The theory laid out in the previous sections can be extended to
non-uniform flux configurations. If the Lorentz force is neglected then
the problem is reduced to finding the flux enclosed by a periodic orbit.
Of course, if the flux distribution is not smooth enough, like the ABFL
case, we have to worry about the diffraction. To estimate the effect of
diffraction consider a finite size flux line of radius $\rho $ and flux
$\phi .$ An orbit that goes through the flux region is subjected to the
Lorentz force produced by the magnetic field $B=\phi /\pi \rho ^2$ and
gets deflected by an angle
\begin{equation}
\delta \theta \sim \frac{\phi /\phi _0}{k\rho }
\end{equation}
where $\phi _0$ is the flux quantum. The diffraction can be neglected if
$\delta \theta \ll 1.$ Clearly, in the zero radius flux line the
diffraction effects cannot be neglected. If we define the perturbation
parameter $\epsilon =\left( \phi /\phi _0\right) /k,$ i.e.\ the ratio of
the perturbed part of the action to the unperturbed action, then the
condition is $\epsilon /\rho \ll 1.$ This means that the limit $\rho
\rightarrow 0$ cannot be taken before $\epsilon \rightarrow 0$ if we want
to neglect the diffraction. Therefore we have to consider a finite size
flux line. In the numerical examples we take $\phi /\phi _0=0.1$ and $\rho
=0.01,$ while $k>120.$ Instead of one flux line, four quarter-strength
lines symmetrically located were used in order to reduce the amount of
calculations. Within the approximation of our theory it gives essentially
the same result as a single line. Note that for the ideal ABFL the matrix
elements of the Hamiltonian are infinite if the basis functions do not
vanish on the flux line. One can avoid this problem by making the
substitution $\Psi \rightarrow r^{\left| \phi /\phi _0\right| }\bar \Psi $
where $r$ is the distance from the flux line \cite{Nar,Pra}. The new
wavefunction $\bar \Psi $ satisfies the equation
$\overline{\mathcal{H}}\bar \Psi =E\bar \Psi $ with the same energy as
$\bar \Psi .$ The most singular term of the non-hermitian Hamiltonian
$\overline{\mathcal{H}}$ is of order $r^{-1}$ instead of $r^{-2}.$

We consider the states near the $\left( 1,1\right) $ resonance. A periodic
orbit of this family may either enclose the flux line and have $\pm 2\pi
\phi /\phi _0$ added to its action, or pass by the flux line and have no
extra action. Hence the one-dimensional wavefunction satisfies the
Schr\"odinger equation (\ref{epeSch}) with a step-wise effective
potential. Strictly speaking, the walls of the potential will have a
finite width $\rho ,$ because when the orbit goes \emph{through} the flux
line there is an additional action between 0 and $\pm 2\pi \phi /\phi _0.$
Of course, the profile of the walls depends on the flux distribution
within the line. If we are not going beyond the second order of the
perturbation theory (Sec.\ \ref {fott}), we may disregard the finite width
of the walls, since the wavefunction does not change much. This way the
results would be independent of $\rho .$ The third order contains the
derivative of the potential. In order to neglect it we must require
$k\epsilon ^{3/2}/\rho \ll 1.$ Since $\epsilon /\rho \ll 1,$ we have the
condition $k\sqrt{\epsilon }\lesssim 1,$ or $k\epsilon \ll 1.$

The width and position of the square well and barrier depend on the location
of the flux. If the flux is located at $x=0,$ $y=-\frac 12+a,$ where $0\leq
a\leq 1/2,$ the potential is 
\begin{equation}
V\left( x\right) =\left\{ 
\begin{array}{l}
-2\pi \frac \phi {\phi _0}\frac k{\mathcal{L}},\quad x\in \left[ -a,a\right]
\\ 
+2\pi \frac \phi {\phi _0}\frac k{\mathcal{L}},\quad x\in \left[
1-a,1+a\right] \\ 
0,\quad x\in \left[ a,1-a\right]
\end{array}
\right.  \label{abflV}
\end{equation}
extended periodically by $V\left( x+2\right) =V\left( x\right) .$ Hence,
assuming $\phi >0,$ the potential consists of the well of width $2a$ and
depth $2\pi \frac \phi {\phi _0}\frac k{\mathcal{L}}$ in the interval
$\left[ -\frac 12,\frac 12\right] $ (counter-clockwise orbits) and the
same size barrier in $\left[ \frac 12,\frac 32\right] $ (clockwise
orbits). For sufficiently large $\frac \phi {\phi _0}\frac k{\mathcal{L}}$
the eigenfunctions are approximately $\hat \psi _m\left( x\right) =\cos
\left[ \left( m+1\right) \pi x/2a+m\pi /2\right] ,$ $\left| x\right| <a,$
and zero elsewhere. This expression holds for sufficiently small $m.$ The
energy $E_m\approx $ $-2\pi \frac \phi {\phi _0}\frac
k{\mathcal{L}}+\frac{\pi ^2\left( m+1\right) ^2}{4a^2}.$

\begin{figure}[tbp]
{\hspace*{2.7cm} 
\psfig{figure=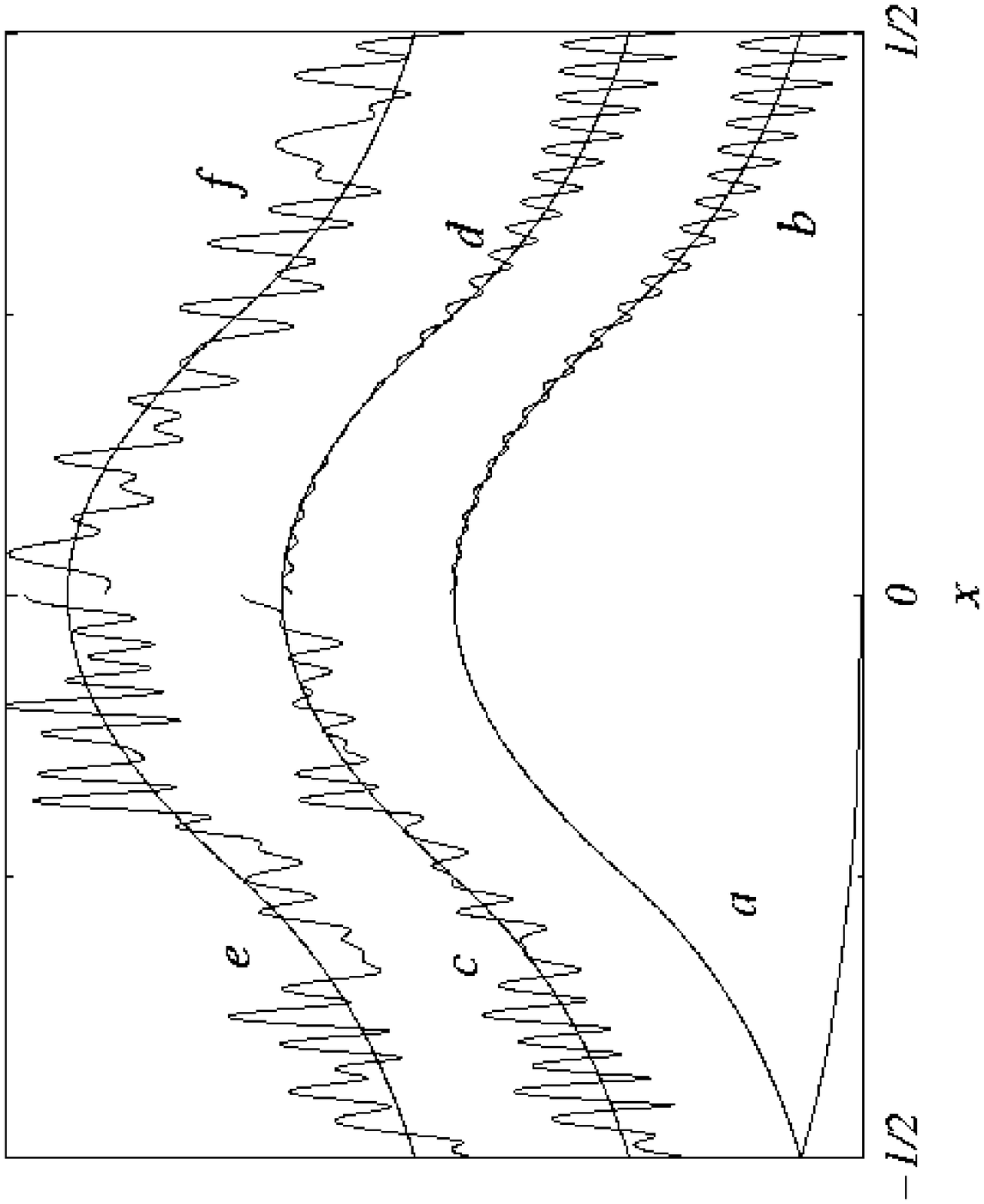,height=7.35cm,width=9cm,angle=270}}
{\vspace*{.13in}}
\caption[State $m=0$, $n=86$ in the square with the flux line, $a= \frac 1
4$, $\phi = 0.1 \phi_0$, $\rho = 0.01$.]
{State $m=0$, $n=86$ in the square with the flux line, $a= \frac 1 4$,
$\phi = 0.1 \phi_0$, $\rho = 0.01$. (a) Upper curve $\hat \psi_0 (x)$,
lower curve $\hat \psi_0 (-1-x)$. This function is superimposed on the
other plots that show the normal derivative $| \partial \Psi / \partial
n|_{y=-1/2}$: (b) is the theoretical result from Eq.\ (4.8), (c) is from
the full numerical diagonalization, (d) is from the diagonalization in the
reduced basis that consists of the states along the diagonal $p=\frac
n2+l,$ $q=\frac n2-l$ (Sec.\ 4.1.4). For comparison, curves (e) and (f)
show the numerical results for the ideal single flux line with $n=82$
and $n=70$, respectively. Although there are strong diffraction effects,
the overall shape is given correctly by the theory.
\label{4_15}}
{\vspace{1.2 cm}}
\end{figure}

\begin{figure}[tbp]
{\hspace*{2.7cm} 
\psfig{figure=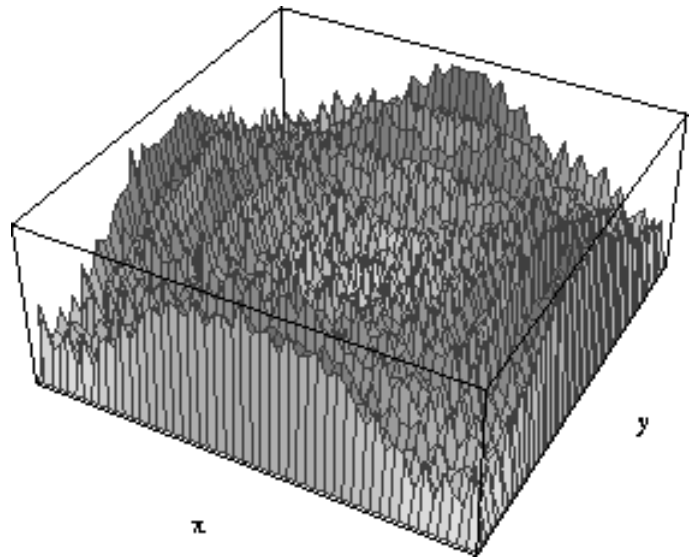,height=7.5cm,width=9cm,angle=0}}
{\vspace*{.13in}}
\caption[Absolute value of the wavefunction, $n=58$, $m=0$, for an ideal
single flux line, $a= \frac 1 4$, $\phi = 0.1 \phi_0$.]
{Absolute value of the wavefunction, $n=58$, $m=0$, for an ideal single
flux line, $a= \frac 1 4$, $\phi = 0.1 \phi_0$. Note that $| \Psi (x,y)|$
is almost but not strictly symmetric under the $90^\circ$ rotation due to
the higher order effects. 
\label{4_16}}
{\vspace{1.2 cm}}
\end{figure}

The function $\hat \psi _0$ for $a=1/4$ and $n=86$ is shown in Fig.\
\ref{4_15} (a). We include the exponentially small portion of the function
for $\left| x\right| >1/4.$ The upper curve is $\hat \psi _0\left(
x\right) $ and the lower curve is $\hat \psi _0\left( -1-x\right) .$ The
PSS wavefunction $\psi \left( x\right) =e^{i\kappa x}\hat \psi \left(
x\right) $ should be compared to the normal derivative of the
two-dimensional wavefunction $\left| \partial \Psi /\partial n\right| $ at
$y=-1/2$ [curve (b), Eq.\ (\ref{epePsi})]. The oscillations appear
because of interference of $\psi \left( x\right) $ and $\psi \left(
-1-x\right) ,$ they scale as $\hat \psi _0\left( -1-x\right) .$ The other
four curves show the normal derivative calculated numerically. Curve (c)
is from full numerical diagonalization, (d) is from the diagonalization in
the reduced basis that includes only the unperturbed states along the
diagonal $p=\frac n2+l,$ $q=\frac n2-l$ (Sec.\ \ref{vroe}). The latter
result is very close to the theoretical. Curves (e) and (f) are the
numerical results for the ideal single ABFL ($\rho =0$) with $n=82$ and
$n=70.$ Although there are strong diffraction corrections, the overall
shape is given well by the theory. The two-dimensional wavefunction is
localized along the stable orbit. The strength of localization is
characterized by $\hat \psi _0.$ Figure \ref{4_16} shows $\left| \Psi \left(
x,y\right) \right| $ for an ideal ABFL, $n=58.$ Although the localization
is not very strong, the wavefunction has little support in the center and
the corners of the square. 

\begin{figure}[tbp]
{\hspace*{2.7cm} 
\psfig{figure=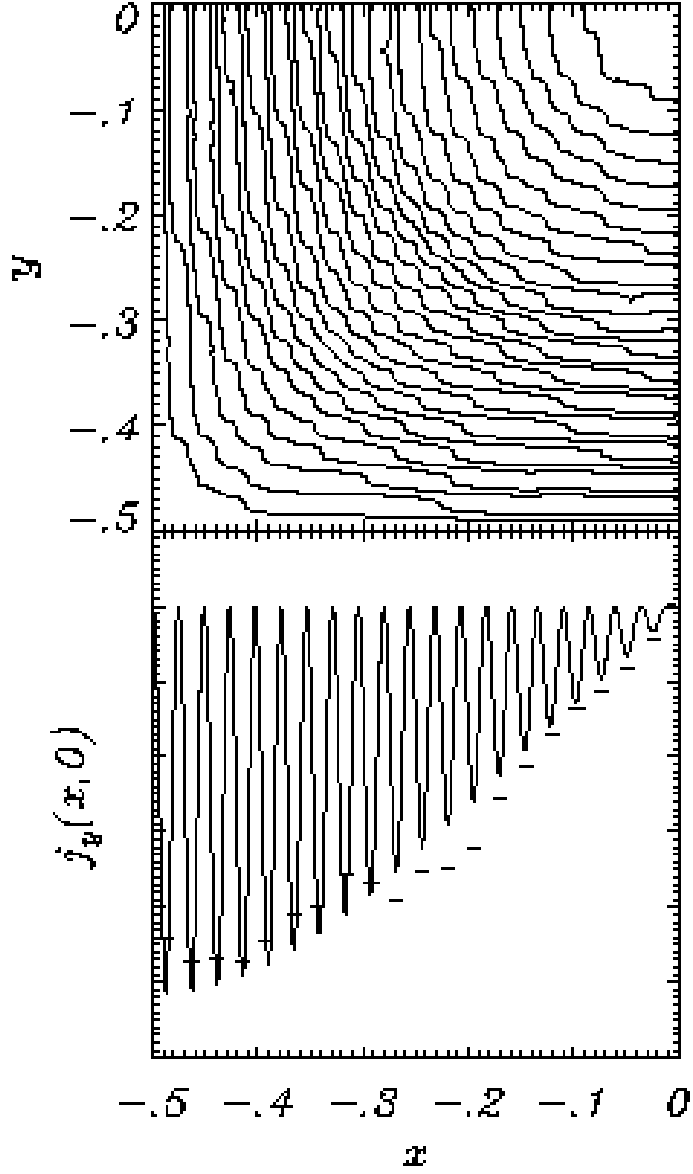,height=15.7cm,width=9cm,angle=0}}
{\vspace*{.13in}}
\caption[Ideal flux line at the center of the square, $\phi = 0.1 \phi_0$.
Current in the state $n=82$, $m=0$, symmetric under the $90^\circ$
rotation.]
{Ideal flux line at the center of the square, $\phi = 0.1 \phi_0$. 
Current in the state $n=82$, $m=0$, symmetric under the $90^\circ$
rotation. Upper figure: numerical streamlines for a quarter of the square. 
Lower figure: theoretical current density $j_y(x,0)$, dashes indicate the 
numerical minima. 
\label{4_17}}
{\vspace{1.2 cm}}
\end{figure}

If the flux is located at the center of the square ($a=1/2$), the
potential well extends to the whole side of the square and there is no
spatial localization. For small $m$ the counter-clockwise orbits are
preferred and there is a strong paramagnetic current. The upper part of
Fig.\ \ref{4_17} shows the streamlines for the state $n=82,$ $m=0,$ symmetric
under the $90^{\circ } $ rotation. The lower part of the figure gives the
current density along the line $y=0,$ $x\in \left[ -\frac 12,0\right] .$
Neglecting the $A$-term in Eq.\ (\ref{vroej}) we have an approximate
formula for it, $j_y\left( x,0\right) \propto -\left( \cos \pi nx/2\right)
^2\left[ \hat \psi _m\left( -x-\frac 12\right) ^2-\hat \psi _m\left(
x-\frac 12\right) ^2\right] .$ For low $m$ the second term in the square
brackets is small when $x<0,$ so $j_y\leq 0.$ It has equally spaced double
zeros at $x=\left( 2l+1\right) /n.$ The factor depending on $\hat \psi _m$
has a zero at $x=0.$

It is interesting to compare the square with ABFL with the ``step''
billiard. This is a square with the step-wise boundary perturbation. For
example, suppose the lower side is perturbed by $\xi \left( x\right)
=-\epsilon ,$ $\left| x\right| <a,$ and $\xi \left( x\right) =0,$ $\frac
12>\left| x\right| >a.$ Then for the $\left( 0,1\right) $ resonance
(bouncing ball states with the large $y$-momentum) the effective potential
is proportional to Eq.\ (\ref{abflV}) limited to $x\in \left[ -\frac
12,\frac 12\right] $ with the periodicity $V\left( x+1\right) =V\left(
x\right) $ (the system has the time-reversal symmetry). When $k\epsilon
\ll 1$ and $k\sqrt{\epsilon }\sim 1$, our theory predicts localization
within $\left| x\right| <a.$ If $k\epsilon =\pi /2,$ i.e.\ $2\epsilon $ is
a half-wavelength, the classical action for the $\left( 0,1\right) $
orbits within $\left| x\right| <a$ and outside of this region differ by
$\pi .$ This is equivalent to the case $\phi /\phi _0=1/2.$ Since the
action enters the phase of the $T$-operator, this difference becomes
ambiguous, it can be made $-\pi $ by adding the phase $2\pi $ for $\left|
x\right| >a.$ (This would be impossible in the case of continuous
perturbation.) Thus the effective potential is undetermined when $k\epsilon
\gtrsim 1.$ The preliminary numerical results \cite{Pra} show that in this
case there are states localized within $\left| x\right| <a$ \emph{and} the
states localized outside of this region. 

\section{Experimental suggestions}

\label{pe2}

Some of the experimental methods mentioned in Sec.\ \ref{pe} could
conceivably be adapted to the billiard with a magnetic flux. The mesoscopic
systems like the quantum corrals \cite{CroLutEig} or the GaAs squares \cite
{LevReiPfe} are directly related to our theoretical model, though accurately
measuring the wavefunction may be a challenge.

In the liquid surface wave experiments the effect of the flux can be
modeled by the flow of the medium \cite{BerChaLar}. If $\mathbf{V}\left(
\mathbf{r}\right) $ is the velocity of the liquid then there is a
correspondence
\begin{equation}
\frac{e\mathbf{A}\left( \mathbf{r}\right) }{c\hbar }\longleftrightarrow 
\frac{-k\mathbf{V}\left( \mathbf{r}\right) }{v_g\left( k,\mathbf{r}\right) }
\end{equation}
where the group velocity $v_g\ll \left| \mathbf{V}\right| .$ (This analogy
neglects the $A^2$ term in the Hamiltonian.) Thus, rotating the tank with a
constant angular velocity is equivalent to applying the uniform field to a
quantum billiard. The flux 
\begin{equation}
2\pi \frac \phi {\phi _0}\longleftrightarrow -\frac k{v_g}\oint
\mathbf{V\cdot dr}
\end{equation}
in the case of homogeneous medium. A flux line would be analogous to a
vortex formed by the water pouring through a small hole on the bottom of the
tank. Reference \cite{BerChaLar} reports the experiments on scattering of
the surface waves on a vortex.

\begin{figure}[tbp]
{\hspace*{2.7cm} \psfig{figure=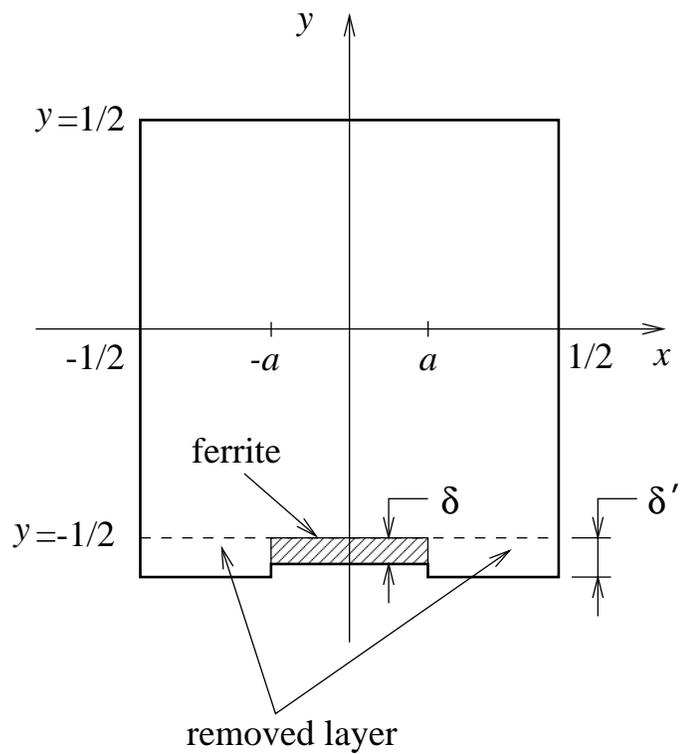,height=9.96cm,width=9cm,angle=0}}
{\vspace*{.13in}}
\caption[Possible microwave experiment with ferrite.]
{Possible microwave experiment with ferrite. The ferrite layer of width
$\delta$ and length $2a$ is mounted on the wall. Part of the wall of width
$\delta^\prime$ is removed to compensate for the width of the ferrite. The
dashed line indicates the boundary of the original square. 
\label{4_18}}
{\vspace{1.2 cm}}
\end{figure}

In the microwave field experiments \cite{SoAnlOtt,GokWuBri,So} a ferrite
strip of length $2a$ can be embedded in the wall $y=-1/2$ (Fig.\
\ref{4_18}). If one applies the static magnetic field to the ferrite, the
wave will acquire an additional phase upon reflection from the strip. The
phase is different for the forward and backward directions, i.e.\ the time
reversal symmetry is broken. This system is analogous to the ABFL square
billiard with the flux line located distance $a$ from the bottom. Suppose
the square cavity is in $xy$-plane, and the ferrite is under the magnetic
field in $z$-direction. Then the permeability of the ferrite in the
absence of losses is
\begin{equation}
\hat \mu =\left[ 
\begin{array}{ccc}
\mu _{\Vert } & -i\varkappa  & 0 \\ 
i\varkappa  & \mu _{\Vert } & 0 \\ 
0 & 0 & \mu _z
\end{array}
\right] .
\end{equation}
Consider a plane wave with electric field
$\mathbf{E}_i\mathbf{=}E\mathbf{\hat z}e^{i\left( k_xx-k_yy\right) }$
($k_y>0$) propagating in the cavity. Suppose this wave is incident on the
ferrite layer of width $\delta .$ Assuming the Dirichlet conditions on the
metal wall to which this layer is attached, one can show that upon
reflection the wave becomes $\mathbf{E}_r\mathbf{=-}E\mathbf{\hat
z}e^{i\left( k_xx+k_yy+\phi \right) }$ where the phase\footnote{This
differs from the result reported in Ref.\ \cite{So}.}
\begin{equation}
\phi =2\tan ^{-1}\left[ \frac{k_y\left( \mu _{\Vert }^2-\varkappa ^2\right)
\sin \left( k_y^f\delta \right) }{k_y^f\mu _{\Vert }\cos \left( k_y^f\delta
\right) -k_x\varkappa \sin \left( k_y^f\delta \right) }\right] .
\end{equation}
Here $k_y^f=\sqrt{k_f^2-k_x^2}$ where $k_f=k\sqrt{\mu _{\Vert }-\varkappa
^2/\mu _{\Vert }}$ is the wavenumber inside the ferrite. The time-reversal
is achieved by reversing the sign of $k_x.$ Then $\phi $ changes. In the
quantum billiard the phase changes sign but keeps the magnitude when time
is reversed. Here it is not the case because $\phi $ includes the phase
that the wave acquires by traveling through the bulk of the ferrite. In
order to separate the effect of the magnetic field from the effect of the
width we may remove a metal layer from the wall outside of the strip, i.e.\
for $\left| x\right| >a$ (Fig.\ \ref{4_18}). The width of the removed layer
$\delta ^{\prime }$ is determined by the condition
\begin{equation}
2k_y\delta ^{\prime }=\frac 12\left[ \phi \left( k_x\right) +\phi \left(
-k_x\right) \right] 
\end{equation}
meaning that the phase the wave acquires outside of the ferrite strip should
be equal to the average phase on the strip. Then the effective phase 
$\phi_{\mathrm{eff}} = \phi - 2k_y \delta^\prime$ is odd in $k_x$. In the 
experiments the wavenumber in the ferrite $k_f$ was about ten times 
larger than $k$, so we expect $\delta^\prime > \delta$ to compensate for 
the phase.

\section{Conclusions}

The classically weak magnetic field acts as a perturbation in an integrable
square billiard. In this system there are special states that are localized
near the short stable periodic orbits that enclose a finite flux. The
perturbation breaks the time-reversal symmetry and there is a preferred
current direction. Not surprisingly, the states carry persistent currents.
We observed a variety of probability and current distributions. The overall
magnetic response ranges from the paramagnetic for the well-localized states
to the diamagnetic. The short periodic orbits dominate the susceptibility.
The localization also takes place for an off-center Aharonov-Bohm flux line.
Although there is no direct classical effect in this case, the vector
potential changes the action and enters the phase of the wavefunction. The
diffraction effects are quite important for the flux line and limit the
preciseness of the semiclassical results. 

\chapter{Quasiclassical Born-Oppenheimer approximations}

\label{qboa}

Up to now all our results followed from the Bogomolny equation. It gives the
natural semiclassical formulation and reduces the dimensionality of the
problem through the use of Poincar\'e's surface of section (PSS). The
perturbation theory allows one to make an asymptotic expansion in the phase
of the wavefunction in a controllable way and relate the quantum states to
the classical phase space. The method has certain shortcomings. To find the
full two-dimensional wavefunction an additional trivial, but lengthy, step
is required. The resulting function is inaccurate near the caustics and in
the classically forbidden regions.

There are cases, on the other hand, when the equivalent results can be
obtained by the adiabatic, or Born-Oppenheimer, approximation (BOA) \cite
{BorOpp}. In general, the BOA is appropriate when a subset of the system's
coordinates varies in time slower than the remaining coordinates. This
allows an approximate separation of variables in the partial differential
equation describing the system. We will see in the examples below that such
separation of variables is possible sometimes even if the notion of ``fast''
and ``slow'' is not well defined (Sec.\ \ref{wgm2}). In any case, the
conditions for the separation always follow from the differential equation
itself, and we refer to all such cases as the BOA.

The relationship between the BOA and the $T$-operator method is quite
complex \cite{ZaiNarPra}. Both methods are usually equivalent in the
leading order. Unlike the $T$-operator, the BOA directly generates the
two-dimensional wavefunction. If the PSS is chosen along the slow
direction, the one-dimensional PSS wavefunction will be a part of the BOA
result. The $T $-operator method often produces an asymptotic form of the
BOA\ solution. It is not always clear whether the BOA can be constructed
for a given problem. For instance, we were unable to formulate the BOA in
a simple form for a perturbed circle, apart from the states near the
$\left( 1,2\right) $ resonance (Sec.\ \ref{bita}). The $T$-operator, once
constructed, treats all resonances uniformly. The purpose of this chapter
is to illustrate the above remarks on several examples. 

\section{Textbook example}

We remind the reader of the standard example where the BOA can be used
\cite {Bay}. In the molecules or solids the electrons' positions
$\mathbf{r}_e$ can be treated as ``fast'' compared to the ``slow'' ionic
positions $\mathbf{R}_i.$ This is based on the small electron-ion mass
ratio $m_e/M_i.$ In the Schr\"odinger equation
\begin{equation}
\left[ -\frac{\hbar ^2}{2m_e}\nabla _e^2-\frac{\hbar ^2}{2M_i}\nabla
_i^2+V\left( \mathbf{r}_e,\mathbf{R}_i\right) \right] \Psi \left(
\mathbf{r}_e,\mathbf{R}_i\right) =E\Psi \left(
\mathbf{r}_e,\mathbf{R}_i\right)
\end{equation}
we make the Born-Oppenheimer ansatz 
\begin{equation}
\Psi \left( \mathbf{r}_e,\mathbf{R}_i\right) =\Phi \left(
\mathbf{r}_e|\mathbf{R}_i\right) \psi \left( \mathbf{R}_i\right) . 
\end{equation}
Here we assume that $\Phi \left( \mathbf{r}_e|\mathbf{R}_i\right) $ is the
$N$th electronic eigenstate for fixed ionic variables which solves
\begin{equation}
\left[ -\frac{\hbar ^2}{2m_e}\nabla _e^2+V\left(
\mathbf{r}_e,\mathbf{R}_i\right) \right] \Phi \left(
\mathbf{r}_e|\mathbf{R}_i\right) =U\left( \mathbf{R}_i\right) \Phi \left(
\mathbf{r}_e|\mathbf{R}_i\right)
\end{equation}
where $\mathbf{R}_i$ is treated as a parameter. The eigenenergy $U\left( 
\mathbf{R}_i\right) $ then acts as a potential for the slow variable: 
\begin{equation}
\left[ -\frac{\hbar ^2}{2M_i}\nabla _i^2+U\left( \mathbf{R}_i\right) \right]
\psi \left( \mathbf{R}_i\right) =E\psi \left( \mathbf{R}_i\right) .
\end{equation}
The adiabatic invariance provides that the electronic eigenstate label $N$
does not change as $\mathbf{R}_i$ is slowly varied. The approximation made
to the Schr\"odinger equation implies that 
\begin{equation}
\left| \frac{\nabla _i\psi \nabla _i\Phi }{M_i}\right| ,\left| \frac{\psi
\nabla _i^2\Phi }{M_i}\right| \ll \left| \frac{\psi \nabla _e^2\Phi
}{m_e}\right| .  \label{teest}
\end{equation}
In the case of the ground state of hydrogen molecule one can estimate
$\left| \nabla _i\Phi \right| \sim \left| \nabla _e\Phi \right| \sim
\left| \Phi \right| /a_B$ where $a_B$ is the Bohr radius. The ions will
oscillate with frequency $\omega \sim e/\sqrt{M_ia_B^3}$ and amplitude
$\delta R_i\sim \hbar /M\omega \sim \left( m_e/M_i\right) ^{1/4}a_B.$
Hence we estimate $\left| \nabla _i\psi \right| \sim \left| \psi \right|
/\delta R_i\sim \left| \psi \right| \left( M_i/m_e\right) ^{1/4}/a_B.$
Therefore the first term on the left of Eq.\ (\ref{teest}) is the biggest
and it is $\left( M_i/m_e\right) ^{3/4}$ times smaller than the
\emph{r.h.s}. This example shows that the BOA amounts to neglecting
certain derivatives in the differential equation. The higher order
corrections can also be written down. 

It was shown that the interaction between the fast quantum electronic and
the slow classical ionic degrees of freedom may lead to chaotic behavior 
\cite{BluEss}.

\section{Bouncing ball states}

\label{bbs}

In Sec.\ \ref{bwsp} we have shown how to derive the bouncing ball (BB)
states in a Bunimovich stadium (Fig.\ \ref{3_10}) using the $T$-operator.
We obtained an almost square well effective potential which made the
states localized within the straight part of the billiard. Equivalently,
we could use the BOA \cite{BaiHosSte} since there is a separation between
the fast and slow motion. We have also mentioned a generalization of this
problem: the stadium with the endcaps of slightly different radii (Fig.\
\ref{3_11}). In the latter case the effective potential is a square well
with a sloped bottom, and the states are spatially shifted towards the
wider part of the billiard. In both cases the wavefunction stays away from
the semicircular regions, so the exact shape of the boundary there is of
little consequence for the states with the low transverse quantum number
$m.$ Another interesting example is the $\pi/3$-rhombus billiard
\cite{BisJai}, which has two degenerate families of the BB states. 

In this section we apply the BOA method to the $\left( 0,1\right) $
resonance states in the tilted unit square introduced in Sec.\ \ref{ets}. The
lower side of the square has a slope, $y=\epsilon x,$ and the states of
interest have a small $x$-momentum compared to the $y$-momentum. These are,
of course, the BB modes, and, as we just explained, they are similar to the
BB states in a tilted stadium (the effective potential is twice as large in
the latter case, because two sides are tilted).

We solve the Helmholtz equation $\left( \nabla ^2+k^2\right) \Psi =0$ with
the BOA ansatz $\Psi \left( x,y\right) =\Phi \left( y|x\right) \psi \left(
x\right) .$ The fast equation is 
\begin{equation}
\frac{\partial ^2\Phi }{\partial y^2}=-U\left( x\right) \Phi .
\end{equation}
The solution 
\begin{equation}
\Phi \left( y|x\right) =\sqrt{\frac 2{1-\epsilon x}}\sin \left( \pi
n\frac{1-y}{1-\epsilon x}\right)
\end{equation}
vanishes at $y=\epsilon x$ and $y=1$ and is normalized. From this we find
the function $U\left( x\right) =\left[ \pi n/\left( 1-\epsilon x\right)
\right] ^2\approx \pi ^2n^2+2\epsilon \pi ^2n^2x.$ Then the slow function
satisfies the equation 
\begin{equation}
-\psi ^{\prime \prime }+\alpha ^3x\psi =\mathcal{E}_m\psi  \label{bbsAi}
\end{equation}
where $\mathcal{E}_m=k^2-\pi ^2n^2$ and $\alpha =\left( 2\epsilon \pi
^2n^2\right) ^{1/3}.$ Note that $\alpha ^3x$ is in the leading order
$k^2\epsilon $ times the effective potential $2x$ of the $T$-operator
method if the lower side is chosen as the PSS. Hence $\psi \left( x\right)
$ is the surface of section wavefunction given by Eq.\ (\ref{ptSch}) with
$E_m=\mathcal{E}_m/k^2\epsilon $. The energy $k^2$ and the two-dimensional
wavefunction are also given correctly. 

Equation (\ref{bbsAi}) has a solution 
\begin{equation}
\psi \left( x\right) =\mathrm{Ai}\left( \alpha x-z_m\right)
\end{equation}
for large $\alpha .$ Here $z_m$ is a root of $\mathrm{Ai}\left( -z\right)
$, so that $\psi \left( 0\right) =0.$ If $\alpha -z_m\gg 1,$ $\psi $ also
effectively vanishes at $x=1.$ The transverse energy $\mathcal{E}_m=\alpha
^2z_m$ must be much smaller than the maximum potential $\alpha ^3,$ i.e.\
$\alpha \sim \left( k^2\epsilon \right) ^{1/3}\gg z_m.$ The wavefunction
$\Psi \left( x,y\right) $ is localized within $x\lesssim z_m/\alpha .$ If
this condition is not satisfied, Eq.\ (\ref{bbsAi}) still can be solved by
a linear combination of $\mathrm{Ai}$ and $\mathrm{Bi}$ functions. In the
opposite case, $k^2\epsilon \ll 1,$ the slope of the side can be neglected
in the leading order. This is the situation when the non-resonant
perturbation theory applies and there is no localization. The BOA is
applicable if $\left| \Phi ^{-1}\left( \partial \Phi /\partial x\right)
\right| \left| \psi ^{-1}\psi ^{\prime }\right| \ll \left| \Phi
^{-1}\left( \partial ^2\Phi /\partial y^2\right) \right| .$ When
$z_m/\alpha \ll 1,$ we estimate $\left| \psi ^{-1}\psi ^{\prime }\right|
\sim \alpha ,$ and the above condition gives $\epsilon ^4\ll k$, which is
always true. For larger $m $, i.e.\ when $z_m/\alpha \sim \left(
m/k\sqrt{\epsilon }\right) ^{2/3}\gtrsim 1,$ we have $\left| \psi
^{-1}\psi ^{\prime }\right| \sim m,$ so the condition becomes $\epsilon
m\ll k\sim n.$ This shows that the BOA is valid even if $m\sim n.$

\begin{figure}[tbp]
{\hspace*{2.7cm} 
\psfig{figure=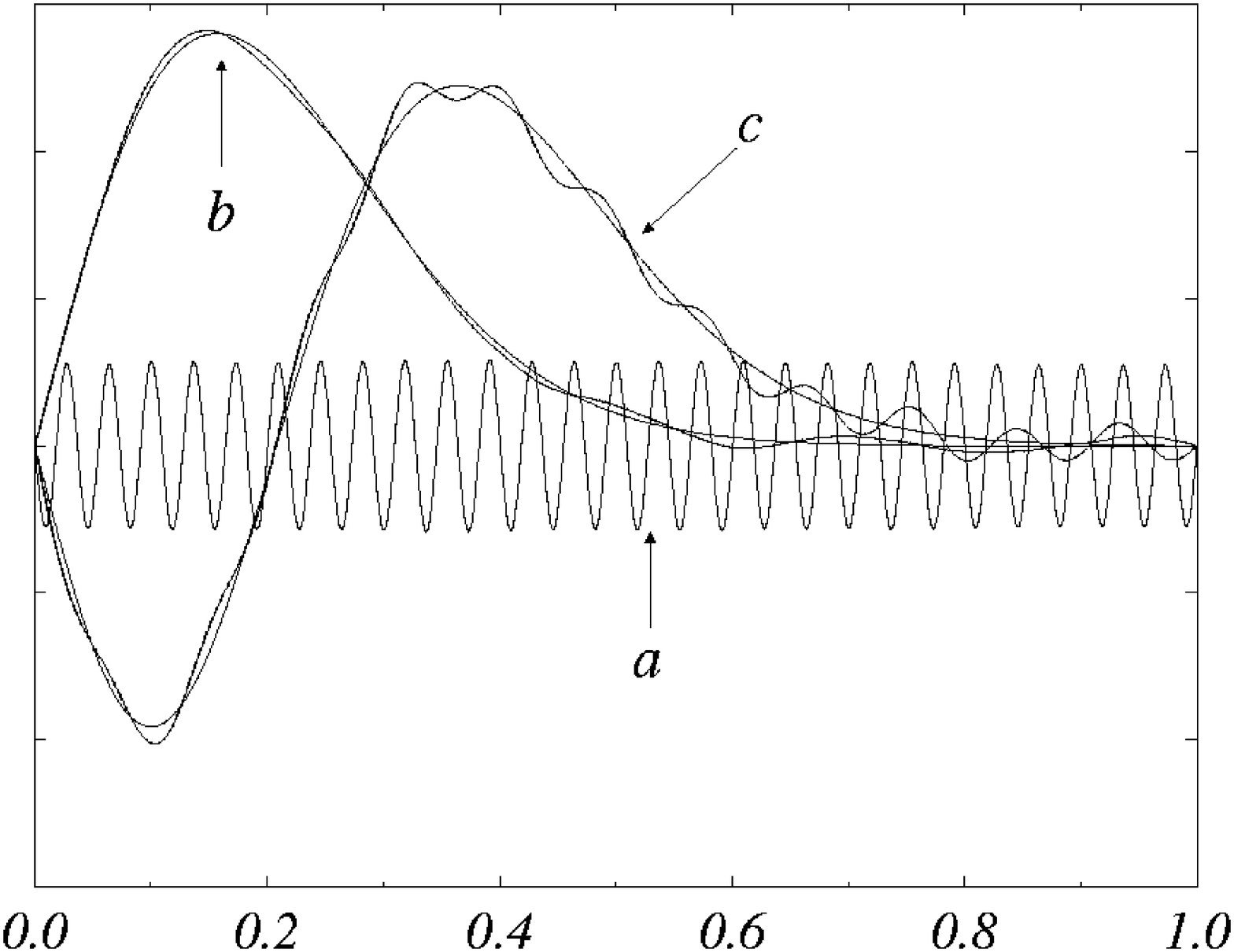,height=7cm,width=9cm,angle=0}}
{\vspace*{.13in}}
\caption[Cross-sections of the numerical wavefunctions for the $(0,1)$ 
resonance in the tilted square.]
{Cross-sections of the numerical wavefunctions $\Psi_{nm}$ for the $(0,1)$ 
resonance in the tilted square, $\epsilon = 0.01$. (a) $\Psi_{55,1} 
(x=0.01,y)$; (b) $\Psi_{55,1} (x,y=0.99)$ compared with $\mathrm{Ai} ( 
\alpha x-z_1)$; (c) $\Psi _{55,2}( x,y=0.99)$ compared with $\mathrm{Ai} 
(\alpha x-z_2)$. The magnitude of the states has been normalized.
\label{5_1}}
{\vspace{1 cm}}
\end{figure}

\begin{figure}[tbp]
{\hspace*{2.7cm} 
\psfig{figure=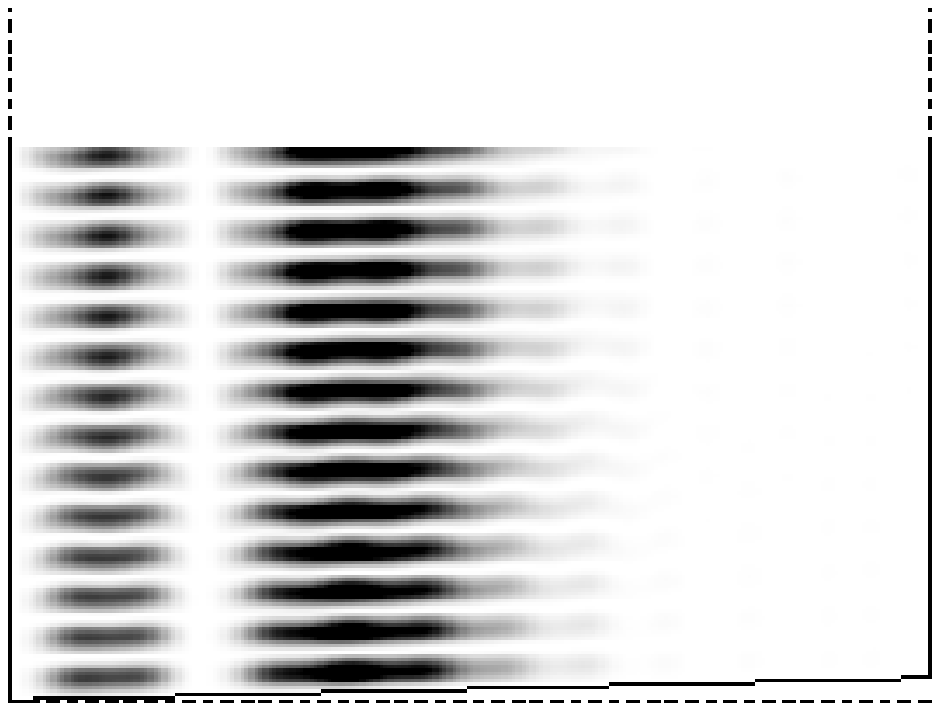,height=6.97cm,width=9cm,angle=0}}
{\vspace*{.13in}}
\caption[Density plot of $|\Psi_{55,2}(x,y)|^2$, the state of Fig.\ 5.1 (c).]
{Density plot of $|\Psi_{55,2}(x,y)|^2$, the state of Fig.\ 5.1 (c). The
lower part of the billiard ($0<y<0.25$) is shown and the graph is expanded
in the $y$ direction by a factor of three in order to display more details
of the wavefunction. The dashed line is at $y=0$. 
\label{5_2}}
{\vspace{1.2 cm}}
\end{figure}

Figure \ref{5_1} shows the cross-sections of the numerically obtained
wavefunctions $\Psi _{nm}:$ (a) $\Psi _{55,1}\left( x=0.01,y\right) $
[this is proportional to $\Phi \left( y|x=0.01\right) $], (b) $\Psi
_{55,1}\left( x,y=0.99\right) $ compared with $\mathrm{Ai}\left( \alpha
x-z_1\right) ,$ and (c) $\Psi _{55,2}\left( x,y=0.99\right) $ compared
with $\mathrm{Ai}\left( \alpha x-z_2\right) .$ Here $\epsilon =0.01$ is
less than the wavelength $\lambda =0.036$ which in turn is less than
$\sqrt{\epsilon }=0.1. $ This results in sufficiently strong localization.
The two-dimensional representation of $\left| \Psi _{55,2}\left(
x,y\right) \right| ^2$ is shown in Fig.\ \ref{5_2}. 

One concludes that in the case of the $\left( 0,1\right) $ resonance in the
tilted square the BOA is straightforward and readily provides the
two-dimensional wavefunction, while the $T$-operator method would require
more work. The BOA method is not, however, directly generalizable to the
higher resonances (Sec.\ \ref{ets}), although the next section suggests a
possible approach.

\section{Channeling approximation}

\label{chap}

The necessity to find the coordinates in which the fast and slow motion
separate may prevent us from treating generic resonances by the BOA. In
some cases the method of images resolves this problem. Consider, for
example, a rectangular billiard perturbed by a potential or a magnetic
flux. The copies of the rectangle obtained by reflection about its sides
will cover the whole two-dimensional plane, forming a lattice. (The
magnetic flux has opposite signs in the neighboring rectangles.) Suppose
that the perturbation is classically weak, that is the classical orbit
does not deviate much from the straight line after one passage across the
billiard. (For the estimate in uniform magnetic field see Sec.\
\ref{loa}.) An orbit in the original rectangle extends along a straight
line with minor deviations in the extended scheme (cf.\ Fig.\ \ref{4_1}).
Now the fast and slow directions separate and the BOA can be used. The BOA
in this case is related to the channeling method for the energetic
particles traveling through the crystal lattice \cite{Lin}. 

For definitiveness, consider the uniform magnetic field $B$ as a
perturbation. We start with the $\left( 1,1\right) $ orbits. We rotate the
coordinate system to make one variable $\xi =\left( ax+by\right) /d$ run
along the fast $\left( 1,1\right) $ direction and the other variable $\eta
=\left( -bx+ay\right) /d$ along the slow perpendicular direction. Here
$a\times b$ are the dimensions of the rectangle and $d=\sqrt{a^2+b^2}.$
The Schr\"odinger equation in these coordinates in the units $\hbar
=c=e=2m=1$ is
\begin{equation}
\left\{ \left[ -i\partial _\xi -A_\xi \left( \xi ,\eta \right) \right]
^2+\left[ -i\partial _\eta -A_\eta \left( \xi ,\eta \right) \right]
^2\right\} \Psi \left( \xi ,\eta \right) =E\Psi \left( \xi ,\eta \right) .
\label{caSch}
\end{equation}
In the channeling approximation we assume that the particle moving fast
in the $\xi $ direction ``sees'' only the averaged vector potential. We
may neglect $A^2$ comparing to the linear $kA$ terms (since $\epsilon
=B/k\ll 1$) and substitute the average vector potential $\bar A_\xi \left(
\eta \right) $ for $A_\xi \left( \xi ,\eta \right) ,$ where
\begin{equation}
\bar A_\xi \left( \eta \right) =L_\xi ^{-1}\int_0^{L_\xi }d\xi A_\xi \left(
\xi ,\eta \right)
\end{equation}
and $L_\xi =2d,$ the period in $\xi .$ Note that $\bar A_\xi L_\xi =\oint
\mathbf{A\cdot dl}$ is the flux enclosed by the periodic orbit in the
original billiard. Similarly, $\bar A_\eta L_\xi =\oint \mathbf{dl\times
A}=0 $ in the gauge where $\mathrm{div}\mathbf{A}=0.$ Now Eq.\
(\ref{caSch}) becomes separable. 

\begin{sloppypar}
The same result can be obtained on a more formal basis. Namely, we make an
ansatz $\Psi \left( \xi ,\eta \right) =\Phi \left( \xi |\eta \right) \psi
\left( \eta \right) $ and consider the vector potential as a small
perturbation in the fast equation for $\Phi $. Then, according to the
elementary perturbation theory, we can approximate $\Phi \left( \xi |\eta
\right) =e^{ik\xi }$ and find the fast eigenenergy $U\left( \eta \right) $
after multiplying the fast equation by $\Phi ^{*}$ and integrating over $\xi
.$ Again neglecting $A^2$ we find that 
\begin{equation}
U\left( \eta \right) =k^2-2k\bar A_\xi \left( \eta \right) .
\end{equation}
The $A_\eta \partial _\eta $ term automatically disappears in the BOA
approach. The slow equation is 
\begin{equation}
-\psi ^{\prime \prime }-2k\bar A_\xi \left( \eta \right) \psi =\left(
E-k^2\right) \psi .
\end{equation}
Now we can rewrite the results in the $x,y$ coordinates. Suppose the origin
is chosen at the center. Define the function $\hat \psi \left(
x-ay/b-a/2\right) =\hat \psi \left( -d\eta /b-a/2\right) =\psi \left( \eta
\right) .$ This definition relates $\hat \psi $ to the function in Eq.\ (\ref
{epeP0}) when $a=b=1.$ The above equation turns into Eq.\ (\ref{epeSch}), 
\begin{equation}
\hat \psi _m^{\prime \prime }\left( x\right) +\left[ E_m-V\left( x\right)
\right] \hat \psi _m\left( x\right) =0,
\end{equation}
with $V\left( x\right) =-2\frac{b^2}{d^2}k\bar A_\xi \left(
-\frac{bx}d-\frac{ab}{2d}\right) .$ It has the symmetries $V\left(
x\right) =V\left( -x\right) $ and $V\left( -a/2+x\right) =-V\left(
-a/2-x\right) $ and has a period $2a.$ For uniform field
\begin{equation}
\bar A_\xi \left( \eta \right) =\left\{ 
\begin{array}{l}
-B\eta \left( 1+\frac d{ab}\eta \right) ,\quad 0\leq \eta \leq \frac{ab}d,
\\ 
B\left( \frac{2ab}d+\eta \right) \left( 1+\frac d{ab}\eta \right) ,\quad 
\frac{ab}d\leq \eta \leq 2\frac{ab}d,
\end{array}
\right.
\end{equation}
with $\bar A_\xi \left( \eta +2ab/d\right) =\bar A_\xi \left( \eta \right) ,$
and $V\left( x\right) =-\left( Bkab^3/d^3\right) \left[ \frac 12-2\left(
\frac xa\right) ^2\right] $ for $\left| x\right| <a/2$ repeated
antiperiodically outside this region. An eigenfunction in this potential
satisfies the Bloch condition 
\begin{equation}
\hat \psi _m\left( x+2a\right) =e^{i\beta }\hat \psi _m\left( x\right)
\end{equation}
where $\beta $ is to be determined.
\end{sloppypar}

The two-dimensional wavefunction has the form 
\begin{equation}
\Psi _0\left( x,y\right) =e^{ik\left( ax+by\right) /d}\hat \psi _m\left(
x-ay/b-a/2\right) .
\end{equation}
This is one of the four degenerate solutions. Another solution is obtained
from this after the rotation by $\pi ,$ $\Psi _2\left( x,y\right) =\Psi
_0\left( -x,-y\right) .$ The other pair comes from shifting to the
neighboring square, where the magnetic field has the opposite direction, and
considering the $\left( -1,1\right) $ orbit. These solutions are $\Psi
_1\left( x,y\right) =\exp \left[ ik\left( -ax+by\right) /d\right] $ $\hat
\psi _m\left( -x-ay/b+a/2\right) $ and $\Psi _3\left( x,y\right) =\Psi
_1\left( -x,-y\right) .$ The wavefunction in the original rectangle is
constructed from these four solutions under the condition that it vanishes
on the boundary. The state can be made even or odd under $\left( x,y\right)
\rightarrow \left( -x,-y\right) .$ With $r=\pm 1$ it has the form 
\begin{equation}
\Psi _{nm}=A\left( \Psi _0+r\Psi _3\right) +B\left( \Psi _2+r\Psi _4\right) .
\end{equation}
Requiring $\Psi _{nm}\left( \pm \frac a2,y\right) =0,$ we find that
$A=-Be^{-i\chi _a}$ and $e^{i\beta }=e^{-i2\chi _a}$ where $\chi
_a=ka^2/d$. And from $\Psi _{nm}\left( x,\pm \frac b2\right) =0$ we get
$A=-rBe^{i\chi _b}$ and $e^{i\beta }=e^{i2\chi _b}$ where $\chi
_b=kb^2/d$. Therefore $\chi _a+\chi _b=\pi n$ where $n$ is even for $r=1$
and odd for $r=-1.$ This determines $k=\pi n/d,$ and the energy can be
expressed as
\begin{equation}
E_{nm}=\left( \pi n/d\right) ^2+d^2E_m/b^2.
\end{equation}
The phase shift $\beta =2\pi n\left( b/d\right) ^2-$ $2\pi n_b.$ If we
assume $\left| \beta \right| \leq \pi $ then $n_b$ is an integer part of
$\left( nb^2/d^2-1/2\right) .$ For a square $\beta =0$ for even $n$ or
$\pi $ for odd $n.$ This is the case considered in Ch.\ \ref{sbim}. It is
clear that the BOA and the $T$-operator theory agree at least to the order
$\sqrt{\epsilon }$ in the\ wavefunction and $\epsilon $ in energy. 

We were able to apply the BOA to the rectangular billiard by extending it to
a bigger system and thus ``unfolding'' the trajectories. The method can be
generalized to include the boundary perturbation. In this case we again
extend the unperturbed rectangle by reflection to the whole plane and assume
that the classical orbits are almost straight lines in the plane (actually,
they will slightly bend on the boundary crossings). Then $U\left( \eta
\right) =k^2-2k^2\delta L\left( \eta \right) /L_\xi ,$ where $\delta L\left(
\eta \right) $ is the additional path due to perturbed boundary for an
unperturbed orbit of period $L_\xi .$ In light of Eq.\ (\ref{caSch}) we might
say that the change in length is compensated by a change of momentum $\delta
k=-k\delta L/L$ that preserves the action. Notice also that the time it
takes the particle to travel distance $L$ with the modified momentum is
equal to the time it travels distance $L+\delta L$ with the old momentum. If
the extension of a system to the whole plane in the original coordinates is
not possible (like for a circle billiard), it can be done in the
action-angle variables.

\section{BOA in the asymptotic region}

\label{bita}

Now we return to the low angular momentum resonance in a perturbed unit
circle (Sec.\ \ref{r12}). In this example the separation of fast and slow
coordinates holds only over a part of the billiard. The BOA wavefunction in
this region is in agreement with the $T$-operator solution.

The Helmholtz equation in cylindrical coordinates has the form 
\begin{equation}
\left( r\frac \partial {\partial r}r\frac \partial {\partial
r}+k^2r^2+\frac{\partial ^2}{\partial \theta ^2}\right) \Psi =0. 
\end{equation}
For the low angular momentum $l$ states the angular coordinate is slow and
the radial coordinate is fast everywhere except the region near the center
$r\lesssim l/k\sim \sqrt{\epsilon }.$ Outside of this area we can make the
BOA ansatz $\Psi \left( r,\theta \right) =\Phi \left( r|\theta \right)
\psi \left( \theta \right) .$ The fast equation
\begin{equation}
r\frac \partial {\partial r}r\frac \partial {\partial r}\Phi +\left[
k^2r^2-l^2\left( \theta \right) \right] \Phi =0
\end{equation}
is the Bessel equation with the variable order $l\left( \theta \right) .$
Since it is invalid for $r\rightarrow 0$ the solution may include both
Bessel and Neumann functions. It is convenient to invoke the asymptotic
expansion for large $kr$ and $l$ but small $l/kr$ \cite{AbrSte} 
\begin{equation}
\Phi \left( r|\theta \right) \approx \frac 1{\sqrt{kr}}\cos \left[
kr+\frac{l^2\left( \theta \right) -\frac 14}{2kr}+\alpha \left( \theta
\right) \right] \label{bitaPhi}
\end{equation}
where $\alpha \left( \theta \right) $ is the phase that mixes Bessel and
Neumann functions. Function $\Phi $ must vanish at the boundary $r\left(
\theta \right) =1+\epsilon \Delta R\left( \theta \right) $ which requires 
\begin{equation}
k+k\epsilon \Delta R\left( \theta \right) +\frac{l^2\left( \theta \right)
-\frac 14}{2k}+\alpha \left( \theta \right) =\pi \left( n-\frac 12\right) .
\label{bitabc}
\end{equation}
Both functions $l\left( \theta \right) $ and $\alpha \left( \theta \right) $
are unknown, and we need some additional information to determine them.
Consider the slow equation 
\begin{equation}
\psi ^{\prime \prime }+l^2\left( \theta \right) \psi =0.  \label{bitasl}
\end{equation}
The radial wavefunction $\psi \left( \theta \right) $ is expected to coincide
with the PSS wavefunction that solves Bogomolny's equation. Equation (\ref
{r12Sch}) implies that 
\begin{equation}
l^2\left( \theta \right) =k^2\epsilon \left[ E_m-V\left( \theta \right)
\right]
\end{equation}
where $V\left( \theta \right) =\Delta R\left( \theta \right) +\Delta R\left(
\theta -\pi \right).$ Since $V\left( \theta \right) $ has period $\pi,$
Eq.\ (\ref{bitasl}) is solved with the boundary condition $\psi \left( \theta
-\pi \right) =\left( -1\right) ^m\psi \left( \theta \right) $, and the
eigenvalues $E_m$ are determined. Now Eq.\ (\ref{bitabc}) enables us to find 
\begin{equation}
\alpha \left( \theta \right) =-\frac 12k\epsilon \left[ \Delta R\left(
\theta \right) -\Delta R\left( \theta -\pi \right) \right] +\alpha _0
\end{equation}
and the quantization condition 
\begin{equation}
k-\frac 1{8k}+\frac 12k\epsilon E_m+\alpha _0=\pi \left( n-\frac 12\right) .
\end{equation}
The constant $\alpha _0$ is determined from the special case $\Delta R=0.$
Then $l=m$ and $\Phi \left( r|\theta \right) \propto J_m\left( kr\right),$ 
which implies $\alpha _0=-\frac \pi 2\left( m\limfunc{mod}2\right)
-\frac \pi 4.$ Figures \ref{2_5}, \ref{2_6} show the states with $m=2,3$,
respectively, in the short stadium. 

In the considered example the BOA is not self-contained --- we could not
complete the solution without knowing the behavior of the wavefunction at
the origin. Having obtained the radial wavefunction by other means (e.g.
from Bogomolny's equation), we reconstructed the two-dimensional
wavefunction [cf.\ Eq.\ (\ref{tdwl0})] and the quantization condition [cf.\
Eq.\ (\ref{r12qc})] of the perturbation theory. It is remarkable that the
fast wavefunction $\Phi \left( r|\theta \right) $ becomes asymptotically
the Bessel function $J_{l\left( \theta \right) }\left( kr\right) $ only in
an unperturbed circle, in all other cases it is mixed with the Neumann
function $N_{l\left( \theta \right) }\left( kr\right) .$ 

In this section the phase of the wavefunction (\ref{bitaPhi}) and the
quantization condition (\ref{bitabc}) contain a $k^{-1}$ order corrections 
that were absent in the $T$-operator theory. This corrections are normally
small in the semiclassical regime although they may begin to play role
when the angular momentum $l \sim k \sqrt{\epsilon} \lesssim 1$. There are
several sources of the $k^{-1}$ corrections in the $T$-operator method.
First, the $k^{-1}$ terms can be added to the phase of the $T$-operator
itself.  For example, in the boundary integral method these terms
originate in the derivative of Hankel's function [cf.\ Eq.\ (\ref{bimH0})].
Second, the Bogomolny equation $\psi = T \psi$ should be solved to higher
precision, namely, one has to retain the next order terms in the expansion
of the unperturbed action and the prefactor near the stationary point
[cf.\ Eq.\ (\ref{r12Texp})]. Similarly, the two-dimensional wavefunction 
can be found from Eq.\ (\ref{Psiint}) with the Hankel function as the 
kernel. Note that the more general kernel (\ref{tdwGtil}) is valid only in 
the leading order in $k^{-1}$.

The low angular momentum resonance in the perturbed sphere of Sec.\ \ref{psb}
can be described by a similar technique. We start with a three-dimensional
Helmholtz equation 
\begin{equation}
\left( \frac \partial {\partial r}r^2\frac \partial {\partial r}+k^2r^2-\hat
l^2\right) \Psi =0,
\end{equation}
where $\hat l^2\equiv -\frac 1{\sin \theta }\frac \partial {\partial
\theta }\sin \theta \frac \partial {\partial \theta }$ $-\frac 1{\sin
^2\theta }\frac{\partial ^2}{\partial \varphi ^2}.$ The equation in the
fast radial direction
\begin{equation}
\frac \partial {\partial r}r^2\frac \partial {\partial r}\Phi +\left[
k^2r^2-l\left( l+1\right) \right] \Phi =0
\end{equation}
is an equation for a spherical Bessel function of order $l=l\left( \theta
,\varphi \right) $ \cite{AbrSte}. Again, since the BOA is not valid near the
center, the solution will include both Bessel and Neumann spherical
functions. Asymptotically, 
\begin{equation}
\Phi \left( r|\theta ,\varphi \right) \approx \frac 1{kr}\sin \left[
kr+\frac{l\left( l+1\right) }{2kr}+\alpha \left( \theta ,\varphi \right)
\right]
\end{equation}
where $\alpha \left( \theta ,\varphi \right) $ is the phase to be
determined. The slow equation 
\begin{equation}
-\hat l^2\psi +l\left( l+1\right) \psi =0
\end{equation}
implies that 
\begin{equation}
l\left( l+1\right) =k^2\epsilon \left[ E_{lm}-V\left( \theta ,\varphi
\right) \right] ,
\end{equation}
where $V\left( \theta ,\varphi \right) =\Delta R\left( \theta ,\varphi
\right) +\Delta R\left( \pi -\theta ,\varphi +\pi \right) .$ The
eigenfunctions $\psi _{lm}$ [the quantum number $l$ should not be confused
with the function $l\left( \theta ,\varphi \right) $] will be even or odd
under inversion, $\psi _{lm}\left( \theta ,\varphi \right) =\left(
-1\right) ^l\psi _{lm}\left( \pi -\theta ,\varphi +\pi \right) ,$ where we
assume the state $\psi _{lm}$ has the same symmetry as the unperturbed
state $Y_{lm}\left( \theta ,\varphi \right) .$ As before, the Dirichlet
condition for $\Phi $ gives
\begin{equation}
\alpha \left( \theta ,\varphi \right) =-\frac 12k\epsilon \left[ \Delta
R\left( \theta ,\varphi \right) -\Delta R\left( \pi -\theta ,\varphi +\pi
\right) \right] -\frac \pi 2\left( l\limfunc{mod}2\right)
\end{equation}
and the quantization for $k$
\begin{equation}
k+\frac 12k\epsilon E_{lm}=\pi n+\frac \pi 2\left( l\limfunc{mod}2\right) .
\end{equation}
Clearly, the BOA is consistent with the $T$-operator approach and even
improves on the angular differential equation.

\section{Whispering gallery modes}

\label{wgm2}

The whispering gallery modes discussed in Sec.\ \ref{wgm} can also be
derived with the BOA. The adiabaticity comes from the slow variation of
curvature of the boundary. This approach is related to the parabolic
equation and the etalon methods \cite{BabBul} in the sense that in the
former some derivatives in the partial differential equation are neglected
and in the latter an ansatz involving the Bessel function with variable
order is made. All these methods give the same leading order results. 

The billiard boundary is locally a circle of radius $R\left( s\right) $
where the variable $s$ runs along the perimeter. Therefore the whispering
gallery wavefunction is locally a wavefunction for a circle, 
\begin{equation}
\Psi \left( \rho ,s\right) =\Phi \left( \rho |s\right) \psi \left( s\right)
=\alpha \left( s\right) J_{l\left( s\right) }\left( kr_s\right) \psi \left(
s\right) ,
\end{equation}
where $\rho \ll R$ is the distance from the boundary, $r_s=R\left(
s\right) -\rho $ is the radius measured from the local center of curvature
and $\alpha \left( s\right) $ is the normalization for $\Phi $. The slowly
changing angular momentum $l\left( s\right) $ does not have to be integer
and will be determined by the boundary conditions. The ``slow''
wavefunction $\psi \left( s\right) $ satisfies the equation similar to
Eq.\ (\ref{bitasl}) and is equal to
\begin{equation}
\psi \left( s\right) =\frac{\exp i\int^sl\left( s^{\prime }\right)
ds^{\prime }/R\left( s^{\prime }\right) }{\sqrt{l\left( s\right) /R\left(
s\right) }}
\end{equation}
where $ds/R$ plays a role of $d\theta .$ Since $l\sim kR,$ the function
$\psi \left( s\right) $ is not really slow, varying approximately as
$e^{iks}. $ However, $\Phi \left( \rho |s\right) $ depends only on the
slow varying functions of $s$, like $l\left( s\right) $ and $R\left(
s\right) ,$ which justifies the BOA\ ansatz. Since the order of the Bessel
function is large and close to its argument, it can be approximated by an
Airy function. If we define a function $f\left( s\right) $ by $l=kR\left(
1-f^{\prime }\right) $ then \cite{AbrSte}
\begin{eqnarray}
J_{l\left( s\right) }\left\{ k\left[ R\left( s\right) -\rho \right] \right\}
&=&J_l\left[ l+l^{1/3}\left( \frac{kRf^{\prime }-k\rho }{l^{1/3}}\right)
\right]  \nonumber \\
\ &\simeq &\left( \frac 2{kR}\right) ^{1/3}\mathrm{Ai}\left[
-2^{1/3}\frac{kRf^{\prime }-k\rho }{\left( kR\right) ^{1/3}}\right] . 
\end{eqnarray}
($l$ can be replaced by $kR$ in the leading order in $f^{\prime }$ and $\rho
/R.$) The Airy function must vanish at $\rho =0$ which makes 
\begin{equation}
f^{\prime }\left( s\right) =1-\frac l{kR}=\frac{z_n}{2^{1/3}\left( kR\right)
^{2/3}}  \label{wgm2f}
\end{equation}
where $z_n$ is the $n$th root of $\mathrm{Ai}\left( -z\right) .$ The full
wavefunction can now be written down as 
\begin{equation}
\Psi \left( \rho ,s\right) =\left( kR\right) ^{-1/6}\mathrm{Ai}\left(
\frac{2^{1/3}k^{2/3}}{R^{1/3}}\rho -z_n\right) \psi \left( s\right)
\label{wgm2Psi}
\end{equation}
where the normalization factor has been added. Note that the normalization
ensures the same total current through any section $s=\mathrm{const}.$

The $T$-operator theory produces the asymptotic form of this solution,
Eq.\ (\ref{wgmP}).\footnote{Although denoted by the same symbol, the PSS
wavefunction and the slow wavefunction $\psi(s)$ differ by a factor
$R^{-1/3}(s)$.} There the function $f\left( s\right) $ was defined via
Eq.\ (\ref {wgmpsi}) as a part of the phase of the PSS wavefunction.
Clearly, this is consistent with the current definition. The explicit
expression for $f^{\prime },$ Eq.\ (\ref{wgmf}), compares with Eq.\
(\ref{wgm2f}) if $z_n$ is used in its approximate form $\left[ \frac 32\pi
\left( n-\frac 14\right) \right] ^{2/3}$ for large $n$. The BOA solution
works both in the classically allowed and classically forbidden regions
including the caustic. It does not require the large number of wavelengths
in the radial direction, which would be the condition for the asymptotic
expansion of the Airy function. 

It is possible to derive the Airy function solution directly, without
referring to the Bessel function. The Helmholtz equation in the $\left( \rho
,s\right) $ coordinates for small $\rho /R$ is 
\begin{equation}
\left[ \frac{\partial ^2}{\partial \rho ^2}-\frac 1{R(1-\rho /R)}\frac
\partial {\partial \rho }+\frac 1{(1-\rho /R)^2}\frac{\partial ^2}{\partial
s^2}+\frac{\rho \left( R^{-1}\right) ^{\prime }}{(1-\rho /R)^3}\frac
\partial {\partial s}+k^2\right] \Psi =0.
\end{equation}
The second term can be neglected as compared to the first term since $\left|
\Phi ^{-1}\partial \Phi /\partial \rho \right| $ $\sim k^{2/3}R^{-1/3}$
[according to Eq.\ (\ref{wgm2Psi})] and $kR\gg 1.$ The third term can be
expanded in $\rho /R.$ The fourth term can be neglected compared with $\frac
\rho R\frac{\partial ^2}{\partial s^2}$ since $\left| \psi ^{\prime }/\psi
\right| \sim k$ and $kR\gg R^{\prime }.$ The simplified equation 
\begin{equation}
\left[ \left( 1+2\frac \rho R\right) ^{-1}\left( \frac{\partial ^2}{\partial
\rho ^2}+k^2\right) +\frac{\partial ^2}{\partial s^2}\right] \Psi =0,
\end{equation}
when solved by the BOA, yields Eq.\ (\ref{wgm2Psi}). We can also estimate the
limits of applicability of the BOA from this differential equation. When we
make the BOA ansatz, the largest term that we drop is 
\begin{equation}
\left| \frac{\partial \Phi }{\partial s}\psi ^{\prime }\right| \sim \left|
kz_n\frac{R^{\prime }}R\frac{\mathrm{Ai}^{\prime }}{\mathrm{Ai}}\Phi \psi
\right|
\end{equation}
where we take the typical $\rho \sim z_nR^{1/3}k^{-2/3}.$ We should compare
it with 
\begin{equation}
\left| \frac{\partial ^2\Phi }{\partial \rho ^2}\psi \right| \sim \left| 
\frac{k^{4/3}}{R^{2/3}}\frac{\mathrm{Ai}^{\prime \prime }}{\mathrm{Ai}}\Phi
\psi \right| .
\end{equation}
Estimating $\left| \mathrm{Ai}^{\prime \prime }/\mathrm{Ai}^{\prime }\right|
\sim n/z_n$ we obtain the condition 
\begin{equation}
\left( kR\right) ^{1/3}\gg \left| R^{\prime }\right| z_n^2/n\sim \left|
R^{\prime }\right| n^{1/3}
\end{equation}
which is the same as the requirement (\ref{wgmapp}) for the simple
$T$-operator theory to work. In Sec.\ \ref{sp} we consider the case when
this condition is not satisfied. 

The classical caustic is given by the turning point of the radial
equation, or a point where the argument of the Airy function vanishes,
$\rho \left( s\right) =\frac{z_nR^{1/3}\left( s\right) }{2^{1/3}k^{2/3}}.$
If the energy $ k^2$ is fixed, the possible caustics are quantized by
$z_n.$ In the classical picture the product $\epsilon \left( s\right)
R^{1/3}\left( s\right) \equiv I$, where $\epsilon \left( s\right) $ is an
angle that the classical trajectory makes with the boundary, is an
adiabatic invariant in the limit $\epsilon \rightarrow 0$ \cite{BabBul}
(see Sec.\ \ref{sp}). Semiclassically, $I=2^{1/3} \sqrt{z_n}/k^{1/3}$ and,
therefore, the caustics can be labeled solely by this parameter. This is
not surprising since, apart from the energy which does not change the
geometry of the orbits, this adiabatic invariant is the only (approximate)
integral of motion. 

\section{Conclusions}

The Born-Oppenheimer approximation can be used alongside the $T$-operator
and other techniques to describe the perturbed integrable systems. When the
separation of the fast and slow variables is possible it provides a
convenient way to find a two-dimensional approximate wavefunction and energy
levels. The BOA solution is usually valid near caustics and in the
classically forbidden (shadow) regions where the semiclassical approximation
often fails or needs modification. Whether the natural coordinates, in which
the separation of fast and slow motion is possible, exist, depends on the
geometry of the system or a family of orbits. In principle, the separation
should always be possible in the action-angle variables, but then again the
shadow regions are off limits. The adiabatic approximation may also be
possible when there is a slowly changing parameter instead of the slowly
changing position of the particle, as in the whispering gallery case.

\chapter{Trace formulas}

\label{trform}

The \emph{trace formulas} are used to describe the density of states in
terms of the classical periodic orbits of the system. The Gutzwiller trace
formula \cite{Gut} applies to the hard chaotic systems and the Berry-Tabor
trace formula \cite{BerTab,BerTab2,Gut2} works for the integrable systems.
In the former case the periodic orbits are unstable and isolated, while in
the latter case they form families. The intermediate situations include
the perturbed integrable systems \cite{Ozo} and mixed systems. 

The $T$-operator formalism (Sec.\ \ref{Tgen}) adequately describes a system
on a semiclassical level. In particular, one should be able to derive the
trace formulas directly from the $T$-operator. We show in Sec.\ \ref{gdot}
how this can be done. In the following sections we apply the general
formalism to the Gutzwiller and Berry-Tabor cases. Then we consider a
perturbed integrable system, which interpolates between the two extremes.
The interpolation formula can be parametrized. At least \emph{four}
parameters are needed to describe the perturbed family of periodic orbits
correctly \cite{ZaiNarPra2}, not three, as Ref.\ \cite{UllGriTom} claims.
We illustrate these points with the example of the coupled quartic
oscillators in Sec.\ \ref{pqo}. 

\section{General derivation of the trace formula}
\label{gdot}

The energy spectrum of a system described by the Bogomolny operator $T(E)$
is given by the zeros of the Fredholm determinant [cf.\ Eq.\
(\ref{totgdet})]
\begin{equation}
D(E) = \det \left[1 - T(E) \right].
\end{equation}
The oscillatory part of the density of states can be expressed as a 
logarithmic derivative
\begin{equation}
d_{\mathrm{osc}}(E) = d(E) - \bar d(E) = \frac{-1} \pi \mathrm{Im} \left[ 
\frac {d \ln D(E + i\epsilon)} {dE} \right] \label{gdotdos}
\end{equation}
where $d(E)=\sum_a \delta(E-E_a)$ and $\bar d(E)$ is the smoothed (Weyl) 
density of states. To justify this result \cite{Pra2} we write $D(E) = 
\prod_n \left[1 - e^{i\theta_n(E)} \right]$. Here $e^{i\theta_n(E)}$ are 
the eigenvalues of the $T$-operator which is approximately unitary. Then
Eq.\ (\ref{gdotdos}) yields
\begin{equation}
d_{\mathrm{osc}}= - \frac 1 {2\pi} \sum_n \theta^{\prime}_n \left(1 +
\mathrm{Im} \cot \frac {\theta_n} 2 \right)
= \sum_n \left[ - \frac 1 {2\pi} \theta^{\prime}_n + \left|
\theta^{\prime}_n \right| \delta (\theta_n (E)) \right].
\end{equation}
With $\theta^{\prime}_n$ approximately independent of $n$ the first sum 
gives the smoothed part $-\bar d = -\frac N {2\pi} \theta^{\prime}$ where 
$N$ is the size of the $T$-matrix. The second sum is $d(E)$. 

\begin{sloppypar}
Using the relationship $\ln \det (1 - T) = \mathrm{Tr} \ln (1 - T)$ in 
Eq.\ (\ref{gdotdos}) and expanding the logarithm in powers of $T$, we find 
\begin{equation}
d_{\mathrm{osc}}(E) = \frac 1 \pi \mathrm{Im} \sum_{n=1}^\infty \frac 1 n 
\frac {d \tau_n (E)} {dE} \label{gdotdosc}
\end{equation}
where 
\begin{equation}
\tau_n (E) = \mathrm{Tr} T^n (E) = \int dq_1 \cdots dq_n T(q_1,q_2) 
\cdots T(q_n,q_1). \label{gdottaun}
\end{equation}
The composition property for the $T$-operator (Sec.\ \ref{propT}) gives for
a two-dimensional system
\begin{equation}
\tau_n (E) = \int dq \sum_p \left( \frac 1 {2\pi i \hbar} \left| \frac 
{\partial^2 S_p (q,q^\prime;E)} {\partial q \partial q^\prime} \right| 
\right)^{1/2}_{q=q^\prime} \exp \left[ \frac i \hbar S_p (q,q;E) \right]. 
\label{gdottau} 
\end{equation}
Here $p$ denotes a ``closed'' orbit that leaves the Poincar\'e surface of
section (PSS) from point $q$ and arrives at the same point on its $n$th 
crossing of the PSS, $S_p (q,q;E)$ is the reduced action for this orbit, 
and the Maslov index was omitted. In effect, we calculated $n-1$ out of 
$n$ integrals in Eq.\ (\ref{gdottaun}) by the stationary phase ($S\Phi$). 
Whether the last integral integral can be done by the $S\Phi$ as well, 
depends on the $S_p (q,q;E)$.
\end{sloppypar}

Since the exact density of states is a collection of $\delta$-functions,
the formal series (\ref{gdotdosc}) diverges when $E$ is on the spectrum.
In practice, one includes only the first few terms that account for the
short periodic orbits. This produces the density of states smoothed over
some energy scale ($\hbar/\mathrm{time}$). Such averaging can be
experimentally relevant, say, due to the finite resolution of the spectrum
because of non-zero temperature. For example, the susceptibility of a
square in magnetic field is determined mostly by the shortest periodic
orbit that encloses flux (Sec.\ \ref{mres}). Thus, only one term in the
trace formula is needed in this case \cite{RicUllJal}. When one is
interested in the energy level correlations on the scale of the average
spacing, the orbits with periods up to the \emph{Heisenberg time} $\hbar
\bar d$ should be included. 

\section{Gutzwiller trace formula}

In the Gutzwiller case the periodic orbits are isolated, which means that
$S_p (q,q;$ $E)$ has the well defined stationary points. [$\partial
S_p(q,q^\prime) / \partial q + \partial S_p(q,q^\prime) / \partial
q^\prime = p(q) - p^\prime (q^\prime) = 0$ at $q = q^\prime = q^*$, a
$S\Phi$ point of the integral (\ref{gdottau}).] Let $q$, $q^\prime$ be in
the vicinity of $q^*$ and expand in $\delta q = q - q^*$ and $\delta
q^\prime = q^\prime - q^*$
\begin{equation}
S_p(q,q^\prime) \simeq S_p (E) + p^* (\delta q - \delta q^\prime) + \frac 
1 2 V_{11} \delta q^2 + V_{12}\delta q \delta q^\prime + \frac 1 2 V_{22} 
\delta q^{\prime 2}
\end{equation}
where $p^*$ is the momentum of the periodic orbit. The matrix of second
derivatives $V$ depends on $q^*$ and the energy. Suppose the orbit $p$ of
``length'' $n$ (i.e.\ returning to the surface of section $n$ times)
consists of $r$ repetitions of a \emph{primitive} periodic orbit of length
$s$, with $n = rs$. Then the contribution of orbit $p$ to the integral
(\ref{gdottau}) is
\begin{equation}
\tau_p = \left( \sum_1^s \left| \frac {V_{12}} {V_{11} + 2V_{12} + V_{22}} 
\right|^{1/2} \right) \exp\left[ \frac i \hbar S_p (E) \right].
\end{equation}
The summation is over the $s$ stationary points, where the orbit 
crosses the PSS (each of them can be chosen as a starting/ending point 
for the periodic orbit). 

The prefactor can be expressed in terms of the \emph{monodromy matrix}
$M_p$ of orbit $p$, which relates the final momentum $\delta p = p - p^* =
V_{11} \delta q + V_{12} \delta q^\prime$ and position $\delta q$ to the
initial $\delta p^\prime = - V_{12} \delta q - V_{22} \delta q^\prime$ and
$\delta q^\prime$. With 
\begin{equation}
M_p = \left( 
  \begin{array}{cc}
  - \frac {V_{11}} {V_{12}} & \frac {V_{12}^2 - V_{11} V_{22}} {V_{12}} \\
  - \frac 1 {V_{12}} & - \frac {V_{22}} {V_{12}}
  \end{array}
\right)
\end{equation}
the prefactor can be written as $s |\det (M_p - 1)|^{-1/2}$. The sum over
the crossing points appears as a factor. [To see that $\det (M_p - 1)$ is
independent of the choice of the initial point $q^*$, express $M_p =
M_{12} M_{23} \cdots M_{n1}$, where $M_{i,i+1}$ is the monodromy matrix
between the two consecutive crossings. Then $\det (M_p - 1) = \det (M_{23}
\cdots M_{n1}M_{12} - 1)$.]

For a repeated orbit, $S_p = r S_s$ and $M_p = (M_s)^r$. Only the rapidly
varying phase needs to be differentiated when evaluating the derivative in
Eq.\ (\ref{gdotdosc}). With $dS_s(E)/dE = T_s$, the period of the
primitive orbit, the density of states
\begin{equation}
d_{\mathrm{osc}} = \sum_{s,r} \frac {T_s/ \hbar} {|\det(M_s^r - 
1)|^{1/2}} \cos \left\{r \left[ \frac {S_s(E)} \hbar + \frac \pi 2 \nu_s 
\right] \right\}
\end{equation}
where we restore the Maslov index. 

\section{Berry-Tabor formula}

In the integrable systems the periodic orbits form continuous families,
and therefore the $S\Phi$ cannot be applied. On the other hand, since the
action is constant within a family, the integral (\ref{gdottau}) can be
easily done in the action-angle variables. For a two-dimensional system we
define the variables $I$, $\theta$, $J$, $\Theta$, and choose the surface
of section $\Theta = 0$. The action after the first return $S(\theta -
\theta^\prime,E)$ depends only on the difference of the angles (cf.\ Sec.\
\ref{tdss}). For an orbit $p$ of length $n$ the action $S_p (\Delta
\theta) = nS(\Delta \theta / n)$. If $p$ is a periodic orbit with a
frequency ratio $\omega_I/\omega_J = m/n$ then $\Delta \theta = \theta -
\theta^\prime = 2\pi m$. The derivative $S_p^{\prime\prime} = -(2\pi n
g_E^{\prime\prime})^{-1}$, where $J=g_E(I)$ (Sec.\ \ref{tdss}). Using these
results in Eqs.\ (\ref{gdottau}) and (\ref{gdotdosc}) we arrive to the
Berry-Tabor trace formula \cite{BerTab,BerTab2}
\begin{equation}
d_{\mathrm{osc}} = \sum_p \frac {T_p} {\pi \hbar^{3/2} n^{3/2} 
|g_E^{\prime\prime}|^{1/2}} \cos \left( \frac {S_p} \hbar + \frac \pi 2 
\nu_p -\frac \pi 4\right). \label{btfd}
\end{equation}

Note that a chaotic system may have a family of non-isolated periodic
orbits. Among the examples are the bouncing ball orbits in the stadium
\cite{Tan,SieSmiCre,PriSmi} or Sinai billiard \cite{Ber,SiePriSmi}. 

\section{Perturbed Berry-Tabor formula}

\label{pbtf}

In a perturbed integrable system the families of periodic orbits are
broken with only a few isolated orbits remaining (Sec.\ \ref{pis}). 
However, the action $S_p (q, q^\prime; E)$ in the integral (\ref{gdottau})
varies too slowly for the $S\Phi$ to be applied. It may be convenient to 
express this one-dimensional integral in the action-angle variables. For 
an orbit of length $n$ 
\begin{equation}
S_p (\theta, \theta^\prime) = nS(\Delta \theta / n) + \epsilon W(\theta,
\theta^\prime)
\end{equation}
where $\epsilon W(\theta, \theta^\prime)$ is the perturbed part of the
action that can be calculated by the standard technique. We may neglect
the order $\epsilon$ terms in the prefactor and take $\theta^\prime =
\theta - 2\pi m$ for a broken family with the winding number $m/n$. Then
the integral that remains is 
\begin{equation}
I_W = \frac 1 {2\pi} \int d\theta \exp \left[ \frac {i\epsilon} \hbar 
\hat W(\theta) \right] \label{pbtfI}
\end{equation}
where $\hat W(\theta) = W(\theta, \theta - 2\pi m)$.\footnote{Note that
$\hat W(\theta) = r \bar V_q(\theta)$ for a periodic orbit $(rp,rq)$,
where $\bar V_q(\theta)$ is just the effective potential of the
$T$-operator perturbation theory (cf.\ Sec.\ \ref{dqoc}). $p$, $q$ are
relatively prime.} The perturbed result for the $d_{\mathrm{osc}}$ is just
the Berry-Tabor formula with the substitution $\cos \phi \rightarrow
\mathrm{Re}[I_W \exp (i\phi)]$ where $\phi$ is the argument of cosine in
Eq.\ (\ref{btfd}) \cite{Ozo}.  One might say that the perturbed trace
formula interpolates between the Berry-Tabor case, when $\epsilon = 0$ and
$I_W = 1$, and the Gutzwiller case, when $\epsilon \hat W/ \hbar$ is large
and $I_W$ can be done by the $S\Phi$. 

When a family of periodic orbits is broken by the perturbation, at least
one stable and one unstable periodic orbits remain, which means that $\hat
W(\theta)$ has at least one minimum and one maximum. Assume, for example,
that $\hat W(\theta)$ has a single maximum and minimum at $\theta = 0,
\pi$, respectively. Often the perturbed action $\hat W(\theta)$ can be
approximated by a simple parametric expression that retains the essential
information about its behavior. In the first attempts of this sort the
function was approximated by the first terms of its Fourier series
\cite{Ozo}, e.g., $\hat W(\theta) \approx w_0 + w_1 \cos (\theta -
\theta_0)$. Then $I_W \approx \exp(i \epsilon w_0/ \hbar) J_0 (\epsilon
w_1/ \hbar)$ where $J_0$ is the Bessel function. However, this expression
is, in general, incorrect in the limit of large $\epsilon \hat W/ \hbar$.
Indeed, the integral (\ref{pbtfI}) can be done by the $S\Phi$ and,
therefore, depends on the values of $\hat W$ and $\hat W^{\prime\prime}$
at the extrema. This means that $I_W$ must depend on four parameters. 

Instead of increasing the number of terms in the Fourier expansion, it is
more efficient to parametrize \cite{UllGriTom} $\hat W(\theta) = W_0 + W_1
\cos [\xi (\theta)]$ where $\xi (0) = 0$ and $\xi (\pi) = \pi$. The
unknown function $\theta (\xi)$ can be approximated by the first two terms
of its Fourier expansion $\theta = \xi - A \sin \xi - B \sin 2\xi$. With 
$d \theta / d\xi = 1 - A \cos \xi -2B \cos 2\xi$ the integral 
(\ref{pbtfI}) becomes 
\begin{equation}
I_W = \frac {I_0} {2\pi} \int d\xi (1 - A \cos \xi -2B \cos 2\xi) \exp 
\left[ \frac {i \epsilon} \hbar W_1 \cos \xi \right]
\end{equation}
where $I_0 = \exp[i (\epsilon/ \hbar) W_0]$. The integrals may be 
expressed in terms of Bessel functions $J_m(\epsilon W_1/ \hbar)$, giving 
\begin{equation}
I_W = I_0 (J_0 - iAJ_1 + 2BJ_2). \label{pqoI}
\end{equation}

The four parameters $W_0$, $W_1$, $A$, $B$, can be related to $\hat W$ and
$\hat W^{\prime\prime}$ evaluated at the extrema by
\begin{eqnarray}
&& \hat W (0) = W_0 + W_1, \; \hat W (\pi) = W_0 - W_1, \nonumber \\
&& \hat W^{\prime\prime}(0) = - \frac {W_1} {(1 - A - 2B)^2}, \;  \hat
W^{\prime\prime} (\pi) = \frac {W_1} {(1 + A - 2B)^2}. 
\end{eqnarray}
Thus, the above interpolation formula is correct in the limit of large
$\epsilon \hat W/ \hbar$, as well as for $\epsilon \rightarrow 0$. One can
show that $\det(M_p -1)$ is proportional to $\epsilon \hat
W^{\prime\prime}$. 

The authors of Ref.\ \cite{UllGriTom} (UGT) take $B=0$. They compensate the
lack of another parameter by letting the function $g_E^{\prime\prime}$
[Eq.\ (\ref{btfd})] depend on $\epsilon$. Although formally this method 
seems to be correct, it is physically misleading. The function $g_E$ 
describes the constant energy surface in the action space $(I,J)$. When 
the system is perturbed, the topology of the invariant tori changes and 
the old actions are no longer the integrals of motion. Therefore 
$g_E^{\prime\prime}$ does not have a clear physical interpretation 
for finite $\epsilon$. Secondly, UGT argue that ``the independent evaluation 
of $g_E^{\prime\prime}$ can be rather laborious and time 
consuming.'' However, in order to explicitly evaluate this function in 
the limit $\epsilon \rightarrow 0$ (which must be equal to the standard 
$g_E^{\prime\prime}$) UGT would need to know the parameter $W_1$ (in our 
notation) and the values of $\det(M_p - 1)$ for the periodic orbits in 
this limit, which in effect requires the solution of the unperturbed 
problem. With this solution on hand it should not be hard to find the 
standard $g_E^{\prime\prime}$.

\section{Example: coupled quartic oscillators}
\label{pqo}

We study the correction to the trace formula for the coupled quartic 
oscillators, which are defined by the Hamiltonian 
\begin{equation}
H = \frac 1 2 (p_x^2 + p_y^2) + ax^4 + by^4 + \epsilon x^2 y^2.
\end{equation}
When $\epsilon = 0$ we may introduce the action-angle variables
$(I,\theta)$ in the $x$ direction and $(J,\Theta)$ in $y$ direction. For a
given energy $E$ and winding number $\alpha = \omega_I/ \omega_J$
\begin{eqnarray}
&&x = \frac 1 {\sqrt 2} \left( \frac {E/a} {1 + \frac a b \alpha^{-4}} 
\right)^{1/4} \mathrm{sd} (\kappa \theta), \nonumber \\
&&I = \frac 2 3 \kappa \sqrt a \left( \frac {E/a} {1 + \frac a b 
\alpha^{-4}}\right)^{-3/4} 
\end{eqnarray}
and similar for $\Theta$ and $J$. Here $\kappa = 2K/ \pi$, where $K 
\equiv K(m = 1/2)$ is the complete elliptic integral of the first kind, and
$\mathrm{sd}$ is one of the Jacobi elliptic functions \cite{AbrSte}. The 
constant energy surface is described by 
\begin{equation}
g_E(I) = J = \left[ \left( \frac 2 3 \kappa \right)^{4/3} E - \left( 
\frac a b \right)^{1/3} I^{4/3} \right]^{3/4}.
\end{equation}

The action after the first return to the PSS $\Theta = 0$ is 
\begin{equation}
S(\theta - \theta^\prime, E) = \frac 2 3 \kappa E^{3/4} \left[ \frac 
{(\theta - \theta^\prime)^4} a + \frac {(2\pi)^4} b \right]^{1/4}.
\end{equation}
The perturbed part of the action $\delta S = - \int \delta H dt$ in the
leading order in $\epsilon$. The integral of the perturbed Hamiltonian 
can be evaluated along the unperturbed orbit. For a family of periodic 
orbits of winding number $\alpha = m/n$ we find
\begin{equation}
\hat W(\theta) = - \frac {\kappa E^{3/4}} {8 (ab)^{1/4}} \frac \alpha {[a 
+ b \alpha^4]^{3/4}} \int_{-2\pi m}^0 d \theta^{\prime\prime} 
\mathrm{sd}^2 [\kappa (\theta^{\prime\prime} + \theta)] \mathrm{sd}^2 
[\kappa \theta^{\prime\prime}/ \alpha].
\end{equation}
Note that the shape of $\hat W(\theta)$ is independent of $a$, $b$, and
$E$, and each is taken to be unity. 

\begin{figure}[tbp]
{\hspace*{2.7cm} \psfig{figure=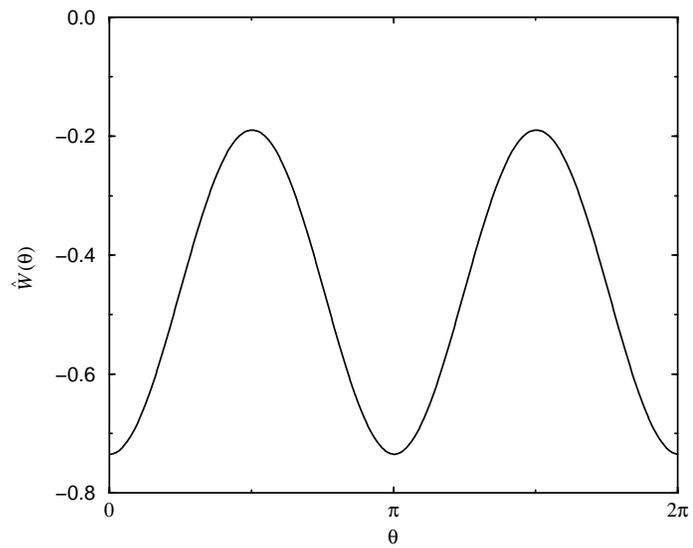,height=7.36cm,width=9cm,angle=270}}
{\vspace*{.13in}}
\caption[$\hat W (\theta)$ for the $(1,1)$ family of orbits of the quartic
oscillators with $x^2y^2$ coupling.]
{$\hat W (\theta)$ for the $(1,1)$ family of orbits ($\alpha=1$) of the
quartic oscillators with $x^2y^2$ coupling. The parameters $a=b=E=1$. 
\label{6_1}}
{\vspace{1.2 cm}}
\end{figure}

\begin{figure}[tbp]
{\hspace*{2.7cm} \psfig{figure=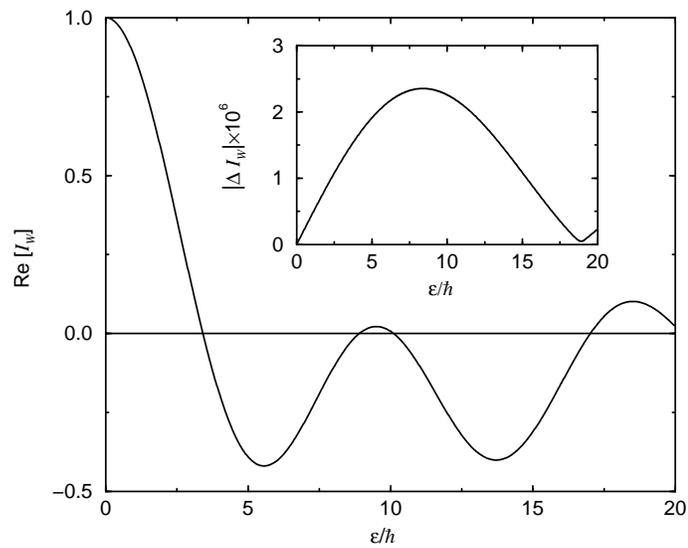,height=7.32cm,width=9cm,angle=270}}
{\vspace*{.13in}}
\caption[Real part of $I_W$ as a function of $\epsilon / \hbar$.]
{Real part of $I_W$ as a function of $\epsilon / \hbar$. The difference
between this function computed by numerical quadrature and from the
interpolation formula, $| \Delta I_W|$, is shown in the inset.
\label{6_2}}
{\vspace{1.2 cm}}
\end{figure}

\begin{figure}[tbp]
{\hspace*{2.7cm} \psfig{figure=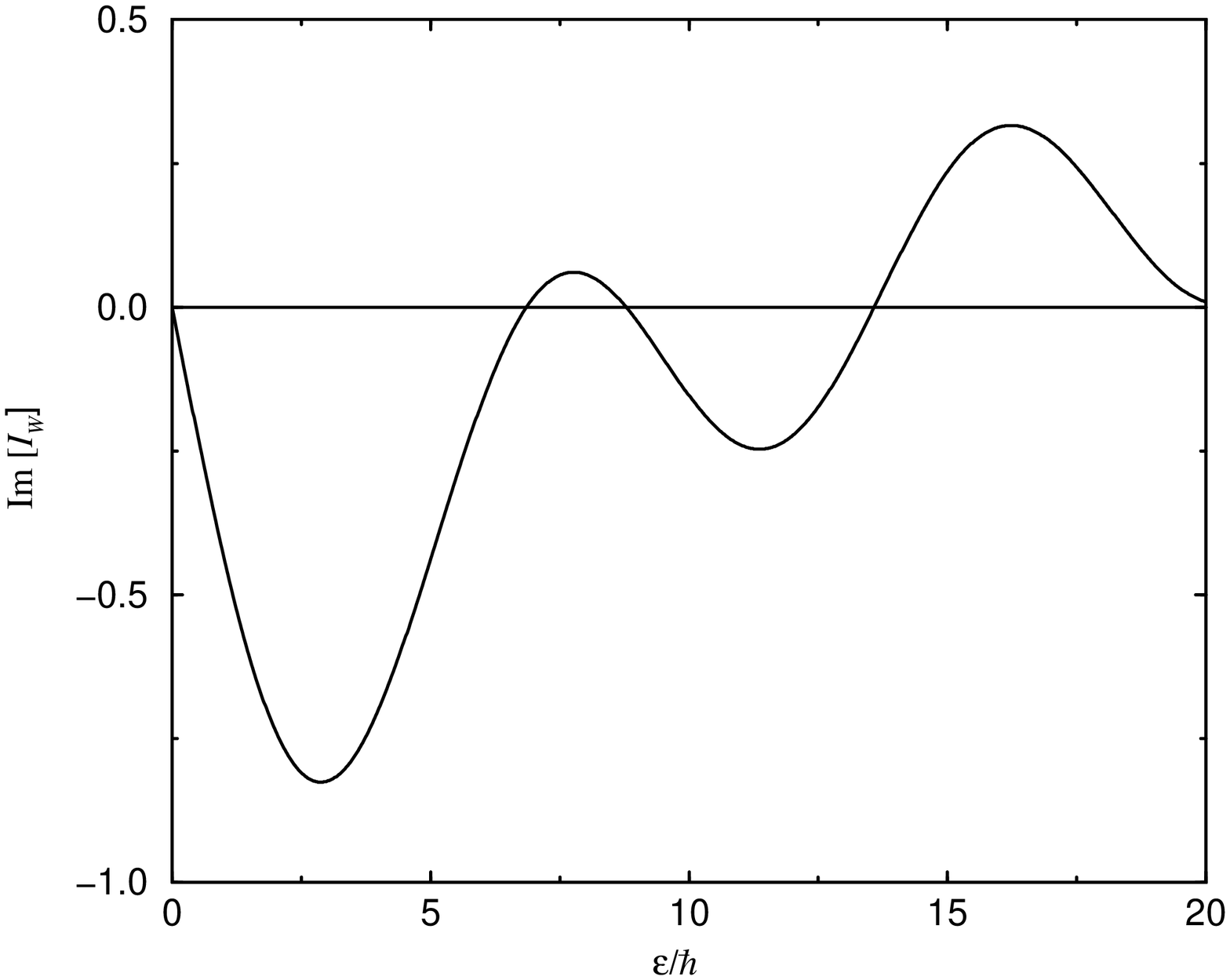,height=7.32cm,width=9cm,angle=270}}
{\vspace*{.13in}}
\caption{Imaginary part of $I_W$ as a function of $\epsilon / \hbar$.
\label{6_3}}
{\vspace{1.2 cm}}
\end{figure}

Figure \ref{6_1} depicts $\hat W(\theta)$ for the $(1,1)$ family, $\alpha
= 1$. It is periodic with period $\pi$ due to symmetries. The
interpolation formulas of the previous section can be adjusted, when the
period of $\hat W(\theta)$ is $2\pi/ r$, by changing $\theta \mapsto
r\theta$. From the numerical values of $\hat W$ and $\hat
W^{\prime\prime}$ at the extrema we find the interpolation parameters $A =
-1.49 \times 10^{-2}$, $B = -1.04 \times 10^{-4}$ (independent of $a$,
$b$, $E$) and $W_0 = -0.46$, $W_1 = 0.27$. The real part of $I_W$ as a
function of $\epsilon/ \hbar$ is shown in Fig.\ \ref{6_2} and the
imaginary part appears in Fig.\ \ref{6_3}. The deviation of the
interpolation formula from the exact integral is given in the inset of
Fig.\ \ref{6_2}. They agree to within a part in $10^{-5}$. 

Note that the parameter $B$ is relatively small. This may have worked in 
favor of UGT who assumed $B=0$. In general, however, this assumption 
implies a relation between the curvatures at minimum and maximum of 
$\hat W(\theta)$, namely,
\begin{equation}
\sqrt {W_1/ \hat W^{\prime\prime}_{\min}} + \sqrt {W_1/ \left| 
W^{\prime\prime}_{\max} \right|} = \frac 2 r
\end{equation}  
(in this case $r=2$). We find numerically that the \emph{l.h.s.} of this 
equation is equal to $1.00021$ for the $(1,1)$ resonance. 

\section{Conclusions}

The oscillatory part of the density of states can be expressed in terms of
traces of the powers of $T$-operator. The powers can be evaluated by the
stationary phase. For their traces the $S\Phi$ can be used only when the
periodic orbits are isolated and the action of the closed orbits in their
neighborhoods divided by Planck's constant changes by order of one on the
scale shorter than the distance between periodic orbits (Gutzwiller case).
For an (almost) integrable system the whole family of (almost) periodic
orbits contributes to the trace (Berry-Tabor case). When the perturbed
action becomes large compared to the Planck constant, the surviving
periodic orbits can be considered well isolated and the Gutzwiller formula
is recovered. 

\chapter{Summary}

In this work we formulated the semiclassical theory for perturbed
integrable systems, applied this theory in a number of cases, and compared
it with other methods. The use of the Poincar\'e surface of section (PSS)
allows to reduce the spatial dimensions by one. The Bogomolny $T$-operator
connects the classical and semiclassical pictures on the surface of
section. It is a kernel of an integral equation $T\psi = \psi$ that
determines the PSS wavefunction. It has a solution when the energy is on
the spectrum, i.e.\ when $\det(1-T)=0$. The $T$-operator is unitary in the
stationary phase approximation. When the boundary of a billiard is chosen
as a surface of section, Bogomolny's equation follows from the boundary
integral method in the semiclassical approximation. The $T$-operator
contains the Maslov phases that come from the regions where the
semiclassical approximation is not valid. 

A classical resonance in a perturbed integrable system manifests itself
in the quantum picture if the size of the resonant island is greater than
the Planck constant. In a billiard with a perturbed boundary this
translates to the condition $\sqrt {L \delta L} \gtrsim \lambda$ where $L$
is the size of the billiard, $\delta L \sim \epsilon L$ is the magnitude
of the perturbation, and $\lambda=1/k$ is the wavelength. When this is the
case, the resonant perturbation theory is required. The theory can be
constructed by expanding the phase of the PSS wavefunction $\psi$ in $k
\epsilon^M$, starting with $M = \frac 1 2$. In the leading order the
perturbed part of the wavefunction satisfies the one-dimensional
Schr\"odinger equation with the effective potential that scales as $k^2
\epsilon$. Because of the above condition, the equation has bound state
solutions that indicate the localized two-dimensional wavefunction. 
Usually the localization is in the neighborhood of a stable periodic
orbit. The two-dimensional wavefunction can be found by propagating $\psi$
from the PSS into the bulk of the system, basically along the classical
paths. The spread of the wavefunction in the momentum space scales as $k
\sqrt{\epsilon}$. The non-resonant states are delocalized in the 
coordinate space, but are better localized (as $k \epsilon$) in the 
momentum space. Our theory automatically takes into account the whole 
perturbed family of periodic orbits, not just the neighborhood of a 
stable orbit. The solutions of the one-dimensional Schr\"odinger equation 
include both bound and unbound and thus cover the transition from 
the resonant to non-resonant states, as long as the influence of other 
resonances can be neglected. In this work we considered two examples of 
billiards with a perturbed boundary: the circle, in particular, the short 
stadium, and the rectangle, in particular, the tilted square. 

In some instances the perturbation theory can be applied to the
non-pertur\-bative systems. This can be done if the system is close to
some integrable system in the regions where the wavefunction is large. We
mentioned the ice-cream cone billiard that has the localized states in the
circular part, the bouncing ball states in the stadium localized outside
of the semicircles, the bouncing ball states near a period-2 orbit, and
the whispering gallery modes. The system may have an additional
perturbation. For example, the bouncing ball modes in the slanted stadium
are shifted towards the wider end. The whispering gallery mode in a convex
billiard follows from the standard EBK quantization procedure of the
respective classical motion. However, when the boundary has a point or a
region of zero curvature, there is no classical adiabatic invariant and
the classical particle is not localized near the boundary. Still, the
localized quantum state may exist. Outside of the region of zero curvature
it is a superposition of the standard whispering gallery modes that are
well-defined. The singular region acts like a scatterer that mixes the
standard modes. The scattering matrix can be approximately expressed in
terms of the $T$-operator. If the eigenstates of the scattering propagator
are composed of the small number of the standard modes, they are localized
near the boundary. 

The magnetic field is a small perturbation for charged particle in a
square billiard if its momentum $p \gg eA/c$, where $A$ is the vector
potential. Then, in the uniform field case, the cyclotron radius is
greater than the size of the square. Thus, in the leading order, the shape
of the orbits is unchanged. The resonant perturbation theory should be
applied when $kL \sqrt{\epsilon} \equiv \sqrt {(e/c \hbar) kBL^3} \gtrsim
1$ where $B$ is the magnetic field. In this case there exist states that
are localized near the stable periodic orbits. The unstable periodic
orbits are the time-reversals of stable orbits. The states associated with
the $(1,1)$ resonance dominate the susceptibility. Since the time-reversal
symmetry is broken, the states carry the persistent currents. The low
transverse modes are paramagnetic and the higher modes are diamagnetic. 
In the square with a flux line the localization also takes place but the 
diffraction effects are strong. The diffraction is smaller for a finite 
size tube. 

When one of the system's coordinates changes much faster than the other,
one can solve the two-dimensional Schr\"odinger equation in the
Born-Oppenheimer approximation (BOA). The solution usually agrees with the
perturbation theory in the leading order. The standard example are the
bouncing ball states where there is a natural separation of the fast and
slow motion. The rectangular geometry is easy to deal with as well, if one
employs the method of images. In this case the BOA is similar to the
channeling approximation. On the other hand, in the circular geometry
only the low angular momentum states are suitable for the BOA. Even then
it fails in the polar coordinates near the center. Thus the BOA solution
cannot be completed in this case since it lacks the boundary condition at
the center. At the same time, it agrees with the perturbation theory in
the asymptotic region if the missing information is provided. In the
whispering gallery mode the slow variable is the curvature of the
boundary. In this instance the BOA solution is valid also near the caustic
and in the classically forbidden region where the perturbation theory is
not applicable. 

The $T$-operator contains all the information about the energy spectrum in
the semiclassical approximation. In particular, the oscillatory part of
the density of states can be expressed in terms of the traces of the
powers of $T$-operator. Depending on the dynamics of the system the traces
can be evaluated to yield the Gutzwiller trace formula in the hard chaotic
case or the Berry-Tabor formula in the integrable case. In the former the
periodic orbits are isolated and in the latter they form families. The
perturbed integrable system connects the opposite situations and provides 
an interpolation formula. The interpolation formula can be parametrized. 
The parametrization depends on four parameters: the perturbed part of the 
action and the monodromy determinant for the stable and unstable periodic 
orbits. When the action difference between the stable and unstable orbits 
is greater than $\hbar$, these orbits are well isolated, when the actions 
are equal, the system is integrable. In the case of the coupled quartic 
oscillators the parametrization works very well. 

The results of this work may be hard to verify experimentally, since it is
the energy spectrum, not the probability density, that is usually
measured. Nevertheless, there are experimental techniques available. The
simplest would be to measure the electric field distribution in a
microwave cavity. The experiments have already been done for a chaotic
shape cavity, so there is no principal difficulty to make it almost
integrable. The electric field in a cavity and the electron wavefunction
in a billiard satisfy the same differential equation. The magnetic
billiard can be modeled by adding a ferrite strip on the walls of the
cavity. In this case the analogy holds only in the leading order in vector
potential. It should also be possible to measure the electron wavefunction
in quantum corrals, but the effect may be diminished due to a substantial
leakage through the walls. Another class of experiments involves the 
surface waves in water, although here the dissipation may be a big 
problem.

\end{document}